# Cuffless, calibration-free hemodynamic monitoring with physics-informed machine learning models

Henry Crandall[1,*], Tyler Schuessler[2,*], Filip Bělík[2,3,*], Albert Fabregas[4], Barry M. Stults[5], Alexandra Boyadzhiev[6], Huanan Zhang[6], Jim S. Wu[7], Aylin R. Rodan[5,8,9], Stephen P. Juraschek[10], Ramakrishna Mukkamala[11,12], Alfred K. Cheung[5], Stavros G. Drakos[5,13], Christel Hohenegger[2], Braxton Osting[2], and Benjamin Sanchez[4,14]

[1]Department of Electrical and Computer Engineering, University of Utah, Salt Lake City, UT, USA

[2]Department of Mathematics, University of Utah, Salt Lake City, UT, USA

[3]Scientific Computing and Imaging Institute, University of Utah, Salt Lake City, UT, USA

[4]Department of Electrical and Computer Engineering, University of Illinois Chicago, Chicago, IL, USA

[5]Department of Internal Medicine, Spencer Fox Eccles School of Medicine, University of Utah Health, Salt Lake City, UT, USA

[6]Department of Chemical Engineering, University of Utah, Salt Lake City, UT, USA

[7]Department of Radiology, Brigham and Women's Hospital, Harvard Medical School, Boston, MA, USA

[8]Molecular Medicine Program, University of Utah Health, Salt Lake City, UT, USA

[9]Medical Service, Veterans Affairs Salt Lake City Health Care System, Salt Lake City, UT, USA

[10]Department of Medicine, Beth Israel Deaconess Medical Center, Harvard Medical School, Boston, MA, USA

[11]Department of Bioengineering, University of Pittsburgh, PA, USA

[12]Department of Anesthesiology & Perioperative Medicine, University of Pittsburgh, PA, USA

[13]Division of Cardiovascular Medicine and Nora Eccles Harrison CVRTI, University of Utah Health, Salt Lake City, UT, USA

[14]Richard and Loan Hill Department of Biomedical Engineering, University of Illinois Chicago, Chicago, IL, USA

[*]These authors contributed equally to this work

Corresponding author: Benjamin Sanchez, 851 S. Morgan St., Office 1104 SEO, Chicago, IL 60607. Email: bst@uic.edu. Phone: 312-996-5847.

Number of Words Main Text: 2,750 (maximum: 2,500)

Number of Figures: 4 (maximum: 4)

Number of Words Methods: 3,254 (maximum: 3,000)

Number of Extended Data Tables: 4

Number of Extended Data Figures: 5

Number of Extended Figures and Tables combined: 9 (maximum: 10)

Number of References: 36 (maximum: 50)

Supplementary Information includes:

- Number of Supplementary Figures: 34
- Number of Supplementary Tables: 11
- Number of Supplementary Videos: 11
- Number of Supplementary References: 222

Supplementary Statistical Table: 1 (Supplementary Table 12)

# Abstract

Wearable technologies have the potential to transform ambulatory and at-home hemodynamic monitoring by providing continuous assessments of cardiovascular health metrics and guiding clinical management. However, existing cuffless wearable devices for blood pressure (BP) monitoring often rely on methods lacking theoretical foundations, such as pulse wave analysis or pulse arrival time, making them vulnerable to physiological and experimental confounders that undermine their accuracy and clinical utility. Here, we developed a smartwatch device with real-time electrical bioimpedance (BioZ) sensing for cuffless hemodynamic monitoring. We elucidate the biophysical relationship between BioZ and BP via a multiscale analytical and computational modeling framework, and identify physiological, anatomical, and experimental parameters that influence the pulsatile BioZ signal at the wrist. A signal-tagged physics-informed neural network incorporating fluid dynamics principles enables calibration-free estimation of BP and radial and axial blood velocity. We successfully tested our approach with healthy individuals at rest and after physical activity including physical and autonomic challenges, and with patients with hypertension and cardiovascular disease in outpatient and intensive care settings. Our findings demonstrate the feasibility of BioZ technology for cuffless BP and blood velocity monitoring, addressing critical limitations of existing cuffless technologies.

Cardiovascular disease (CVD) remains the leading cause of mortality worldwide, despite decades of progress in diagnosis and treatment.[1,2] Many CVDs are preventable, yet current strategies for early detection and management are limited by the constraints of existing hemodynamic monitoring technologies.[3–5] While ambulatory and at-home blood pressure (BP) monitoring devices are crucial for identifying at-risk hypertensive patients, guiding management, and preventing CVDs (Supplementary Discussion 1), these devices are obtrusive and impractical for preventive and long-term use due to discomfort (Supplementary Discussion 2).[6] Moreover, as cuff-based technologies must be implemented within controlled environments under near-rest conditions, pathologic BP information in dynamic states associated with acute CVD events (e.g., acute plaque rupture or burst cerebral aneurysm) may be missed.[7,8] Given the significant healthcare burden and mortality associated with CVD,[9] smart wearable devices have emerged as promising solutions for tracking BP (Supplementary Discussion 3).[10–12] However, many existing approaches rely on methods that require calibration and lack consistent scientific foundations, thus remaining susceptible to poorly understood confounders and limiting their clinical utility.[13]

Here, we present a smartwatch equipped with electrical bioimpedance (BioZ) sensing capable of continuously monitoring pulsatile resistance and reactance changes at the wrist (Fig.1a). To address key scientific gaps of existing approaches, we develop a multiscale modeling framework that mechanistically links BP to BioZ signals. Experimental BioZ signals are analyzed using a signal-tagged physics-informed neural network (sPINN) that integrates Navier-Stokes equations to enable cuffless, calibration-free estimation of BP and radial and axial blood velocity (V). Finally, we evaluate the system's performance in healthy individuals, patients with hypertension (HTN), and those with CVD. This technology enables advanced hemodynamic monitoring in real-world settings and represents a potential paradigm shift in wearable cardiovascular assessment and personalized HTN and CVD care.

## Wearable smartwatch

We built a smartwatch equipped with BioZ sensing for measuring pulsatile resistance and reactance changes at the wrist. To quantify the device's precision (Fig.1b), we first performed measurements on two resistors of 49.9 and 51.1 ohms representative of wrist values with 1 and 3.14 kohms contact impedance at 51.2 kHz, respectively (Fig.1c i–ii). Driving 100 $\mu A_{rms}$, the rolling standard deviation over 100 samples obtained was 3.22±0.23 and 4.53±0.23 mohms, respectively (Supplementary Fig.1). This level of precision allows us to detect pulsatile electrical changes associated with radial BP.

We then integrated stainless steel electrodes into the smartwatch to support BioZ measurements on participants. The electrodes exhibited low and stable impedance magnitude (13.4±3.2 kohms) and phase (-15.9±6 degrees) at 46.4 kHz (Fig.1d i–ii). These values indicate a good electrode–skin contact and support reliable current injection and voltage sensing in the BioZ frequency of interest. Cyclic voltammetry testing on the electrodes revealed broad, hysteretic currents, consistent with pseudocapacitive charging and irreversible surface reactions (Fig.1d iii). Repeated cyclic voltammetry test over 500 cycles demonstrated electrochemical stability, with minimal hysteresis drift and preservation of capacitive behavior (Fig.1d iv). The near-overlap between the first and 500th cycles indicates robust interfacial integrity and negligible electrode degradation necessary to enable long term hemodynamic monitoring.

## Fluid parameters and blood conductivity

We developed a multiscale model of the human arm to establish the relationship between BP and BioZ through changes in blood volume and blood conductivity across the cardiac cycle. The first two stages of the model combine the Navier-Stokes and Maxwell-Fricke theories for fluids with immersed particles in a branched arterial tree with elastic wall motion and structured tree outlet conditions (Fig.2a and Supplementary Discussion 4).[14–18]

To evaluate the influence of fluid, anatomical, and physiological parameters on radial BP, wall shear stress magnitude, and radial blood conductivity, we conducted local sensitivity analyses (LSA) under physiologically realistic conditions (Fig.2b, Supplementary Table 1 and Supplementary Video 1-8). Specifically, variations in the systolic peak of the brachial BP waveform modulated the corresponding peaks in both wall shear stress and radial blood conductivity waveforms (Fig.2b i). Increasing the mean wall shear stress had little effect on BP, but amplified the peaks and led to an overall increase in the baseline shear stress and radial blood conductivity (Fig.2b ii). As the radii of all arteries were increased, the radial systolic pressure and mean radial blood conductivity increased (Fig.2b iii). Hematocrit and plasma conductivity had no influence on the fluid parameters or waveform morphology; however, hematocrit levels were inversely related to mean radial blood conductivity (Fig.2b iv), whereas increased plasma conductivity increased the mean radial blood conductivity (Fig.2b v). The red blood cells (RBCs) orientation had a greater contribution to radial blood conductivity than RBC deformation (15.5% vs 2.8%, Fig.2c).

To account for combined interactions among physiological factors, we conducted a global sensitivity analyses (GSA) by jointly perturbing multiple sources and quantifying their collective contributions. Specifically, BP, the radial artery radius, and radial mean shear stress had the greatest effects on the peak-to-peak radial blood conductivity (Fig.2d and Supplementary Table 2). Likewise, the BP, radial artery radius, and radial mean shear stress primarily determined the time of minimum radial blood conductivity, whereas BP, the radial artery radius, and heart rate had the greatest impacts on time the maximum radial blood conductivity. Hematocrit and plasma conductivity influenced only the mean radial blood conductivity, without altering any other features of the conductivity waveform. Since the hematocrit and plasma conductivity do not change per cardiac cycle, the GSA demonstrates that the BP and underlying fluid dynamics heavily impact the resulting blood conductivity.

To incorporate patient-specific geometries, we extended our LSA and GSA study by performing com-putational fluid dynamics simulations on a human arterial tree model. In the specific model simulated, we found that velocity and wall shear stress peaked immediately proximal to the brachial bifurcation, where arterial segments were compressed, while pressure decreased uniformly in the distal branches (Fig.2e).[19]

## Sources of bioimpedance signals

The final stage of our multiscale framework solves Maxwell's equations for electromagnetism and translates these blood changes into a measurable electrical signal at the skin. This enables end-to-end conversion of BP–driven variations in blood volume and conductivity within the radial artery into a pulsatile BioZ signal (Fig.3a, Supplementary Fig.2-3 and Supplementary Discussion 5).

With this model, we investigated the relationships among radial blood conductivity, the arterial radius, and resistance (Fig.3b,c). Our GSA demonstrated that BP and fluid parameters drive pulsatile resistance. In particular, radial BP and arterial radius had the greatest impacts on the times of the minimum and maximum resistances (Supplementary Table 3). In comparison, the peak-to-peak resistance and mean resistance were the most sensitive to electrode offset and radial arterial radius sources (Fig.3d). Additionally, LSA revealed how the wrist surface resistance signal changes with the brachial BP (Fig.3e i), mean wall shear stress (Fig.3e ii), arterial radius (Fig.3e iii), hematocrit (Fig.3e iv), and plasma conductivity (Fig.3e v). The brachial BP and mean wall shear stress altered the morphology of the resistance waveform, whereas the arterial radius to length ratio, hematocrit, and plasma conductivity affected the mean resistance value (Fig.3e, Supplementary Fig.4).

We also investigated the influence of patient-specific anatomical, physiological, and biological variability by performing finite element method simulations on computable human phantoms (Fig.3f,g, Extended Data Fig.1, Supplementary Fig.5–8 and Supplementary Table 4). The results confirmed that our electrode configuration and measurement frequency are sensitive to blood volume and conductiv-ity changes. We first examined the current density field in the tissues directly underneath the electrodes (Fig.3h). The movement of the electrodes' position was simulated to account for experimental variability between different placements when wearing the smartwatch, resulting in resistance and reactance changes of 10.2%±14.2% and 8.4%±14.3%, respectively (Fig.3i, Extended Data Fig.1 and Supplementary Fig.9). Anatomic variation in the depth of the radial artery changed the resistance and reactance by 0.74%±1.61% and 0.51%±0.92%, respectively (Fig.3j and Extended Data Fig.1). Furthermore, changes in the radius (Fig.3k) and blood conductivity (Fig.3l) of the radial artery had negative linear relationships with resistance and reactance (Extended Data Fig.1). Additionally, to quantify the contribution of individual tissues to the measured resistance and reactance data, we simulated the sentivity regions underneath the electrodes (Supplementary Discussion 5). The resistance and reactance volumes were (55.2±7) × (57.3±3.1) × (45.7±2) $mm^3$ and (49.2±2.5) × (54.3±2.6) × (45±1.2) $mm^3$, respectively (Fig.3m, Extended Data Fig.1 and Supplementary Fig.10–11). The radial artery was the fourth-largest contributor to baseline resistance and the third-largest contributor to baseline reactance (Fig.3n, Extended Data Fig.1, Supplementary Fig.12–14 and Supplementary Table 5).

## Physics-informed machine learning

To predict both BP and V fields from BioZ, we developed our sPINN model, augmenting the space-time inputs with a learned encoding of the measured resistance signal including optional anthropometric and physiological metadata (Fig.4a and Extended Data Fig.2).[20,21] With our population-wide trained sPINN model, we evaluated the performance of our calibration-free model while accounting for both intra- and interindividual variability (Extended Data Table 1-2 and Supplementary Fig.15–17).

We first validated the sPINN architecture with a hybrid dataset comprising experimental arterial line BP waveforms from the PulseDB database[22] and synthetic resistance signals generated by our multiscale model (Supplementary Fig.18). The sPINN generated interpretable, spatiotemporal predictions of BP and V fields that were informed both by measured data and constrained by the governing fluid-dynamics equations in a compliant arterial domain (Extended Data Fig.3, Supplementary Discussion 6–7 and Supplementary Fig.19–20). The model achieved high prediction precision for systolic BP (SBP) and diastolic BP (DBP), with coefficients of determination of $r^2$ = 0.998 and $r^2$ = 0.974, respectively (Extended Data Fig.3). The axial V predictions were very accurate, with an error of $<$1.8 cm/s. Moreover, the predicted versus and ground-truth Vs showed excellent agreement for both the systolic and diastolic average V ($r^2$ = 0.994 and $r^2$ = 0.888, respectively), further validating the model's ability to learn the mapping from BioZ to the underlying fluid dynamics (Extended Data Fig.3).

Next, we evaluated the generalizability of the sPINN model across diverse simulated physiological conditions, informed by our LSA and GSA studies, using an expanded PulseDB synthetic dataset. The sPINN maintained strong SBP prediction accuracy despite the imposed biological variability in the resistance signals (Extended Data Fig.3). The precision of the DBP predictions decreased owing to limitations associated with the generation of the biological variability dataset (Supplementary Discussion 8); indeed, the covariance between the resistance and SBP values was $\approx 10\times$ greater than that between the resistance and DBP values (Supplementary Fig.21).

## Hemodynamic monitoring

We successfully tested our smartwatch device and sPINN BP and V measurement capabilities in $N$=75 healthy individuals (Group 1) before and after walking, running, and cycling, assessing exercise-induced hemodynamic changes (Supplementary Fig.22–24). We also examined responses to different autonomic challenges (Extended Data Fig.4 and Supplementary Table 6–9). The participants wore the device on their left wrist for continuous recording, and the measurements were sensitive to the radial pulsatile BioZ signal (Supplementary Fig.25). After training, we benchmarked the sPINN's V predictions with a test set against reference ultrasound Doppler flowmetry measurements (Supplementary Video 9–11). The measured peak systolic velocity (PSV) fell within the sPINN-predicted range and predicted interquartile range for 76% and 38% of the participants, respectively (Extended Data Fig.4). The distributions of the measured PSV and sPINN-predicted PSV overlap closely across all the subjects, indicating excellent agreement (Extended Data Fig.4).

We also evaluated patients in outpatient ($N$=85) and intensive care unit settings ($N$=3), consisting of three cohorts of patients (Group 2) with HTN (both controlled and uncontrolled), CVD, or other conditions (Supplementary Discussion 6, Extended Data Table 3 and Supplementary Fig.26–29).

A calibration-free sPINN model tackles a more challenging prediction problem yet offers greater translational potential than subject-specific, calibration-dependent approaches by eliminating the need for subject- or session-specific tuning. Based on test data (Extended Data Table 4), calibration-free sPINN models demonstrated strong predictive performance and achieved coefficients of determination of $r^2$ = 0.774 for SBP and $r^2$ = 0.807 for DBP predictions compared with the cuff reference BP (Fig.4b,c). Stratified analyses revealed even greater precision within specific subgroups: $r^2$ = 0.752 and $r^2$ = 0.854 SBP and DBP predictions for patients with HTN (Fig.4d,e), $r^2$ = 0.859 and $r^2$ = 0.811 for those with CVD (Fig.4f,g), and $r^2$ = 0.588 and $r^2$ = 0.642 for patients with other conditions (Fig.4h,i). This likely reflects the broader variability and higher clinical complexity introduced in populations with more comorbidities (47%) than the HTN (19%) and CVD (23%) cohorts (Supplementary Discussion 6).

To further evaluate the limits of prediction precision of our calibration-free sPINN models, we trained subject-specific sPINN models (Extended Data Table 4 and Supplementary Table 10-11). At the cost of fine-tuning, calibrated models outperformed the population-wide approach and ahieved higher precision across all cohorts. When all patients were grouped, these models yielded $r^2$ = 0.860 and $r^2$ = 0.895 for SBP and DBP predictions, respectively. Model performance was further improved in the clinical subgroups, reaching $r^2$ = 0.861 and $r^2$ = 0.922 for SBP and DBP predictions for patients with HTN, $r^2$ = 0.944 and $r^2$ = 0.913 for those with CVD, and $r^2$ = 0.662 and $r^2$ = 0.836 for patients with other condi-tions, respectively (Extended Data Fig.5).

These results demonstrate the potential clinical utility of the sPINN as a robust method for calibration-free and cuffless hemodynamic monitoring.

## Discussion

Wearable devices hold promise for preventive and continuous monitoring in normotensive individuals at risk of HTN, as well as patients with HTN and CVDs.[23] However, many existing methods depend on calibration and utilize pulse wave analysis or pulse transit time approaches that lack robust scientific foundations, leaving them vulnerable to poorly understood physiological and experimental confounders.[24–27] To address these critical limitations, we present a smartwatch with BioZ sensing capability and demonstrate, through modeling and sensitivity analyses, that radial BP is a primary contributor to wrist BioZ signals. These findings are experimentally validated with a calibration-free sPINN model, successfully monitoring BP and V in healthy individuals and patients with HTN and CVD.

The biophysical model couples the Navier-Stokes equations for blood flow with an elastic arterial wall model, the Maxwell-Fricke theory for blood conductivity, and Maxwell's equations of electromagnetism. This end-to-end framework provides a mechanistic link between fluid dynamics governing arterial BP, blood volume, blood conductivity, and the BioZ signal recorded at the wrist (Supplementary Discussion 8). Our approach overcomes the limitations of existing wearable devices for cuffless BP monitoring that rely on Moens-Korteweg pulse wave velocity approximation;[16] and models the influence of blood viscosity, arterial wall density and elasticity, Poisson's ratio, and heart rate on BioZ signals. By using structured-tree outlet boundary conditions, we model BP-driven elastic and pulsatile effects with complex wave numbers at and throughout the outlet, resulting in consistent volumetric changes in the radial artery and RBC reorientation that generate the BioZ signals measured at the wrist. We also developed a sPINN algorithm to provide physically interpretable BP and axial and radial V field predictions.

This monitoring capability is unprecedented among conventional ambulatory BP monitors and experimental wearable devices for cuffless BP monitoring.

Our calibration-free sPINN models demonstrate strong BioZ predictive performance for both SBP and DBP. Notably, this performance is further enhanced with subject-specific models (Extended Data Table 4, Supplementary Fig.30–31 and Supplementary Table 10–11). Beyond these two discrete values, our sPINN platform also accurately reconstructs the full BP waveform. In contrast to arterial dynamics methods for estimating BP (Supplementary Discussion 3 and Supplementary Fig.32–34),[23] the sPINN achieves lower average root mean square error of 6.23 mm Hg in healthy individuals, 6.39 mm Hg in patients with HTN, 5 mm Hg in patients with CVD, and 4.93 mm Hg in patients with other conditions (Supplementary Discussion 8 and Extended Data Table 4). Remarkably, the predicted PSVs strongly agree with the measured Vs across a broad cohort of healthy individuals, despite the model being trained without access to any reference V data (Extended Data Fig.4). This high level of agreement achieved in a fully self-supervised manner demonstrates the ability of the sPINN to capture the underlying fluid dynamics governing blood flow, increasing its physical interpretability and predictive robustness.

Integrating cuffless hemodynamic monitoring capabilities into a smartwatch addresses key limitations of current ambulatory and home-based cuff BP monitors. It also bypasses the obtrusiveness of experimental cuffless wearable devices that require direct skin attachment, which can hinder both usability and long-term patient adherence. However, future work will be crucial to enabling the widespread clinical adoption of our hemodynamic monitoring technology in both inpatient and outpatient care (Supplementary Discussion 9).[28,29] First, a clinical study following established protocols is needed to evaluate real-world performance at the bedside and in at-home settings. Second, device performance should be assessed across a larger cohort of individuals with uncontrolled HTN and those with periodic or orthostatic HTN. Third, the longitudinal accuracy of the calibration-free sPINN models must be evaluated. Our findings establish the foundations of calibration-free, cuffless wearable BioZ hemodynamic monitor-ing, warranting this future work to evaluate and refine this approach for clinical deployment.

# Methods

## Smartwatch

The smartwatch includes a display for real-time visualization of the resistance and reactance values at the measured frequency, and indicates the battery level and wireless connection status (Fig.1b). The smartwatch can store 32 GB of data for up to 1 month and has a type-A USB port for downloading data and recharging a 3.7 V battery. The watch's battery supports up to 14 days of operation and, when continuously transmitting resistance and reactance values at 100 samples/s via Bluetooth Low Energy, has a runtime of ≈6 hours. The electrodes have an outer diameter of 30 mm, inner diameter of 18 mm, electrode gap of 2.5 mm, and area 392 $mm^2$.

## Fluid dynamics

### Multiscale coupled Navier-Stokes and Maxwell-Fricke model of blood conductivity

In the model, analytical solutions of the reduced Navier-Stokes equations with an elastic tube boundary are used to characterize blood fluid dynamics in the brachial, radial, and ulnar arteries (Fig.2a). Under geometric assumptions of the arterial tree, a brachial BP waveform can propagate to the radial artery at the wrist. According to the shear stresses within the artery, red blood cells (RBCs) are oriented and deformed with the flow, and Maxwell-Fricke theory is used to compute the radial arterial blood conductivity (Supplementary Discussion 4).

The Navier-Stokes equations are first linearized according to the assumption that the radius of the artery is small compared with the wavelength of the propagating pressure wave.[15,30] Solutions are assumed to be time-periodic and space-varying with complex wavenumbers. The general solutions to these equations are coupled with the stress-strain equations of motion of a thin elastic tube wall.[16] After cou-pling these equations, we solve for the complex wavenumbers, which yield explicit solutions for the BP, the pulsatile radial and axial fluid velocities, the arterial wall displacements, and the shear stresses within the artery.[16]

Additionally, these fluid solutions provide an elastic relationship between arterial BP and the volumetric flow rate that differs from the relationship obtained under the assumption of an inviscid Moens-Korteweg wavespeed.[16,31,32] The linearized Navier-Stokes equation for axial momentum is integrated to obtain an equation describing the relationships among the rate of change in the flow rate, the pressure gradient, and the wall shear stress.[16] Then, without relying on a Moens-Korteweg wave speed approximation,[16,31,32] the explicit fluid solutions are used to relate the wall shear stress to the flow rate, and the equation is manipulated to derive complex characteristic admittance coefficients, which are defined as the ratios of the Fourier coefficients for the flow rate and BP.

We form our arterial tree under the assumption that the arteries are elastic cylinders that bifurcate without loss of energy (Fig.2a). Our arterial tree consists of a distal portion of the brachial artery that bifurcates into the radial and ulnar arteries. The ulnar artery then bifurcates into the common interosseous artery and the distal ulnar artery. At the outlets of the radial, interosseous, and distal ulnar arteries, we implement structured tree outlet conditions.[18] In the structured tree, the arteries are repeatedly and consistently branched until the vessels reach a minimum radii. Additionally, we set a structured tree length-to-radius ratio of 25.[18] The wall thickness ratio in the structured tree is chosen to match the corresponding outlet artery wall thickness ratio. Finally, distal stiffening occurs according to a power law based on the Young's moduli of the structured tree vessels with respect to their radii. Under the assumptions of continuous of pressure and mass conservation, with the admittance coefficients in each artery and at the outlets derived from the structured tree, complex reflection coefficients are computed at each bifurcation to produce primary wave reflections.[16]

Shear stresses within the radial artery are input into a model in which ellipsoid RBCs experience shear stress-induced orientation and deformation, and Maxwell-Fricke theory is used to tie this to arterial blood conductivity (Fig.2a).[15,17] Since our fluid solutions vary both radially and axially, the conductivity calculation is performed not only over an arterial cross-section but also along an axial segment of interest. Finally, the time-dependent conductivity and spatially-averaged radial artery radius are obtained and input into a three-layer half-space model (skin, subcutaneous adipose tissue, muscle, and radial artery) to compute the BioZ signal at the wrist surface over time (Fig.3a).

We utilized this model to isolate the impact of RBC orientation and deformation by modeling radial blood conductivity in three consecutive scenarios: stationary blood with no RBC deformation or orientation, pulsatile flow with RBCs undergoing orientation, and finally pulsatile flow with both RBC deformation and orientation. The deformation impact is measured by subtracting the results of the full scenario from those of the scenario without deformation. Similarly, the orientation impact is measured by comparing the results of the deformation-free scenario with those of the stationary blood scenario.

**Experimental radial blood pressure and synthetic wrist surface bioimpedance dataset**

We leveraged the PulseDB source dataset to create our own training and test datasets of experimental radial BP signals and corresponding synthetic wrist surface resistance signals using our physiological modeling framework (Extended Data Fig.3).[22] Experimental BP recordings were obtained in the radial artery; therefore, our physiological model used to generate the corresponding synthetic wrist surface resistance signals did not consider the bifurcation of the brachial or ulnar arteries. We first isolated radial BP data from a total of 2,423 participants (age: 63.42±15.8 years, ratio male:female ratio, 1,405:1,018) and split these signals into individual radial BP periods on the basis of the maximum in the waveform to ensure no signal overlap.[33] Next, we ran these radial BP periods through our forward biophysical model without branching to synthetically generate the corresponding wrist surface resistance signals while considering nominal values for the fluid and electrical parameters (Extended Data Fig.3 and Supplementary Table 1). In total, we generated over 1.2M pairs of coupled experimental radial BP and synthetic BioZ signals. The SBP and DBP are 126.7±25.2 mm Hg and 61.6±14.2 mm Hg, respectively, 48% of the samples have DBP<60 mm Hg and 18% of the samples have SBP>150 mm Hg. We then sampled 100k periods of radial BP signals that met the following criteria: SBP/DBP>90/60, SBP/DBP<140/90, and heart rate between 40 and 200 beats per minute. We randomly sampled a subset of 90k pairs for training and 10k pairs for testing.

**Sensitivity analysis**

We performed a sensitivity analysis to identify the fluid dynamic parameters that most influenced the radial blood conductivity. We first performed a local sensitivity analysis by visualizing the outputs of various stages of the forward physiological model when only one parameter was perturbed (Fig.2b). Physiologically relevant ranges for each parameter were determined from a broad literature search (Supplementary Table 1). We applied a variance-based approach to quantify the global sensitivity over the entire parameter range.[17,34] Except for BP, we assumed that each parameter followed a uniform distribution (Supplementary Table 1). The BP is a uniformly randomly selected waveform from the 100k sampled signals (Supplementary Discussion 4).

The model output is a radial blood conductivity or wrist surface resistance waveform; however, we measured the sensitivities with respect to the scalar quantities of interest: peak-to-peak conductivity, average conductivity over one period, and times of the maximum and minimum conductivity (Fig.2c). We used a quasi-Monte-Carlo sampling approach and ran 1.5M iterations to approximate first-order and total-order Sobol indices along with standard deviations for each. We then measured the sensitivities in terms of first and total order Sobol indices (Supplementary Table 2).[34] Additionally, we performed a global sensitivity analysis on the impacts of the BioZ model (Fig.3d). We sampled BP periods uniformly from our dataset and used quasi-Monte-Carlo sampling for the remaining parameter domain (Supple-mentary Table 1). For each sample drawn, we calculated the resulting mean resistance, peak-to-peak resistance, and times of maximum and minimum resistance. First- and total-order Sobol indices were computed for each parameter over a sample of 700k parameter values (Supplementary Table 3).

For the parameters of radial artery radius, heart rate, hematocrit, Young's modulus at the wall, mean wall shear stress, skin conductivity, and fat layer thickness, we illustrate the impacts of local physiological variability (Extended Data Fig.3). For each parameter, we determined a nominal, minimum and maximum values via a literature search (Supplementary Table 1). To assess the impact of biological variability on our synthetic BP predictions, we created normal distributions with the mean given by the nominal value and standard deviation chosen such that three standard deviations would leave the minimum to maximum interval. Rejection sampling was used to ensure that all samples were within the minimum to maximum range. We then generated one dataset of size 500k, where all seven parameters were randomly selected over all 100k BP periods, which was repeated five times (Extended Data Fig.3). Additionally, we created seven individual datasets of size 50k each in which one parameter was randomly selected while the others were held fixed.

## Bioimpedance

### Three-layer cylindrical bioimpedance wrist model

We developed a four-terminal analytical BioZ model of the wrist to relate radial volume and blood conductivity with surface BioZ measurements. The model consists of an infinite layered half-space with isotropic skin, subcutaneous adipose tissue, and skeletal muscle tissues (Fig.3a, Supplementary Discussion 5, Supplementary Fig.2 and Supplementary Table 1). The radial artery is modeled as an infinitely long conductive cylinder embedded within the muscle layer. We then solved Maxwell's equations under a quasistatic approximation and derived the apparent electrical resistance arising from the superposition of a primary and secondary electrical potential. The primary and secondary potentials were computed via an $h$-adaptive numerical integration scheme with tolerances set to ensure at least three decimal digits of accuracy. This model was implemented in Julia 1.10 and validated with finite element method simulations in COMSOL Multiphysics 5.5 (COMSOL Inc, Burlington, MA) (Supplementary Fig.3 and Supplementary Discussion 5). The time-varying radial arterial radius and blood conductivity computations from our fluid model are passed into our BioZ model to infer the wrist surface resistance signals (Fig.3b). We examined the individual effects of artery radius and blood conductivity on BioZ by varying one parameter while holding the other at its nominal value. (Fig.3c).

**Computational bioimpedance model**

We performed finite-element method electrical simulations of BioZ measurements collected by the smartwatch at the left wrist in Sim4Life V 7.2.1 (Zurich MedTech, Zurich Switzerland). We used the left forearms of the Ella, Morphed Ella, Fats, and Glenn version 3.1 human phantoms to investigate the impact of subject-specific anatomy (mean±standard deviation age 43±27.7 years, body mass index 25±4.9 kg/m$^2$, Fig.3f-n, Supplementary Table 4).[19] The smartwatch electrodes were modeled as perfect electric conductors on the left anterior wrist directly above the radial artery (Fig.3f,g). The electrical properties of the tissues were assigned using the IT'IS database version 4.1 at the corresponding frequency.[19,35] The simulation was performed using the electro-quasi-static module with 1 V Dirichlet boundary conditions, $10^{-12}$ relative convergence tolerance, and a limit of 100k iterations.

We studied the electrical resistance and reactance behavior of the left wrist in human computable models by performing five parameter sweeps: frequency, electrode position, radial artery depth, radial artery radius, and radial blood conductivity (Fig.3i-l and Extended Data Fig.1). For the frequency sweep simulations, we included 25 logarithmically spaced frequencies between 1 kHz and 1 MHz. For the electrode position sweep (Fig.3i and Extended Data Fig.1), we moved the electrode configuration 10 mm left or right of the nominal position directly above the radial artery in 2.5 mm increments. For the radial artery depth (Fig.3j), we moved the radial artery 2 mm above and below the nominal location in the human phantoms in 0.5 mm increments. To study the impact of the radial artery radius (Fig.3k and Extended Data Fig.1), we incrementally inflated and deflated the radius at 17 values between 1.5 and 3 mm. For the conductivity sweep (Fig.3l and Extended Data Fig.1), we simulated 25 linearly-spaced blood conductivity values between 0.3 and 1.1 S/m; this range is centered on the nominal conductivity of blood.[19,35] Unless otherwise noted, tissue electrical properties (i.e., conductivity and relative permittivity) were determined by the frequency of the electrical current and the electrode position was fixed at the nominal location directly above the radial artery. Furthermore, we visualized the isopotential lines on the wrist (Supplementary Fig.5-8) and performed an impedance sensitivity distribution analysis (Supplementary Discussion 5).

We calculated the volume within the wrist responsible for 95% of the measured resistance and reactance (Fig.3m, Extended Data Fig.1, Supplementary Fig.10-11) and quantified the contribution of individual tissues (Fig.3n, Extended Data Fig.1, Supplementary Fig.12, and Supplementary Table 5). Single frequency simulations were performed at 50 kHz.

## Machine learning

### Signal-tagged physics-informed neural network

We implemented a dual component machine learning model with three training phases (Fig.4a, Extended Data Fig.5 and Supplementary Discussion 7). The model maps a point in space-time, a resistance signal, and (optional) participant anthropometric and anatomical input metadata information to pressure and axial-radial velocity fields evaluated at that point. The resistance signal and individual participant data are passed through a 1D convolutional neural network signal encoder (Fig.4a ii); the reduced-dimensional encoded signal along with the space-time point, is passed to a novel sPINN model (Fig.4a i). We modified a standard physics informed neural network architecture[20,21] by augmenting the space-time input with an additional encoded resistance signal so that the model can provide interpretable, space-time blood flow field predictions that: (i) are dependent on the resistance data sample and (ii) are grounded in physics. In particular, we require that the fields satisfy pulsatile fluid flow in an elastic tube, described by the incompressible Navier-Stokes equations in a cylindrical tube with elastic boundary conditions (Supplementary Discussion 4).

We trained each model component separately and then in tandem. This tri-phase training scheme achieves multiple goals: (i) the signal encoder must learn a low-dimensional representation of the resistance signal while capturing (theoretically-grounded) BP-features; (ii) the sPINN is constrained to produce physically realistic blood flow predictions; and (iii) both components are fine-tuned to increase accuracy. Our model and training scheme address the inverse/prediction problem and allow several different types of model losses to be evaluated (physics, predictive error, adaptive filtering, and autoencoder reconstruction).

**Model physics, loss, and training**

We first pretrain the signal encoder (Fig.4a ii). A resistance signal and (optional) participant metadata information are encoded via the signal encoder, then passed to a decoder network and adaptive filter network. The decoder reconstructs the input resistance signal and the adaptive filter predicts a paired simultaneous BP signal. All three networks' parameters are updated to minimize a compound reconstruction and adaptive filter loss function (Fig.4a, Extended Data Fig.5, and Supplementary Discussion 7). This modified task-informed autoencoder training method enables extractions of feature-rich representations of the input resistance signal for the encoding, via the encoder-decoder network. In addition, this approach enables (i) adaptively filtering of the input resistance such that its encoding contains expressive

BP features; and (ii) shaping of the encoding space so that resistance signals that are similar to BP signals are grouped together, leading to better intersubject generalizability.

We then train the sPINN, with static input resistance encodings by freezing the signal encoder parameters. The sPINN model parameters are updated to minimize a compound loss function consisting of partial differential equation residuals, initial conditions, and boundary conditions. The combination of these loss terms produces accurate solutions to the Navier-Stokes equations in a thin elastic tube. We introduce a novel iterative procedure to update the arterial wall boundary position during training to accommodate arterial wall elasticity and consequent radial inflation/deflation; wall positions are iteratively updated on the basis of a governing elasticity ordinary differential equation and the sPINN's evaluated pressure and velocity (Supplementary Fig.20 and Supplementary Discussion 7).

The use of the Navier-Stokes equations as governing equations for the residual calculations ensures that our learned pressure and velocity fields model fluid motion in an artery. However, we cannot express an explicit input resistance signal dependent physical relationship within the Navier-Stokes equations, e.g., with a forcing term. Instead, our reference boundary and initial conditions are input dependent, allowing our model to learn our established biophysical relationship between BioZ and BP/V.

To train the sPINN, we randomly sample a batch of collocation space-time points from the arterial domain: a subset of the batch is sampled from the domain such that it is relevant to one of the five loss terms (Supplementary Fig.19). Each of these space-time points is combined with a resistance signal sampled from the training data set. The resistance signal is passed through the signal encoder to produce an encoded resistance signal, which is augmented with the space-time point and input into the sPINN model, yielding predicted arterial fluid field samples. These field samples are used to evaluate the loss functions and derivatives with respect to model inputs/parameters, and update the model parameters. The reference fields used in the loss computation (Supplementary Discussion 7) vary with the input resistance/pressure pair, ensuring that our model can learn the relevant boundary and initial conditions on the basis of the input measured resistance.

For the final phase of training, the signal encoder parameters and sPINN parameters are unfrozen and updated to minimize the same loss functions as those for the sPINN. This tandem training strategy further improves the extracted resistance features (via the encoder), increasing pressure prediction accuracy while still adhering to blood-flow physics.

### Data sources

All data were collected under protocols approved by the institutional review board, with written informed consent obtained from all participants. The sPINN was implemented with several datasets (Supplementary Discussion 6 and Supplementary Table 8–9), including three experimental datasets (Supplementary Discussion 6). The Group 1 dataset includes data from $N$ = 75 healthy participants with reference NOVA Plus (Finapress Medical Systems, Enschede, Netherlands). The Group 2 dataset includes data from patients with HTN ($N$ = 32), CVD ($N$ = 22), or other conditions ($N$ = 32) with reference VitalStream 5 (CareTaker Medical, Charlottesville, VA) . We also used the Graphene-HGCPT dataset.[36] In addition, we considered two hybrid experimental-synthetic datasets: PulseDB[22] and PulseDB[22] with biological variability. For all the datasets, an $\approx$90/10 training/testing split was used for the data of each individual participant to construct the training/testing sets for the whole dataset (Supplementary Discussion 7). Furthermore, for the Group 1 and Group 2 datasets, $N$ = 4 and $N$ = 6 random participants, respectively, were considered test-exclusive subjects, with none of their data in the training set, for use in an unseen-subject generalizability evaluation for the sPINN (Supplementary Discussion 8). For both the Group 1 and Group 2 datasets, we trained calibration-free, population-wide and subject-specific sPINN models without any metadata. For the Group 1 dataset, we also trained a calibration-free, population-wide sPINN including anthropometric and physiological metadata consisting of eight measured parameters (wrist circumference, forearm length, radial artery depth, radial artery major diameter, radial artery mi-nor diameter, brachial artery depth, brachial artery major diameter, and brachial artery minor diameter) and one derived parameter (average radial artery radius); we did not include any demographic metadata. For the PulseDB[22] and PulseDB[22] with biological variability datasets, we trained population-wide sPINN models. The model trained with the PulseDB[22] dataset with biological variability includes the biologically variable parameters as metadata.

## Statistical analyses

The data were analyzed using Prism 10 (Dotmatics, Boston, MA) to evaluate statistical significance (Supplementary Discussion 10).

## Data availability

Restrictions apply to the availability of the data that support the findings of this study, which were collected as part of an industry-funded project. As a result, the data are not publicly available. Access to specific data may be granted upon reasonable request and is subject to relevant licensing agreements.

## Code availability

The code used in this study was developed as part of an industry-funded project and is protected as intellectual property under a patent. As a result, it is not publicly available. Access specific parts of the code may be granted upon reasonable request and is subject to relevant licensing agreements.

## Acknowledgments

This material is partially based upon work supported by the National Science Foundation (NSF) GRF under Grant No. 2139322 (H.C). C.H. and B.O. acknowledge partial support by NSF 2136198 and NSF 2529648. B.S. acknowledges the direct financial support for the research reported in this publication provided by B-Secur, Ltd (Belfast, United Kingdom); University of Illinois System & Universidad Nacional Autónoma de México Seed Funding Initiative; NSF under Award numbers 2136198 and 2529648; the National Cancer Institute of the National Institutes of Health (NIH) under Award Number 1R21CA273984-01A1, 1P01CA285249-01A1, and 1R21CA289101-01A1; and the National Institute on Minority Health and Health Disparities under Award Number 1R21MD018488-01A1. The content is solely the responsibility of the authors and does not necessarily represent the official views of the NSF, NIH, or Veterans Affairs.

# Author Contributions

Conceptualization - B.S.; Data curation - H.C., T.S., F.B., A.F., A.B.; Formal Analysis - C.H., B.O., B.S.; Funding acquisition - C.H., B.O., B.S.; Investigation - all; Methodology - B.M.S, H.Z., J.W., R.M., A.R.R., A.K.C., S.G.D., C.H., B.O., B.S.; Project administration - C.H., B.O., B.S.; Resources - B.S.; Software - H.C., T.S., F.B., A.F.; Supervision - C.H., B.O., B.S.; Validation - all; Visualization - H.C., T.S., F.B., A.F., A.B.; Writing – original draft - all; Writing – review & editing – all.

# Competing Interests

Dr. Sanchez is a co-founder of and holds equity in Haystack Diagnostics, Inc. He holds equity and serves as Scientific Advisor to B-Secur, Ltd., and Sobr Safe, Inc. He holds equity and serves as a Chief Scientific Officer of Hemodynamiq, Inc. He serves as a Chief Scientific Advisor to First Capital Ventures, LLC, and Promptus, LLC. The other authors have no conflicts of interest to declare.

# Main Figures

Fig.1. **Wearable smartwatch for continuous cuffless blood pressure and blood velocity monitoring on the basis of electrical bioimpedance**. **a**, Underlying measurement principles. **b**, **i**. Exploded view of the computer-aided smartwatch design model with key components labeled; **ii**. Smartwatch on the anterior forearm in the measurement configuration. **c**, Rolling standard deviation. **d**, Electrical impedance spectroscopy **i**. Magnitude and **ii**. Phase; **iii**. Cyclic voltammetry, mean (solid black line) and two standard deviation (shaded color) plotted; **iv**. Repeated cyclic voltammetry test. SD, standard deviation. Scale bars, 1 s.

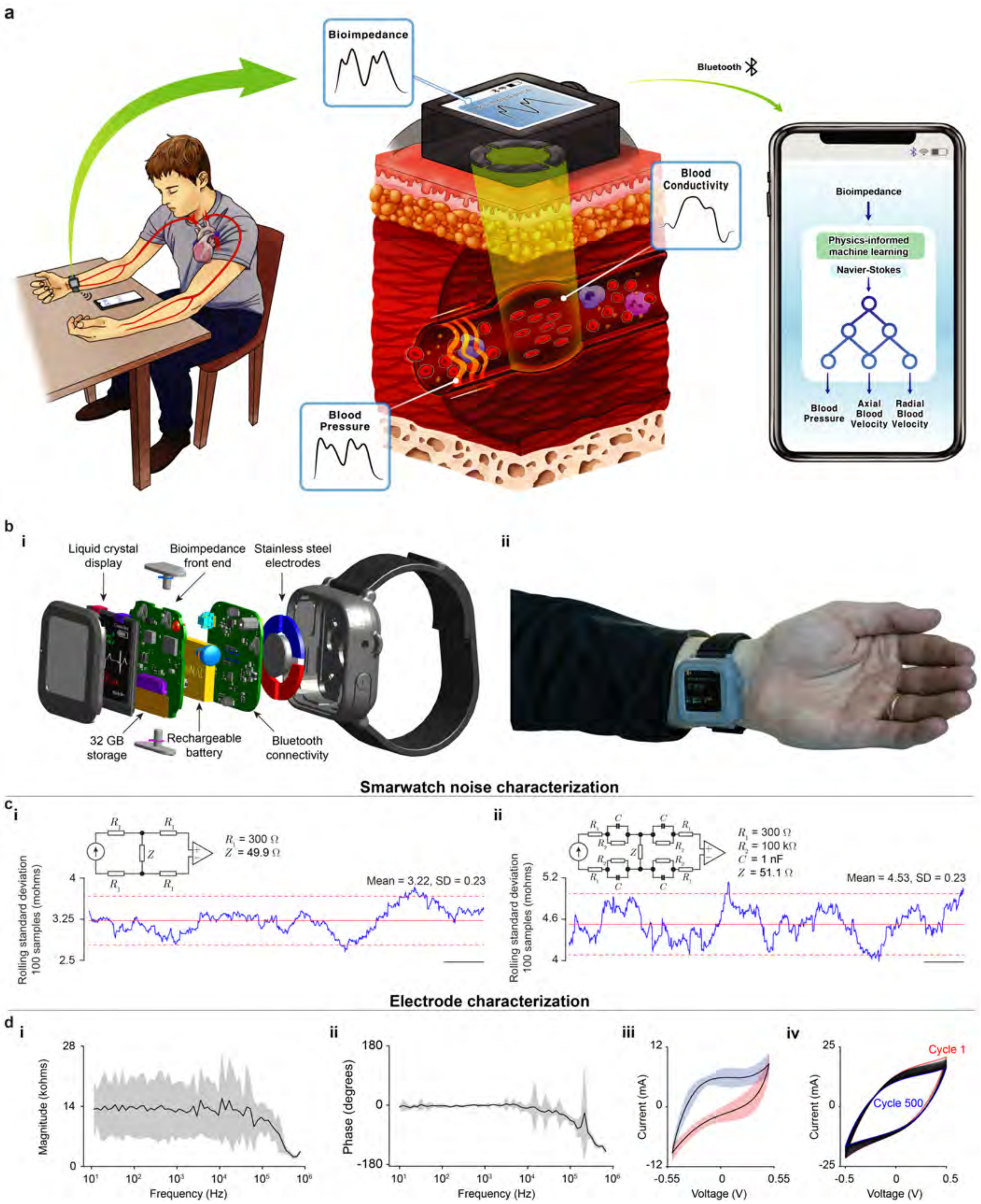

Fig.2. **Influence of brachial blood pressure, wall shear stress, and radial arterial radius impact radial blood volume and conductivity**. **a**, Multiscale coupled Navier-Stokes and Maxwell-Fricke modeling framework for immersed particle fluid dynamics featuring a branched elastic wall artery and structured tree outlet conditions. **b**, Local sensitivity analysis considering: **i**. Brachial mean arterial pressure (MAP); **ii**. Mean wall shear stress; **iii**. Radial artery length scaling; **iv**. Hematocrit; and **v**. Plasma conductivity. **c**, Contribution of red blood cell (RBC) **i**. Deformation; and **ii**. Orientation to the blood conductivity. **d**, Global sensitivity analysis. **e**, Computational fluid dynamic simulations in a subject-specific arm arterial model. Error bars, means and standard deviations. Scale bars, one-quarter period.

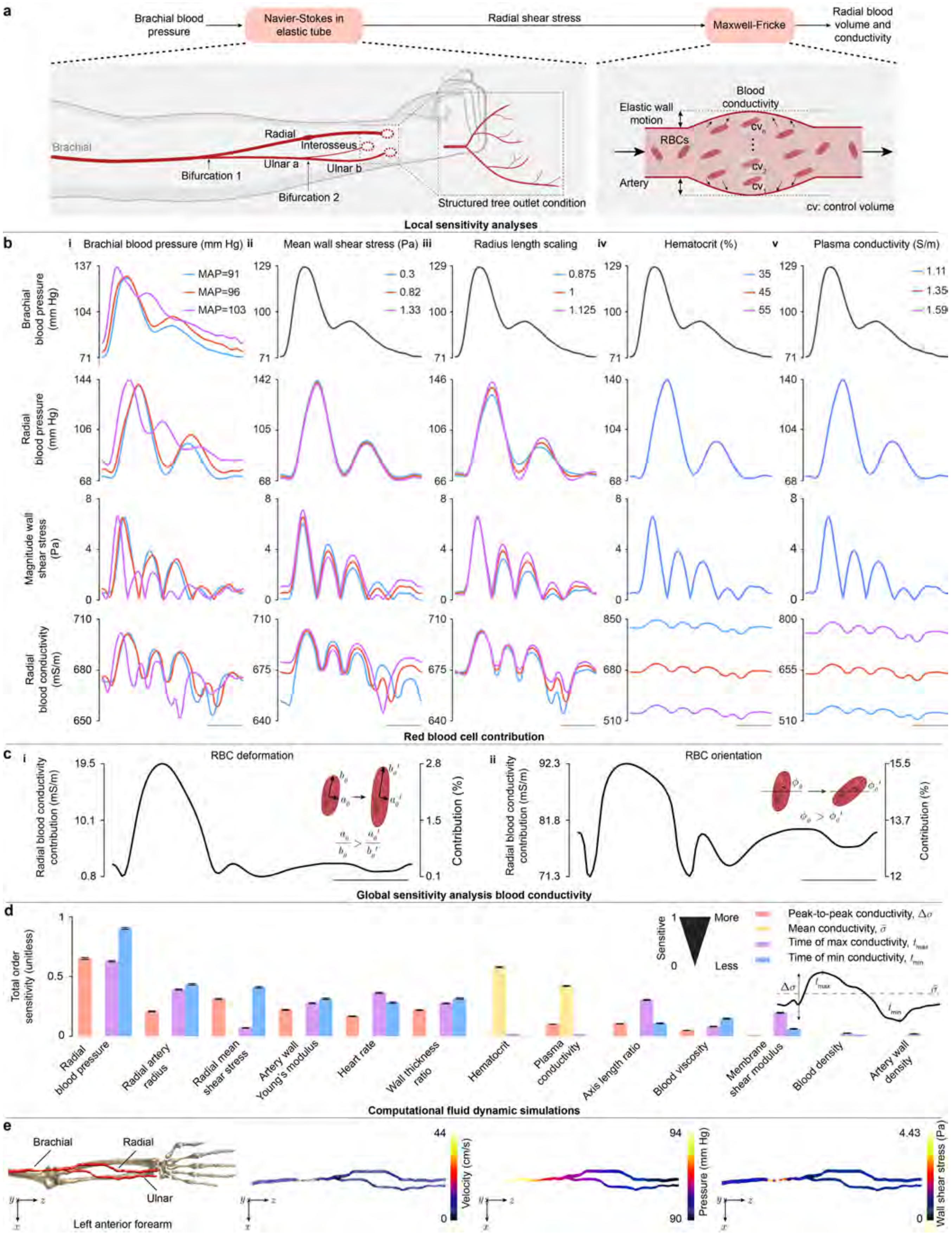
a
Brachial blood pressure
Navier-Stokes in elastic tube
Radial shear stress
Maxwell-Fricke
Radial blood volume and conductivity
Brachial
Radial
Interosseus
Ulnar a
Ulnar b
Bifurcation 1
Bifurcation 2
Structured tree outlet condition
Elastic wall motion
Blood conductivity
RBCs
Artery
cv: control volume
Local sensitivity analyses
b
i Brachial blood pressure (mm Hg)
ii Mean wall shear stress (Pa)
iii Radius length scaling
iv Hematocrit (%)
v Plasma conductivity (S/m)
MAP=91
MAP=96
MAP=103
Brachial blood pressure (mm Hg)
Radial blood pressure (mm Hg)
Magnitude wall shear stress (Pa)
Radial blood conductivity (mS/m)
Red blood cell contribution
c
i RBC deformation
ii RBC orientation
Radial blood conductivity contribution (mS/m)
Contribution (%)
Global sensitivity analysis blood conductivity
d
Total order sensitivity (unitless)
Sensitive
More
Less
Peak-to-peak conductivity, Δσ
Mean conductivity
Time of max conductivity
Time of min conductivity
Radial blood pressure
Radial artery radius
Radial mean shear stress
Artery wall Young's modulus
Heart rate
Wall thickness ratio
Hematocrit
Plasma conductivity
Axis length ratio
Blood viscosity
Membrane shear modulus
Blood density
Artery wall density
Computational fluid dynamic simulations
e
Brachial
Radial
Ulnar
Left anterior forearm
Velocity (cm/s)
Pressure (mm Hg)
Wall shear stress (Pa)

Fig.3. **Influence of anatomical, physiological, and experimental factors on electrical resistance a**, Three-layer cylindrical bioimpedance model of the wrist. **b**, Arterial blood conductivity and resultant wrist surface resistance signal over a representative period. **c**, Effects of the radius of the radial artery and blood conductivity on resistance. **d**, Global sensitivity analysis. **e**, Local sensitivity analysis. **f**, Subject-specific model based on the finite element method (FEM). **g**, Anatomical features of the FEM left wrist and placement of current-injecting (red) and voltage-sensing (blue) electrodes. **h**, Current density vector field. Local FEM sensitivity analyses: **i**, Electrode location; **j**, Artery depth; **k**, Artery radius; and **l**, Blood conductivity. **m**, Volume resistance density (VRD). **n**, Tissue and fluid contributions to the resistance at the wrist. Error bars, means and standard deviations. Scale bars in **f-k, m**, 1 cm. Scale bars in **b, e**, one-quarter period. Resistance data were acquired at 50 kHz.

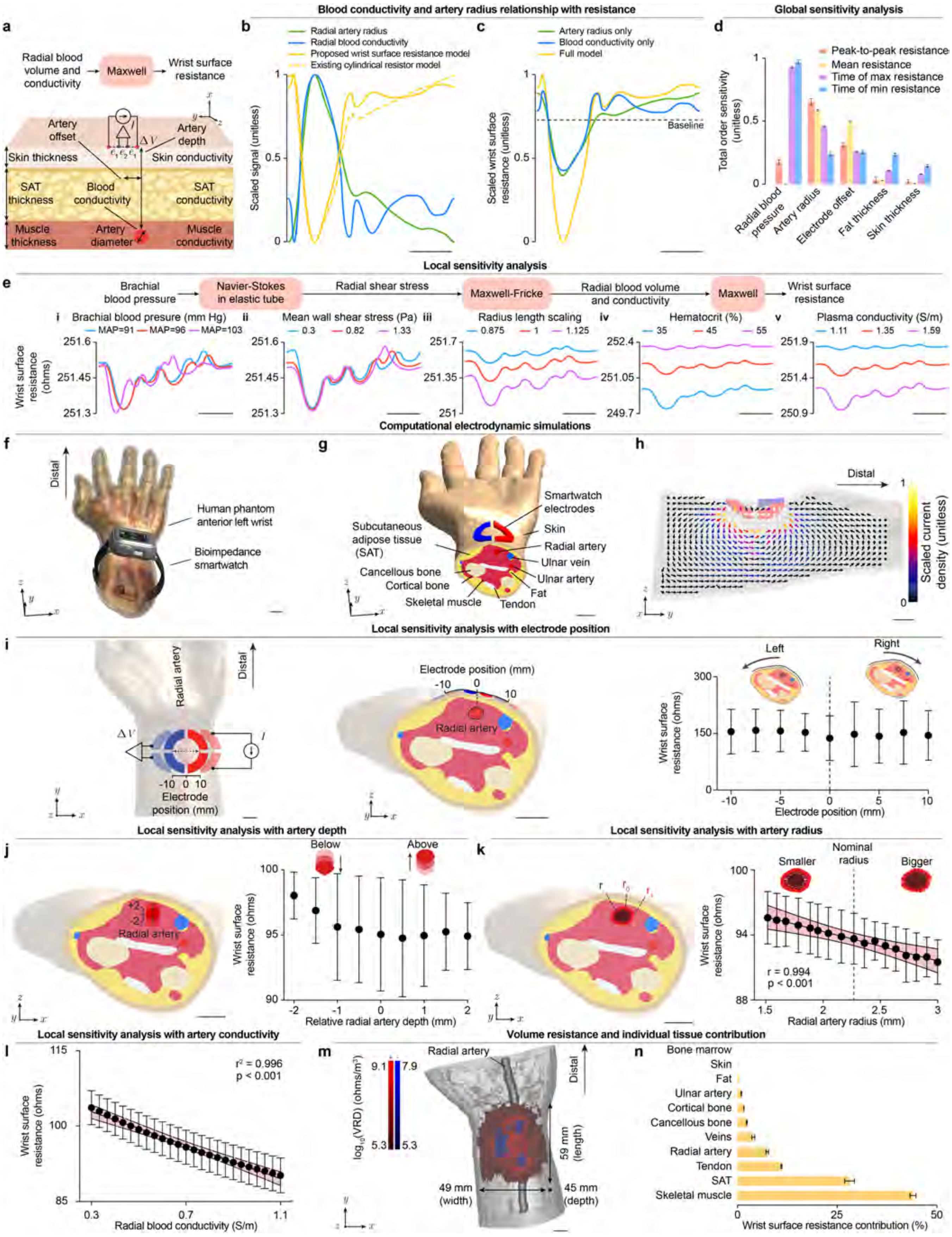

Blood conductivity and artery radius relationship with resistance
Global sensitivity analysis
Radial blood volume and conductivity
Maxwell
Wrist surface resistance
Artery offset
Artery depth
Skin thickness
Skin conductivity
SAT thickness
Blood conductivity
SAT conductivity
Muscle thickness
Artery diameter
Muscle conductivity
Radial artery radius
Radial blood conductivity
Proposed wrist surface resistance model
Existing cylindrical resistor model
Scaled signal (unitless)
Artery radius only
Blood conductivity only
Full model
Baseline
Scaled wrist surface resistance (unitless)
Peak-to-peak resistance
Mean resistance
Time of max resistance
Time of min resistance
Total order sensitivity (unitless)
Radial blood pressure
Artery radius
Electrode offset
Fat thickness
Skin thickness
Local sensitivity analysis
Brachial blood pressure
Navier-Stokes in elastic tube
Radial shear stress
Maxwell-Fricke
Radial blood volume and conductivity
Maxwell
Wrist surface resistance
Brachial blood presure (mm Hg)
MAP=91 MAP=96 MAP=103
Mean wall shear stress (Pa)
Radius length scaling
Hematocrit (%)
Plasma conductivity (S/m)
Wrist surface resistance (ohms)
Computational electrodynamic simulations
Distal
Human phantom anterior left wrist
Bioimpedance smartwatch
Smartwatch electrodes
Skin
Subcutaneous adipose tissue (SAT)
Radial artery
Ulnar vein
Cancellous bone
Ulnar artery
Cortical bone
Fat
Skeletal muscle
Tendon
Scaled current density (unitless)
Local sensitivity analysis with electrode position
Radial artery
Electrode position (mm)
Left
Right
Wrist surface resistance (ohms)
Local sensitivity analysis with artery depth
Local sensitivity analysis with artery radius
Below
Above
Relative radial artery depth (mm)
Smaller
Nominal radius
Bigger
r = 0.994
p < 0.001
Radial artery radius (mm)
Local sensitivity analysis with artery conductivity
Volume resistance and individual tissue contribution
r² = 0.996
p < 0.001
Radial blood conductivity (S/m)
log10(VRD) (ohms/m³)
49 mm (width)
59 mm (length)
45 mm (depth)
Bone marrow
Skin
Fat
Ulnar artery
Cortical bone
Cancellous bone
Veins
Radial artery
Tendon
SAT
Skeletal muscle
Wrist surface resistance contribution (%)

Fig.4. **Cuffless blood pressure monitoring performed in the clinic with no cuff and no self-reported data for calibration**. **a**, Architecture of the population-wide signal-tagged physics-informed neural network (sPINN). Systolic and diastolic blood pressure (SBP and DBP) calibration-free prediction results across **b, c**, All patients, **d, e**, hypertensive patients, **f, g**, cardiovascular disease patients, and **h, i**, patients with other conditions. AE, absolute error; SD, standard deviation; SBP, systolic blood pressure; DBP, diastolic blood pressure; CNN, convolutional neural network. All results are shown without test-exclusive patients.

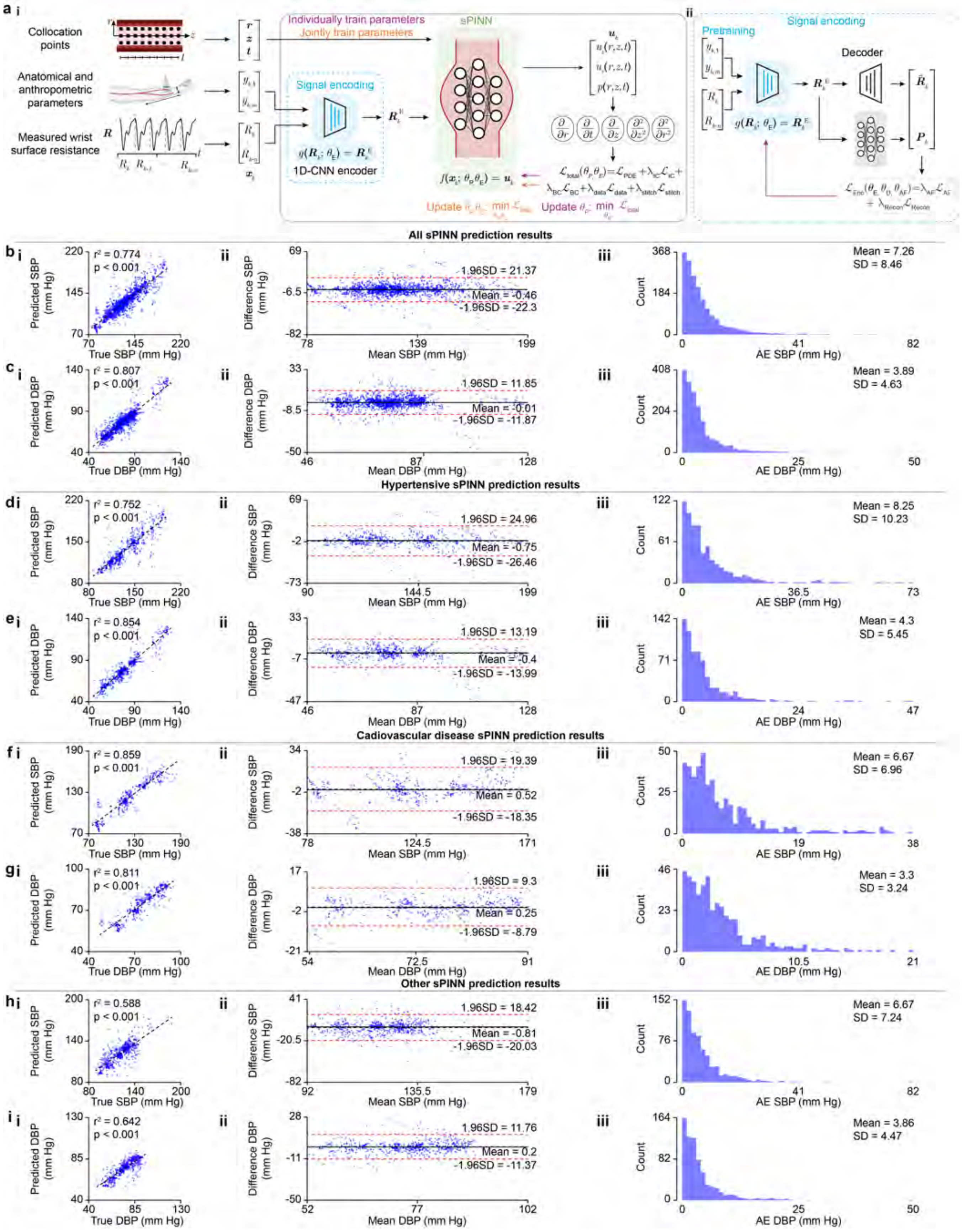

Collocation points
Anatomical and anthropometric parameters
Measured wrist surface resistance
Individually train parameters
Jointly train parameters
sPINN
Signal encoding
1D-CNN encoder
Pretraining
Decoder
All sPINN prediction results
r² = 0.774
p < 0.001
True SBP (mm Hg)
Predicted SBP (mm Hg)
Difference SBP (mm Hg)
Mean SBP (mm Hg)
1.96SD = 21.37
Mean = -0.46
-1.96SD = -22.3
Count
AE SBP (mm Hg)
Mean = 7.26
SD = 8.46
r² = 0.807
p < 0.001
True DBP (mm Hg)
Predicted DBP (mm Hg)
Difference DBP (mm Hg)
Mean DBP (mm Hg)
1.96SD = 11.85
Mean = -0.01
-1.96SD = -11.87
AE DBP (mm Hg)
Mean = 3.89
SD = 4.63
Hypertensive sPINN prediction results
r² = 0.752
1.96SD = 24.96
Mean = -0.75
-1.96SD = -26.46
Mean = 8.25
SD = 10.23
r² = 0.854
1.96SD = 13.19
Mean = -0.4
-1.96SD = -13.99
Mean = 4.3
SD = 5.45
Cardiovascular disease sPINN prediction results
r² = 0.859
1.96SD = 19.39
Mean = 0.52
-1.96SD = -18.35
Mean = 6.67
SD = 6.96
r² = 0.811
1.96SD = 9.3
Mean = 0.25
-1.96SD = -8.79
Mean = 3.3
SD = 3.24
Other sPINN prediction results
r² = 0.588
1.96SD = 18.42
Mean = -0.81
-1.96SD = -20.03
Mean = 6.67
SD = 7.24
r² = 0.642
1.96SD = 11.76
Mean = 0.2
-1.96SD = -11.37
Mean = 3.86
SD = 4.47

# Extended data figures

Extended Data Fig.1. **Anatomical, physiological, and experimental factors influence electrical reactance**. Local reactance sensitivity considering: **a**, Electrode placement, radial artery **b**, Depth, **c**, Radius, and **d**, Blood conductivity. **e**, Volume reactance density. **f**, Individual tissue contribution to the baseline reactance. Error bars, means and standard deviations. Single frequency simulations were performed at 50 kHz. SAT, subcutaneous adipose tissue; VXD, volume reactance density. Scale bar, 1 cm.

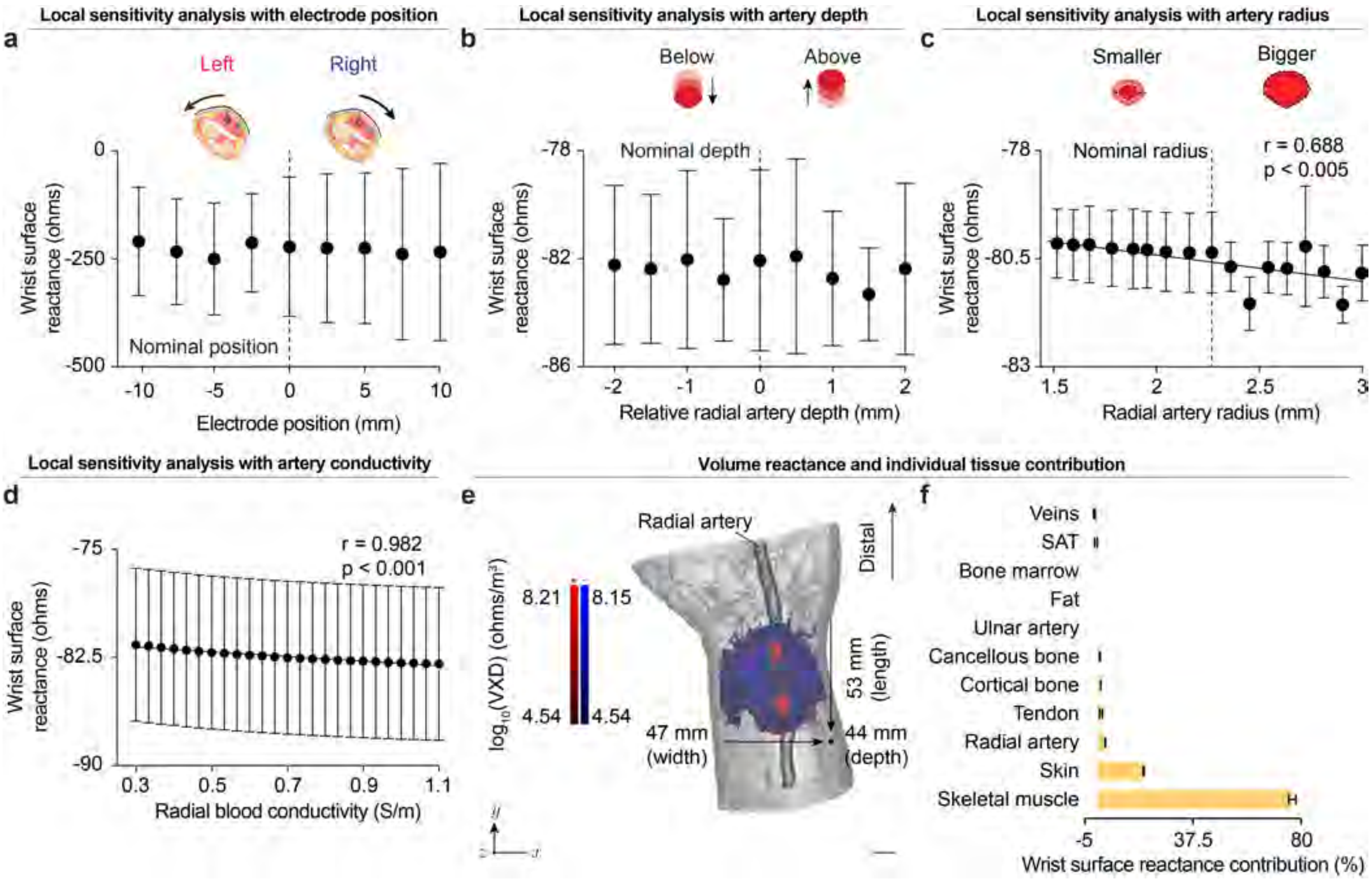

Extended Data Fig.2. **Anatomical analysis of the upper limb arteries among the Group 1 participants.** **a**, Doppler flowmetry and B-mode imaging with depth and diameter of the left distal **i, ii**. Radial; **iii, iv**. Ulnar; and **v, vi**. Brachial arteries. **b**, Radial artery **i**. Depth; **ii**. Major diameter; **iii**. Minor diameter; and **iv**. Peak systolic velocity. **c**, Ulnar artery **i**. Depth; **ii**. Major diameter; **iii**. Minor diameter; and **iv**. Peak systolic velocity. **d**, Brachial artery **i**. Depth; **ii**. Major diameter; **iii**. Minor diameter; and **iv**. Peak systolic velocity. **e,** Arteries **i**. Depth; **ii**. Major diameter; **iii**. Minor diameter; and **iv**. Peak systolic velocity.

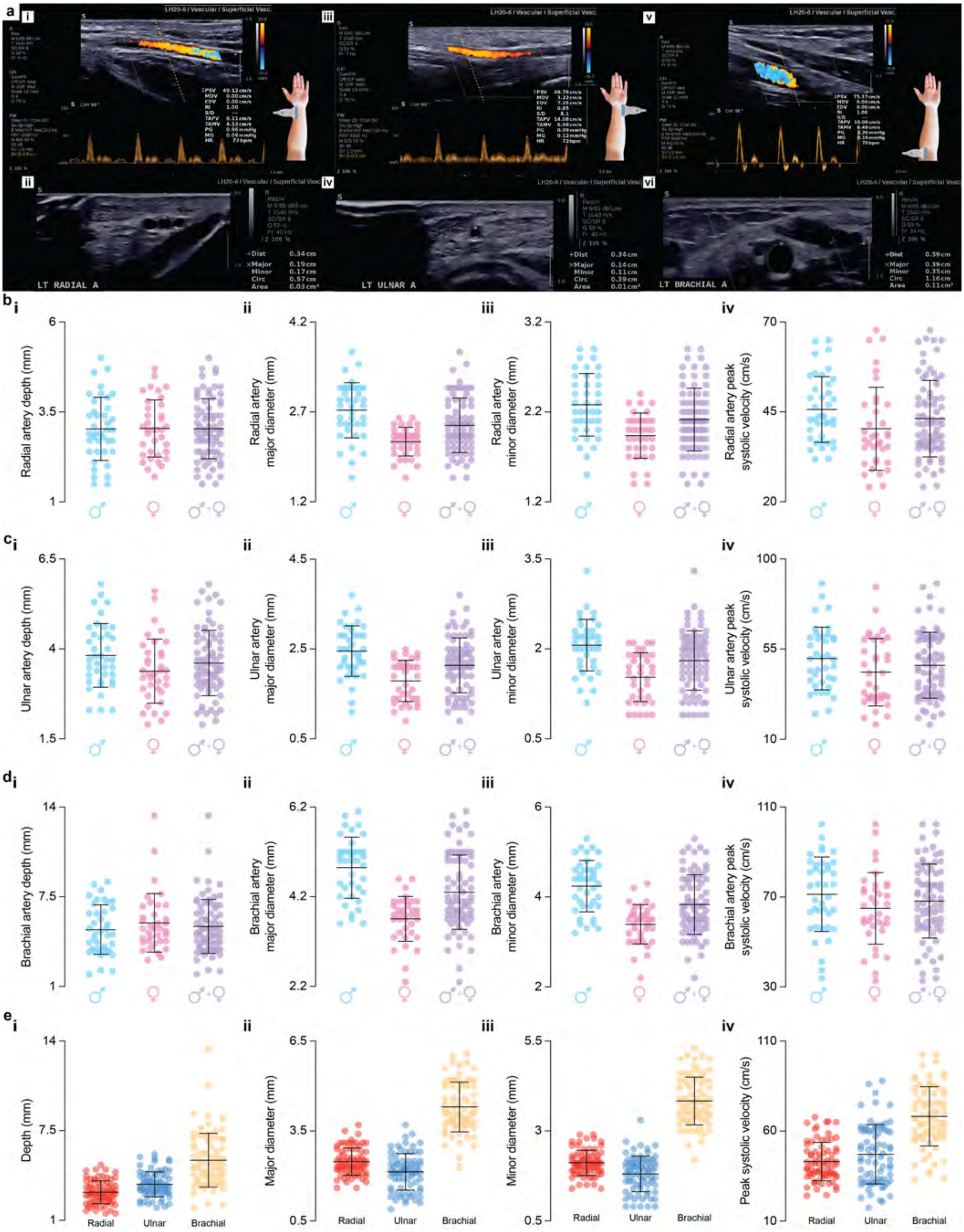

a
i
ii
iii
iv
v
vi
LT RADIAL A
LT ULNAR A
LT BRACHIAL A
b
c
d
e
Radial artery depth (mm)
Radial artery major diameter (mm)
Radial artery minor diameter (mm)
Radial artery peak systolic velocity (cm/s)
Ulnar artery depth (mm)
Ulnar artery major diameter (mm)
Ulnar artery minor diameter (mm)
Ulnar artery peak systolic velocity (cm/s)
Brachial artery depth (mm)
Brachial artery major diameter (mm)
Brachial artery minor diameter (mm)
Brachial artery peak systolic velocity (cm/s)
Depth (mm)
Major diameter (mm)
Minor diameter (mm)
Peak systolic velocity (cm/s)
Radial
Ulnar
Brachial

Extended Data Fig.3. **Hemodynamic predictions with a signal-tagged physics-informed neural network (sPINN) model on the basis of synthetic data**. **a**, Flowchart illustrating the generation of synthetic resistance signals. **b, i**. Predicted and modeled arterial velocity fields plotted for radial and axial locations in the artery at 5 different indicated times. **ii**. Axial velocity profiles evaluated at inlet. **c**, Example experimental and predicted BP waveforms. Analysis of BP and blood velocity predictions at **d**, systole and **e**, diastole. **i**. Correlation; **ii**. Bland-Altman plots; **iii**. Absolute prediction error histogram plots; and **iv**. Correlation plots for the artery cross-sectional mean velocity. **f**, Framework to generate synthetic data with biological variability. **g**, Systolic blood pressure (SBP) correlations. **h**, Diastolic blood pressure (DBP) correlations. The $x$- and $y$- axes are consistent across columns. Scale bars in **b i** and **c**, 1 cm and one-quarter period, respectively. AE, absolute error; RMSE, root mean squared error; SD, standard deviation; SAT, subcutaneous adipose tissue.

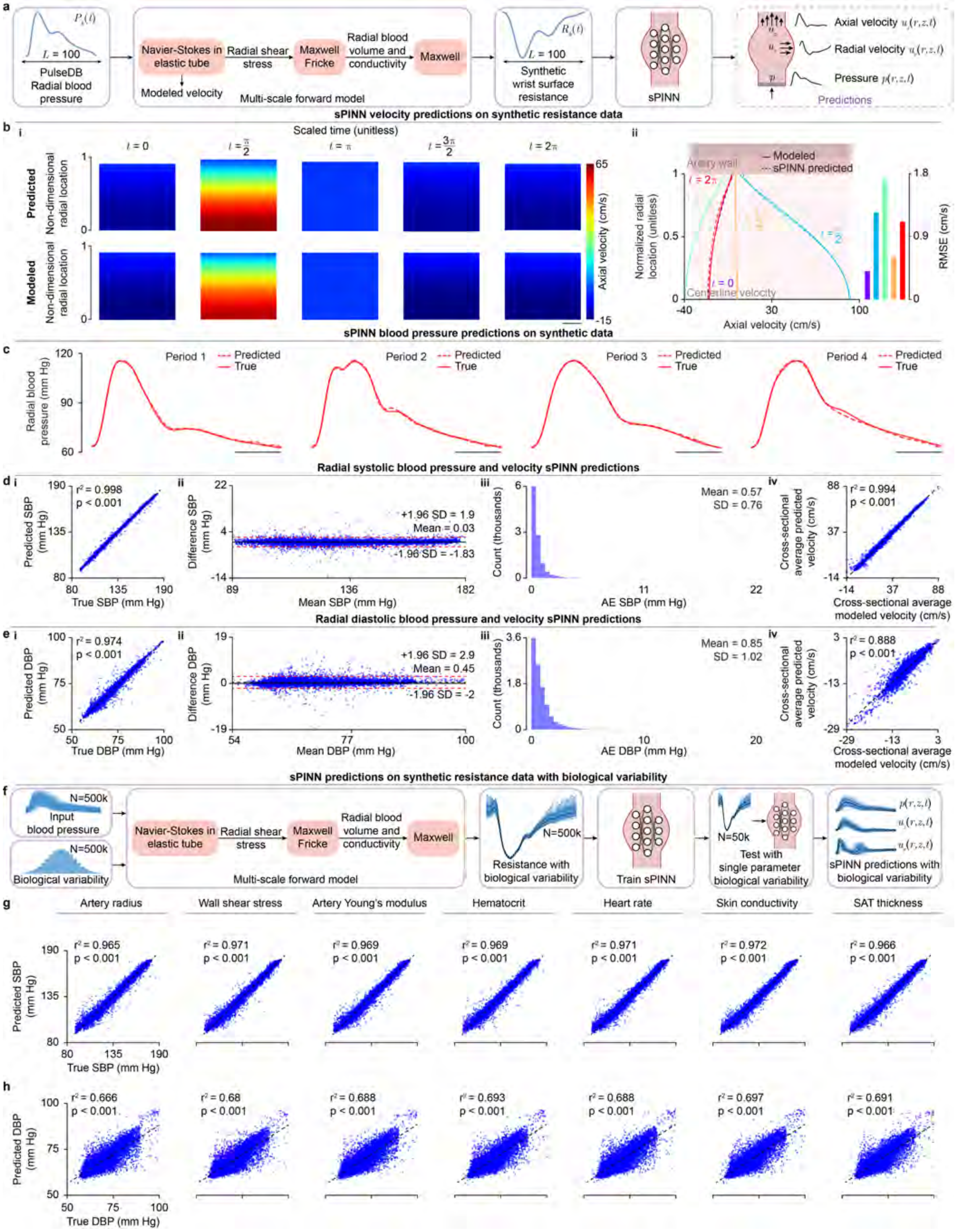

a
PulseDB Radial blood pressure
Navier-Stokes in elastic tube
Modeled velocity
Radial shear stress
Maxwell Fricke
Radial blood volume and conductivity
Maxwell
Multi-scale forward model
Synthetic wrist surface resistance
sPINN
Axial velocity
Radial velocity
Pressure
Predictions
sPINN velocity predictions on synthetic resistance data
b
Scaled time (unitless)
Predicted
Modeled
Non-dimensional radial location
Axial velocity (cm/s)
Modeled
sPINN predicted
Artery wall
Centerline velocity
Normalized radial location (unitless)
Axial velocity (cm/s)
RMSE (cm/s)
sPINN blood pressure predictions on synthetic data
c
Radial blood pressure (mm Hg)
Period 1
Period 2
Period 3
Period 4
Predicted
True
Radial systolic blood pressure and velocity sPINN predictions
d
Predicted SBP (mm Hg)
True SBP (mm Hg)
Difference SBP (mm Hg)
Mean SBP (mm Hg)
Count (thousands)
AE SBP (mm Hg)
Mean = 0.57
SD = 0.76
Cross-sectional average predicted velocity (cm/s)
Cross-sectional average modeled velocity (cm/s)
Radial diastolic blood pressure and velocity sPINN predictions
e
Predicted DBP (mm Hg)
True DBP (mm Hg)
Difference DBP (mm Hg)
Mean DBP (mm Hg)
AE DBP (mm Hg)
Mean = 0.85
SD = 1.02
sPINN predictions on synthetic resistance data with biological variability
f
Input blood pressure
Biological variability
Resistance with biological variability
Train sPINN
Test with single parameter biological variability
sPINN predictions with biological variability
g
Artery radius
Wall shear stress
Artery Young's modulus
Hematocrit
Heart rate
Skin conductivity
SAT thickness
h

Extended Data Fig.4. **Vital data for the healthy cohort vitals and blood velocity predictions by the signal-tagged physics-informed neural network.** **a**, Ensemble beat-to-beat heart rate, blood pressure (BP) and mean-centered resistance for the healthy participants in Group 1. **b**, Comparison between the peak systolic velocity (PSV) distribution predicted by the signal-tagged physics-informed neural network (sPINN) based on the test set for Group 1 (containing data from all participants) and a reference per-subject Doppler flowmetry PSV in the brachial artery. **c**, Overlays of the distributions of the sPINN PSV and Doppler flowmetry measured PSV. Scale bars, one-quarter period. BP, blood pressure; DBP, diastolic blood pressure; P, number of periods; SBP, systolic blood pressure.

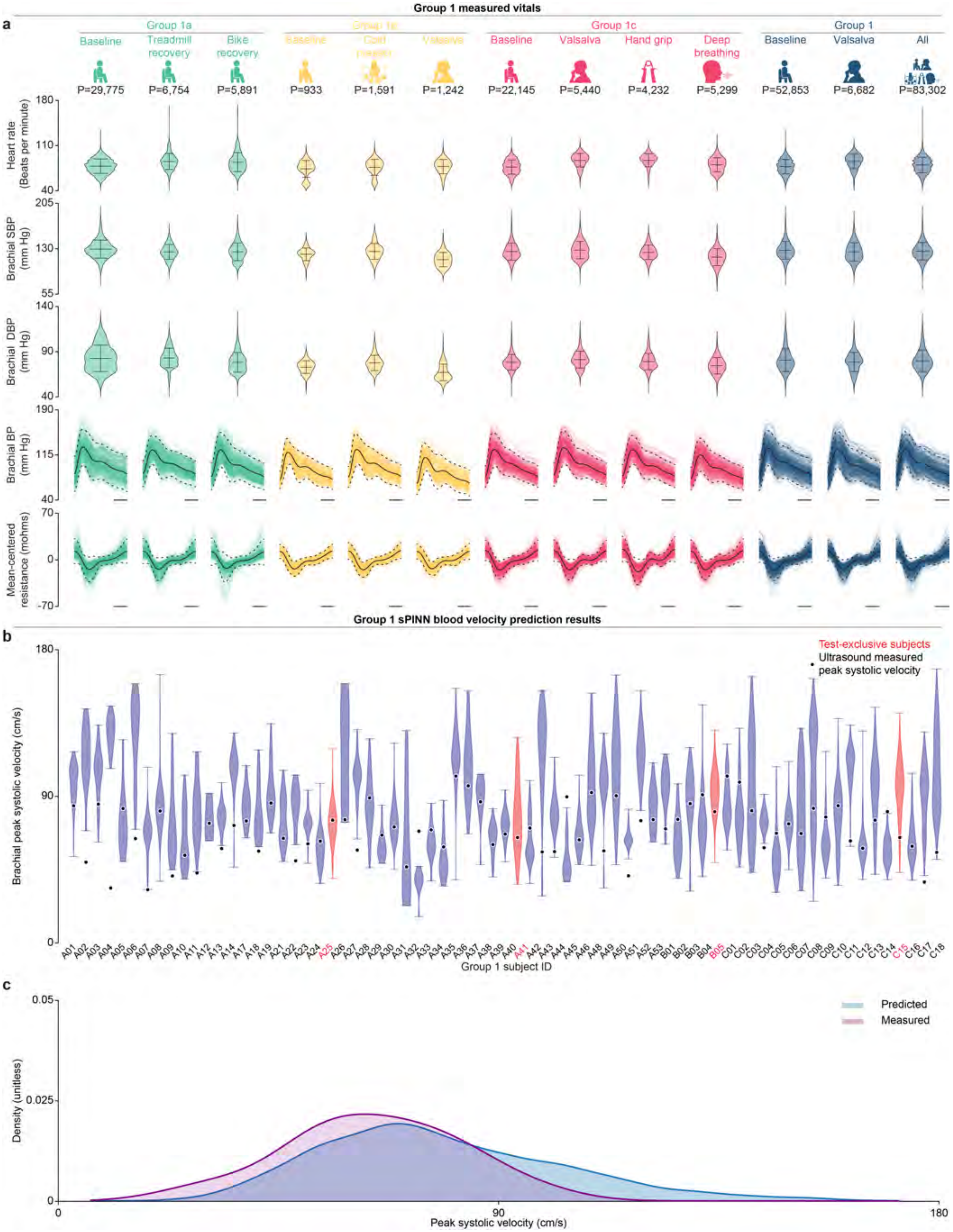

Group 1 measured vitals
a
Group 1a
Baseline
Treadmill recovery
Bike recovery
Baseline
Valsalva
Group 1c
Baseline
Valsalva
Hand grip
Deep breathing
Group 1
Baseline
Valsalva
All
P=29,775
P=6,754
P=5,891
P=933
P=1,591
P=1,242
P=22,145
P=5,440
P=4,232
P=5,299
P=52,853
P=6,682
P=83,302
Heart rate (Beats per minute)
Brachial SBP (mm Hg)
Brachial DBP (mm Hg)
Brachial BP (mm Hg)
Mean-centered resistance (mohms)
Group 1 sPINN blood velocity prediction results
b
Test-exclusive subjects
Ultrasound measured peak systolic velocity
Brachial peak systolic velocity (cm/s)
Group 1 subject ID
c
Predicted
Measured
Density (unitless)
Peak systolic velocity (cm/s)

Extended Data Fig.5. **Cuffless blood pressure monitoring using subject-specific signal-tagged physics-informed neural network models in a clinical cohort.** **a**, Subject-specific signal-tagged physics-informed neural network (sPINN) architecture. sPINN systolic and diastolic blood pressure (SBP and DBP) prediction results in **b, c**, All patients, **d, e**, Hypertensive patients, **f, g**, Cardiovascular disease patients, and **h, i**, Patients with other conditions. AE, absolute error; SD, standard deviation; SBP, systolic blood pressure; DBP, diastolic blood pressure; CNN, convolutional neural network. All results are shown without text-exclusive patients.

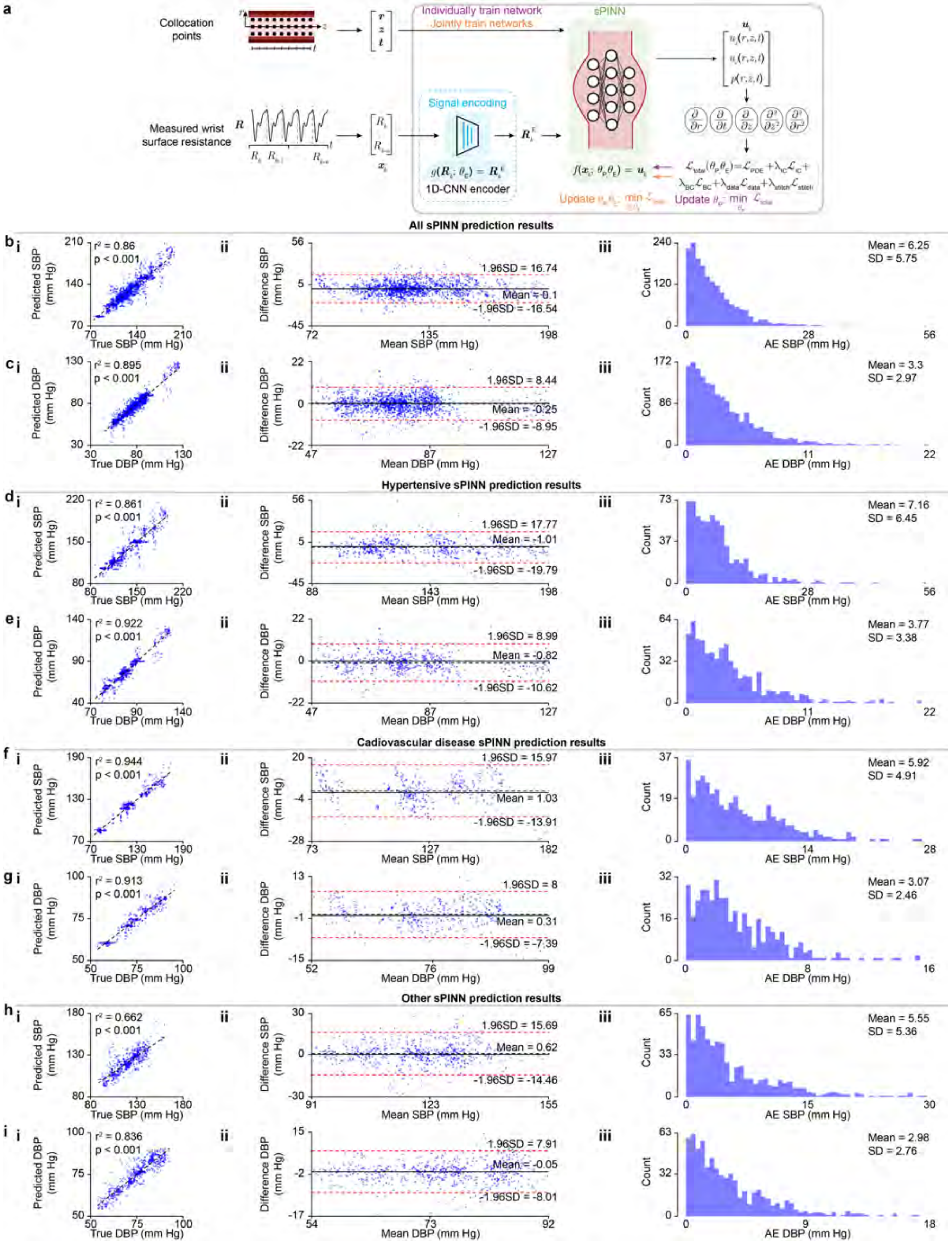

# Extended Data Tables

Extended Data Table 1. **Demographic and anthropometric data for cohort of healthy individuals who participated in laboratory experiments**. The data are shown as male:female (total). M, Male; F, Female; BMI, body mass index.

| Parameter (units) | Mean | Standard deviation | Minimum | First quartile | Median | Third quartile | Maximum |
|---|---|---|---|---|---|---|---|
| **All (M:F 39:36)** | | | | | | | |
| Age (years) | 25.62:25.75 (25.68) | 5.95:6.69 (6.28) | 18:18 (18) | 21:21.5 (21) | 23:25.5 (25) | 29:28 (28) | 47:48 (48) |
| Height (m) | 1.8:1.66 (1.73) | 0.07:0.07 (0.1) | 1.68:1.55 (1.55) | 1.73:1.6 (1.66) | 1.8:1.65 (1.73) | 1.85:1.73 (1.8) | 1.93:1.8 (1.93) |
| Weight (kg) | 75.92:66.24 (71.27) | 9.16:11.82 (11.53) | 53.5:48.07 (48.07) | 70.86:56.24 (61.79) | 77.1:65.08 (72.56) | 80.39:74.15 (79.37) | 102.04:97.51 (102.04) |
| BMI (kg/m$^2$) | 23.54:23.88 (23.7) | 2.72:3.83 (3.28) | 17.88:18.24 (17.88) | 21.88:21.25 (21.59) | 23.11:23.74 (23.49) | 25.11:25.29 (25.11) | 31.47:36.9 (36.9) |
| Wrist circumference (cm) | 16.28:14.66 (15.5) | 0.72:1.9 (1.62) | 14.8:4.7 (4.7) | 15.83:14.25 (14.93) | 16.4:15 (15.7) | 16.7:15.6 (16.4) | 17.7:16.5 (17.7) |
| Forearm length (cm) | 26.36:24.39 (25.41) | 2.83:1.5 (2.48) | 16.8:22 (16.8) | 25.53:23 (23.78) | 27:24.25 (25.5) | 28:25.4 (27) | 30:28 (30) |
| Upper arm length (cm) | 33.97:31.54 (32.8) | 3.94:2.75 (3.61) | 27:25 (25) | 30.73:29.25 (30.03) | 33.2:32 (32.5) | 37:34 (35) | 41.2:36 (41.2) |
| **Group 1a (M:F 26:26)** | | | | | | | |
| Age (years) | 25.77:25.46 (25.62) | 6.81:7.61 (7.15) | 18:18 (18) | 21:21 (21) | 23:23.5 (23) | 30:26 (28.5) | 47:48 (48) |
| Height (m) | 1.81:1.66 (1.73) | 0.07:0.07 (0.1) | 1.7:1.55 (1.55) | 1.75:1.6 (1.65) | 1.8:1.65 (1.73) | 1.88:1.73 (1.8) | 1.93:1.75 (1.93) |
| Weight (kg) | 76.96:64.05 (70.5) | 9.25:10.65 (11.83) | 58.96:48.07 (48.07) | 72.56:55.78 (60.09) | 78.23:61.22 (70.98) | 81.63:68.03 (79.37) | 102.04:97.51 (102.04) |
| BMI (kg/m$^2$) | 23.58:23.3 (23.44) | 2.23:3.69 (3.02) | 20.36:18.24 (18.24) | 21.87:21.25 (21.58) | 23.29:23.28 (23.28) | 25.11:24.69 (25.1) | 28.69:36.9 (36.9) |
| Wrist circumference (cm) | 16.46:14.97 (15.72) | 0.74:0.88 (1.1) | 14.9:13 (13) | 16.3:14.5 (15) | 16.5:15 (16) | 17:15.6 (16.5) | 17.7:16.5 (17.7) |
| Forearm length (cm) | 26.73:24.36 (25.54) | 2.62:1.49 (2.43) | 17.8:22 (17.8) | 25.5:23 (24) | 27:24.25 (25.45) | 28:25.4 (27.25) | 30:28 (30) |
| Upper arm length (cm) | 35.1:31.81 (33.45) | 4:2.98 (3.87) | 28:25 (25) | 32:29 (30.5) | 35.7:32.25 (34) | 38:34 (36.2) | 41.2:36 (41.2) |
| **Group 1b (M:F 4:1)** | | | | | | | |
| Age (years) | 25.75:24 (25.4) | 3.77:0 (3.36) | 21:24 (21) | 23:24 (23.25) | 26:24 (25) | 28.5:24 (27.75) | 30:24 (30) |
| Height (m) | 1.8:1.65 (1.77) | 0.09:0 (0.1) | 1.68:1.65 (1.65) | 1.74:1.65 (1.67) | 1.83:1.65 (1.8) | 1.87:1.65 (1.86) | 1.88:1.65 (1.88) |
| Weight (kg) | 74.83:68.03 (73.47) | 12.14:0 (10.95) | 63.49:68.03 (63.49) | 64.63:68.03 (65.19) | 73.7:68.03 (68.03) | 85.03:68.03 (83.33) | 88.44:68.03 (88.44) |
| BMI (kg/m$^2$) | 23.32:24.96 (23.64) | 5.76:0 (5.04) | 18.47:24.96 (18.47) | 19.34:24.96 (19.78) | 21.66:24.96 (23.11) | 27.29:24.96 (26.58) | 31.47:24.96 (31.47) |
| Wrist circumference (cm) | 15.93:15.5 (15.84) | 0.48:0 (0.46) | 15.5:15.5 (15.5) | 15.6:15.5 (15.5) | 15.8:15.5 (15.7) | 16.25:15.5 (16.08) | 16.6:15.5 (16.6) |
| Forearm length (cm) | 25.85:23.7 (25.42) | 2.9:0 (2.69) | 21.7:23.7 (21.7) | 23.85:23.7 (23.2) | 26.85:23.7 (26) | 27.85:23.7 (27.78) | 28:23.7 (28) |
| Upper arm length (cm) | 31.43:33.4 (31.82) | 3.37:0 (3.05) | 27:33.4 (27) | 29.25:33.4 (30.38) | 31.75:33.4 (32) | 33.6:33.4 (33.85) | 35.2:33.4 (35.2) |
| **Group 1c (M:F 9:9)** | | | | | | | |
| Age (years) | 25.11:26.78 (25.94) | 4.17:3.63 (3.89) | 21:20 (20) | 22.5:25 (23) | 23:28 (26.5) | 27.25:29.25 (28) | 34:31 (34) |
| Height (m) | 1.77:1.68 (1.73) | 0.06:0.08 (0.08) | 1.7:1.6 (1.6) | 1.72:1.62 (1.68) | 1.74:1.68 (1.73) | 1.82:1.77 (1.8) | 1.85:1.8 (1.85) |
| Weight (kg) | 73.39:72.35 (72.87) | 8.02:14.1 (11.14) | 53.5:50 (50) | 72.9:61.61 (69.84) | 75:73.47 (74.91) | 78.23:81.54 (78.91) | 79.37:92.97 (92.97) |
| BMI (kg/m$^2$) | 23.52:25.42 (24.47) | 2.67:4.23 (3.57) | 17.88:19.05 (17.88) | 22.73:22.93 (23.01) | 23.61:25 (24.15) | 25.66:28.17 (26.14) | 26.93:33.08 (33.08) |
| Wrist circumference (cm) | 15.93:13.67 (14.8) | 0.6:3.44 (2.66) | 14.8:4.7 (4.7) | 15.65:14 (14.5) | 15.9:14.5 (15.55) | 16.4:15.53 (15.9) | 16.7:15.7 (16.7) |
| Forearm length (cm) | 25.53:24.54 (25.04) | 3.48:1.69 (2.7) | 16.8:22.4 (16.8) | 25.53:23.28 (23.5) | 26.1:24.7 (25.5) | 27.35:25.45 (26.3) | 28.6:28 (28.6) |
| Upper arm length (cm) | 31.82:30.56 (31.19) | 2.59:1.89 (2.29) | 27:27.7 (27) | 30.4:29.33 (29.6) | 31.8:30.3 (31.2) | 33.05:32.15 (33) | 36.4:33.5 (36.4) |

Extended Data Table 2. **Anatomical and hemodynamic data of the brachial, radial, and ulnar arteries for the cohort of healthy individuals who participated in laboratory experiments.** The data shown for the healthy lab subjects are reported as male:female (total). $^*$, $N$=25 males and $N$=26 females. $^\dagger$, $N$=38 males and $N$=36 females.

| Parameter (units) | Mean | Standard deviation | Minimum | First quartile | Median | Third quartile | Maximum |
|---|---|---|---|---|---|---|---|
| **All (M:F 39:36)** | | | | | | | |
| Radial artery depth (cm) | 0.3:0.3 (0.3) | 0.09:0.08 (0.08) | 0.15:0.17 (0.15) | 0.24:0.24 (0.24) | 0.29:0.3 (0.3) | 0.36:0.37 (0.36) | 0.5:0.47 (0.5) |
| Radial artery major diameter (cm) | 0.27:0.22 (0.25) | 0.05:0.02 (0.05) | 0.16:0.16 (0.16) | 0.25:0.21 (0.22) | 0.28:0.23 (0.24) | 0.31:0.24 (0.28) | 0.37:0.26 (0.37) |
| Radial artery minor diameter (cm) | 0.23:0.19 (0.21) | 0.03:0.03 (0.03) | 0.15:0.14 (0.14) | 0.2:0.18 (0.19) | 0.22:0.2 (0.21) | 0.26:0.21 (0.23) | 0.29:0.24 (0.29) |
| Radial peak systolic velocity (cm/s) | 45.62:40.26 (43.05) | 9.18:11.56 (10.67) | 31.69:24 (24) | 37.68:31.27 (35.39) | 43.63:37.59 (41.71) | 52.3:46.95 (48.92) | 64.9:67.75 (67.75) |
| Ulnar artery depth (cm) | 0.38:0.34 (0.36) | 0.09:0.09 (0.09) | 0.23:0.19 (0.19) | 0.32:0.29 (0.3) | 0.37:0.33 (0.35) | 0.44:0.39 (0.42) | 0.58:0.56 (0.58) |
| Ulnar artery major diameter (cm) | 0.25:0.18 (0.21) | 0.06:0.05 (0.06) | 0.11:0.09 (0.09) | 0.21:0.14 (0.16) | 0.25:0.19 (0.21) | 0.28:0.22 (0.25) | 0.37:0.25 (0.37) |
| Ulnar artery minor diameter (cm) | 0.21:0.15 (0.18) | 0.04:0.04 (0.05) | 0.11:0.09 (0.09) | 0.18:0.12 (0.15) | 0.21:0.16 (0.19) | 0.23:0.19 (0.21) | 0.33:0.21 (0.33) |
| Ulnar peak systolic velocity (cm/s)† | 50.43:43.51 (47.06) | 15.67:16.83 (16.5) | 22.71:17.39 (17.39) | 38.49:30.68 (32.21) | 50.12:40.26 (46.06) | 60.88:59.6 (60.23) | 87.93:86.16 (87.93) |
| Brachial artery depth (cm) | 0.51:0.56 (0.54) | 0.18:0.21 (0.2) | 0.19:0.29 (0.19) | 0.37:0.43 (0.4) | 0.48:0.52 (0.5) | 0.66:0.65 (0.66) | 0.86:1.34 (1.34) |
| Brachial artery major diameter (cm) | 0.48:0.37 (0.43) | 0.07:0.05 (0.08) | 0.36:0.23 (0.23) | 0.43:0.34 (0.37) | 0.5:0.38 (0.41) | 0.52:0.41 (0.5) | 0.61:0.46 (0.61) |
| Brachial artery minor diameter (cm) | 0.42:0.34 (0.38) | 0.06:0.04 (0.07) | 0.32:0.22 (0.22) | 0.38:0.31 (0.34) | 0.44:0.35 (0.38) | 0.46:0.37 (0.44) | 0.53:0.43 (0.53) |
| Brachial peak systolic velocity (cm/s) | 71.25:64.93 (68.21) | 16.65:15.96 (16.52) | 33.76:32.72 (32.72) | 58.06:57.5 (57.25) | 75.05:64.34 (67.32) | 84.22:74.17 (80.93) | 102.43:102.43 (102.43) |
| **Group 1a (M:F 26:26)** | | | | | | | |
| Radial artery depth (cm) | 0.32:0.29 (0.3) | 0.09:0.08 (0.08) | 0.17:0.17 (0.17) | 0.24:0.22 (0.24) | 0.32:0.29 (0.3) | 0.39:0.36 (0.38) | 0.5:0.42 (0.5) |
| Radial artery major diameter (cm) | 0.28:0.22 (0.25) | 0.04:0.03 (0.05) | 0.2:0.16 (0.16) | 0.26:0.2 (0.23) | 0.29:0.23 (0.25) | 0.31:0.24 (0.29) | 0.37:0.26 (0.37) |
| Radial artery minor diameter (cm) | 0.24:0.19 (0.21) | 0.03:0.02 (0.04) | 0.18:0.14 (0.14) | 0.21:0.17 (0.19) | 0.23:0.2 (0.21) | 0.27:0.21 (0.24) | 0.29:0.24 (0.29) |
| Radial peak systolic velocity (cm/s) | 46.22:38.26 (42.24) | 8.84:8.92 (9.67) | 31.69:24.32 (24.32) | 39.94:31.56 (35.47) | 45.05:36.87 (40.69) | 52.66:41.44 (48.64) | 64.9:64.9 (64.9) |
| Ulnar artery depth (cm) | 0.39:0.34 (0.37) | 0.08:0.1 (0.09) | 0.23:0.19 (0.19) | 0.33:0.25 (0.31) | 0.39:0.33 (0.36) | 0.46:0.39 (0.44) | 0.58:0.56 (0.58) |
| Ulnar artery major diameter (cm) | 0.25:0.16 (0.21) | 0.06:0.05 (0.07) | 0.11:0.09 (0.09) | 0.22:0.13 (0.15) | 0.25:0.15 (0.21) | 0.28:0.21 (0.25) | 0.37:0.25 (0.37) |
| Ulnar artery minor diameter (cm) | 0.21:0.14 (0.18) | 0.04:0.04 (0.05) | 0.11:0.09 (0.09) | 0.19:0.11 (0.13) | 0.22:0.14 (0.19) | 0.23:0.19 (0.22) | 0.33:0.21 (0.33) |
| Ulnar peak systolic velocity (cm/s)* | 50.27:39.09 (44.57) | 12.3:16.04 (15.27) | 22.71:17.39 (17.39) | 42.03:28.99 (31.28) | 50.25:32.41 (44.29) | 60.37:49.6 (58.46) | 74.56:86.16 (86.16) |
| Brachial artery depth (cm) | 0.53:0.57 (0.55) | 0.18:0.23 (0.21) | 0.19:0.29 (0.19) | 0.37:0.43 (0.4) | 0.52:0.53 (0.52) | 0.7:0.62 (0.68) | 0.84:1.34 (1.34) |
| Brachial artery major diameter (cm) | 0.51:0.37 (0.44) | 0.06:0.05 (0.09) | 0.36:0.23 (0.23) | 0.5:0.34 (0.38) | 0.52:0.38 (0.44) | 0.56:0.41 (0.52) | 0.61:0.46 (0.61) |
| Brachial artery minor diameter (cm) | 0.45:0.35 (0.4) | 0.05:0.04 (0.07) | 0.32:0.22 (0.22) | 0.42:0.32 (0.35) | 0.45:0.35 (0.39) | 0.48:0.37 (0.45) | 0.53:0.43 (0.53) |
| Brachial peak systolic velocity (cm/s) | 70.79:61 (65.9) | 17.79:14.5 (16.81) | 33.76:32.72 (32.72) | 56.37:53.8 (55.96) | 71.83:60.64 (64.5) | 85.23:71.18 (75.69) | 102.43:89.7 (102.43) |
| **Group 1b (M:F 4:1)** | | | | | | | |
| Radial artery depth (cm) | 0.26:0.32 (0.27) | 0.07:0 (0.06) | 0.17:0.32 (0.17) | 0.22:0.32 (0.25) | 0.28:0.32 (0.28) | 0.31:0.32 (0.32) | 0.33:0.32 (0.33) |
| Radial artery major diameter (cm) | 0.23:0.22 (0.23) | 0.06:0 (0.05) | 0.16:0.22 (0.16) | 0.19:0.22 (0.2) | 0.23:0.22 (0.22) | 0.27:0.22 (0.26) | 0.3:0.22 (0.3) |
| Radial artery minor diameter (cm) | 0.2:0.21 (0.2) | 0.03:0 (0.03) | 0.15:0.21 (0.15) | 0.17:0.21 (0.18) | 0.2:0.21 (0.21) | 0.22:0.21 (0.22) | 0.23:0.21 (0.23) |
| Radial peak systolic velocity (cm/s) | 38.99:65.38 (44.27) | 4.43:0 (12.41) | 34.3:65.38 (34.3) | 35.27:65.38 (35.76) | 39.02:65.38 (41.79) | 42.71:65.38 (49.07) | 43.63:65.38 (65.38) |
| Ulnar artery depth (cm) | 0.37:0.48 (0.39) | 0.12:0 (0.11) | 0.3:0.48 (0.3) | 0.31:0.48 (0.31) | 0.32:0.48 (0.33) | 0.44:0.48 (0.5) | 0.55:0.48 (0.55) |
| Ulnar artery major diameter (cm) | 0.2:0.24 (0.21) | 0.03:0 (0.03) | 0.18:0.24 (0.18) | 0.19:0.24 (0.19) | 0.19:0.24 (0.19) | 0.22:0.24 (0.24) | 0.25:0.24 (0.25) |
| Ulnar artery minor diameter (cm) | 0.17:0.17 (0.17) | 0.04:0 (0.03) | 0.13:0.17 (0.13) | 0.14:0.17 (0.15) | 0.16:0.17 (0.16) | 0.19:0.17 (0.18) | 0.22:0.17 (0.22) |
| Ulnar peak systolic velocity (cm/s) | 42.58:61.04 (46.27) | 12.51:0 (13.62) | 33.58:61.04 (33.58) | 34.63:61.04 (35.15) | 37.92:61.04 (40.17) | 50.53:61.04 (60.92) | 60.88:61.04 (61.04) |
| Brachial artery depth (cm) | 0.52:0.71 (0.56) | 0.23:0 (0.21) | 0.36:0.71 (0.36) | 0.39:0.71 (0.41) | 0.44:0.71 (0.45) | 0.66:0.71 (0.75) | 0.86:0.71 (0.86) |
| Brachial artery major diameter (cm) | 0.42:0.37 (0.41) | 0.07:0 (0.06) | 0.37:0.37 (0.37) | 0.38:0.37 (0.37) | 0.4:0.37 (0.39) | 0.46:0.37 (0.43) | 0.52:0.37 (0.52) |
| Brachial artery minor diameter (cm) | 0.37:0.32 (0.36) | 0.04:0 (0.04) | 0.33:0.32 (0.32) | 0.34:0.32 (0.33) | 0.37:0.32 (0.35) | 0.41:0.32 (0.39) | 0.43:0.32 (0.43) |
| Brachial peak systolic velocity (cm/s) | 81.77:75.85 (80.58) | 8.9:0 (8.15) | 70.05:75.85 (70.05) | 75.29:75.85 (74.4) | 83.02:75.85 (80.52) | 88.25:75.85 (86.89) | 90.98:75.85 (90.98) |
| **Group 1c (M:F 9:9)** | | | | | | | |
| Radial artery depth (cm) | 0.28:0.34 (0.31) | 0.09:0.08 (0.09) | 0.15:0.23 (0.15) | 0.22:0.29 (0.28) | 0.29:0.31 (0.3) | 0.31:0.42 (0.35) | 0.42:0.47 (0.47) |
| Radial artery major diameter (cm) | 0.26:0.22 (0.24) | 0.05:0.02 (0.05) | 0.19:0.18 (0.18) | 0.22:0.21 (0.21) | 0.25:0.22 (0.23) | 0.3:0.23 (0.25) | 0.35:0.25 (0.35) |
| Radial artery minor diameter (cm) | 0.22:0.2 (0.21) | 0.03:0.03 (0.03) | 0.17:0.14 (0.14) | 0.2:0.19 (0.19) | 0.23:0.2 (0.22) | 0.24:0.22 (0.23) | 0.26:0.23 (0.26) |
| Radial peak systolic velocity (cm/s) | 46.8:43.26 (45.03) | 11.15:15.3 (13.11) | 31.89:24 (24) | 37.77:28.63 (35.27) | 42.77:44.93 (43.85) | 57.09:55.52 (55.72) | 62.32:67.75 (67.75) |
| Ulnar artery depth (cm) | 0.36:0.32 (0.34) | 0.09:0.04 (0.07) | 0.23:0.28 (0.23) | 0.31:0.29 (0.29) | 0.37:0.3 (0.34) | 0.39:0.34 (0.37) | 0.53:0.41 (0.53) |
| Ulnar artery major diameter (cm) | 0.25:0.21 (0.23) | 0.06:0.02 (0.05) | 0.13:0.19 (0.13) | 0.21:0.21 (0.21) | 0.26:0.21 (0.22) | 0.3:0.22 (0.26) | 0.33:0.24 (0.33) |
| Ulnar artery minor diameter (cm) | 0.21:0.18 (0.19) | 0.04:0.02 (0.03) | 0.13:0.14 (0.13) | 0.2:0.17 (0.18) | 0.21:0.18 (0.2) | 0.23:0.2 (0.21) | 0.26:0.21 (0.26) |
| Ulnar peak systolic velocity (cm/s) | 54.34:54.32 (54.33) | 23.98:14.2 (19.12) | 24.32:31.73 (24.32) | 32.66:46.95 (33.34) | 51.99:59.44 (56.68) | 77.22:62.52 (65.22) | 87.93:74.56 (87.93) |
| Brachial artery depth (cm) | 0.47:0.52 (0.49) | 0.16:0.15 (0.15) | 0.22:0.31 (0.22) | 0.35:0.44 (0.37) | 0.47:0.49 (0.49) | 0.6:0.58 (0.59) | 0.72:0.77 (0.77) |
| Brachial artery major diameter (cm) | 0.43:0.37 (0.4) | 0.05:0.06 (0.06) | 0.36:0.26 (0.26) | 0.4:0.31 (0.38) | 0.42:0.39 (0.41) | 0.45:0.41 (0.43) | 0.52:0.46 (0.52) |
| Brachial artery minor diameter (cm) | 0.39:0.32 (0.36) | 0.05:0.04 (0.05) | 0.33:0.26 (0.26) | 0.34:0.29 (0.33) | 0.38:0.34 (0.35) | 0.41:0.36 (0.38) | 0.47:0.38 (0.47) |
| Brachial peak systolic velocity (cm/s) | 67.88:75.06 (71.47) | 15.15:16.65 (15.88) | 37.36:58.14 (37.36) | 60.9:59.15 (59.38) | 67.16:73.11 (70.22) | 81.53:85.19 (81.17) | 84.23:102.43 (102.43) |

Extended Data Table 3. **Demographic and hemodynamic data for the clinical cohort.** The data are shown as male:female (total). HTN, hypertension; CVD, cardiovascular disease; SBP, systolic blood pressure; DBP, diastolic blood pressure; HR, heart rate; BPM, beats per minute.

| Parameter (units) | Mean | Standard deviation | Minimum | First quartile | Median | Third quartile | Maximum |
|---|---|---|---|---|---|---|---|
| **All (M:F 51:35)** | | | | | | | |
| Age (years) | 67.22:68.06 (67.66) | 13.83:16.44 (14.79) | 28:24 (24) | 58.5:55 (58) | 71:72 (71) | 76:80 (77) | 93:90 (93) |
| Office brachial SBP (mm Hg) | 128.53:126.44 (128.1) | 17.06:19.12 (18.13) | 101:91 (91) | 116:114 (115) | 125:120.5 (124) | 139.5:145 (142) | 186:165 (186) |
| Office brachial DBP (mm Hg) | 76.02:72.06 (74.51) | 10.2:9.25 (9.94) | 48:53 (48) | 71:65 (69) | 76:74 (75.5) | 80.75:80 (80) | 109:87 (109) |
| Office HR (BPM) | 71.84:74.06 (72.75) | 14.29:13.61 (13.89) | 40:52 (40) | 63:64 (63.75) | 69.5:72.5 (72) | 80:84 (80.5) | 113:105 (113) |
| Baseline resistance (ohms) | 22.55:26.6 (24.37) | 2.69:4.03 (4.06) | 13.59:18.34 (12.64) | 20.79:23.83 (21.11) | 21.87:27.6 (23.8) | 24.01:29.5 (27.53) | 30.3:34.47 (36.13) |
| Peak-to-peak resistance (mohms) | 26.29:29.32 (28.21) | 15.17:13.71 (15.55) | 7.01:7.31 (7.01) | 14.23:18.36 (15.85) | 21.76:25.21 (23.72) | 34.79:39.5 (37.61) | 70.82:74.06 (74.56) |
| **HTN Cohort (M:F 23:9)** | | | | | | | |
| Age (years) | 62.96:77.33 (67) | 15.44:6.91 (14.99) | 28:66 (28) | 55.75:72.5 (60) | 68:78 (71) | 75.25:82 (77.5) | 82:88 (88) |
| Office brachial SBP (mm Hg) | 133.83:138 (135) | 19.59:18.34 (19.05) | 101:114 (101) | 121.5:119.75 (121) | 132:145 (134.5) | 145.25:151 (149.5) | 186:165 (186) |
| Office brachial DBP (mm Hg) | 79.48:69.89 (76.78) | 11.41:9.75 (11.67) | 56:54 (54) | 74.25:65.25 (68.5) | 80:70 (75.5) | 84:75.25 (83.5) | 109:87 (109) |
| Office HR (BPM) | 75.23:75.56 (75.32) | 9.52:15.32 (11.23) | 60:52 (52) | 68:63.75 (68) | 75:79 (76) | 82:81.5 (81.5) | 95:105 (105) |
| Baseline resistance (ohms) | 23.5:27.08 (24.42) | 3.53:2.63 (3.72) | 13.35:20.77 (13.66) | 21.63:25.19 (21.69) | 23.23:27.51 (23.97) | 25.87:28.47 (26.51) | 33.92:32.29 (34.29) |
| Peak-to-peak resistance (mohms) | 29.11:29.34 (29.15) | 13.24:12.13 (12.94) | 7.04:9.35 (7.04) | 18.28:18.92 (18.47) | 27.96:27.08 (27.69) | 38.08:38.52 (38.16) | 72.38:68.43 (71.53) |
| **CVD Cohort (M:F 15:7)** | | | | | | | |
| Age (years) | 74.87:79.86 (76.45) | 7.92:6.34 (7.68) | 58:71 (58) | 71:75.5 (71) | 75:80 (76) | 79.5:84 (81) | 93:90 (93) |
| Office brachial SBP (mm Hg) | 129.07:130.57 (129.55) | 15.07:18.36 (15.76) | 103:108 (103) | 117:115.25 (116) | 130:125 (128.5) | 141:149.75 (145) | 154:151 (154) |
| Office brachial DBP (mm Hg) | 73:72.71 (72.91) | 9.09:10.78 (9.4) | 48:53 (48) | 70.25:66.25 (70) | 73:78 (74.5) | 79:80.5 (79) | 85:83 (85) |
| Office HR (BPM) | 65.13:73.71 (67.86) | 13.53:13.79 (13.9) | 40:60 (40) | 57.5:64.25 (59) | 63:68 (66) | 75.25:82 (76) | 87:99 (99) |
| Baseline resistance (ohms) | 21.49:32.88 (21.68) | 1.52:0.78 (1.78) | 19.42:30.89 (18.38) | 20.38:32.28 (20.5) | 20.88:33.29 (21.04) | 23.32:33.48 (23.33) | 25.07:33.71 (26.76) |
| Peak-to-peak resistance (mohms) | 18.08:31 (20.07) | 7.06:11.54 (8.83) | 7.01:7.58 (7.01) | 12.91:23.73 (13.48) | 16.25:29.94 (17.27) | 21.76:36.28 (25.82) | 41.47:81.52 (51.75) |
| **Other Cohort (M:F 13:19)** | | | | | | | |
| Age (years) | 65.92:58.83 (62.28) | 13.32:17.18 (15.81) | 46:24 (24) | 54.25:49 (49.5) | 71:59.5 (68) | 75.25:70 (75.5) | 85:83 (85) |
| Office brachial SBP (mm Hg) | 118.54:119.06 (120.22) | 9.07:17.22 (15.99) | 102:91 (91) | 112.75:111 (112) | 118:119 (119) | 125.25:128 (127) | 133:165 (165) |
| Office brachial DBP (mm Hg) | 73.38:72.89 (73.34) | 7.54:8.77 (8.14) | 56:59 (56) | 69.75:64 (67) | 76:74.5 (75.5) | 78.25:80 (80) | 83:84 (84) |
| Office HR (BPM) | 73.85:73.44 (73.63) | 19.5:13.43 (15.69) | 48:53 (48) | 61:64 (64) | 68:72.5 (72.5) | 86.5:84 (84.5) | 113:103 (113) |
| Baseline resistance (ohms) | 21.76:26.02 (24.53) | 1.96:3.87 (3.88) | 18.26:18.34 (13.33) | 21.22:23.77 (21.42) | 21.53:27.53 (24.03) | 23.89:29.47 (28.17) | 30.02:34.47 (34.47) |
| Peak-to-peak resistance (mohms) | 46.04:28.69 (35.11) | 35.62:14.23 (22.98) | 7.01:7.31 (7.01) | 13.01:17.73 (16.53) | 39.59:23.31 (25.09) | 74.5:41.63 (50.42) | 165.48:69.97 (93.54) |

Extended Data Table 4. **Summary of signal-tagged physics-informed neural network results.** Prediction accuracies for each model trained and tested with various datasets. MAE, mean absolute error; AMAE, average MAE; RMSE, root mean square error; ARMSE, average RMSE; SD, standard deviation; HTN, hypertension; CVD, cardiovascular disease; BP, blood pressure; SBP, systolic blood pressure; DBP, diastolic blood pressure; P, population-wide; PM, population-wide + metadata; S, subject-specific. All the results are shown without test-exclusive individuals.

| **Dataset** | **Test data included** | **Model type** | **Signal-tagged physics-informed neural network** | | | | | | | |
|---|---|---|---|---|---|---|---|---|---|---|
| | | | **SBP** | | | **DBP** | | | **BP waveform** | |
| | | | $r^2$ | MAE (mm Hg) | SD (mm Hg) | $r^2$ | MAE (mm Hg) | SD (mm Hg) | AMAE (mm Hg) | ARMSE (mm Hg) |
| Group 1 | All | PM | 0.575 | 7.24 | 6.98 | 0.629 | 5.45 | 5.19 | 6.29 | 6.64 |
| | All | P | 0.428 | 8.59 | 8.34 | 0.448 | 6.67 | 6.56 | 7.56 | 8.02 |
| | All | S | 0.626 | 6.62 | 6.42 | 0.680 | 4.99 | 4.78 | 5.85 | 6.23 |
| Group 2 | All | P | 0.774 | 7.26 | 8.46 | 0.807 | 3.89 | 4.63 | 5.54 | 6.14 |
| | HTN | P | 0.752 | 8.25 | 10.23 | 0.854 | 4.30 | 5.45 | 6.32 | 7.02 |
| | CVD | P | 0.859 | 6.67 | 6.96 | 0.811 | 3.30 | 3.24 | 4.87 | 5.42 |
| | Other | P | 0.588 | 6.67 | 7.24 | 0.642 | 3.86 | 4.47 | 5.21 | 5.73 |
| Group 2 | All | S | 0.860 | 6.25 | 5.75 | 0.895 | 3.30 | 2.97 | 4.94 | 5.48 |
| | HTN | S | 0.861 | 7.16 | 6.45 | 0.922 | 3.77 | 3.38 | 5.74 | 6.39 |
| | CVD | S | 0.944 | 5.92 | 4.91 | 0.913 | 3.07 | 2.46 | 4.51 | 5.00 |
| | Other | S | 0.662 | 5.55 | 5.36 | 0.836 | 2.98 | 2.76 | 4.47 | 4.93 |
| PulseDB[22] synthetic | All | P | 0.998 | 0.57 | 0.76 | 0.974 | 0.85 | 1.02 | 0.76 | 0.95 |
| | Biological variability | PM | 0.969 | 2.43 | 2.54 | 0.686 | 3.27 | 3.04 | 3.13 | 3.52 |

# Supplementary Information for

## Cuffless, calibration-free hemodynamic monitoring with physics-informed machine learning models

Henry Crandall[1,*], Tyler Schuessler[2,*], Filip Bělík[2,3,*], Albert Fabregas[4], Barry M. Stults[5], Alexandra Boyadzhiev[6], Huanan Zhang[6], Jim S. Wu[7], Aylin R. Rodan[5,8,9], Stephen P. Juraschek[10], Ramakrishna Mukkamala[11,12], Alfred K. Cheung[5], Stavros G. Drakos[5,13], Christel Hohenegger[2], Braxton Osting[2], and Benjamin Sanchez[4,14]

[1]Department of Electrical and Computer Engineering, University of Utah, Salt Lake City, UT, USA
[2]Department of Mathematics, University of Utah, Salt Lake City, UT, USA
[3]Scientific Computing and Imaging Institute, University of Utah, Salt Lake City, UT, USA
[4]Department of Electrical and Computer Engineering, University of Illinois Chicago, Chicago, IL, USA
[5]Department of Internal Medicine, Spencer Fox Eccles School of Medicine, University of Utah Health, Salt Lake City, UT, USA
[6]Department of Chemical Engineering, University of Utah, Salt Lake City, UT, USA
[7]Department of Radiology, Brigham and Women's Hospital, Harvard Medical School, Boston, MA, USA
[8]Molecular Medicine Program, University of Utah Health, Salt Lake City, UT, USA
[9]Medical Service, Veterans Affairs Salt Lake City Health Care System, Salt Lake City, UT, USA
[10]Department of Medicine, Beth Israel Deaconess Medical Center, Harvard Medical School, Boston, MA, USA
[11]Department of Bioengineering, University of Pittsburgh, PA, USA
[12]Department of Anesthesiology & Perioperative Medicine, University of Pittsburgh, PA, USA
[13]Division of Cardiovascular Medicine and Nora Eccles Harrison CVRTI, University of Utah Health, Salt Lake City, UT, USA
[14]Richard and Loan Hill Department of Biomedical Engineering, University of Illinois Chicago, Chicago, IL, USA
[*]These authors contributed equally to this work
Corresponding author: Benjamin Sanchez, 851 S. Morgan St., Office 1104 SEO, Chicago, IL 60607. Email: bst@uic.edu. Phone: 312-996-5847.

# Contents

# Supplementary Discussions

## 1 Supplementary Discussion 1. Clinical significance of hemodynamic monitoring

Hemodynamic monitoring provides critical insights into the pathophysiology and management of several cardiovascular and renal diseases; these are briefly discussed next.

### 1.1 Hypertension

Elevated blood pressure (BP) and hypertension (HTN) are associated with a high burden of death and disability worldwide due to their contribution to diseases like ischemic heart disease, heart failure (HF), and stroke.[1,2] Globally, HTN is one of the largest public health epidemics, reducing quality of life, burdening healthcare systems, and endangering over one billion people.[3] Despite proven procedures for diagnosing, treating, and controlling the disease, awareness (<65%) and control (<25%) remain low.[4] Even with advances in diagnosis and treatments, the prevalence of HTN and the disability-adjusted life years due to HTN increased considerably over the last 30 years.[2] Substantial evidence links elevated BP to severe cardiovascular diseases (CVDs), including stroke, myocardial infarction, heart failure, peripheral artery disease, renal disease, atrial fibrillation, and dementia.[5–8]

Measuring BP is critical for the early diagnosis and on-going treatment of HTN. Accordingly, BP measurement is a routine part of most clinic visits,[2] especially for those under treatment for HTN, CVD and kidney disease. The proper preparation of the patient and the proper procedure to measure BP in the outpatient clinic remain the gold standard for BP management and are recommended by most practice guidelines. Large randomized outcome trials on BP levels also target BP obtained in such standardized manner. However, because of the excellent correlation between out-of-office BP and clinical outcomes, the empowerment of the patient in managing their own BP that also leads to better adherence in their treatment, as well as other benefits, out-of-office BP monitoring is commonly recommended as complements to standard BP measurements.[9–11]

Two modalities exist for out-of-office BP measurements: ambulatory BP monitoring (ABPM) and home BP monitoring (HBPM). ABPM involves wearing an automated oscillometric device for a defined period, typically 24 hours. The device is programmed to measure at fixed intervals, usually every 15-30 minutes. Averaged values are then calculated for waking, resting, and various sub-intervals to better reflect BP during daily life. HBPM refers to regular measurements taken at home, preferably in the morning and evening. Both methods provide data that are complementary to office BP.[2] Similar to office-based BP measurements, HBPM is intermittent and depends on the patient or family to perform the measurements correctly. As proper technique may be unreliable in a significant portion of the population, ongoing training and retraining are necessary to ensure accuracy. ABPM provides numerous measurements over a 24-hour period, offering detailed temporal data. However, it is associated with patient discomfort, inconvenience, poor reproducibility, high cost, and is restricted to the duration of the monitoring period.[12–14] Moreover, ABPM requires one to be relatively motionless at the time of assessment. While this can still be used to assess BP in a broader range of body positions and settings, it is unable to capture dynamic states. In addition, its discomfort precludes frequent measurement (typical is 15 to 20 minute intervals). As part of their position statement, the European Society of Hypertension identified the need to develop new technologies to make out-of-office BP measurements

more user-friendly.[12]

Out-of-office BP measurements are also essential for measuring BP in patients with white coat HTN and masked HTN. Although estimates vary depending on the population studied, ≈20%-30% of patients seen in clinic have these conditions. In patients with white coat HTN, office BP is higher in the office than at home. Conversely, patients with masked HTN have lower BP in the office than at home. Importantly, both conditions are associated with increased risks of CVD and mortality, indicating the importance of understanding which patients may be affected by these entities and what their BP is outside the office. However, understanding of how to treat these populations has been hampered by the difficulty of consistently obtaining out-of-office BP measurements.[8,15–17]

There is increasing recognition of the importance of detecting extreme fluctuations in BP. This is particularly relevant for older adults with treated HTN, who often have more variable BP and are at greater risk for low BP events. Indeed, syncope and hypotensive events were more common among participants of the Systolic Blood Pressure Intervention Trial (SPRINT), who were assigned a lower treatment goal. While more intensive BP treatment is consistently better for preventing CVD, intermittent low BP is a limiting factor and being able to avoid complications of low BP is critical for optimizing intensive therapy.

## 1.2 Heart failure

Heart failure (HF) is a type of CVD in which there is functional limitation due to impairment of ventricular filling or ejection of blood, or a combination of both. It can cause a range of symptoms, such as shortness of breath, fatigue, swelling in the legs and ankles, and difficulty performing everyday activities. Once developed, HF results in significant morbidity and mortality, with a 1-year mortality rate of 7.2% and a 1-year hospitalization rate of 31.9% in patients with chronic HF, and in patients hospitalized for acute HF, these rates increase to 17.4% and 43.9%.[18] Prevention of HF continues to be a primary target for health-care systems.[19,20] One of the most crucial aspects of preventing HF is BP management.[15,21] HTN promotes pathological changes at the cellular level in the heart muscle (cardiac remodeling) due to increased pressure overload that the heart must pump against. Over time, this remodeling drives the progression to impaired systolic and/or diastolic function. Moreover, HTN can indirectly increase the risk for HF, by increasing the risk for myocardial infarction. Therefore, monitoring BP in hypertensive subjects and those at-risk of developing HF is essential for early diagnosis, appropriate management, and improved clinical outcomes. BP monitoring helps detect changes over time, enabling clinicians to adjust treatment as needed and optimize BP control. Early detection of BP changes in HF patients can help prevent both acute and long-term complications through timely medication adjustments and clinical intervention. When HF symptoms become refractory to optimal medical therapy – a condition known as advanced HF – devices that provide long-term mechanical circulatory support, such as left ventricular assist devices (LVADs), are often used as a treatment option.[22] In essence, LVADs are mechanical pumps comprised of an inflow cannula (inserted in the left ventricle) and an outflow graft (connected to the aorta). By drawing blood from the heart and pumping it to the rest of the body, LVADs unload the failing heart and undertake part of its systolic function.[22] BP monitoring in LVAD patients is of paramount importance because deviations from normal BP ranges have been consistently linked to adverse LVAD-related events (such as strokes, organ hypoperfusion and other).[23] However, contemporary LVADs provide continuous flow circulatory support and this significantly blunts the natural variation between systolic and diastolic pressures (called pulse pressure) and the corresponding oscillometric changes in arterial walls. Of note, both manual auscultation and traditional oscillometric BP monitors rely on these pressure fluctuations to measure BP. As a result, standard BP measurement methods are not reliable in LVAD patients.[22] Doppler ultrasound, on the other hand, has been shown to effectively measure BP even if pulse pressure is minimal and, thus, is considered the most reliable method for this patient population.[24]

## 1.3 Kidney disease

Kidney disease is an important contributor to HTN, and HTN in turn contributes to the progression of kidney disease. Among other mechanisms such as activation of the renin-angiotensin system and the sympathetic nervous system, the kidney is often plays an essential pathogenic role in HTN by retention of salt. Conversely, moderate and severe HTN, with systolic BP (SBP) >140 mm Hg may accelerate kidney function decline, presumably from intrarenal vascular damage by the high BP. Once chronic kidney disease (CKD) develops, the kidney's capability of excreting salt is further impaired, thus establishing a vicious cycle.[25] HTN is highly prevalent in CKD.[26] As a result, BP monitoring and management are essential components of care for patients with CKD.[27]

A transient decrease in BP is common in healthy individuals, a phenomenon known as the "nocturnal dip". This dip is often diminished or absent in patients in CKD and has been postulated to contribute to their prevalent cardiac disease.[28–30] An additional unique feature in patients with end-stage kidney disease (ESKD) requiring maintenance intermittent hemodialysis is the often drastic fluctuation in BP, characterized by a gradual increase in BP in the inter-dialytic interval as a result of gradual fluid retention and the sudden decrease in BP during the 4-hour intra-dialytic period. This rather sudden decrease in BP not infrequently results in shock and hypovolemic signs and symptoms, with often diminished perfusion to the coronary arteries and myocardial stunning.[31,32]

It can be readily appreciated that BP management in CKD and ESKD patients can be complicated and can benefit from a convenient mode of continuous hemodynamic monitoring.

## 1.4 Atherosclerosis

Atherosclerosis, where plaque buildup in the arteries narrow the blood vessels and slows down blood velocity, often remains undetected until it reaches advanced stages.[33] Despite being a preventable disease, monitoring blood velocity in patients with atherosclerosis presents unique challenges. First, the need for frequent testing to monitor disease progression and treatment efficacy can be burdensome for patients and healthcare systems. Non-invasive imaging techniques, such as ultrasound Doppler flowmetry and computed tomography (CT) angiography,[34,35] while useful for visualizing atherosclerotic plaques, have important limitations including cost, expertise required, radiation exposure, and difficulty detecting smaller or unstable plaques. Magnetic resonance angiography (MRA) or phase-contrast magnetic resonance imaging (MRI) can provide detailed images of blood vessels and can visualize the flow of blood within arteries and veins without the need for contrast agents.[36] Unlike CT angiography, MRI has the advantage of being non-invasive and free of ionizing radiation; however, MRA and MRI equipment is more expensive, time-consuming, and less accessible than other modalities such as ultrasound for routine blood velocity monitoring. Finally, traditional hemodynamic monitoring with ABPMs and HBPMs does not fully capture the complexity of arterial stiffness and endothelial dysfunction in patients with atherosclerosis.[37]

# 2 Supplementary Discussion 2. Clinical challenges of blood pressure monitoring

There are several methods to measure BP. Clinically accurate methods available to measure BP include continuous invasive measurements using an arterial catheter (also known as A-line),[38] intermittent non-invasive measurements using oscillometry,[39] continuous non-invasive measurements such as the vascular unloading technique (also known as finger-cuff technology),[40,41] Doppler opening pressure ultrasound, and arterial tonometry.

## 2.1 Arterial line

The gold standard clinical procedure to measure BP is arterial catheterization, where a thin, hollow tube is placed typically into the radial artery.[42] The success of catheter placement is largely dependent on palpation of the radial artery's point of maximal pulsation with the fingertips followed by insertion of a separate needle or a needle-guided catheter immediately beneath the fingertips toward the pulsation and into the artery. Complications associated with arterial cannulation include temporary vascular occlusion, thrombosis, ischemia, hematoma formation, and local and catheter-related infection and sepsis.[43] A-line, although the gold standard to monitor continuous BP, is not an option for preventive and routine outpatient use.

## 2.2 Oscillometry

Intermittent measurements (e.g., taken at 30 min intervals) using oscillometry is measured noninvasively by occluding a major artery (typically the brachial artery in the arm) with an external pneumatic cuff.[44] Intermittent oscillometry still remains the cornerstone of non-invasive clinical assessment. When the pressure in the cuff is higher than the BP inside the artery, the artery collapses. As the pressure in the external cuff is slowly decreased by venting through a bleed valve, cuff pressure drops below systolic BP, and blood will begin to spurt through the artery. The pressure in the cuff when blood first passes through the cuffed region of the artery is an estimate of systolic pressure. The pressure in the cuff when blood first starts to flow continuously is an estimate of diastolic pressure.

## 2.3 Finger-cuff

Finger-cuff technology is an indirect measure of intra-arterial BP, employing a vascular unloading technique at the fingers using a double finger cuff.[45] These cuffs consists of an inflatable bladder and a photoplethysmograph (PPG). The light is absorbed by the blood and the pulsation of arterial diameter during a heartbeat causes a pulsation in the light detector signal. The first step in this method is determining the proper unloaded diameter of the finger arteries. This is the point at which finger cuff pressure and intra-arterial pressure are equal and at which the transmural pressure across the finger arterial walls is zero. Next, the arteries are clamped, to keep them at this unloaded diameter, by varying the pressure of the inflatable bladder using the fast cuff pressure control system. The control system incorporates a servo-controller, which defines a set point and a measured value that is compared with this set point. In this case, the set point is the signal of the plethysmograph (unloaded diameter of the arteries) that must be clamped, while the measured value comes from the light detector. The amplified difference between the set point and measured value, "*the error signal*", is used to control a fast pneumatic valve. This valve modulates the air pressure generated by the air compressor. This causes changes in the finger cuff pressure, in parallel with intra-arterial BP in the finger, to dynamically unload the arterial walls in the finger and thus estimate the BP signal. This approach needs to be calibrated using a standard oscillometry. During the calibration process, the device locates the pulse at the finger and performs a partial occlusion. Then it switches from one finger to the next during the course of the recording to relieve the pressure from the occluded finger. Experimental finger-cuff technology offers a non-invasive alternative to A-line to monitor continuous BP, but the accuracy is largely impacted by cuff insertion and anatomic position.

## 2.4 Doppler opening pressure

Doppler opening pressure offers an alternative method to estimate BP, particularly in patients where BP readings with sphygmomanometer and stethoscope are difficult or unreliable. With the patient positioned comfortably, a BP cuff is placed around the arm, just as with traditional BP measurement. Then, a Doppler device with a probe is used to detect blood flow in the arteries. Next, the cuff is inflated to a pressure higher than the expected systolic pressure, occluding the artery and stopping blood flow. This causes the Doppler signal to disappear, indicating that no blood is flowing past the cuff. The cuff pressure is then slowly released. As the pressure drops, blood flow resumes when the systolic pressure exceeds the cuff pressure. The Doppler device will detect the return of blood flow. The pressure at which the Doppler signal reappears corresponds to the systolic BP or Doppler opening pressure.

## 2.5 Arterial tonometry

Arterial tonometry is performed by placing a strain gauge pressure sensor over the radial artery and applying mild pressure to partially flatten the artery.[46,47] Ideally, this results in a transmural pressure of zero on the radial artery and the maximum pulse pressure can be acquired by the sensor. The mean arterial pressure is determined from the maximum pulse pressure. Systolic and diastolic pressure are scaled through calibration. This technique is rarely used in practice since it is very sensitive to motion artifacts and is unable to measure continuously.

# 3 Supplementary Discussion 3. Cuffless blood pressure measurement with wearables

This section describes diverse technologies used to measure BP, from ABPMs to emerging optical, mechanical, and ultrasound-based methods.

## 3.1 Ambulatory blood pressure monitors

Out-of-office BP measurements are essential for measuring BP in patients with white coat HTN and masked HTN. Although estimates vary depending on the population studied, 20-30% of patients seen in clinic have these conditions. In patients with white coat HTN, office BP is higher in the office than at home. Conversely, patients with masked HTN have lower BP in the office than at home. Importantly, both conditions are associated with increased risks of CVD and mortality, indicating the importance of understanding which patients may be affected by these entities and what their BP is outside the office.[8,15–17] However, understanding of how to treat these populations has been hampered by the difficulty of consistently obtaining out-of-office BP measurements.

ABPMs remain the cornerstone of outpatient BP monitoring. However, their use is restricted to intermittent measurements only (e.g., taken at 15-30 min intervals during the day, 60 mins at night) during few days since the cuff is uncomfortable to wear. Furthermore, the inflation of the cuff can trigger higher BP readings, cause sleep disturbances, and increase stress during night measurements. HBPM have shown to be a superior predictor of cardiovascular events compared to clinic BP measurements.[48] Additionally, HBPM is capable of identifying phenotypes that may be obscured in clinical settings, such as white coat HTN and masked HTN.[49] Consequently, the European Society of Hypertension now recommends out-of-office BP measurements to confirm a diagnosis of HTN.[8] However, HBPMs remain challenged by limited data, patient discomfort, poor reproducibility, and patient stigmatization.[12–14] The cost of testing, patient compliance, technology accessibility, technology inconvenience, and accuracy of devices have been identified as barriers for widespread adoption of HBPM and ABPM.[50]

Healthcare providers are, thus, in need of new BP monitoring solutions to expand the armamentarium to combat BP and CVD more broadly. Despite the value of existing clinical tools, novel wearables technologies have the potential to offer an unprecedented opportunity to enable convenient outpatient hemodynamic monitoring and reduce hospitalizations associated to CVD. Such wearables could have widespread applications in preventive care, enabling early detection of individuals at risk for CVD, as well as in patient management by facilitating timely interventions and monitoring treatment efficacy.

Cuffless BP devices attempt to solve the problems mentioned above with existing BP measurement techniques.[51,52] The topic has received significant interest, yet many cuffless BP monitors have not been formally validated using proper standards.[53–56] Consequently, as of 2021, the European Society of Hypertension opposed the use of existing cuffless BP devices for diagnostic or therapeutic decisions.[49]

Cuffless BP measurement principles are diverse, but generally rely on a combination of wearable sensors, arterial mechanics models, and signal processing techniques.[52,55,57] A variety of sensing methods have been developed using mechanoelectrical, optical, and acoustic physical principles.[57,58] Data acquired from one or a combination of sensors are used to extract signal features and provide an estimate for BP–through empirical pulse wave analysis (PWA), pulse wave velocity (PWV), or arterial wall motion–oftentimes combined with data-driven machine learning (ML) models.[58]

## 3.2 Optical

Optical sensors rely on one or more light emitting source and corresponding photo-detectors to measure changes in peripheral blood volume. They are widely deployed for monitoring heart rate (HR) and blood oxygen saturation and have been explored for measuring BP.[59] Optical sensors, like PPG, have been incorporated into conformable patches,[60] flexible circuits,[61,62] wrist bands,[63,64] and finger probes.[65] The signal acquired from the optical sensors are typically paired with PWA, pulse transit time (PTT) or an ML model to estimate BP. The sensors are cheap, widely adopted, and have high signal quality, but suffer from poor penetration depth and interference from ambient light sources.[59,66] Importantly, PPG data lack a solid theoretical background that causally relates PPG to BP.[51]

## 3.3 Mechanical

Mechanical sensors transduce fluctuations in applied pressure into an electric signal. Mechanical sensing methods include piezoelectric, piezoresistive, piezocapacitive, and triboelectric.[57,66] Piezoelectric sensors are self-powered, feature high sensitivity, and quick response times, but they suffer from poor sensitivity to low frequency stimuli.[57,66,67] Piezoresistive sensors have high pressure sensitivities, but feature a trade-off between sensitivity and stretchability, and require an external power source.[57,66] Piezocapacitive sensors feature high sensitivity, low power consumption, and strong stability to environmental factors, but require an external power source.[57,66,68] Triboelectric sensors are self-powered and feature better sensitivity than piezoelectric sensors, but suffer from envrionmental interference and inability to measure low frequency pressures.[57,69,70] Each method is being explored for incorporation into flexible substrates to continuously measure pulsatile blood waveforms.[57,67,68] However, like all noninvasive sensing methodologies, they cannot directly measure arterial pressure and, thus, still rely on ML, PTT, or mechanics models to relate the signal to an arterial pressure waveform.

## 3.4 Ultrasound

Ultrasound imaging is a widely used, non-invasive medical imaging technique that utilizes high-frequency sound waves to create real-time images of internal body structures. Advancements in Doppler ultrasound have enabled real-time assessment of blood flow, aiding in the diagnosis of vascular conditions.[51] Recently, miniature wearable ultrasound sensors have been explored for pulse wave monitoring.[57,71,72]

State-of-the-art ultrasound sensor research such as Dr. Xu's lab at UC San Diego has largely focused on developing flexible, stretchable, conformable materials that can be worn on the skin. The approach implemented by the group at UC San Diego to estimate SBP and diastolic BP (DBP) is based on measuring changes in arterial diameter within the cardiac cycle. The approach relies on an empirical relationship between the arterial cross-section and BP in the form of exponential function fit,[72,73] namely

$$\text{SBP} = P_\text{d} e^{\alpha\left(\frac{A_{\max}}{A_\text{d}} - 1\right)} \quad \text{and} \quad \text{DBP} = P_\text{d} e^{\alpha\left(\frac{A_{\min}}{A_\text{d}} - 1\right)},$$

where $A_{\max}$ and $A_{\min}$ represent the artery cross section at systole and diastole respectively measured with ultrasound, $P_\text{d}$ is the reference pressure at diastole, $A_\text{d}$ is the corresponding reference artery cross section at diastole measured with ultrasound, and $\alpha$ is the vessel stiffness coefficient calculated as $\alpha = A_\text{d} \ln(P_\text{s}/P_\text{d})/(A_\text{s} - A_\text{d})$, with $A_\text{s}$ the systolic arterial cross-section measured with ultrasound and $P_s$ the reference SBP. The artery cross-section is assumed rotationally symmetric and defined as $A = \pi d^2/4,$ where $d$ is the artery diameter measured with ultrasound.

While this approach shows accurate prediction of SBP and DBP values with mean $\pm$ standard deviation (SD) error values of 3.12$\pm$2.55 and 2.93$\pm$1.92 mm Hg (Supplementary Fig.34), respectively, a linear (Supplementary Fig.33) or exponential fit on the arterial wall motion fails to capture the temporal dynamics of the BP waveform with limits of agreement of 29.46 and -21.51 mm Hg, and mean and SD error values of 8.29 and 10.54 mm Hg (Supplementary Fig.34). Additionally, the exponential model is derived from experiments on canine and ex vivo human arteries[74–76] but, to our knowledge, it has never been evaluated with in vivo human arteries. While ex vivo studies may provide insights into arteries' mechanical properties, they might be limited in their generalizability to in vivo human conditions and lead to divergent BP predictions.

Alternatively, recently proposed resonance sonomanometry measures arterial dimensions and wall resonances elicited by acoustic stimulation, enabling BP estimation through a fully determined physical model.[77] Through the inviscid assumptions of the Moens-Korteweg and Bramwell-Hill equations the artery's Young's modulus can be directly related to changes in arterial pressure. Additionally, under thin-wall and long-cylinder assumptions, the resonant frequency can be related to the arterial pressure, Young's modulus, and arterial radius.[78] Through an iterative approach, upon measuring resonant frequencies and arterial radius displacements, dynamic BP and Young's modulus can be computed as functions of time.[77]

Conventional ultrasound might face additional limitations, primarily due to the challenge of impedance matching between the transducer and the skin. For long-term monitoring, skin conditions will change due to factors such as hydration, sweat, and movement, making it difficult to maintain consistent impedance matching over time. Traditional ultrasound gels used to improve conductivity might not be suitable for long-term wear, as they could dry out quickly and require frequent reapplication.

# 4 Supplementary Discussion 4. Multiscale coupled fluid and blood conductivity model

Here we describe the full forward model from BP to electrical conductivity. This model can be described in two parts, the fluids modeling of the blood within the arm's arterial tree, and then the blood conductivity modeling that depends on the resulting shear stresses. With this model, we establish how BP, along with several other physiological parameters, impact the electrical properties of the blood in the radial artery and of the entire arm.

## 4.1 Navier-Stokes linearization

Begin by considering irrotational fluid flow within a tube. This can be modeled by the Navier-Stokes (N.S.) equations in cylindrical coordinates

$$\rho\left(\frac{\partial u_z}{\partial t}+u_z\frac{\partial u_z}{\partial z}+u_r\frac{\partial u_z}{\partial r}\right)+\frac{\partial p}{\partial z}=\mu\left(\frac{\partial^2 u_z}{\partial z^2}+\frac{\partial^2 u_z}{\partial r^2}+\frac{1}{r}\frac{\partial u_z}{\partial r}\right), \tag{1a}$$

$$\rho\left(\frac{\partial u_r}{\partial t}+u_z\frac{\partial u_r}{\partial z}+u_r\frac{\partial u_r}{\partial r}\right)+\frac{\partial p}{\partial r}=\mu\left(\frac{\partial^2 u_r}{\partial z^2}+\frac{\partial^2 u_r}{\partial r^2}+\frac{1}{r}\frac{\partial u_r}{\partial r}-\frac{u_r}{r^2}\right), \tag{1b}$$

$$\frac{\partial u_z}{\partial z}+\frac{1}{r}\frac{\partial (r u_r)}{\partial r}=0, \tag{1c}$$

where $u_z$ denotes axial fluid velocity, $u_r$ denotes radial fluid velocity, $p$ denotes pressure, $r$ is the radial coordinate, $z$ is the axial coordinate, $t$ is time, $\rho$ is the fluid density, and $\mu$ is the fluid viscosity. In addition, let $a$ denote the resting radius of the tube. Equations (1a) and (1b) represent conservation of linear momentum and Equation (1c) represents conservation of mass.

We assume the presence of an $\omega$-time-periodic pressure wave that travels down the artery with wavelength $\lambda$ that is much longer than the radius of the artery; so we introduce the small positive quantity $\delta=\frac{a}{\lambda}\ll 1$.[79] We perform a nondimensionalization of the form $\overline{r}=\frac{r}{a}$, $\overline{z}=\frac{z}{\lambda}=\delta\frac{z}{a}$, $\overline{t}=\omega t$, $\overline{u_z}=\frac{u_z}{U}$, $\overline{u_r}=\frac{u_r}{V}$, and $\overline{p}=\frac{p}{P}$ where $U$, $V$, and $P$ are characteristic velocities and pressures yet to be determined. Nondimensionalizing Equation (1c), we must choose $V=\delta U$ in order to allow for nontrivial axial and radial fluid flow. Hence, we obtain the nondimensional equation

$$\frac{\partial \overline{u_z}}{\partial \overline{z}}+\frac{1}{\overline{r}}\frac{\partial(\overline{r}\,\overline{u_r})}{\partial \overline{r}}=0.$$

Nondimensionalizing the linear momentum equations, and introducing the nondimensional Wormersly number $\alpha=\sqrt{\frac{\rho\omega}{\mu}}a$ and the Reynold's number $\text{Re}=\frac{\rho a U}{\mu}$ yields

$$\alpha^2\frac{\partial \overline{u_z}}{\partial \overline{t}}+\delta\text{Re}\,\overline{u_z}\frac{\partial \overline{u_z}}{\partial \overline{z}}+\delta\text{Re}\,\overline{u_r}\frac{\partial \overline{u_z}}{\partial \overline{r}}+\delta\frac{Pa}{U\mu}\frac{\partial \overline{p}}{\partial \overline{z}}=\delta^2\frac{\partial^2 \overline{u_z}}{\partial \overline{z}^2}+\frac{\partial^2 \overline{u_z}}{\partial \overline{r}^2}+\frac{1}{\overline{r}}\frac{\partial \overline{u_z}}{\partial \overline{r}},$$

$$\alpha^2\delta^2\frac{\partial \overline{u_r}}{\partial \overline{t}}+\delta^3\text{Re}\,\overline{u_z}\frac{\partial \overline{u_r}}{\partial \overline{z}}+\delta^3\text{Re}\,\overline{u_r}\frac{\partial \overline{u_r}}{\partial \overline{r}}+\delta\frac{Pa}{U\mu}\frac{\partial \overline{p}}{\partial \overline{r}}=\delta^4\frac{\partial^2 \overline{u_r}}{\partial \overline{z}^2}+\delta^2\frac{\partial^2 \overline{u_r}}{\partial \overline{r}^2}+\delta^2\frac{1}{\overline{r}}\frac{\partial \overline{u_r}}{\partial \overline{r}}-\delta^2\frac{\overline{u_r}}{\overline{r}^2}.$$

Under the additional assumption that $\delta\ll\frac{\alpha^2}{\text{Re}}$, we retain all terms of lower order to obtain the nondimensional equations

$$\alpha^2\frac{\partial \overline{u_z}}{\partial \overline{t}}+\frac{Pa\delta}{U\mu}\frac{\partial \overline{p}}{\partial \overline{z}}=\frac{\partial^2 \overline{u_z}}{\partial \overline{r}^2}+\frac{1}{\overline{r}}\frac{\partial \overline{u_z}}{\partial \overline{r}}, \tag{2a}$$

$$\alpha^2\delta^2\frac{\partial \overline{u_r}}{\partial \overline{t}}+\frac{Pa\delta}{U\mu}\frac{\partial \overline{p}}{\partial \overline{r}}=\delta^2\frac{\partial^2 \overline{u_r}}{\partial \overline{r}^2}+\delta^2\frac{1}{\overline{r}}\frac{\partial \overline{u_r}}{\partial \overline{r}}-\delta^2\frac{\overline{u_r}}{\overline{r}^2}. \tag{2b}$$

Equations (2a) and (2b) give us two choices for $P$. If we choose $P = \frac{\delta U \mu}{a}$, then the radial derivative of pressure will be on the order of the other terms in Equation (2b), but then in Equation (2a), the axial pressure gradient would have to very large, on the order of $\frac{1}{\delta^2}$ to balance the equation. Instead, we choose $P = \frac{U\mu}{\delta a}$. Eliminating terms of higher order, the dimensionless linearized N.S. equations[79] are

$$\alpha^2 \frac{\partial \overline{u_z}}{\partial \overline{t}} + \frac{\partial \overline{p}}{\partial \overline{z}} = \frac{\partial^2 \overline{u_z}}{\partial \overline{r}^2} + \frac{1}{\overline{r}} \frac{\partial \overline{u_z}}{\partial \overline{r}}, \tag{3a}$$

$$\frac{\partial \overline{p}}{\partial \overline{r}} = 0, \tag{3b}$$

$$\frac{\partial \overline{u_z}}{\partial \overline{z}} + \frac{1}{\overline{r}} \frac{\partial (\overline{r}\overline{u_r})}{\partial \overline{r}} = 0. \tag{3c}$$

It is from these equations that we derive the analytical solutions to fluid flow and wall displacement.

## 4.2 Derivation of fluid solutions

We assume time-periodic wave solutions of the form[79]

$$\overline{u_z}(\overline{r}, \overline{z}, \overline{t}) = \sum_{n=0}^{N} \overline{u_{z,n}}(\overline{r}) e^{i(n\overline{t} - \overline{K}_n \overline{z})},$$

$$\overline{u_r}(\overline{r}, \overline{z}, \overline{t}) = \sum_{n=0}^{N} \overline{u_{r,n}}(\overline{r}) e^{i(n\overline{t} - \overline{K}_n \overline{z})},$$

$$\overline{p}(\overline{z}, \overline{t}) = \sum_{n=0}^{N} \overline{B}_n e^{i(n\overline{t} - \overline{K}_n \overline{z})}.$$

Note that the $\overline{B}_n$ coefficients do not vary with $\overline{r}$ in order to satisfy Equation (3b). Fixing an axial location $z$, the $\overline{B}_n$ coefficients can be derived from the Fourier decomposition of a BP waveform.

Inserting these forms into Equations (3a) and (3c), and matching coefficients with like exponentials, we get the equations

$$-i\overline{K}_n \overline{B}_n = \overline{u_{z,n}}''(\overline{r}) + \frac{\overline{u_{z,n}}'(\overline{r})}{\overline{r}} - \alpha^2 in \overline{u_{z,n}}(\overline{r}),$$

$$0 = -i\overline{K}_n \overline{u_{z,n}}(\overline{r}) + \overline{u_{r,n}}'(\overline{r}) + \frac{\overline{u_{r,n}}(\overline{r})}{\overline{r}}.$$

In the case that $n = 0$, we achieve general solutions of the form

$$\overline{u_{z,0}}(\overline{r}) = \frac{-i\overline{K}_0 \overline{B}_0}{4} (\overline{r}^2 - \overline{A}_0),$$

$$\overline{u_{r,0}}(\overline{r}) = \frac{\overline{K}_0^2 \overline{B}_0}{16} \overline{r} (\overline{r}^2 - 2\overline{A}_0),$$

where $\overline{A}_0$ is an undetermined, nondimensional constant.

Now, suppose $n > 0$. Introduce the nondimensional parameter $\Lambda_n = i^{3/2}\sqrt{n}\alpha$, and the nondimensional variable $\zeta_n = \Lambda_n \overline{r}$, perform a change of variables, and the above equations become

$$\frac{-i\overline{K}_n \overline{B}_n \zeta_n^2}{\Lambda_n^2} = \zeta_n^2 \frac{\partial^2 \overline{u_{z,n}}}{\partial \zeta_n^2} + \zeta_n \frac{\partial \overline{u_{z,n}}}{\partial \zeta_n} + \zeta_n^2 \overline{u_{z,n}}, \tag{4a}$$

$$0 = -i\overline{K}_n \overline{u_{z,n}}(\overline{r}) + \overline{u_{r,n}}'(\overline{r}) + \frac{\overline{u_{r,n}}(\overline{r})}{\overline{r}}. \tag{4b}$$

Equation (4a) is of the form of Bessel equations of the zeroth order. It has a general solution (requiring that $\overline{u_{z,n}}(0)$ exists)

$$\overline{u_{z,n}}(\overline{r}) = \overline{A}_n J_0(\zeta_n) + \frac{\overline{K}_n \overline{B}_n}{n\alpha^2}. \tag{5}$$

We then insert Equation (5) into Equation (4b) and solve for $\overline{u_{r,n}}(\overline{r})$,

$$\overline{u_{r,n}}(\overline{r}) = \frac{\overline{A}_n i\overline{K}_n}{\Lambda_n} J_1(\zeta_n) + \frac{i\overline{K}_n^2 \overline{B}_n \overline{r}}{2n\alpha^2}.$$

Finally, we change our solutions back to dimensional form, making note of the conversions $A_i = U\overline{A}_i$, $B_i = P\overline{B}_i$, and $K_n = \frac{1}{\lambda}\overline{K}_n$,

$$u_z(r,z,t) = \frac{-iK_0B_0}{4\mu}(r^2 - A_0)e^{-iK_0z} + \sum_{n=1}^{N}\left(A_nJ_0(\zeta_n) + \frac{K_nB_n}{\rho(n\omega)}\right)e^{i(n\omega t - K_nz)},$$

$$u_r(r,z,t) = \frac{K_0^2B_0r}{16\mu}(r^2 - 2A_0)e^{-iK_0z} + \sum_{n=1}^{N}\left(\frac{aA_niK_n}{\Lambda_n}J_1(\zeta_n) + \frac{iK_n^2B_nr}{2\rho(n\omega)}\right)e^{i(n\omega t - K_nz)},$$

$$p(r,z,t) = \sum_{n=0}^{N} B_n e^{i(n\omega t - K_nz)}.$$

Defining $\tau_w$ to be the shear stress evaluated at the wall, we can show through our nondimensionalization that the only dominant contribution is given by

$$\tau_w = -\tau_{rz} = -\mu\left(\frac{\partial u_z}{\partial r} + \frac{\partial u_r}{\partial z}\right)_{r=a} = -\mu\left(\frac{\partial u_z}{\partial r}\right)_{r=a},$$

where the last inequality follows from the same nondimensionalization and linearization.

## 4.3 Matching with the arterial wall

We denote $\xi(z,t)$ and $\eta(z,t)$ to be the arterial wall displacements in the axial and radial directions respectively. We set up the stress-strain relationships for an elastic artery and make the simplifying assumption that $a/h \gg 1$ and $a/h \gg \sigma$, where $h$ is the thickness of the wall and $\sigma$ is Poisson's ratio. Eliminating small terms from the stress-strain equations, we obtain the equations of motion of the arterial wall,

$$\frac{\partial^2\xi}{\partial t^2} = \frac{E_\sigma}{\rho_w}\left(\frac{\partial^2\xi}{\partial z^2} + \frac{\sigma}{a}\frac{\partial\eta}{\partial z}\right) - \frac{\tau_w}{\rho_w h}, \tag{6a}$$

$$\frac{\partial^2\eta}{\partial t^2} = \frac{p_w}{\rho_w h} - \frac{E_\sigma}{\rho_w a}\left(\frac{\eta}{a} + \sigma\frac{\partial\xi}{\partial z}\right), \tag{6b}$$

where $\rho_w$ is the density of the arterial wall, $E_\sigma = E/(1-\sigma^2)$ where $E$ is Young's Modulus of the arterial wall, and $p_w$ is the pressure at the arterial wall.

We then similarly assume time-periodic wave representations of our displacements by

$$\xi(z,t) = \sum_{n=0}^{N} C_n e^{i(n\omega t - K_nz)}, \quad \eta(z,t) = \sum_{n=0}^{N} D_n e^{i(n\omega t - K_nz)}.$$

Then, we make a linearized wall-matching assumption by matching the fluid velocities at $r = a$ to the rates of changes of the wall displacements,

$$\frac{\partial\xi(z,t)}{\partial t} = u_z(a,z,t), \quad \frac{\partial\eta(z,t)}{\partial t} = u_r(a,z,t).$$

Taking into account Equations (6a) and (6b) by solving for $\tau_w$ and $p_w$, matching the fluid velocities, and matching coefficients with equal exponentials, we obtain the following system of equations for $n = 0$,

$$0 = \frac{E_\sigma}{\rho_w}\left(-K_0^2 C_0 - \frac{i\sigma}{a}K_0 D_0\right) - \frac{iK_0 B_0 a}{2\rho_w h}, \tag{7a}$$

$$0 = \frac{B_0}{\rho_w h} - \frac{E_\sigma}{\rho_w a}\left(\frac{D_0}{a} - i\sigma K_0 C_0\right), \tag{7b}$$

$$0 = \frac{-iK_0 B_0}{4\mu}(a^2 - A_0), \tag{7c}$$

$$0 = \frac{-K_0^2 B_0}{16\mu}a(a^2 - 2A_0). \quad . \tag{7d}$$

Since we cannot choose the constant $A_0$ to satisfy Equations (7c) and (7d) simultaneously, we must have $K_0 = 0$. However, this implies, $u_0 \equiv v_0 \equiv 0$, implying zero net flow per cycle, as the remaining solutions are purely oscillatory. So, we make use of the linearity of our PDE in order to superimpose Poiseuille flow. Given a mean wall shear stress parameter, $\overline{\tau_0}$, we can determine the pressure drop term of the Poiseuille solution,[79] $k_s$, per the relationship

$$k_s = \frac{-2\overline{\tau_0}}{a}.$$

In addition, Equation (7b) provides us an "inflated" wall displacement term given by

$$D_0 = \frac{B_0 a^2}{hE_\sigma},$$

where we see that the resting wall displacement depends linearly on the average pressure, $B_0$. Instead of carrying forward with a zeroth order displacement term and requiring knowledge of an "uninflated arterial radius", we assume that $\bar{a}$ is the arterial radius of an individual under a typical mean BP, $\hat{B}_0$, and set the arterial radius given a BP with average pressure $B_0$ as

$$a = \bar{a} + \frac{(B_0 - \hat{B}_0)\bar{a}^2}{hE_\sigma}, \quad D_0 = 0,$$

so that the radius $a$ is some perturbation of $\bar{a}$ depending on if $B_0$ is above or below $\hat{B}_0$.

The corresponding system of equations for $n \geq 1$ is

$$-(n\omega)^2 C_n = \frac{E_\sigma}{\rho_w}\left(-K_n^2 C_n - \frac{i\sigma}{a}K_n D_n\right) + \frac{\mu A_n \Lambda_n J_1(\Lambda_n)}{\rho_w h a}, \tag{8a}$$

$$-(n\omega)^2 D_n = \frac{B_n}{\rho_w h} - \frac{E_\sigma}{\rho_w a}\left(\frac{D_n}{a} - i\sigma K_n C_n\right), \tag{8b}$$

$$i(n\omega)C_n = A_n J_0(\Lambda_n) + \frac{K_n B_n}{\rho(n\omega)}, \tag{8c}$$

$$i(n\omega)D_n = \frac{A_n i K_n a}{\Lambda_n}J_1(\Lambda_n) + \frac{iK_n^2 B_n a}{2\rho(n\omega)}. \tag{8d}$$

For the oscillatory solutions, we need to solve for the wavenumbers, $K_n$. Equations (8a)–(8d) define a homogeneous linear system of equations. Under the requirement that $a^2 \ll \frac{E_\sigma}{\rho_w(n\omega)^2}$, which can be shown true under our parameter regime with $n \leq 50$, we have that $(n\omega)^2 - \frac{E_\sigma}{\rho_w a^2} \approx -\frac{E_\sigma}{\rho_w a^2}$.[79] Then, $K_n$ is chosen so that the matrix of coefficients for this linear system is nonsingular, providing nontrivial solutions for $A_n$, $B_n$, $C_n$, and $D_n$. This is also found in Zamir Equation (5.7.21),[79] replacing $K_n = n\omega/c$. The solution is found by first setting $z_n = \frac{E_\sigma h K_n^2}{\rho a(n\omega)^2}$, $g_n = \frac{2J_1(\Lambda_n)}{\Lambda_n J_0(\Lambda_n)}$, and finding the (further positive real) solution to the quadratic equation

$$\left[(g_n - 1)(\sigma^2 - 1)\right] z_n^2 + \left[\frac{\rho_w h}{\rho a}(g_n - 1) + (2\sigma - \frac{1}{2})g_n - 2\right] z_n + \frac{2\rho_w h}{\rho a} + g_n = 0.$$

The wavenumbers can then be solved for from the definition of $z_n$. We additionally define the elasticity constants,[79]

$$G_n = \frac{2 + z_n(2\sigma - 1)}{z_n(2\sigma - g_n)}, \quad G_n^{(1)} = \frac{2 - z_n(1 - g_n)}{z_n(2\sigma - g_n)}, \quad G_n^{(2)} = \frac{g_n + \sigma z_n(g_n - 1)}{z_n(g_n - 2\sigma)}.$$

Combining the steady and oscillatory solutions, our final fluids equations are

$$u_z(r,z,t) = \frac{k_s}{4\mu}(r^2 - a^2) + \sum_{n=1}^{N} \frac{K_n B_n}{\rho(n\omega)} \left(1 - G_n \frac{J_0(\zeta_n)}{J_0(\Lambda_n)}\right) e^{i(n\omega t - K_n z)}, \tag{9a}$$

$$u_r(r,z,t) = \sum_{n=1}^{N} \frac{iaK_n^2 B_n}{2\rho(n\omega)} \left(\frac{r}{a} - G_n \frac{2J_1(\zeta_n)}{\Lambda_n J_0(\Lambda_n)}\right) e^{i(n\omega t - K_n z)}, \tag{9b}$$

$$p(r,z,t) = k_s z + \sum_{n=0}^{N} B_n e^{i(n\omega t - K_n z)}, \tag{9c}$$

$$\xi(z,t) = \sum_{n=1}^{N} \frac{iK_n G_n^{(1)}}{\rho(n\omega)^2} B_n e^{i(n\omega t - K_n z)}, \tag{9d}$$

$$\eta(z,t) = \sum_{n=1}^{N} \frac{aK_n^2 G_n^{(2)}}{\rho(n\omega)^2} B_n e^{i(n\omega t - K_n z)}. \tag{9e}$$

The pressure gradient can be computed explicitly by taking the partial derivative of Equation (9c),

$$\frac{\partial p(r,z,t)}{\partial z} = k_s + \sum_{n=0}^{N} -iK_n B_n e^{i(n\omega t - K_n z)}. \tag{9f}$$

From Equation (9a) we compute the shear stress by taking the partial derivative of Equation (9a),

$$\tau(r,z,t) = \frac{k_s r}{2} + \sum_{n=1}^{N} \frac{\mu K_n B_n \Lambda_n}{\rho(n\omega)a} G_n \frac{J_1(\zeta_n)}{J_0(\Lambda_n)} e^{i(n\omega t - K_n z)}. \tag{10}$$

## 4.4 Arterial branching

Consider the dimensionless linearized conservation of axial momentum shown in Equation (3a). Converting this back to dimensional form, we obtain

$$\rho \frac{\partial u_z}{\partial t} + \frac{\partial p}{\partial z} = \frac{\mu}{r} \frac{\partial}{\partial r}\left(r \frac{\partial u_z}{\partial r}\right). \tag{11}$$

Define the flow rate as $q(z,t) = 2\pi \int_0^a r u_z(r,z,t)\, \mathrm{d}r$. From Equation (9a), this is written as

$$q(z,t) = \frac{-k_s \pi a^4}{8\mu} + \sum_{n=1}^{N} \frac{\pi a^2 K_n B_n}{\rho(n\omega)} \left(1 - G_n \frac{2J_1(\Lambda_n)}{\Lambda_n J_0(\Lambda_n)}\right) e^{i(n\omega t - K_n z)}, \tag{12}$$

and make the definition

$$\hat{q}_n = \frac{\pi a^2 K_n B_n}{\rho(n\omega)} \left(1 - G_n \frac{2J_1(\Lambda_n)}{\Lambda_n J_0(\Lambda_n)}\right). \tag{13}$$

Inserting Equation (13) into Equation (10), we obtain the following definition for wall shear stress in terms of the Fourier coefficients for flow rate,

$$\tau_w = \frac{-k_s a}{2} + \sum_{n=1}^{N} \frac{i\rho(n\omega)}{2\pi a} \frac{G_n \frac{2J_1(\Lambda_n)}{\Lambda_n J_0(\Lambda_n)}}{1 - G_n \frac{2J_1(\Lambda_n)}{\Lambda_n J_0(\Lambda_n)}} \hat{q}_n e^{i(n\omega t - K_n z)}. \tag{14}$$

Multiplying Equation (11) by $2\pi r$ and integrating it from $r = 0$ to $r = a$, we obtain

$$\rho \frac{\partial q}{\partial t} + \pi a^2 \frac{\partial p}{\partial z} = -2\pi a \tau_w. \tag{15}$$

For $n \geq 1$, inserting Equations (9c), (12) and (14) into Equation (15), we derive the ratio of the Fourier coefficients of flow and pressure to be

$$\frac{\hat{q}_n}{B_n} = \frac{\pi a^2 K_n}{\rho(n\omega)} \left[1 - G_n \frac{2J_1(\Lambda_n)}{\Lambda_n J_0(\Lambda_n)}\right]. \tag{16}$$

Note that $\frac{\pi a^2 K_n}{\rho(n\omega)}$ defines the characteristic admittance of inviscid flow,[79] while the second term attenuates that value based on the relationship between pulsatile, elastic shear stress and flow rate.

Now, consider a pressure wave traveling through the brachial artery that undergoes a bifurcation into the ulnar and radial arteries. We will only assume the presence of primary wave reflections at this bifurcation as they have been found to be a reasonable modeling simplification and that modeling secondary reflections throughout an arterial tree quickly becomes intractable.[79] Define the reflection constant $R$, which is typically a real number between $-1$ and 1, however, in the case of elastic cylinders, it can be a complex number with magnitude less than or equal to 1. To model primary reflections, consider the forward and backward traveling pressure waves in the brachial artery defined as

$$p(z,t) = p_f(z,t) + p_b(z,t) = Be^{i(\omega t - kz)} + RBe^{i(\omega t + kz - 2kl)}.$$

In addition, there is a flow wave in the brachial artery which we will assume to be of the form

$$q(z,t) = q_f(z,t) + q_b(z,t) = \hat{q}_f e^{i(\omega t - kz)} - R\hat{q}_f e^{i(\omega t + kz - 2kl)}.$$

We also assume forward and backward traveling pressure and flow waves in each of the ulnar and radial arteries, each of which have their own pressure coefficient, $C_j$, wavenumber, $k_j$, and arterial length, $l_j$, for $j = 1, 2$. The pressure and flow waves, $p_j(z,t)$ and $q_j(z,t)$ respectively, are then represented as a sum of forward and backward traveling waves denoted similarly by

$$p_{f,j}(z,t) = C_j e^{i(\omega t - k_j(z-l))}, \quad p_{b,j}(z,t) = R_j C_j e^{i(\omega t + k_j(z-l) - 2k_j l_j)},$$
$$q_{f,j}(z,t) = \hat{q}_{f,j} e^{i(\omega t - k_j(z-l))}, \quad q_{b,j}(z,t) = -R_j \hat{q}_{f,j} e^{i(\omega t + k_j(z-l) - 2k_j l_j)},$$

respectively for $j = 1, 2$.

The characteristic admittance, $Y$, for such an artery is defined as the ratio of flow and pressure at the outlet, which is given by the ratio of forward coefficients of flow rate and pressure.[79] This quantity was derived for our elastic, pulsatile solutions in Equation (16). The effective admittance for such an artery, $Y^e$, is given by the ratio of flow and pressure at the inlet. The forms for characteristic and effective admittance are

$$Y = \frac{\hat{q}_f}{B}, \quad Y_j^e = \frac{\hat{q}_{f,j}}{C_j} \left( \frac{1 - R_j e^{i(\omega t - 2k_j l_j)}}{1 + R_j e^{i(\omega t - 2k_j l_j)}} \right), \tag{17a}$$

for $j = 1, 2$ for each of the ulnar and radial arteries.[79] The effective admittance is a product of the characteristic admittance and an attenuation constant depending on the outlet reflection coefficients, wave numbers, and artery lengths of the ulnar and radial arteries. In the case that the radial and ulnar arteries have no-reflection outlet conditions, their effective admittances are simply their characteristic admittances. Additionally, we require continuity of pressure and conservation of flow at the bifurcation which is represented by

$$p_f(l,t) + p_b(l,t) = p_1(0,t) = p_2(0,t), \tag{17b}$$
$$q_f(l,t) + q_b(l,t) = q_1(0,t) + q_2(0,t). \tag{17c}$$

Equations (17a), (17b), and (17c) allow us to derive the reflection coefficient in terms of the admittance coefficients,

$$R = \frac{Y - Y_1^e - Y_2^e}{Y + Y_1^e + Y_2^e}. \tag{17d}$$

Hence, once we know the reflection coefficients in the ulnar and radial arteries, $R_1$ and $R_2$, we derive the admittances in each artery by Equations (16) and (17a), in order to solve for the reflection coefficient in the brachial artery by Equation (17d). Given a tree with several bifurcations, one must start at the leaf vessels to compute their reflection coefficients before working back up to the most proximal parent vessel. Additionally, this process is repeated to solve for reflection coefficients for each of the $N$ Fourier terms of our solutions.

## 4.5 Outflow boundary condition

It is well studied that the BP waveform is not simply attenuated as it travels from the brachial artery to the radial artery. Through additive reflections, it is common to see "peaking" of the BP waveform where the systolic pressure at the wrist can be higher than in the brachial artery by 10 mm Hg or more.[80–82] In the case of a no reflection outlet condition at the end of the radial artery, the pressure wave would simply decay linearly with the pressure drop and exponentially with rates of the imaginary parts of the wavenumbers. We use a structured tree (ST) outlet condition as it allows us to use our elastic fluids solutions through the entirety of the tree.[83,84]

In the ST outlet condition, the outlet artery is continually branched into smaller and smaller vessels until the vessels reach some minimum radius, $a_{\text{min}}$, which is chosen to be 30 µm.[83,84] The bifurcations are treated the same as is described in the previous section, and the terminal vessels of the ST are prescribed no reflection boundary conditions. The outlet is called a ST because at each bifurcation, one daughter vessel has a radius of $\alpha$ times the parent radius while the other has a radius of $\beta$ times the parent radius. In order to prescribe values to $\alpha$ and $\beta$, we must make two assumptions on the ST.[83,84]

First, a $\xi$-law is prescribed at each bifurcation. At an arbitrary bifurcation, given that the parent artery has radius $a_0$, and the two daughter arteries have radii $\alpha a_0$ and $\beta a_0$, the $\xi$ law tells us that

$$1 = \alpha^{\xi} + \beta^{\xi}.$$

$\xi$ is chosen between 2.33 and 3 depending on the Reynolds number of the flow; a nominal value of $\xi \approx 2.76$ is taken.[83,84] Second, an area ratio is determined,

$$\kappa = \alpha^2 + \beta^2.$$

From studies of the arterial tree, a value of $\kappa = 1.16$ is chosen.[83,84] Once these two values are chosen, the values of $\alpha$ and $\beta$ are determined.

One additional parameter that is required is the length-radius ratio of each of the arteries, $L_{\text{rr}}$, such that a vessel in the ST with radius $a_0$ (mm) will have a length of $L_{\text{rr}} \cdot a_0$ (mm). From arterial tree measurements, we use an intermediate value of $L_{\text{rr}} = 25 \pm 10$.[83,85]

Finally, it has been shown that the vessels of the arterial tree stiffen the further they are from the aorta. Previous sources have used a shifted exponential curve to match this stiffening,[83,84] however, from physiological data collected by Westerhof et al.,[86] plotting admittance on a log-log scale reveals that a power law better fits the data (Supplementary Fig.32). In order to not break our assumptions in deriving our fluids solutions, we keep the wall thickness to artery radius ratio fixed while increasing the Young's modulus by a power law as we travel down the ST. We fix the nondimensional exponent in order to achieve peaking on the order of 10 mm Hg, and at each outlet we choose the coefficient on the fit to require continuity with the Young's modulus of the outlet artery, whether that be the radial, ulnar i., or ulnar b.

## 4.6 Blood electrical conductivity model

Here, we describe the modeling of blood conductivity from shear stress in a cylindrical artery.[87,88] We assume that blood is a dilute suspension of liquid plasma and red blood cells (RBCs) and that the RBCs are uniformly concentrated through the artery. Since the arteries that we are interested in modeling conductivity have diameters larger than 0.3 mm, accumulation of RBCs near the center of the artery, also called the Fåhræus-Lindqvist effect, is negligible.[89,90] A control volume is defined as a small cylindrical volume containing one RBC at its center and otherwise plasma. This is illustrated in Hoetink et al. Figure 1.[87] Further, we assume that RBCs are ellipsoidal particles with one axis of length $2a$, and two axes of lengths $2b$ with $a \leq b$. Assuming that the conductivity of an RBC is negligible in comparison to the surrounding plasma, the Maxwell-Fricke equation for the electrical conductivity of a single control volume is

$$\frac{\sigma_{\text{cv}}}{\sigma_{\text{plasma}}} = \frac{1-H}{1+(C-1)H}, \tag{18}$$

in which $\sigma_{\text{cv}}$ is the conductivity of a single control volume, $\sigma_{\text{plasma}}$ is the conductivity of the plasma, $H$ is the hematocrit, and $C$ is a factor depending on the geometry and orientation of the RBC.[87,91]

Assuming constant volume but deforming the $a$ and $b$-axes, we derive the relation

$$\frac{a}{b} = \frac{a_0}{b_0}\left(1+\frac{\Delta b}{b_0}\right)^{-3},$$

where $a_0$ and $b_0$ are the initial axes lengths, $a$ and $b$ are the deformed axes lengths, and $\Delta b/b_0$ is the strain of the $b$-axis.[87] Further assume the relationship between extension ratio $\lambda_b$ and magnitude of shear stress, $|\tau|$,

$$\lambda_b^2 - \lambda_b^{-2} = \frac{|\tau| b}{\mu_m},$$

where $\mu_m$ is the membrane or surface shear modulus.[87,92] From this, using a linear Taylor's approximation for the extension ratios, we obtain

$$\frac{a}{b} = \frac{a_0}{b_0}\left(1+\frac{|\tau| b_0}{4\mu_m}\right)^{-3}.$$

Assuming that all RBCs are aligned such that either their $a$-axis or $b$-axis is oriented with the flow, define

$$M = \frac{\phi - 0.5\sin(2\phi)}{\sin^3(\phi)}\cos(\phi), \quad \cos(\phi) = \frac{a}{b},$$

and the quantities

$$C_a = \frac{1}{M}, \quad C_b = \frac{2}{2-M}, \quad C_r = \frac{1}{3}C_a + \frac{2}{3}C_b,$$

corresponding to geometry factors for $a$-aligned RBCs, $b$-aligned RBCs, or RBCs that are randomly oriented.[87] Here, random orientation describes an averaging between RBCs that are 1/3 $a$-aligned and 2/3 $b$-aligned.

Next, we make use of the $f$ function first defined by Bitbol and Leterrier to describe how RBCs orient themselves as a function of shear stress.[88,93] This function is given by

$$f(r,z,t) = \frac{\tau_0^{-1}(r,z,t)}{\tau_d^{-1}(r,z,t) + \tau_0^{-1}(r,z,t)},$$

$$\tau_0(r,z,t) = k_0\left|\frac{\partial u_z}{\partial r}\right|^{-1}, \quad \tau_d(r,z,t) = k_d\left|\frac{\partial u_z}{\partial r}\right|^{-1/2},$$

where $k_0$ and $k_d$ are modeling parameters. Since shear stress is represented as $\tau = \mu\frac{\partial u_z}{\partial r}$, we reduce the above to

$$f(r,z,t) = \frac{\sqrt{|\tau(r,z,t)|}k}{1+\sqrt{|\tau(r,z,t)|}k}, \quad \text{and} \quad k = \frac{k_d}{k_0\sqrt{\mu}},$$

where the only required quantity is the ratio of $k_0$ and $k_d$ rather than each independently. With this information, the geometry factor for Maxwell-Fricke is given by

$$C = f(r,z,t)C_b + (1 - f(r,z,t))C_r = \left(\frac{2}{3} + \frac{1}{3}f(r,z,t)\right)C_b + \left(\frac{1}{3} - \frac{1}{3}f(r,z,t)\right)C_a. \tag{19}$$

The $C$-factor from Equation (19) is then inserted into Equation (18) to obtain the conductivity of a control volume.

Finally, through integration over all control volumes, and an extension to variations in the $z$-direction, under the assumption that wall displacements are small, we obtain the blood conductivity of blood given by

$$\sigma_{\text{blood}} = \frac{2}{a^2 L}\int_0^L \int_0^a r\sigma_{\text{cv}}(r,z,t)\,\mathrm{d}r\mathrm{d}z. \tag{20}$$

Equation (20) is equivalent to the equations for blood conductivity by Gaw et al. and Hoetink et al. if it is simplified that $\sigma_{\text{cv}}$ does not vary in the $z$ direction.[87,88]

## 4.7 Computational fluid dynamic model

We performed arterial flow simulations using the computable human Ella phantom and the Flow solver in Sim4Life V 7.2.1 (Zurich MedTech, Zurich Switzerland).[94] We isolated the artery segment from the left upper arm to the forearm including the brachial bifurcation into the ulnar and radial arteries (Fig.2e). The simulation solved the full Navier-Stokes equations with boundary conditions defined as cross-sectional average brachial inlet velocity of 70 cm/s and a constant pressure at the radial and ulnar outlets of 90 mm Hg. The domain was meshed using 10M tetrahedrons. The velocity tolerance was set to 0.01 cm/s and the max number of iterations was limited to 100.

# 5 Supplementary Discussion 5. Bioimpedance

## 5.1 Electrodynamic model of the wrist

We developed three-layer electrodynamic model of the wrist to relate changes in blood volume and electrical in a cylindrical radial artery with a surface BioZ measurement. In this model we solved Maxwell's equations under a quasi-electrostatic approximation with a point-like current excitation and ground at infinity, and derived the apparent electrical resistance arising from the superposition of a primary and secondary electric potential.[95] In our derivation, the primary potential ignores the contribution of the radial artery, and the secondary potential accounts for the artery's contribution.

Our model consists of an infinite three-layered half-space domain $\mathbb{R}^3_- = \{(x,y,z) \in \mathbb{R}^3 \colon x \geq 0\}$ with isotropic layers of skin with thickness $d_1$ and conductivity $\sigma_1$, subcutaneous fat tissue with thickness $d_2$ and conductivity $\sigma_2$, and skeletal muscle tissue with conductivity $\sigma_3$. The radial artery is modeled as an infinitely long conductive cylinder with radius $a$ filled with blood with conductivity as in Equation (20). For notational convenience, we hereafter denote blood conductivity as $\sigma_a$, and the artery is embedded within the muscle layer as shown in Supplementary Fig.2.

The derivation in Wait et al.[95] provides a framework for increasing the number of layers to $N$, but the reflection coefficients for the additional layers are implicitly defined and therefore impractical for computation. Here, we provide the framework for expanding the model to three-layers with explicit reflection coefficients.

Following Wait et al. rationale,[95] we consider the quasi-static Maxwell's equations in the three-layers with an injected current source $I$ located at the origin and sink $-I$ located at infinity. We assume azimuthal and axisymmetric solutions due to the presence and geometry of the cylinder. Neglecting displacement currents and focusing on axial currents on the cable, we only consider transverse electromagnetic (TEM) fields. The Maxwell's equations take the form

$$\nabla \times \vec{E} = -i\mu\omega\vec{H} \tag{21a}$$

$$\nabla \times \vec{H} = (\sigma + i\omega\varepsilon)\vec{I}, \tag{21b}$$

where $\vec{E}$ is the electric field, $\vec{H}$ is the magnetic field, and $\omega$ is the angular frequency. The constant material parameters $\mu, \sigma, \rho$, and $\varepsilon$ are the magnetic permeability, electric conductivity, electric resistivity, and electric permittivity, as defined previously. Introducing the Hertz vector potential, $\vec{\Pi}$ in Equation (21b), so that $i\mu\omega\vec{H} = \gamma^2\nabla \times \vec{\Pi}$,[96] using vector identities to distribute the curl operator, we solve for $\vec{E}$ in terms of $\vec{\Pi}$:

$$\vec{E} = -\tilde{\gamma}^2\vec{\Pi} + \nabla(\nabla \cdot \vec{\Pi}), \tag{22}$$

where $\tilde{\gamma}^2 = i\mu\omega(\sigma + i\omega\varepsilon)$. The electric potential is defined as the divergence of the Hertz vector potential, $\Omega = -\nabla \cdot \vec{\Pi}$, so that if $\gamma = 0$ corresponding to DC current, we recover the usual form $\vec{E} = -\nabla\Omega$. Taking the curl of Equation (21b), using the Hertz potential, the Equation (22), and vector identities, we find that $\vec{\Pi}$ satisfies the Helmholtz equation

$$\nabla^2\vec{\Pi} - \tilde{\gamma}^2\vec{\Pi} = 0. \tag{23}$$

Here $\nabla^2 = \Delta$ is the Laplacian operator acting on a vector field. We seek solutions so that $E_z$, the $z-$component of $\vec{E}$, is odd around the origin, where the injected current is located. Owing to the assumptions of azimuthal and axisymmetric solutions under TEM fields, the vector equations above reduce to scalar equations for $E_z, H_\phi, \Pi_\rho$, where $(\rho, \phi, z)$ are cylindrical coordinates centered around the plane $z = 0$ and the subscripts denote the respective vector components.

The first step consists of finding the impedance of the artery. This is equivalent to solving the equations inside an infinitely long cylinder of conductivity $\sigma_a$. We define the impedance of the artery as

$$Z_a = \left.\frac{\widehat{E_z}(\lambda)}{2\pi\rho\widehat{H_\phi}(\lambda)}\right|_{\rho=a}. \tag{24}$$

Here, $\widehat{\cdot}$ denotes the Fourier sine transform in $z$: $\widehat{f} = \sqrt{\frac{2}{\pi}}\int_0^\infty f(z)\sin(\lambda z)\,\mathrm{d}z$ and $a$ is the radius of the radial artery. Focusing on the non vanishing components only and transforming into Fourier space,[95]

$$H_\phi = -\sigma_a\frac{\partial}{\partial\rho}\widehat{\Pi_z}, \quad E_z = (-\gamma_a^2+\lambda^2)\widehat{\Pi_z}, \quad \left(\frac{1}{\rho}\frac{\partial}{\partial_\rho}\left(\rho\frac{\partial}{\partial_\rho}\right) - \lambda^2 - \gamma_a^2\right)\widehat{\Pi_z} = 0,$$

for the fields inside the artery, denoted by the subscript $a$. The bounded general solution of this equation has the form $\widehat{\Pi_z} = C(\lambda)I_0(v\rho)$, where $C(\lambda)$ is an arbitrary constant, $I_0$ is the Bessel function of the first kind of order zero, and $v^2 = \lambda^2 + \gamma_a^2$. Substituting into Equation (24), we finally find

$$Z_a(\lambda) = \frac{vI_0(va)}{2\pi\sigma_a a I_1(va)}, \tag{25}$$

where $I_1$ is the modified Bessel function of the first kind of order one. The cylinder impedance Equation (25) is necessary to apply the impedance boundary condition at the artery $\widehat{E}_z(\lambda)\big|_{\rho=a} = Z_a(\lambda)\widehat{I}_a(\lambda)$, which requires both the total electric field $E(z)$ and the cylinder current $I_a(z)$.

Considering a single source of injected current at the origin and using the linearity property of the Maxwell equations, we obtain the total electric field as the superposition of the primary field in the layered geometry without the artery, and the secondary field due to the presence of the artery.

In the limit of $\tilde{\gamma} = 0$ corresponding to DC current, Equation (23) reduces to the Laplace equation for the Hertz vector potential or in this azimuthal and axisymmetry geometry, it becomes the Laplace equation for the primary and secondary electric potentials $\Delta\Omega = 0$. Using separation of variables, we express these solutions as Fourier-Bessel sine integrals and compute the total electric field as $E(z) = -\frac{\partial}{\partial z}(\Omega^p + \Omega^s)$. While the solutions are developed when $\tilde{\gamma} = 0$, they form a valid approximation to the solution when $\tilde{\gamma}$ is small corresponding to the low AC frequency employed in our setup, however the artery impedance should be viewed as the effective series impedance per unit length $Z_a$ given in Equation (25). The fundamental solution to Laplace's equation with an injected current $I$ at the origin is inversely proportional to the distance to the origin. As a result, the cylinder current in the artery is given as

$$I_a(z) = \left.\int_0^{2\pi}\rho' H^s_{\phi'}(z)\,\mathrm{d}\phi'\right|_{\rho'=a} = -\sigma_3\left.\int_0^{2\pi}\rho'\frac{\partial\Pi_z^s}{\partial\rho'}\,\mathrm{d}\phi'\right|_{\rho'=a} \approx -2\pi\rho'\sigma_3\frac{\partial}{\partial\rho'}\int_0^\infty F(\lambda)K_0(\lambda\rho')\sin(\lambda z)\,\mathrm{d}\lambda, \tag{26}$$

where $H^s_\phi(z)$ is the Hertz potential associated with the secondary potential due to the presence of the cylinder with centerline $x = -h, y = d$ and $(\rho', \phi', z')$ are shifted cylindrical coordinates. Assuming that $h \gg a$, the last approximation in Equation (26) is obtained by first noting that the secondary Hertz potential must be a superposition of the artery contribution and its image at $x = h, y = d$, namely

$$\Pi_z^s = \int_0^\infty F(\lambda)\left[K_0(\lambda\sqrt{(x+h)^2+(y-d)^2}) + K_0(\lambda\sqrt{(x-h)^2+(y-d)^2})\right]\sin(\lambda z)\,\mathrm{d}\lambda.$$

In the above, we use properties of Fourier-Bessel integrals[95] and the fundamental solutions of Laplace equation to write

$$\frac{1}{\sqrt{x^2+y^2+z^2}} = \frac{2}{\pi}\int_0^\infty K_0\left(\lambda\sqrt{x^2+y^2}\right)\cos(\lambda z)\,\mathrm{d}\lambda = \frac{1}{\pi}\int_0^\infty\int_{-\infty}^\infty \frac{1}{u}e^{-u|x|}e^{-i\beta y}\cos(\lambda z)\,\mathrm{d}\beta\mathrm{d}\lambda, \tag{27}$$

where $u^2 = \beta^2 + \lambda^2$ and $K_0$ is the modified Bessel function of the second kind of order zero.

Finally since $K_0'(x) = -K_1(x)$, the Fourier sine transform of Equation (26) is

$$\widehat{I}_a(\lambda)\sqrt{\frac{2}{\pi}} = 2\pi a\sigma_3 F(\lambda)K_1(\lambda a)\lambda. \tag{28}$$

Motivated by Equation (27), the primary potentials, ignoring the presence of the artery, at each of the three tissues have the general form

$$\Omega_1^p = \frac{I}{2\pi^2\sigma_1}\int_0^\infty\int_{-\infty}^\infty \frac{1}{u}\left[f_1^p(u)e^{ux} + g_1^p(u)e^{-ux}\right]e^{-i\beta y}\cos(\lambda z)\,\mathrm{d}\beta\mathrm{d}\lambda \tag{29a}$$

$$\Omega_2^p = \frac{I}{2\pi^2\sigma_1}\int_0^\infty\int_{-\infty}^\infty \frac{1}{u}\left[f_2^p(u)e^{ux} + g_2^p(u)e^{-ux}\right]e^{-i\beta y}\cos(\lambda z)\,\mathrm{d}\beta\mathrm{d}\lambda \tag{29b}$$

$$\Omega_3^p = \frac{I}{2\pi^2\sigma_1}\int_0^\infty\int_{-\infty}^\infty \frac{1}{u}\left[f_3^p(u)e^{ux}\right]e^{-i\beta y}\cos(\lambda z)\,\mathrm{d}\beta\mathrm{d}\lambda. \tag{29c}$$

Recalling from Supplementary Fig.2 that $x \le 0$, terms of the form $e^{ux}$ are decaying and therefore guarantee a bounded solution in the muscle (third layer). The forms of the potentials in Equations (29a) and (29b) can be thought of as accounting for the infinite reflections between the tissues. The unknown functions, $f_1^p(u)$, $f_2^p(u)$, $f_3^p(u)$, $g_1^p(u)$, and $g_2^p(u)$ are to be determined from applying voltage continuity, current continuity, and matching the non-singular behavior of the homogeneous half-space solution boundary conditions. These boundary conditions are

$$\begin{cases} \Omega_1^p - \Omega_2^p = 0 & \text{at } x = -s_1 \quad (29\text{d}) \\ \sigma_1\dfrac{\partial\Omega_1^p}{\partial x} - \sigma_2\dfrac{\partial\Omega_2^p}{\partial x} = 0 & \text{at } x = -s_1 \quad (29\text{e}) \\ \Omega_2^p - \Omega_3^p = 0 & \text{at } x = -s_2 \quad (29\text{f}) \\ \sigma_2\dfrac{\partial\Omega_2^p}{\partial x} - \sigma_3\dfrac{\partial\Omega_3^p}{\partial x} = 0 & \text{at } x = -s_2 \quad (29\text{g}) \\ \dfrac{\partial}{\partial x}\left[\Omega_1^p - \dfrac{I}{2\pi\sigma_1\sqrt{x^2+y^2+z^2}}\right] = 0 & \text{at } x = 0\,. \quad (29\text{h}) \end{cases}$$

Here $s_1$ is the thickness of skin tissue and $s_2$ is the thickness of skin and subcutaneous fat tissue (SAT) (Supplementary Fig.2). The last boundary condition in Equation (29h) can be reformulated using Equation (27) as

$$\frac{\partial\Omega_1^p}{\partial x} = \frac{I}{2\pi^2\sigma_1}\frac{\partial}{\partial x}\int_0^\infty\int_{-\infty}^\infty \frac{1}{u}e^{-u|x|}e^{-i\beta y}\cos(\lambda z)\,\mathrm{d}\beta\mathrm{d}\lambda \quad \text{at } x = 0.$$

Next, we apply all boundary conditions in Equations (29d)–(29h) and rewrite the resulting linear system of equations for the unknown functions $f_1^p, g_1^p, f_2^p, g_2^p, f_3^p$ as a matrix system, $A\vec{f^p} = \vec{b^p}$. We find

$$A = \begin{bmatrix} e^{-us_1} & e^{us_1} & -e^{-us_1} & -e^{us_1} & 0 \\ \sigma_1 e^{-us_1} & -\sigma_1 e^{us_1} & -\sigma_2 e^{-us_1} & \sigma_2 e^{us_1} & 0 \\ 0 & 0 & e^{-us_2} & e^{us_2} & -e^{-us_2} \\ 0 & 0 & \sigma_2 e^{-us_2} & -\sigma_2 e^{us_2} & -\sigma_3 e^{-us_2} \\ 1 & -1 & 0 & 0 & 0 \end{bmatrix}, \quad \text{and} \quad \vec{b^p} = \begin{bmatrix} 0 \\ 0 \\ 0 \\ 0 \\ 1 \end{bmatrix}. \tag{30}$$

We introduce the true reflection coefficient $\Gamma_1, \Gamma_2$ between skin and SAT, and SAT and muscle, respectively; and the damped reflection coefficient $\widetilde{\Gamma}_1(u)$ between skin and SAT as follows

$$\Gamma_1 = \frac{\sigma_1 - \sigma_2}{\sigma_1 + \sigma_2}, \quad \Gamma_2 = \frac{\sigma_2 - \sigma_3}{\sigma_2 + \sigma_3}, \quad \text{and} \quad \widetilde{\Gamma}_1(u) = \frac{\sigma_1 - \sigma_2\frac{1-\Gamma_2 e^{-2ud_2}}{1+\Gamma_2 e^{-2ud_2}}}{\sigma_1 + \sigma_2\frac{1-\Gamma_2 e^{-2ud_2}}{1+\Gamma_2 e^{-2ud_2}}} = \frac{e^{-2ud_2}\Gamma_2 + \Gamma_1}{e^{-2ud_2}\Gamma_1\Gamma_2 + 1},$$

where $d_2 = s_2 - s_1$ is the thickness of SAT. We factor the determinant of $A$ in the form

$$\det(A) = (\sigma_2+\sigma_3)(\sigma_1+\sigma_2)(1-\widetilde{\Gamma}_1(u)e^{-2us_1})\left(e^{-2ud_2}\Gamma_1\Gamma_2+1\right).$$

The solution to the linear system is

$$\begin{aligned}
f_1^p(u) &= \frac{(\sigma_2+\sigma_3)(\sigma_1+\sigma_2)}{\det(A)}\left(e^{-2ud_2}\Gamma_1\Gamma_2+1\right) = \frac{1}{1-\widetilde{\Gamma}_1(u)e^{-2us_1}} = 1+g_1^p(u)\\
g_1^p(u) &= \frac{(\sigma_2+\sigma_3)(\sigma_1+\sigma_2)}{\det(A)}e^{-2us_1}\left(e^{-2ud_2}\Gamma_2+\Gamma_1\right) = f_1^p(u)\widetilde{\Gamma}_1(u)e^{-2us_1}\\
f_2^p(u) &= \frac{(\sigma_2+\sigma_3)}{\det(A)}2\sigma_1 = f_1^p(u)\frac{1+\Gamma_1}{e^{-2ud_2}\Gamma_1\Gamma_2+1} = f_1^p(u)\frac{1+\widetilde{\Gamma}_1(u)}{1+\Gamma_2e^{-2ud_2}}\\
g_2^p(u) &= \frac{(\sigma_2+\sigma_3)}{\det(A)}2\sigma_1e^{-2us_2}\Gamma_2 = f_2^p(u)\Gamma_2e^{-2us_2}\\
f_3^p(u) &= \frac{(\sigma_2+\sigma_3)}{\det(A)}2\sigma_1\frac{2\sigma_2}{\sigma_2+\sigma_3} = f_2^p(u)(1-\Gamma_2).
\end{aligned}$$

The primary potential in the skin thus becomes

$$\begin{aligned}
\Omega_1^p(x,y,z) &= \frac{I}{2\pi^2\sigma_1}\int_0^\infty\int_{-\infty}^\infty\frac{1}{u}\left[e^{ux}+\frac{\widetilde{\Gamma}_1(u)e^{-2us_1}}{1-\widetilde{\Gamma}_1(u)e^{-2us_1}}\left(e^{ux}+e^{-ux}\right)\right]e^{-i\beta y}\cos(\lambda z)\,\mathrm{d}\beta\mathrm{d}\lambda\\
&= \frac{I}{2\pi\sigma_1}\frac{1}{\sqrt{x^2+y^2+z^2}}+\frac{2I}{\pi^2\sigma_1}\int_0^\infty\int_0^\infty\frac{1}{u}\frac{\widetilde{\Gamma}_1(u)e^{-2us_1}}{1-\widetilde{\Gamma}_1(u)e^{-2us_1}}\cosh(ux)\cos(\beta y)\cos(\lambda z)\,\mathrm{d}\beta\mathrm{d}\lambda.
\end{aligned} \tag{31}$$

Since the artery is located in the muscle, we need the primary potential there. It has the form

$$\Omega_3^p(x,y,z) = \frac{I}{\pi^2\sigma_1}\int_0^\infty\int_0^\infty\frac{1}{u}\frac{1+\Gamma_1}{1-\widetilde{\Gamma}_1(u)e^{-2us_1}}\frac{1-\Gamma_2}{1+\Gamma_1\Gamma_2e^{-2ud_2}}e^{ux}\cos(\beta y)\cos(\lambda z)\,\mathrm{d}\beta\mathrm{d}\lambda \tag{32}$$

Next, we determine the secondary potential which accounts for the electrical contribution of the artery. Similarly, we assume that the potential in each tissues takes a Fourier-Bessel form[95]

$$\Omega_1^s = -\frac{1}{2}\int_0^\infty\int_{-\infty}^\infty F(\lambda)\frac{1}{u}\left[f_1^s(u)e^{ux}+g_1^s(u)e^{-ux}\right]e^{-i\beta(y-d)}\lambda\cos(\lambda z)\,\mathrm{d}\beta\mathrm{d}\lambda \tag{33a}$$

$$\Omega_2^s = -\frac{1}{2}\int_0^\infty\int_{-\infty}^\infty F(\lambda)\frac{1}{u}\left[f_2^s(u)e^{ux}+g_2^s(u)e^{-ux}\right]e^{-i\beta(y-d)}\lambda\cos(\lambda z)\,\mathrm{d}\beta\mathrm{d}\lambda \tag{33b}$$

$$\Omega_3^s = -\frac{1}{2}\int_0^\infty\int_{-\infty}^\infty F(\lambda)\frac{1}{u}\left[e^{-u|x+h|}+f_3^s(u)e^{ux}\right]e^{-i\beta(y-d)}\lambda\cos(\lambda z)\,\mathrm{d}\beta\mathrm{d}\lambda \tag{33c}$$

where $h$ is the distance of the center of the radial artery from the skin, and $d$ is the offset of the artery from the electrodes (Supplementary Fig.2). Furthermore, $F(\lambda)$ is a function to be determined from the impedance condition at the artery. The unknown functions $f_1^s(u)$, $f_2^s(u)$, $f_3^s(u)$, $g_1^s(u)$, and $g_2^s(u)$ are to be determined by applying voltage continuity, current continuity, and electric insulation boundary conditions. The first exponential term in the integrand in Equation (33c) accounts for the fundamental solution to Laplace's equation due to the presence of the artery and for $x > -h$, it has the form $e^{-ux}e^{-uh}$. The boundary conditions for the secondary potential are

$$\begin{cases}
\Omega_1^s - \Omega_2^s = 0 & \text{at } x=-s_1\,, \quad (33d)\\
\sigma_1\dfrac{\partial\Omega_1^s}{\partial x}-\sigma_2\dfrac{\partial\Omega_2^s}{\partial x}=0 & \text{at } x=-s_1\,, \quad (33e)\\
\Omega_2^s - \Omega_3^s = 0 & \text{at } x=-s_2\,, \quad (33f)\\
\sigma_2\dfrac{\partial\Omega_2^s}{\partial x}-\sigma_3\dfrac{\partial\Omega_3^s}{\partial x}=0 & \text{at } x=-s_2\,, \quad (33g)\\
\dfrac{\partial}{\partial x}\Omega_1^s = 0 & \text{at } x=0\,. \quad (33h)
\end{cases}$$

Applying the boundary conditions (33d)–(33h), we solve for the unknown functions in the secondary potential as a system of linear equations, $A\vec{f^s} = \vec{b^s}$, with $A$ given in Equation (30) and

$$\vec{b^s} = \left(0, 0, e^{us_2}e^{-uh}, -\sigma_3 e^{us_2}e^{-uh}, 0\right)^T$$

with $^T$ the vector transpose. The solution of $A\vec{f^s} = \vec{b^s}$ is

$$\begin{aligned}
f_1^s(u) &= \frac{e^{-uh}}{\det(A)}4\sigma_2\sigma_3 = e^{-uh}\frac{(1-\Gamma_1)(1-\Gamma_2)}{(1-\widetilde{\Gamma}_1(u)e^{-2us_1})(e^{-2ud_2}\Gamma_1\Gamma_2+1)} \\
g_1^s(u) &= f_1^s(u) \\
f_2^s(u) &= \frac{e^{-uh}(\sigma_1+\sigma_2)}{\det(A)}2\sigma_3\left(1-\Gamma_1 e^{2us_1}\right) \\
g_2^s(u) &= \frac{e^{-uh}(\sigma_1+\sigma_2)}{\det(A)}2\sigma_3\left(1-\Gamma_1 e^{-2us_1}\right) \\
f_3^s(u) &= \frac{e^{-uh}(\sigma_2+\sigma_3)(\sigma_1+\sigma_2)}{\det(A)}(1-\widetilde{\Gamma}_1(-u)e^{2us_1})(1+\Gamma_2\Gamma_1 e^{2ud_2})
\end{aligned}$$

As a result, the secondary potential in the skin layer has the form

$$\begin{aligned}
\Omega_1^s(x,y,z) = -2(1-\Gamma_1)(1-\Gamma_2)\int_0^\infty\int_0^\infty F(\lambda)\frac{1}{u}\frac{e^{-uh}\cosh(ux)}{(1-\widetilde{\Gamma}_1(u)e^{-2us_1})(\Gamma_2\Gamma_1 e^{-2ud_2}+1)}\cos(\beta(y-d)) \\
\lambda\cos(\lambda z)\,\mathrm{d}\beta\mathrm{d}\lambda.
\end{aligned} \tag{34}$$

Finally, to find $F(\lambda)$, we match the impedance solution at the boundary of the artery. Therefore, we seek the corresponding approximation of the secondary potential in the muscle. The exact form is

$$\begin{aligned}
\Omega_3^s = &-\int_0^\infty\int_0^\infty F(\lambda)\frac{1}{u}\left[e^{-u|x+h|}+e^{-u(h-x)}\frac{(1-\widetilde{\Gamma}_1(-u)e^{2us_1})}{(1-\widetilde{\Gamma}_1(u)e^{-2us_1})}\frac{(1+\Gamma_2\Gamma_1 e^{2ud_2})}{(1+\Gamma_2\Gamma_1 e^{-2ud_2})}\right] \\
&\cos(\beta(y-d))\lambda\cos(\lambda z)\,\mathrm{d}\beta\mathrm{d}\lambda \\
= &-\int_0^\infty F(\lambda)K_0\left(\lambda\sqrt{(x+h)^2+(y-d)^2}\right)\lambda\cos(\lambda z)\,\mathrm{d}\lambda \\
&-\int_0^\infty\int_0^\infty F(\lambda)\frac{1}{u}e^{-uh}e^{ux}\frac{(1-\widetilde{\Gamma}_1(-u)e^{2us_1})}{(1-\widetilde{\Gamma}_1(u)e^{-2us_1})}\frac{(1+\Gamma_2\Gamma_1 e^{2ud_2})}{(1+\Gamma_2\Gamma_1 e^{-2ud_2})}\cos(\beta(y-d))\lambda\cos(\lambda z)\,\mathrm{d}\beta\mathrm{d}\lambda.
\end{aligned} \tag{35}$$

The impedance condition at the artery $(x+h)^2+(y-d)^2=a^2$ has the form[95]

$$\left[\widehat{E_z^p}(\lambda)+\widehat{E_z^s}(\lambda)\right]\Big|_{\text{artery}} = Z_a(\lambda)\widehat{I_a}(\lambda) = \frac{\lambda^2\sigma_3 I_0(\lambda a)}{\sigma_a I_1(\lambda a)}\sqrt{\frac{\pi}{2}}F(\lambda)K_1(\lambda a)$$

where $Z_a(\lambda)$ is given in Equation (25) with $\gamma_a=0$, i.e $v=\lambda$ and $\widehat{I_a}(\lambda)$ is given in Equation (28). Next, we use the low frequency approximation $E_z = -\frac{\partial\Omega}{\partial z}$ to compute both primary and secondary potentials and their Fourier sine coefficients from Equations (32) and (35)

$$\begin{aligned}
E_z^p &= \sqrt{\frac{2}{\pi}}\int_0^\infty \widehat{E_z^p}\sin(\lambda z)\,\mathrm{d}\lambda = \frac{I}{\pi^2\sigma_1}\int_0^\infty\int_0^\infty\frac{1}{u}\frac{1+\Gamma_1}{1-\widetilde{\Gamma}_1(u)e^{-2us_1}}\frac{1-\Gamma_2}{1+\Gamma_1\Gamma_2 e^{-2ud_2}}e^{ux}\cos(\beta y)\lambda\sin(\lambda z)\,\mathrm{d}\beta\mathrm{d}\lambda \\
E_z^s &= \sqrt{\frac{2}{\pi}}\int_0^\infty \widehat{E_z^s}\sin(\lambda z)\,\mathrm{d}\lambda = -\int_0^\infty F(\lambda)K_0\left(\lambda\sqrt{(x+h)^2+(y-d)^2}\right)\lambda^2\sin(\lambda z)\,\mathrm{d}\lambda \\
&\quad -\int_0^\infty\int_0^\infty F(\lambda)\frac{1}{u}e^{-uh}e^{ux}\frac{(1-\widetilde{\Gamma}_1(-u)e^{2us_1})}{(1-\widetilde{\Gamma}_1(u)e^{-2us_1})}\frac{(1+\Gamma_2\Gamma_1 e^{2ud_2})}{(1+\Gamma_2\Gamma_1 e^{-2ud_2})}\cos(\beta(y-d))\lambda^2\sin(\lambda z)\,\mathrm{d}\beta\mathrm{d}\lambda.
\end{aligned}$$

When evaluating the above expressions at the artery, we make the further simplification that, for the double integrals terms, $x=-h, y=d$, corresponding to the centerline of the artery. Rearranging and solving for $F(\lambda)$ yields

$$\frac{F(\lambda)}{I} = \widetilde{F}(\lambda) = \frac{1}{\pi^2\sigma_1\lambda}\frac{\displaystyle\int_0^\infty \frac{1}{u}\frac{1+\Gamma_1}{1-\widetilde{\Gamma}_1(u)e^{-2us_1}}\frac{1-\Gamma_2}{1+\Gamma_1\Gamma_2 e^{-2ud_2}}e^{-uh}\cos(\beta d)\,\mathrm{d}\beta}{K_0(\lambda a)+\frac{\sigma_3 I_0(\lambda a)K_1(\lambda a)}{\sigma_a I_1(\lambda a)}+\displaystyle\int_0^\infty \frac{e^{-2uh}}{u}\frac{1-\widetilde{\Gamma}_1(-u)e^{2us_1}}{1-\widetilde{\Gamma}_1(u)e^{-2us_1}}\frac{1+\Gamma_1\Gamma_2 e^{2ud_2}}{1+\Gamma_1\Gamma_2 e^{-2ud_2}}\,\mathrm{d}\beta}. \quad (36)$$

## 5.2 Bioimpedance model of the wrist

In a four-terminal BioZ measurement, one pair of electrodes injects an alternating current $I$ and a separate pair of electrodes measures the voltage difference $\Delta\Omega$. The current electrodes are modeled as point current sources with equal and opposite signs. We define the first current source, $\mathcal{E}_1$, with current, $I$ located at $(0,0,0)$. The voltage sensing electrodes, $\mathcal{E}_2$ and $\mathcal{E}_3$ are placed at $(0,0,e_1)$ and $(0,0,e_1+e_2)$, respectively. Finally, the second current source, $\mathcal{E}_4$, located at $(0,0,2e_1+e_2)$ has current $-I$. In this linear electrode configuration, $2e_1+e_2$ is the distance between the current electrodes, with $e_1$ as the distance between the outer current electrode and the voltage electrode and $e_2$ as the distance between the inner voltage electrodes (Supplementary Fig.2).

By virtue of the principle of superposition, the potential generated by the current sources evaluated anywhere in the domain is the sum of the potential generated by each individual current source. Therefore, the potential evaluated at the location of the first voltage electrode, $e_1$, is $\Omega(0,0,e_1) = \Omega_{\mathcal{E}_1}(0,0,e_1)+\Omega_{\mathcal{E}_4}(0,0,e_1)$, where $\Omega_{\mathcal{E}_1}$ and $\Omega_{\mathcal{E}_4}$ are the potentials generated by the first and second current sources, respectively. This can be equivalently represented as $\Omega(0,0,e_1) = \Omega_{\mathcal{E}_1}(0,0,e_1) - \Omega_{\mathcal{E}_1}(0,0,e_1+e_2)$ due to reciprocity. In the same manner, the potential at the second voltage electrode is $\Omega(0,0,e_1+e_2) = \Omega_{\mathcal{E}_1}(0,0,e_1+e_2) - \Omega_{\mathcal{E}_1}(0,0,e_1)$. Taking the difference in potential between the voltage electrodes yields $\Delta\Omega = \Omega(0,0,e_1) - \Omega(0,0,e_1+e_2) = 2\left[\Omega_{\mathcal{E}_1}(0,0,e_1) - \Omega_{\mathcal{E}_1}(0,0,e_1+e_2)\right]$.

Following Ohm's law, the impedance $Z$ is the potential difference by the injected current, namely

$$Z = \frac{\Delta\Omega}{I} = \frac{2\left(\Omega(0,0,e_1) - \Omega(0,0,e_1+e_2)\right)}{I}. \quad (37)$$

Using the three-layer cylindrical model of the wrist as derived in Supplementary Discussion 5.1, the potential is the superposition of the primary and secondary potentials given in Equations (31) and (34) evaluated at the location of the electrodes, that is, $\Omega(0,0,e_1)I^{-1} = (\Omega_1^p(0,0,e_1) + \Omega_1^s(0,0,e_1))I^{-1}$. Plugging in yields

$$\frac{\Omega_1^p(0,0,e_1)}{I} = \frac{1}{2\pi\sigma_1 e_1} + \frac{2}{\pi^2\sigma_1}\int_0^\infty\int_0^\infty \frac{1}{u}\frac{\widetilde{\Gamma}_1(u)e^{-2us_1}}{1-\widetilde{\Gamma}_1(u)e^{-2us_1}}\cos(\lambda e_1)\,\mathrm{d}\beta\mathrm{d}\lambda$$

$$\frac{\Omega_1^s(0,0,e_1)}{I} = -2(1-\Gamma_1)(1-\Gamma_2)\int_0^\infty\int_0^\infty \frac{1}{u}\frac{\widetilde{F}(\lambda)e^{-uh}}{(1-\widetilde{\Gamma}_1(u)e^{-2us_1})(\Gamma_2\Gamma_1 e^{-2ud_2}+1)}\cos(\beta d)\lambda\cos(\lambda e_1)\,\mathrm{d}\beta\mathrm{d}\lambda.$$

with $\widetilde{F}(\lambda)$ given in Equation (36).

## 5.3 Numerical validation

We simulated a three-layered isotropic domain with skin (2.5 mm), subcutaneous adipose tissue (1 mm), and muscle (infinite element domain) in COMSOL (COMSOL Multiphysics, Burlington, MA) to validate our forward electrodynamic and BioZ models in Equations (31), (34), and (37). The radial artery was modeled as a cylinder located 6.5 mm from the surface of the skin layer with a radius of 1.5 mm. Tissue conductivity values were assigned based on the IT'IS database for 50 kHz. The simulations were performed in the electric currents module using a frequency domain study with Dirichlet boundary conditions for the current source (100 $\mu$A) and ground surfaces defined as infinite element domains on the edges of the half space. The relative tolerance was set to $5{\cdot}10^{-4}$ and the maximum number of iterations was limited to $10^4$. We calculated the absolute error between COMSOL and the analytical model along the $z$-direction as $|V_{\text{model}} - V_{\text{COMSOL}}|$ and the relative difference as $|2(V_{\text{model}} - V_{\text{COMSOL}})/(V_{\text{model}} + V_{\text{COMSOL}})|$.

## 5.4 Volume impedance density

The sensitivity $\mathbf{S}$ is defined as $\mathbf{S} = (\mathbf{J_1}/I_1) \cdot (\mathbf{J_2}/I_2)$,[97] where $\mathbf{J_1}$ is the current density field when the current is applied between the current injecting electrodes and $\mathbf{J_2}$ is the current density field when the alternating electrical current is injected between the voltage sensing electrodes. The current density fields are scaled by the injected currents, $I_1$ and $I_2$. Multiplying the sensitivity by the tissue's frequency-dependent admitivitty, $\gamma$, yields the volume impedance density (VID) $\mathbf{K}$,

$$\mathbf{K} = \gamma^{-1}\mathbf{S} = \gamma^{-1}(\mathbf{J_1}/I_1) \cdot (\mathbf{J_2}/I_2). \tag{38}$$

The volume resistance density (VRD) and volume reactance density (VXD) are the real and imaginary parts of $\mathbf{K}$, respectively. Formulating Equation (38) this way enables a seamless extension to complex-valued input currents and tissue electrical properties.

The resultant impedance for a specified region can be approximated by multiplying each applicable voxel's VID by its volume

$$Z_v = \Sigma_{i=1}^{N} K_i v_i. \tag{39}$$

Here, $Z_v$ represents the total impedance for a specific region consisting of $N$ voxels and $v_i$ and $K_i$ are a voxel's volume and volume impedance density, respectively. We leveraged Equation (39) to calculate the impedance contribution of each tissue in the human phantom models (Fig.3m-n, Extended Data Fig.1, Supplementary Fig.10–14 and Supplementary Table 5).

With VRD and VXD values calculated for every tissue in the simulation domain, we then identified the volume that contributed 95% to the overall measurement by summing the largest individual VRD and VXD voxels, regardless of sign, until the threshold was reached. We then calculated the length of the contributing voxels along the $x$, $y$, and $z$ directions to estimate penetration depth. This method identifies the voxels with the highest contribution (positive or negative) to the measurement configuration.

## 5.5 Electrode impedance characterization

We used a three-electrode electrochemical cell to perform electrochemical analysis on the stainless-steel electrodes with an SP-150 potentiostat (Bio-Logic, France). The electrolyte was 0.01M phosphate-buffered saline (1xPBS, Cytiva) diluted to 10 $\mu$S/cm with deionized water to mimic the conductivity of skin. 8"x12" folded stainless steel mesh was used as a counter electrode, and an Ag/AgCl/1M KCl electrode was used as a reference electrode. We took cyclic voltammetry measurements by sweeping the stimulus voltage from -0.5 to 0.5 V at 100 mV/s for three cycles. Cyclic voltammetry stability measurements were taken by sweeping from -0.5 to 0.5 V at 1 V/s for 500 cycles. We performed electrical impedance spectroscopy analysis with a 20 mV amplitude stimulus sine-wave to obtain impedance magnitude and phase at frequencies between 10 Hz and 1 MHz, the typical range of frequencies for BioZ measurements.

# 6 Supplementary Discussion 6. Experimental study

## 6.1 Standard protocol approvals, registrations, and informed consents

The procedures followed were in accordance with the ethics standards of the responsible committee on human experimentation as overseen by Institutional Review Board . After learning of the study, participants completed prescreening questions. They were then contacted by the study coordinator. Subjects meeting inclusion and exclusion criteria were invited to participate in the study. All subjects enrolled provided written informed consent prior any measurement.

## 6.2 Overall study design

This single-time investigation was designed as a prospective feasibility study using our BioZ smartwatch. The goal was to monitor BP and blood velocity in healthy subjects in the lab setting (Group 1), healthy and unhealthy patients in the clinic setting (Group 2), and HF patients in the acute care setting (Group 3). Group 1 subjects were recruited through the University study locator platform. Group 2 patients were scheduled during routine visits. Group 3 patients were scheduled for assessments both before and after left ventricular assist device (LVAD) implantation in the cardiovascular intensive care unit (ICU). The experimental groups are summarized below and in Supplementary Table 6.

### 6.2.1 Group 1

We collected data for Group 1 while subjects were seated comfortably and relaxed with backs and arms supported, the middle of the upper arm at heart level, legs uncrossed, and feet flat on the floor. Group 1 subjects were further divided into three subgroups with each subgroup following a different study protocol intended to induce dynamic BP changes. Group 1a and Group 1c subjects underwent additional anthropometric and ultrasound measurements of the left arm, and 30 minutes of baseline sensor data collection with simultaneous blood oxygen (SpO2), Finger BP, BioZ, PPG, Brachial BP, and electrocardiogram (ECG) (Supplementary Fig.15 and Extended Data Fig.3). A summary of Group 1's vitals during each trial is provided in Supplementary Table 7.

- Group 1a. Subjects also performed an exercise protocol, consisting of three exercise trials (Fig.4b, Extended Data Fig.4 and Supplementary Fig.15). The exercise's resistance was increased on each trial from mild to intense. Specifically, for the treadmill protocol, participants ran on a treadmill at a pace of three, five, and seven mph for three minutes each, with a one minute rest between increments. Similarly, for the stationary bike protocol, participants pedaled at a cadence that produced 50, 75, and 100 W for three minutes each, with a one minute rest between increments. Following completion of each exercise intervention, data was recorded for an additional 10 minutes while the subject was seated and recovering from exercise.
- Group 1b. Subjects performed the Valsalva maneuver and cold pressor test. During the Valsalva trials, subjects plugged their nose with their right hand and exhaled against closed airways for 15 seconds. The maneuver was repeated three times with 3 minute breaks in-between. During the cold pressor test, subjects immersed their right hand in a bucket of ice water for 60 seconds. The test was repeated three times with 10 minute breaks in-between.
- Group 1c. Subjects performed stationary bike exercise, the Valsalva maneuver, hand grip exercise, and a deep breathing routine. The stationary bike exercise was divided into three trials.

Each trial consisted of three minutes of exercise followed by a minute of rest. Subjects' HR was monitored and a cadence was prescribed to elevate the HR 25% beyond their resting HR. The Valsalva maneuver was identical to Group 1b subjects. During the hand grip test, subjects performed an initial exercise consisting of six sets of eight repetitions per hand with a weight at least 30% of their max grip strength. Five minutes after the initial exercise, a maintenance phase of three sets of eight repetitions with the right hand was performed. Only the right hand was used during the maintenance phase to avoid interference with the finger BP monitor and pulse oximeter. Finally, the deep breathing intervention entailed subjects performing a guided 4-7-8 breathing routine.

### 6.2.2 Group 2

Group 2 patients participated in data collection after routine visits (Fig.4c and Extended Data Table 3). Simultaneous finger BP and BioZ were collected while subjects were seated comfortably and relaxed with backs and arms supported, the middle of the upper arm at heart level, legs uncrossed, and feet flat on the floor.

### 6.2.3 Group 3

Group 3 patients underwent simultaneous BioZ and clinical-grade telemetry monitoring in the cardiovascular ICU. Reference BP data was taken using an arterial line (A-line) with the patient in the supine position and as close to euvolemic as possible (Extended Data Fig.5).

## 6.3 Study enrollment

### 6.3.1 Group 1

Inclusion criteria: Adults $>$18 years of age at the time of screening, capable of running at a pace of 7 mph on a treadmill. Exclusion criteria: 1. Diagnosed with cardiovascular disease. 2. Implantable electronic device 3. Pregnant. In total, $N$ = 76 provided written informed consent, $N$ = 53 were enrolled in Group 1a, $N$ = 5 were enrolled in Group 1b, and $N$ = 18 were enrolled in Group 1c. $N$ = 1 Group 1a participant was excluded from the dataset and analysis due to inability to complete the exercise protocol (Supplementary Fig.15 and Extended Data Table 2).

### 6.3.2 Group 2

Inclusion criteria: Adults $>$18 years of age at the time of screening. A total of $N$ = 86 patients provided written informed consent and were enrolled in the study. Among them, $N$ = 32 patients were classified in the hypertensive cohort, $N$ = 22 patients were classified in the CVD cohort, and $N$ = 32 patients were classified in a third cohort because they had conditions that were neither HTN or CVD. For simplicity, hereafter we refer to this latter group as the "other" cohort (Fig.4, Extended Data Fig.5, Extended Data Table 3, Supplementary Fig.26).

### 6.3.3 Group 3

Inclusion criteria: Adults $>$18 years of age at the time of screening, admitted to the ICU at the time of data collection, existing placement of a radial arterial catheter for BP measurement. Exclusion criteria: patient unable to make own medical decisions. For Group 3 participants, $N$ = 3 provided written informed consent and enrolled in the study, $N = 3$ patients completed the study and were included in our data analyses, which did not include machine learning prediction due to the limited sample size (Supplementary Fig.29).

## 6.4 Data collection

### 6.4.1 Group 1

After Group 1 participants provided written informed consent, they completed a two-hour study protocol with ultrasound measurement, anthropometric measurements, and wearable sensor data collection (Supplementary Fig.15). The anthropometric measurements examined left arm wrist circumference, forearm length (from the olecranon to the ulnar styloid), bicep length (from the olecranon to the acromion), and hand to elbow length (from the middle phalanx of the third digit to the cubital fossa) (Supplementary Fig.16). The ultrasound imaging included transverse B-mode sonograms (SuperSonic Mach 30, LH20-6 transducer, Hologic, Marlborough, MA) to record the diameter and depth of the left radial, ulnar, and brachial arteries (Extended Data Fig.2). We did not control for the cardiac phase while measuring the diameter, which could have introduced a source of experimental uncertainty. Additional longitudinal Doppler sonograms recorded blood flow in the left radial, ulnar, and brachial arteries. The radial and ulnar arteries were imaged at the anterior wrist in line with the radial styloid process, while the brachial artery was imaged 3 cm proximal to the cubital fossa. For all ultrasound measurements we applied a light probe pressure to avoid compressing the arteries. During the wearable sensor data collection, we recorded continuous arterial BP using the NOVA Plus (Finapress Medical Systems, Enschede, Netherlands). The Finapress finger cuff was placed on the left third digit, and the Finapress arm cuff was placed around the left bicep to measure BP at the proper palmar digital artery and the brachial artery, respectively. An ECG was recorded with the Nova Plus ECG module using three leads on the left shoulder, right shoulder, and lower left abdomen. Respiration was measured using the NOVA respiration module with a thoracic impedance measurement through lead II of the ECG. SpO2 was recorded using the Nellcor DS100A pulse oximeter (Nellcor Puritan Bennet, Hayward, CA) placed on the left index finger. We measured PPG with the Verity Sense (Polar Electro, Kempele, Finland) placed on the left anterior forearm just below the cubital fossa. Finally, we collected left wrist BioZ using our newly developed smartwatch placed on the anterior left wrist (Bsecur, Belfast, Ireland) (Fig.1 and Supplementary Fig.1). Anthropometric and ultrasound measurements were consistent across subgroups, and data collection hardware was uniform, except for the absence of PPG data in Group 1b.

### 6.4.2 Group 2

Group 2 patients received standard office BP measurements using an automated oscillometric device operated by a trained medical assistant. When patients presented with elevated BP, an automated office BP technique was applied while patients were alone and the resulting SBP and DBP averaged across measurements. Next, we collected simultaneous BP using the VitalStream 5 (CareTaker Medical, Charlottesville, VA) placed on the left third digit and BioZ from the smartwatch placed on the left anterior wrist. Patients were asked to sit still and refrain from talking for the measurement duration (Supplementary Fig.26).

#### 6.4.3 Group 3

Group 3 patients underwent 15-minute measurement sessions with simultaneous ICU-grade patient vitals and BioZ (Supplementary Fig.29). The vital signs recorded included invasive BP, 5-lead ECG, blood oxygen, and respiration. BP was collected using a gold standard radial artery catheter with a pressure transducer. Blood oxygen was derived from a photoplethysmography signal placed on a right digit. Respiration was derived from the 5-lead ECG signal. Finally, BioZ was recorded with our smartwatch placed on the anterior wrist. When a radial artery catheter or other intravenous catheters were placed on the left wrist, we moved the BioZ watch to the right wrist. All vital sign sensors were routed through a Phillips MX70 patient monitor. We used VitalRecorder software to export all patient telemetry data in real time.[98]

## 6.5 Data preprocessing

We developed an application-specific preprocessing pipeline in MATLAB R2024a (Mathworks, Inc., Natick, MA) to prepare the experimental time-series data for analysis and ML tasks (Supplementary Fig.17). Specifically, the experimental data underwent filtering, resampling, peak finding, synchronization, segmentation, and noise rejection.

#### 6.5.1 Filtering

The resistance data was filtered using a 4th-order Butterworth low-pass filter with cutoff frequency at 3.5 Hz to attenuate signals outside the frequency bands with high cardiovascular power. Note the BP signal can be viewed as a DC coupled signal. We found that low-pass filtered data including BioZ baseline trends improved BP estimation as it is related to the mean arterial pressure. The BP data was passed through a 4th-order Butterworth low-pass filter with a cutoff frequency of 10 Hz.

#### 6.5.2 Resampling

After filtering, the signals were resampled to a uniform frequency of 100 Hz using cubic interpolation. The resampling process corrected discrepancies in sensor sampling rates, ensuring consistent signal timestamps, which facilitated accurate peak detection and signal synchronization.

#### 6.5.3 Peak-finding

To aid in synchronization, we required each signal's inter-beat interval (IBI), the time between adjacent heartbeats. We identified systolic peaks (maxima in BP and minima in BioZ) as zero-crossings in the signals' first derivative with hyper-parameters for minimum signal distance and peak prominence. To aid in peak detection, we split the time-series signal into sub-segments of fixed time. We then used an adaptive technique to set the peak finding hyper-parameters based on the heart rate calculated by taking the Fourier transform of a sub-segment and identifying the maximum power at cardiovascular frequencies. We then calculated the IBI as the time difference between adjacent systolic peaks.

#### 6.5.4 Synchronization

Next, the signals required synchronization as they originated from separate sensors with varying start and stop times. With uniform sampling frequencies and reliable timestamps, each signal featured a time shift relative to the other signals. To accurately identify this time shift, we implemented a brute force algorithm that minimized the median absolute error between a fixed signal's IBI and the remaining signals' IBI. Specifically, we fixed one signal and then buffered the other into sub-segments of length $N$ with overlap $N-1$. We then computed the median absolute difference between the sub-segment at every possible location in the fixed signal. We repeated the process for every sub-segment and determined the optimal shift by selecting the shifted sub-segment that produced the lowest median absolute difference. This method works well in IBI signals containing noise from imperfect peak-finding or experimental noise.

### 6.5.5 Segmentation

After synchronization, the signals were segmented into individual periods based on the feet of the systolic peaks in the BioZ signal. We ensured pressure periods strictly started and ended at a systolic peak foot by padding the period with a fixed number of adjacent points and then cutting the period based on the maxima in the second derivative. Next, the segments were resampled using cubic interpolation to a uniform length of $n$=100 samples in preparation for machine learning analysis.

### 6.5.6 Signal quality assessment

The segmented periods were individually examined for signal quality. Resistance periods were rejected as noisy if they met any of the following criteria: heart rate (HR)$<$40 beats per minute (BPM), HR$>$120 BPM, peak-to-peak resistance $<$7 mohms, and peak-to-peak resistance $>$1 ohm. BP periods were rejected as noise if they met any of the following criteria: HR $<$40 BPM, HR $>$120 BPM, pressure $>$220 mm Hg, pressure $<$40 mm Hg, pulse pressure $<$15 mm Hg, pulse pressure $>$80 mm Hg. The systolic peak location in the BP and resistance signals was evaluated, and segments with peaks occurring beyond the 35th index were rejected. Additional preprocessing steps were applied prior to machine learning as described in Supplementary Discussion 7.

After processing the experimental data, we built two datasets from the Group 1 and Group 2 trials with 223,156 and 45,568 raw segments, respectively. In total, we collected over 265k paired simultaneous pressure and resistance periods from $N$ = 161 unique subjects in Group 1 and Group 2. The datasets are summarized in Supplementary Table 8.

# 7 Supplementary Discussion 7. Signal-tagged physics-informed neural network

## 7.1 Data sources and preprocessing

### 7.1.1 Data sources

We consider five different base datasets: three experimental and two hybrid experimental-synthetic, summarized in Supplementary Table 8. We collected two experimental datasets; see Group 1 and Group 2 in Supplementary Discussion 6. The other experimental dataset is the Graphene-HGCPT[99] dataset. For Group 1 and Group 2 base datasets, we created different datasets for population-wide and subject-specific models (Supplementary Table 8). For the hybrid experimental-synthetic datasets, we use PulseDB,[100] an online database of experimental radial BP signals, and our forward model to generate a set of corresponding synthetic wrist surface resistance signals (Extended Data Fig.2). We also created an additional dataset by varying biological parameters within the forward model. Details on the generation of these two hybrid experimental-synthetic datasets are described below.

Each dataset consists of a set of paired BP, $\{\mathbf{p}_k\}_{k\in[N_d]}$, and resistance signals, $\{\mathbf{R}_k\}_{k\in[N_d]}$. $\mathbf{p}_k$ is a length $n$ discretization of a time-dependent brachial or radial (depending on the dataset) BP signal corresponding to a single heartbeat and $\mathbf{R}_k$ is the time-synchronized length $n$ resistance signal either measured or synthesized at the wrist. Here, $N_d$ is the number of signals in the dataset, prior to any filtering (Supplementary Table 8).

### 7.1.2 Hybrid experimental-synthetic dataset generation

From PulseDB,[100] we uniformly sampled 100k periods of radial BP signals that met the following criteria: 90 mm Hg $<$ SBP $<$140 mm Hg, 60 mm Hg $<$ DBP $<$90 mm Hg, and 40 BPM $<$ HR $<$ 200 BPM. When referring to the forward model (or synthetically generated results), we consider a single elastic artery with no-reflection outflow boundary conditions (Supplementary Discussion 4–5). When working with synthetic data and referring to the forward model, we consider resistance at the same location as BP.

For the first PulseDB[100] synthetic dataset, all parameter values are set at their nominal values besides the BP waveform and the period of the waveform which are provided by PulseDB.[100] This results in a dataset of size 100k with each radial BP waveform corresponding to one resistance waveform. For the PulseDB[100] synthetic with biological variability datasets, biological variability is considered over the parameters of radial artery radius, radial artery wall Young's modulus, blood hematocrit, heart rate, radial mean wall shear stress, subcutaneous adipose fat thickness, and skin conductivity. Random parameter values are sampled with rejection sampling (to ensure they remain in their range) from a normal distribution with mean at their nominal values and standard deviation such that three standard deviations would leave their parameter range (Supplementary Table 1). First, for each radial BP waveform, all parameters are randomly sampled and corresponding resistance waveforms are generated. This is repeated five times to build a training dataset with 500k BP-resistance pairs. Second, seven individual testing datasets of size 50k are built in each of which one parameter is randomly sampled while the others are fixed at their nominal values (including the period/heart rate) and the first 50k BPs are used from PulseDB. In all synthetic results, the centerline artery depth is kept constant while the artery radius oscillates about its instantiated value and the blood conductivity oscillates by the Maxwell-Fricke equations.

### 7.1.3 Data preprocessing

Here we describe the data preprocessing performed for the three experimental datasets; these filtering steps were not used for the synthetic data.

We first apply the BP signal quality assessments described in Supplementary Discussion 6 (Supplementary Fig.17). Then using the following procedure, we identify "clean" blocks of $m$ consecutive resistance periods and paired "clean" BP signals.

1. We find all possible consecutive blocks of resistance that are $m$ periods long, forming: $\{\tilde{\mathbf{r}}_k\}_{k\in[\tilde{N}_d]}$, where $\tilde{\mathbf{r}}_k = [\mathbf{R}_k, \mathbf{R}_{k-1}, \ldots, \mathbf{R}_{k-(m-1)}]$. The $\tilde{\mathbf{r}}_k$ can be overlapping; each $\mathbf{R}_k$ could appear in $m$ different $\tilde{\mathbf{r}}_k$. Requiring $m$ consecutive resistance periods significantly reduces the data size (i.e. $\tilde{N}_d < N_d$). We now take as data the BP/resistance pairs, $\{(\mathbf{p}_k, \tilde{\mathbf{r}}_k)\}_{k\in[\tilde{N}_d]}$. In practice, $m = 5$ in all cases except forward synthesized data where $m = 1$.

2. We split the BP/resistance data into initial training and test sets as follows. We first select a number of subjects whose data will be entirely in the test set. The number of subjects, whose data is entirely within the test set, and the manner in which they are chosen, are dataset dependent. For Group 1 data, we randomly select 2 test-exclusive subjects from Group 1a, 1 test-exclusive from Group 1b, and 1 test-exclusive from Group 1c. For Group 2 data, we choose test-exclusive subjects in a random manner that excludes subjects with the least correlated $\{(\mathbf{p}_k, \tilde{\mathbf{r}}_k)\}_{k\in[\tilde{N}_d]}$ pairs. We first fit a linear regression model, $\mathbf{A}$, to the entire data set, $\{(\mathbf{p}_k, \tilde{\mathbf{r}}_k - \bar{r}_k)\}_{k\in[\tilde{N}_d]}$. We then evaluate the linear fit for each subject, and compute average subject error, $e_i = \frac{1}{\tilde{N}_{s_i}} \sum_k^{\tilde{N}_{s_i}} \|\mathbf{A}(\tilde{\mathbf{r}}_k - \bar{r}_k) - \mathbf{p}_k\|_2^2$, here $\tilde{N}_{s_i}$ is the $i$-th subject's number of data points. We then randomly select 6 subjects from the bottom 50th percentile of $\{e_i\}_{i\in[S_n]}$, where $S_n$ is the number of Group 2 subjects. These 6 subjects constitute the test-exclusive subjects for the Group 2 dataset. For the Graphene-HGCPT[99] dataset, there are only 7 subjects in total, so we do not select any test-exclusive subjects.

   For the remaining subjects, not relegated to the test set, we select ≈10% of their data to be in the test set and the other ≈90% for the training set. We avoid intra-subject data leakage by preventing any overlap between the resistance signals $\tilde{\mathbf{r}}_k$ in the training and test sets. To limit the amount of data lost while preventing intra-subject data leakage, we first break the data into thirds and select a contiguous ≈10% block from each third for the test set, resulting in three contiguous ≈3% blocks in the test set.

3. We perform a 10 dimensional principal component analysis (PCA) on the resistance training data. Each resistance signal, $\tilde{\mathbf{r}}_k$, in both the training and test sets are transformed to the 10 dimensional latent space and then reconstructed using the inverse PCA transform to obtain a reconstructed resistance, $\hat{\mathbf{r}}_k$. We compute each signal's relative reconstruction error, $E_k = \frac{\|\tilde{\mathbf{r}}_k - \hat{\mathbf{r}}_k\|}{\|\tilde{\mathbf{r}}_k\|}$. In both the training and test sets, we discard the resistance signals with reconstruction error, $E_k$, above the $95$-th percentile.

4. We repeat Step 3 for the BP training and test sets, using an 8 dimensional PCA.

5. Finally, we compute the mean of $\tilde{\mathbf{r}}_k$, $\bar{r}_k$, and form the augment resistance $\mathbf{r}_k = [\tilde{\mathbf{r}}_k - \bar{r}_k, \mathbf{y}]$, where $\mathbf{y} = [y_1, \ldots, y_p]$ is augmented data, which is dataset dependent. For all datasets, $y_1$ is the period (60/heart rate) of $\mathbf{R}_k$ and $y_2 = \bar{r}_k$. The remainder of the supplementary inputs, $\{y_i\}_{i=3}^p$, are subject dependent physiological parameters. For the Group 1 data, we include eight physiological parameters: wrist circumference, forearm length, radial artery depth, radial artery major diameter, radial artery minor diameter, brachial artery depth, brachial artery major diameter, and brachial artery minor diameter and one derived parameter: average radial artery radius. For Group 2 and the Graphene-HGCPT[99] dataset, there are no available physiological parameters. For the synthetic dataset, we trained a base model with no physiological parameters, but to study biological variability, we trained additional models with Young's Modulus, artery radius, mean wall shear stress, hematocrit, subcutaneous fat thickness, and skin conductivity.

The resulting filtered BP / augmented resistance pairs, $\{(\mathbf{p}_k, \mathbf{r}_k)\}_{k\in[\hat{N}_d]}$ constitute the final training and test data. For the datasets considered in Supplementary Table 8, we report the number of samples before and after filtering by the above process. The Group 1 dataset has a total of 84,044 cleaned BP/resistance pairs, the Group 2 dataset has 30,656, and the Graphene-HGCPT[99] dataset has 8,928. All reported model accuracies are in regard to the entire or a subset of the test dataset.

## 7.2 Overview of our machine learning approach

In its simplest form, our objective is to solve the prediction/inverse problem: predict an entire BP vector $\mathbf{p}_k$ from an augmented resistance signal vector $\mathbf{r}_k$. Our predictive model has two components (Fig.4a and Supplementary Discussion 7.3).

The first component of our model is a 1D-CNN signal encoder, designed to encode an augmented resistance signal, $\mathbf{r}_k$, into a reduced-dimensional representation, $\mathbf{R}_k^E$. This encoder is first trained in isolation with a constituent decoder and supplemental adaptive filtering (AF) network,[101–103] to minimize a combination of reconstruction and adaptive-filtering loss. The signal encoder is then affixed to the second component of our model: a novel signal-tagged physics-informed neural network (sPINN).

The second component of our model maps a space-time vector, $(r, z, t)$, and encoded resistance information, $\mathbf{R}_k^E$, to pressure and axial-radial velocity fields $[p, u_z, u_r]_k\,(r, z, t)$, evaluated at $(r, z, t)$. By modifying a traditional PINN[104,105] via augmenting the space-time input with an additional (encoded) resistance signal, $\mathbf{R}_k^E$, the model can make interpretable, space-time blood flow-field predictions that (i) are dependent on the data sample $k$ and (ii) are grounded in physics; in particular, we require that the fields satisfy pulsatile fluid flow in an elastic tube, described by the incompressible N.S. equations in a cylindrical tube with elastic boundary conditions (Supplementary Discussion 7.5.3).

Together, the two components comprise a novel model architecture that solves a physics prediction/inverse problem and allows several different types of model loss to be evaluated (physics, predictive error, autoencoder reconstruction, adaptive filtering/contrastive) as described in Supplementary Discussion 7.5. To train the model, we break our training into three phases: two phases of each component training individually, then a final phase of the components training in conjunction (Supplementary Discussion 7.6).

We further distinguish between two kinds of models. Subject-specific models are trained on the data from only a single subject (Extended Data Fig.5). Population-wide models take data from a population of subjects; we view this as calibration-free models and a more challenging prediction problem given inter-subject variability (Fig.4a).

Below, we describe the model components and training strategy in detail.

## 7.3 Model description

### 7.3.1 Signal encoder component

The first component of our model is a signal encoder (Fig.4a), designed to extract dimension-reduced features from an input resistance signal in a manner that is faithful to both its own reconstruction and constructive for the prediction objective. The signal encoder is given by $g(\mathbf{r}_k; \theta_E) = \mathbf{R}_k^E$. Here, $\theta_E$ are the encoder parameters, $\mathbf{r}_k$ is the augmented resistance signal, and $\mathbf{R}_k^E$ is the resultant $M$-dimensional output encoded signal. In practice, we take $M = 47$ in all models.

We train the signal encoder (update $\theta_E$) during training phase 1 and phase 3. During phase 1, the input signal $\mathbf{r}_k$ is encoded into $\mathbf{R}_k^E$. $\mathbf{R}_k^E$ is then fed into both a decoder and an AF network (Fig.4a). The decoder and AF networks are only used for the purpose of pretraining the signal encoder and are discarded after the phase 1 training step. Phase 1 and phase 3 training steps are further described in Supplementary Discussion 7.6.

The signal encoder can be viewed as a task-informed autoencoder or a supervised autoencoder, as it is an autoencoder with the addition of a supervised loss on the representation layer (AF network).[103,106] As in a traditional autoencoder, the decoder mirrors the encoder, to produce a reconstructed augmented resistance signal, $\hat{\tilde{\mathbf{r}}}_k$. This encoder-decoder relationship helps to structure the encoding space, ensuring that the reduced-dimensional $\mathbf{R}_k^E$ is a feature-rich representation of the original resistance signal.[107,108] The AF neural network (NN) maps $\mathbf{R}_k^E$ to a (time-dependent) BP prediction, $\hat{\mathbf{p}}_k$. It is intended to further shape the encoding space, and hence $\mathbf{R}_k^E$, to contain information conducive for predicting $\mathbf{p}_k$. We use the terminology "AF network" as it is similar to (nonlinear) adaptive filtering;[101,102] the encoded signal, $\mathbf{R}_k^E$, is "adapted" to encode information for predicting BP, $\hat{\mathbf{p}}_k$.

#### Implementation details

The signal encoder is a convolutional NN with three 1D convolutional layers, each followed by a $\mathrm{ReLu}$ activation function. The input to the signal encoder is the resistance signal, $\tilde{\mathbf{r}}_k - \bar{r}_k \in \mathbb{R}^{m \cdot n}$. The output of the covolutional block is concatenated with the augmented data, $\mathbf{y}$ and fed into a fully connected linear layer block with three linear layers and $\mathrm{ReLU}$ activation functions between them. The final output, $\mathbf{R}_k^E \in \mathbb{R}^M$, represents the encoding of the augmented resistance data.

The decoder's architecture is a largely inverse of the encoder's: beginning with a linear layer block with three linear layers each followed by a $\mathrm{ReLU}$ activation function. The output of this linear layer block is then fed into a convolution transpose block, which consists of three transposed 1D convolutional layers with $\mathrm{ReLU}$ activation functions between the inner layers. The input to the decoder is $\mathbf{R}_k^E$, and the output is the reconstruction, $\hat{\tilde{\mathbf{r}}}_k$.

The AF network is a fully connected NN with four linear layers and $\mathrm{ReLU}$ activation functions. The input is $\mathbf{R}_k^E$, and the output is a predicted BP vector, $\hat{\mathbf{p}}_k \in \mathbb{R}^n$.

### 7.3.2 Signal-tagged physics-informed neural network model component

The second component of our model is the sPINN. Employing PINN methods allows us to impose physics on our BP predictions–specifically incompressible N.S. in cylindrical coordinates.[104,105] We combine standard PINN space-time coordinate inputs and loss objectives with supplementary features, allowing the resultant learned fluid-flow solutions to be dependent on both the governing physics and accompanying input variables.

We attach our first component: signal encoder, to our second component: signal-tagged physics-informed neural network (sPINN, Fig.4a). Let $f$ represent this combined model (signal encoder + sPINN), with the signal encoder still parameterized by $\theta_E$ and the sPINN parameterized by $\theta_P$, where $f\colon \mathbb{R}^{m\cdot n+p+3} \to \mathbb{R}^3$. In particular, $f(\mathbf{x}_k; \theta_P, \theta_E) = \mathbf{u}_k$, where

$$\mathbf{x}_k = \begin{bmatrix} r, & z, & t, & \mathbf{r}_k \end{bmatrix}, \quad \text{and} \quad \mathbf{u}_k = \begin{bmatrix} p(r,z,t), & u_z(r,z,t), & u_r(r,z,t) \end{bmatrix}.$$

Here, $\mathbf{x}_k$ is comprised of a point in space and time $(r, z, t)$ —like a classical PINN— and the augmented resistance $\mathbf{r}_k$. This augmented signal is what we identify as tagging the solution $\mathbf{u}_k$. Our model's ability to specify $\mathbf{u}_k$, while still satisfying the governing physical equations, from a broad set of input resistance signals is what distinguishes it from a standard PINN framework.

The augmented resistance signal, $\mathbf{r}_k$, is first partitioned from $\mathbf{x}_k$ and passed through the signal encoder. The signal encoder processes $\mathbf{r}_k$, encoding the signal into $\mathbf{R}_k^E$, see Supplementary Discussion 7.3.1 and Fig.4a. For this second component of the model, there are also two distinct training phases: phase 2 and phase 3. During phase 2, the parameters of the signal encoder ($\theta_E$) are fixed. Then during phase 3, the signal learns with respect to the same objectives as the sPINN, allowing the encoding to generalize to extracting features relevant for "tagging" $\mathbf{u}_k$.

In both phases 2 and 3, after encoding, the augmented resistance signal, $\mathbf{R}_k^E$, is concatenated with the spatiotemporal variables, $(r, z, t)$, and input to the sPINN. We use a rotationally symmetric cylindrical domain to describe the radial artery. The inputs $r$ and $z$ represent points in the spatial domain and $t$ is a time during the period of interest. The output of the sPINN, $\mathbf{u}_k = \begin{bmatrix} p(r,z,t), & u_z(r,z,t), & u_r(r,z,t) \end{bmatrix}$, predicts the BP, axial velocity, and radial velocity in the artery. Analogous to a traditional PINN, this characterization is a solution to the governing physical equation, the N.S. partial differential equation. Our sPINN learns a set of solutions to the N.S. with each pertaining to the input augmented resistance data.

**Implementation details**

The sPINN is a fully connected NN with six hidden layers, each with 100 neurons. The input and output layers have $3 + M + |y|$ and 3 neurons, respectively. We use the $\tanh$ activation function between each hidden layer, so that it is possible to compute second order derivatives of the output with respect to the input.

## 7.4 Summary of biophysical model enforced by the signal-tagged physics-informed neural network

To model blood flow in a section of the radial artery, we consider the problem of fully developed fluid flow in a flexible cylindrical domain. Here, we describe the N.S. equations to model fluid flow, boundary conditions, and initial conditions. We use a non-dimensionalization $\tilde{r} = r/\bar{a}$, $\tilde{z} = z/\bar{a}$, $\tilde{t} = 2\pi t/T$, $\tilde{u}_z = u_z/\bar{U}$, $\tilde{u}_r = u_r/\bar{U}$, and $\tilde{p} = p/\bar{P}$, and going forward drop the tilde on the variables e.g. $r = \tilde{r}$. Here, $\bar{a}$ is an average arterial radius, $\bar{U}$ is a characteristic velocity, and $\bar{P} = \bar{U}^2\rho$ is a characteristic pressure. The non-dimensionalization we choose for our model is slightly different from that used in Supplementary Discussion 4; this is because it is preferred to have $O(1)$ values for $r$, $z$, $t$, $u_z$, $u_r$, and $p$ for training a NN.

In this section, we denote the solutions generated by our forward fluid model, $\mathcal{F}$, by

$$\begin{bmatrix} p^{\mathcal{F}}, u_z^{\mathcal{F}}, u_r^{\mathcal{F}} \end{bmatrix}(r, z, t) = \mathcal{F}(r, z, t, \mathbf{p}_k).$$

Where the inputs to $\mathcal{F}$ are a space-time point, $(r, z, t)$, and BP vector, $\mathbf{p}_k$, and the outputs are the pressure, axial velocity, and radial velocity evaluated at $(r, z, t)$.

Breaking our domain into pieces, let the interior of the artery be represented by $\mathcal{D}(t)$, the arterial wall by $\partial\mathcal{D}_1(t)$, and the inlet and outlet of the tube by $\partial\mathcal{D}_2(t)$ (Supplementary Fig.19),

$$\begin{aligned}\mathcal{D}(t) &= \{(r, z) \colon r \in [0, 1 + \eta(z, t)),\ z \in (0, L_{\mathcal{D}}/\bar{a})\} \\ \partial\mathcal{D}_1(t) &= \{(r, z) \colon r = 1 + \eta(z, t),\ z \in (0, L_{\mathcal{D}}/\bar{a})\} \\ \partial\mathcal{D}_2(t) &= \{(r, z) \colon r \in [0, 1 + \eta(z, t)),\ z \in \{0, L_{\mathcal{D}}/\bar{a}\}\}.\end{aligned}$$

Since we are trying to model fluid motion for a single period of BP, we consider $t \in [0, 2\pi]$. Additionally, we let $L_{\mathcal{D}} = 0.1L$, again for NN training purposes.

The non-dimensional N.S. equations — relevant to our model — for $r, z \in \mathcal{D}(t)$, $t \in (0, 2\pi]$ are

$$\alpha^2 \frac{\partial u_z}{\partial t} + Re\left(u_z \frac{\partial u_z}{\partial z} + u_r \frac{\partial u_z}{\partial r} + \frac{\partial p}{\partial z}\right) - \left(\frac{\partial^2 u_z}{\partial z^2} + \frac{\partial^2 u_z}{\partial r^2} + \frac{1}{r}\frac{\partial u_z}{\partial r}\right) = 0 \tag{40a}$$

$$\alpha^2 \frac{\partial u_r}{\partial t} + Re\left(u_z \frac{\partial u_r}{\partial z} + u_r \frac{\partial u_r}{\partial r} + \frac{\partial p}{\partial r}\right) - \left(\frac{\partial^2 u_r}{\partial z^2} + \frac{\partial^2 u_r}{\partial r^2} + \frac{1}{r}\frac{\partial u_r}{\partial r} - \frac{u_r}{r^2}\right) = 0 \tag{40b}$$

$$\frac{\partial u_z}{\partial z} + \frac{u_r}{r} + \bar{a}\frac{\partial u_r}{\partial r} = 0, \tag{40c}$$

where, $\alpha = \bar{a}\sqrt{\frac{\omega}{\nu}}$, $Re = \frac{\bar{a}\bar{U}}{\nu}$ and $\nu = \frac{\mu}{\rho}$ ($\mu$, $\rho$, and $\omega$ are defined in Supplementary Discussion 4). On the arterial wall, we consider two stipulations on the velocity field which constitute our wall boundary conditions. For axial velocity, we impose the no-slip condition: axial velocity is zero at the wall. For the radial velocity, we impose the no-penetration condition: radial velocity at the boundary matches the speed of the wall. This allows our model to capture inflation/deflation of arterial wall: maximizing physical coherence and alignment with the forward fluid model. A more in depth discussion of this is found in Sections 7.5 and 7.6. Concretely,

$$u_z(r, z, t) = 0, \qquad (r, z, t) \in \partial\mathcal{D}_1(t) \times (0, 2\pi] \tag{40d}$$

$$u_r(r, z, t) = \frac{\alpha^2}{Re}\frac{\partial \eta}{\partial t}(z, t), \qquad (r, z, t) \in \partial\mathcal{D}_1(t) \times (0, 2\pi]. \tag{40e}$$

At the inlet and outlet we have time dependent boundary conditions which drive the flow in the artery,

$$p(r, z, t) = p^{\mathcal{F}}(z, t) \qquad (r, z, t) \in \partial\mathcal{D}_2(t) \times (0, 2\pi] \tag{40f}$$

$$u_z(r, z, t) = u_z^{\mathcal{F}}(r, z, t) \qquad (r, z, t) \in \partial\mathcal{D}_2(t) \times (0, 2\pi]. \tag{40g}$$

Here, the forward model constrains the pressure at the inlet and outlet to be radially constant. We don't specify a boundary condition for pressure, $p$ on $\partial\mathcal{D}_1(t) \times (0, 2\pi]$ because the reference pressure is only valid for $r \in [0, 1]$. We don't specify a boundary condition for the radial velocity $u_r$ on $\partial\mathcal{D}_2(t) \times (0, 2\pi]$ as our model is able to learn radial velocity at the inlet and outlet. Finally, the initial conditions are given by

$$p(r, z, t) = p^{\mathcal{F}}(r, z, t) \qquad (r, z, t) \in \mathcal{D}(t) \times \{0\} \tag{40h}$$

$$u_z(r, z, t) = u_z^{\mathcal{F}}(r, z, t) \qquad (r, z, t) \in \mathcal{D}(t) \times \{0\}. \tag{40i}$$

Similar to the boundary conditions, we do not consider an initial condition for $u_r$ at $t = 0$.

## 7.5 Loss functions

### 7.5.1 Loss for first component, signal encoder

During phase 1 training, the signal encoder, decoder, and AF network parameters ($\theta_E$, $\theta_D$, and $\theta_{AF}$, respectively) are updated to minimize a compound loss function,

$$\mathcal{L}_{\text{Enc}}(\theta_E, \theta_D, \theta_{AF}) = \lambda_{\text{Recon}}\mathcal{L}_{\text{Recon}}(\theta_E, \theta_D) + \lambda_{\text{AF}}\mathcal{L}_{\text{AF}}(\theta_E, \theta_{AF}) \tag{41a}$$

consisting of the reconstruction and adaptive filtering terms,

$$\mathcal{L}_{\text{Recon}}(\theta_E, \theta_D) = \sum_{k\in[K]} \|\hat{\tilde{\mathbf{r}}}_k - \tilde{\mathbf{r}}_k\|^2 \tag{41b}$$

$$\mathcal{L}_{\text{AF}}(\theta_E, \theta_{AF}) = \sum_{k\in[K]} \|\hat{\mathbf{p}}_k - \mathbf{p}_k\|^2. \tag{41c}$$

Here, $\hat{\tilde{\mathbf{r}}}_k$ is the decoder's reconstructed augmented resistance signal and $\tilde{\mathbf{r}}_k$ is the reference augmented resistance signal. Similarly, $\hat{\mathbf{p}}_k$ is the AF network's predicted BP signal and $\mathbf{p}_k$ is the reference BP signal. We choose the loss weights to be $\lambda_{\text{Recon}} = 0.65$, and $\lambda_{\text{AF}} = 0.35$.

### 7.5.2 Loss for second component, signal encoder + signal-tagged physics-informed neural network

The signal encoder + sPINN model ($f$) parameters ($\theta_P$ during phase 2; $\theta_E$ and $\theta_P$ during phase 3) are updated to minimize a PINN-like compound loss function consisting five terms: partial differential equation (PDE) residuals, initial conditions (IC), boundary conditions (BC), data, and stitching. The combination the first four of these loss terms constitute the well posed trio required for $f$ to produce solutions to the N.S. equations. The stitching term is added for practicality to ensure solution continuity due to challenges arising within our dynamic domain. Utilizing the N.S. equations as governing equations for the residual calculations, ensures our predicted pressure and velocity fields are consistent with the physical principles of fluid motion in an artery. We cannot, however, establish an explicit input $\mathbf{R}_k$ dependent physical relationship within the N.S. equations, e.g. with a forcing term. Consequently, we do not consider them to be directly parameterized by the input resistance features. Rather, we view the boundary and initial conditions to be input dependent and drive solution variability. Concretely, our model learns the extant relationship between $\mathbf{R}_k$ and $\mathbf{p}_k$ (with additional information present in $\mathbf{r}_k$), using boundary and initial conditions as a proxy.

A collocation point is defined as a point in the arterial spatio-temporal domain: $(r_i, z_i, t_i) \in (\mathcal{D}(t) \cup \partial\mathcal{D}_1(t) \cup \partial\mathcal{D}_2(t)) \times [0, 2\pi]$ (Supplementary Fig.19). Each collocation point is combined with a resistance-physiological signal, $\mathbf{r}_k$, and input to $f$ to generate a value for the velocity and pressure fields. Note that collocation point sampling is independent of sample $k$ selection. The resultant values–$[p, u_z, u_r]\,(r, z, t)$–are then compared with a reference field at the same $(r, z, t)$, or used in the PDE residual calculation. The boundary conditions and initial conditions are prescribed via this point-wise comparison with $\mathbf{p}_k$ informed reference fields, generated using the fluid portion of our forward physics model. The comparison between the reference velocity and pressure fields from each $\mathbf{p}_k$ and the modeled fields using the corresponding $\mathbf{r}_k$, is the explicit learnable contrast from which the sPINN infers feasible initial and boundary conditions using the input $\mathbf{R}^E_k$.

The loss function is given by

$$\mathcal{L}_{\text{total}}(\theta_P, \theta_E) = \mathcal{L}_{\text{PDE}} + \lambda_{\text{IC}}\mathcal{L}_{\text{IC}} + \lambda_{\text{BC}}\mathcal{L}_{\text{BC}} + \lambda_{\text{data}}\mathcal{L}_{\text{data}} + \lambda_{\text{stitch}}\mathcal{L}_{\text{stitch}}, \tag{42a}$$

where $\lambda_i$, $i \in \{\text{IC}, \text{BC}, \text{data}, \text{stitch}\}$ are adaptive weights applied during computation, and

$$\mathcal{L}_{\text{PDE}}(\theta_P, \theta_E) = \sum_{i\in[N_{\text{PDE}}]} \sum_{j\leq 3} \left(e_j(u_{z,i}, u_{r,i}, p_i, r_i) r_i^2\right)^2 \tag{42b}$$

$$\mathcal{L}_{\text{IC}}(\theta_P, \theta_E) = \sum_{i\in[N_{\text{IC}}]} \left[p(r_i, z_i, t_i = 0) - p^{\mathcal{F}}(r_i, z_i, t_i = 0)\right]^2 + \left[u_z(r_i, z_i, t_i = 0) - u_z^{\mathcal{F}}(r_i, z_i, t_i = 0)\right]^2 \tag{42c}$$

$$\mathcal{L}_{\text{BC}}(\theta_P, \theta_E) = \sum_{i\in[N_{\text{BC}}]} u_z^2(r_i, z_i, t_i) + \left[u_r(r_i, z_i, t_i) - (\alpha^2/Re)\partial_t \eta(z_i, t_i)\right]^2 \tag{42d}$$

$$\mathcal{L}_{\text{data}}(\theta_P, \theta_E) = \sum_{i\in[N_{\text{data}}]} \left[p(r_i, z_i, t_i) - p^{\mathcal{F}}(r_i, z_i, t_i)\right]^2 + \left[u_z(r_i, z_i, t_i) - u_z^{\mathcal{F}}(r_i, z_i, t_i)\right]^2 \tag{42e}$$

$$\mathcal{L}_{\text{stitch}}(\theta_P, \theta_E) = \sum_{i\in[N_{\text{stitch}}]} \left[p(r_i, z_i, t_i) - p(r = 1, z_i, t_i)\right]^2 . \tag{42f}$$

Here, $\mathcal{L}_{\text{PDE}}$ is a squared sum of N.S. residuals $e_j(u_{z,i}, u_{r,i}, p_i, r_i)$ (given by Equations (40a), (40b), and (40c) for $j = 1, 2, 3$, respectively) and multiplied by $r_i^2$ for normalization, evaluated at a batch of $N_{\text{PDE}}$ interior collocation points (points in $\mathcal{D}(t)$),

$$\{(u_{z,i} = u_z(r_i, z_i, t_i),\ u_{r,i} = u_r(r_i, z_i, t_i), p_i = p(r_i, z_i, t_i), r_i)\}_{i\in[N_{\text{PDE}}]} .$$

$\mathcal{L}_{\text{IC}}$ is a squared sum of the discrepancy between the predicted and prescribed initial conditions, evaluated at a batch of $N_{\text{IC}}$ collocation points with $t_i = 0$. $\mathcal{L}_{\text{data}}$ is a squared sum of the discrepancy between the predicted and prescribed inlet and outlet boundary conditions, evaluated at a batch of $N_{\text{data}}$ collocation points (points in $\partial\mathcal{D}_2(t)$). Again, the prescribed reference pressure and velocity fields in $\mathcal{L}_{\text{IC}}$ and $\mathcal{L}_{\text{data}}$ are generated using the forward fluids model and $p^{\mathcal{F}}(r_i, z = 0, t_i)$ is simply measured data. $\mathcal{L}_{\text{BC}}$ is a squared sum of the discrepancy between the predicted and prescribed (no-slip, no-penetration) boundary conditions, evaluated at a batch of $N_{\text{BC}}$ collocation points (points in $\partial\mathcal{D}_1(t)$). The no-penetration condition is $(\alpha^2/Re)\partial_t\eta(z, t) = u_r(r = 1 + \eta, z, t)$, where $\eta$ is the arterial wall displacement from a base $\bar{a}$ arterial radius, and $u_r$ is the radial velocity. In order to enforce these boundary conditions at the arterial wall ($r = 1 + \eta$), we first must model the wall location itself. This dynamic wall method causes the spatial arterial domain to change with $t$ and $z$, and necessitates our last loss term, $\mathcal{L}_{\text{stitch}}$. $\mathcal{L}_{\text{stitch}}$ is a squared sum of the discrepancy between the modeled pressure near the found wall position and the modeled pressure at the base wall position. This ensures radial continuity of the modeled pressure between the base cylindrical arterial domain and the space created (or removed), by the moving wall method.

### 7.5.3 Elastic wall motion method

To accommodate arterial wall elasticity and subsequent radial inflation/deflation (i.e. radial location change), we introduce a novel dynamic range for the sPINN's boundary points at the wall. Our method iteratively considers proposed wall positions, then based on the sPINN's own modeled pressure and velocity determines a new satisfactory candidate wall location. We model this elastic wall motion to agree as much as possible with our forward fluid model for maximum consistency and physiological relevance without introducing axial displacement (Supplementary Discussion 4); thus, wall displacement satisfies the following non-dimensional ordinary differential equation (Equation (6b) in Supplementary Discussion 4 with $\xi = 0$ and $a = \bar{a}$):

$$\alpha^4 \frac{\partial^2 \eta}{\partial t^2} = \frac{\rho}{\rho_w} \left[ Re^2 \frac{\bar{a}}{h} p_w - \frac{E_\sigma \bar{a}^2}{\mu\nu} \eta \right] , \tag{43}$$

where $p_w = p(r = 1 + \eta, z, t) - \hat{B}_0/\bar{P}$ is the difference between the pressure at the current wall position and the baseline inflated artery pressure, $\hat{B}_0$.

Exploiting the no penetration boundary condition (40e) and NN derivative functionality, we can represent

$$\frac{\alpha^2}{Re}\frac{\partial^2\eta}{\partial t^2}(z,t) = \frac{\partial u_r}{\partial t}(r,z,t), \qquad (r,z,t) \in \partial\mathcal{D}_1(t) \times (0, 2\pi]$$

and rearrange Equation (43) to find

$$\eta = \frac{\mu\nu}{E_\sigma \bar{a}^2} Re \left( Re\frac{\bar{a}}{h}p_w - \alpha^2 \frac{\rho_w}{\rho}\frac{\partial u_r}{\partial t}\right). \tag{44}$$

Using Equation (44), we iteratively update the wall position via, $r = 1 + \eta$, until convergence (Supplementary Fig.20). In practice, we update the wall position five times as we empirically determined that this was sufficient. Once a final wall position is obtained, we can evaluate $f$ on $\mathcal{D}(t)$ (in particular, near $r(z,t) = 1 + \eta(z,t)$ where the domain has changed), calculate the no-penetration boundary condition using Equation (44), and enforce arterial wall boundary conditions with $\mathcal{L}_{\text{BC}}$.

## 7.6 Training methods

The two components of our model (signal encoder + sPINN) are trained individually, during phase 1 and phase 2, and then together, during phase 3. Here we describe the training process for phases 1, 2, and 3. We emphasize that, unlike a traditional PINN, we learn many solutions to the N.S. PDE system in Equation (40), each corresponding to an input signal, $\mathbf{r}_k$. The reference signals for the initial and inlet/outlet boundary conditions are dependent on both the input $(r, z, t)$ and $\mathbf{p}_k$. Thus, the sPINN learns an $\mathbf{r}_k$ input dependent set of fluid solutions, for each ($\mathbf{p}_k$, $\mathbf{r}_k$) pair in the training set based on the varying initial and inlet/outlet reference conditions.

The signal encoder + sPINN model are implemented in pytorch.[109] Pytorch's autograd functionality is used not just to differentiate the loss functions in Equation (42), but also to compute the derivatives that appear in the N.S. residuals and the wall motion method. Some Jacobian computation implementation is adapted from Lu et al.[110] All model parameters ($\theta_E$, $\theta_D$, $\theta_{AF}$ and $\theta_P$) are updated using NADAM.

There are some differences in the training methodology depending on the dataset and model choice; these are summarized in Supplementary Table 9. For example, for synthetic and subject-specific models, we only train the model using phase 3. The number of epochs for each model differs based on dataset and model type, as the datasets differ significantly in both size and problem complexity. The number of total model parameters is the total number of trainable parameters that constitute both the signal encoder and feed forward neural network sPINN part the model. The number of trainable parameters in the signal encoder depends on the number of input resistance signals ($m$), the number of physiological parameters ($p$), and the complexity of the problem (for example, we increased the size of the signal encoder's fully connected block for the Group 1 dataset population model). The number of trainable parameters for the sPINN, excluding the signal encoder, is the same across all trained models (Supplementary Discussion 7.3.2).

All model training was performed using single GPUs on GPU servers . These servers have an assortment of NVIDIA H100NVL, L40, P40, V100, A100, and RTX2080Ti GPUs. The training times (Supplementary Table 9) are approximate and depend on a variety of model and server factors: the number of epochs and size of the training dataset, the number of trainable parameters, shared compute/memory slowdowns at CHPC, and the type of GPU used during training. The training times for the subject-specific models are an approximate per-subject model training time multiplied by the number of subjects; in practice many of the component subject models are trained simultaneously, requiring access to multiple GPUs. In our implementation, we did not optimize for the GPU memory bandwidth of individual GPUs, so a significant amount of the allocated GPU memory on the newer GPUs (H100NVLs and L40s) was not used. Additionally, the data generation process (Supplementary Discussion 7.6.2) was performed using CPUs without GPU memory pinning; this certainly contributed to idle GPU memory bandwidth and longer training times, especially for larger training datasets.

### 7.6.1 Phase 1, pretraining the signal encoder

In phase 1, we pretrain the first component (signal encoder) of our model, described in Supplementary Discussion 7.3.1 independently from the second component (sPINN) to minimize $\mathcal{L}_{\text{Enc}}$ in Equation (41) (Fig.4a).

To train, first, a minibatch of $K$, $(\mathbf{p}_k, \mathbf{r}_k)$ pairs are selected from the training data set. The $\{\mathbf{r}_k\}_{k=1}^{K}$ are encoded into $\{\mathbf{R}_k^E\}_{k=1}^{K}$, via the signal encoder. $\{\mathbf{R}_k^E\}_{k=1}^{K}$ are then input into both the decoder and the AF network. We then use the decoder and AF network outputs ($\{[\mathbf{R}_k, \ldots, \mathbf{R}_{k-m}]\}_{k=1}^{K}$ and $\{\mathbf{p}_k\}_{k=1}^{K}$, respectively) to compute $\mathcal{L}_{\text{Enc}}$ in Equation (41). The model parameters, $\theta_E$, $\theta_D$, and $\theta_{AF}$, are updated to minimize $\mathcal{L}_{\text{Enc}}$.

The decoder and AF are only used for pretraining the signal encoder and are not used after phase 1 training. The signal encoder is attached to the sPINN, and $\theta_E$ is further updated during phase 3 training.

### 7.6.2 Phase 2, pretraining the signal-tagged physics-informed neural network using a frozen pretrained signal encoder

In phase 2, we pretrain the second component (sPINN) of our model, described in Supplementary Discussion 7.3.2 to minimize $\mathcal{L}_{\text{total}}$ in Equation (42) (Fig.4a). We attach the pretrained signal encoder to the sPINN and freeze its parameters ($\theta_E$) to prevent them from updating. Since the signal encoder is pretrained to extract features that encode both resistance information and the BP-resistance relationship, we think of the sPINN as having a baseline of informative signal “tagging”, which it can use to learn to emulate blood flow, without the additional burden of learning this feature extraction, simultaneously.

To train, we first build a minibatch of inputs for the first and second components of the model (signal encoder + sPINN), defined as: $\{\mathbf{x}_j\}_{j=1}^{N} = \{(r_j, z_j, t_j, \mathbf{r}_j)\}_{j=1}^{N}$. Here, $N = N_{\text{BC}} + N_{\text{IC}} + N_{\text{data}} + N_{\text{PDE}} + N_{\text{stitch}}$ is the size of a minibatch and is comprised of model input data corresponding to the five different loss terms in Supplementary Discussion 7.5.2. To describe the sampling process, we denote $N'_{\text{BC}} = N_{\text{BC}}$, $N'_{\text{IC}} = N_{\text{IC}}$, $N'_{\text{data}} = N_{\text{data}}/2$, $N'_{\text{PDE}} = N_{\text{PDE}}$, and $N'_{\text{stitch}} = N_{\text{stitch}}$. To build a minibatch of model inputs, $\{\mathbf{x}_n\}_{n=1}^{N}$, we perform the following.

1. We sample $K$ pairs ($\{\mathbf{r}_k\}_{k=1}^{K}$, $\{\mathbf{p}_k\}_{k=1}^{K}$) from the training data set.
2. For each signal pair ($\mathbf{r}_k$, $\mathbf{p}_k$), $\sum_{i\in\{\text{BC,IC,data,PDE}\}} N'_i$ collocation points, $(r, z, t)$, are sampled from the relevant domains (Supplementary Fig.19), i.e.,

$$(r, z, t) \in \begin{cases} \partial\mathcal{D}'_1 \times (0, 2\pi], & i = \text{BC} \\ \mathcal{D}' \times \{0\}, & i = \text{IC} \\ \partial\mathcal{D}'_2 \times (0, 2\pi], & i = \text{data} \\ \mathcal{D}' \times (0, 2\pi], & i = \text{PDE} \end{cases},$$

where

$$\mathcal{D}' = \{(r,z) \colon r \in [0,1),\ z \in (0, L_{\mathcal{D}}/\bar{a})\} \tag{45a}$$
$$\partial\mathcal{D}'_1 = \{(r,z) \colon r = 1,\ z \in (0, L_{\mathcal{D}}/\bar{a})\} \tag{45b}$$
$$\partial\mathcal{D}'_2 = \{(r,z) \colon r \in [0,1),\ z \in \{0, L_{\mathcal{D}}/\bar{a}\}\}\,. \tag{45c}$$

3. We sample $N'_{\text{data}}$ supplementary data points, $(r,z,t) \in \partial\mathcal{D}'_2 \times \{t_{\text{dia}}, t_{\text{sys}}\}$, where $t_{\text{dia}}$, $t_{\text{sys}}$ are the times at which the diastole and systole pressure occur for the pressure signal $\mathbf{p}_k$. These collocation points are added to the data points to enhance the fluid solutions predictions at systole and diastole; the total number of data collocation points is $N_{\text{data}} = 2N'_{\text{data}}$.

4. For the ic and data collocation points, reference signals are generated using the forward model, i.e. $\{[p^{\mathcal{F}}, u_z^{\mathcal{F}}, u_r^{\mathcal{F}}]\,(r_j, z_j, t_j) = \mathcal{F}(r_j, z_j, t_j, \mathbf{p}_k)\}_j$, where $j \in [N_{\text{IC}}]$ and $j \in [N_{\text{data}}]$. These collocation points and reference values are used to evaluate $\mathcal{L}_{\text{IC}}$ and $\mathcal{L}_{\text{data}}$ in Equation (42).

5. Each collocation point, $(r_j, z_j, t_j)$, is then concatenated with $\mathbf{r}_k$ to form model inputs: $\{\mathbf{x}_j\}_{j\in[N_i]} = \{(r_j, z_j, t_j, \mathbf{r}_n)\}_{j\in[N_i]}$.

6. The model is evaluated at $\mathbf{x}_{\text{BC}} = \{\mathbf{x}_j\}_{j\in[N_{\text{BC}}]}$ to obtain $\mathbf{u}_{\text{BC}} = \{\mathbf{u}_j\}_{j\in[N_{\text{BC}}]}$, where $\mathbf{u}_j = [p, u_z, u_r]_j = f(\mathbf{x}_j; \theta_E, \theta_P)$. The pressures, $\{p(r=1, z_j, t_j)\}_{j\in[N_{\text{BC}}]}$, extracted from $\mathbf{u}_{\text{BC}}$ are reference pressures used in step 8 below.

7. For each of the $N_{\text{BC}}$ fluid solutions generated by the model, $\mathbf{u}_{\text{BC}}$, we use the elastic wall motion method described in Supplementary Discussion 7.5.3 to update the wall location (i.e. replace $r = 1$ values in $\mathbf{x}_{\text{BC}}$ by $r = 1 + \eta(z,t)$). These updated $r$ values replace those in $\mathbf{x}_{\text{BC}}$. We iterate this procedure five times, and the fifth set of $r$ values define the model's artery wall; that is, $(r,z) \in \partial D_1(t)$. The resulting $\mathbf{x}_{\text{BC}}$ inputs are used to evaluate $\mathcal{L}_{\text{BC}}$ in Equation (42).

8. We then sample two new sets of $r$ values ($\{r_j\}_{j\in[2N_{\text{BC}}]}$), uniformly between the original ($r = 1$) wall position and the final wall location ($r = 1 + \eta(z,t)$).

   For the first set of new $r$ values, $\{r_j\}_{j\in[N_{\text{BC}}]}$, we create a copy of $\mathbf{x}_{\text{BC}}$, and replace its $r$-values with this new set of $r$-values (keeping everything else the same). These points $(r,z)$ are in $\mathcal{D}(t)$. We then relabel this set of model inputs as PDE inputs and add them to the set of $\mathbf{x}_{\text{PDE}}$ inputs. There are now $N_{\text{PDE}} = N'_{\text{PDE}} + N'_{\text{BC}}$, total inputs: $\{\mathbf{x}_j\}_{j\in[N_{\text{PDE}}]}$.

   The second set of $r$ values is similarly inserted into a different copy of $\mathbf{x}_{\text{BC}}$. These points $(r,z)$ are also in $\mathcal{D}(t)$. These inputs are relabeled to become the stitch points, $\mathbf{x}_{\text{stitch}}$, and serve to stitch the domains $\mathcal{D}'$ and $\mathcal{D}(t)$. $\mathbf{x}_{\text{stitch}}$ and their paired reference pressures, $\{p(r=1, z_j, t_j)\}_{j\in[N_{\text{stitch}}]}$, are used to evaluate $\mathcal{L}_{\text{stitch}}$ in Equation (42).

   See Supplementary Fig.19 for an example illustration of a set of BC, IC, data, PDE, stitch collocation points associated with a particular $(\mathbf{p}_k, \mathbf{r}_k)$ pair.

9. The above process yields $K$ minibatches each of size $N$. These minibatches are combined and we resample $K$ final minibatches of size $N$, ensuring that there are $N_i$ of each collocation point type $i \in \{\text{PDE}, \text{IC}, \text{BC}, \text{data}, \text{stitch}\}$.

Each minibatch $\{\mathbf{x}_j\}_{j\in[N]}$ is input to the model, and the loss terms in Supplementary Discussion 7.5.2 are evaluated, using the model output, $\{\mathbf{u}_j\}_{j\in[N]} = f(\{\mathbf{x}_j\}_{j\in[N]}; \theta_P, \theta_E)$, and the relevant reference solution points. During this phase (phase 2) of training, only $\theta_P$ is updated. The loss weights $\lambda_{\text{IC}}$, $\lambda_{\text{IC}}$, $\lambda_{\text{data}}$, and $\lambda_{\text{stitch}}$ are updated prior to the loss computation as prescribed by Jin et al.[111] and Wang et al.[112] We iterate this process, choosing batches of $K$ pairs ($\{\mathbf{r}_k\}_{k=1}^{K}$, $\{\mathbf{p}_k\}_{k=1}^{K}$) until the model has seen all training pairs, which constitutes one epoch. Further details for the learning rate scheduler, number of epochs trained, and starting learning rate can be found in Supplementary Table 9.

### 7.6.3 Phase 3, full model training: signal-tagged physics-informed neural network + signal encoder

After pretraining the sPINN, via phase 2, we unfreeze $\theta_E$, and proceed in the same manner as in phase 2 to update both $\theta_E$ and $\theta_P$. This allows the signal encoder to learn with respect to the same objectives as the sPINN; thus, maximizing the relevance of the features extracted from the resistance/physiological signal, $\mathbf{r}_k$, for the purpose of predicting accurate and signal-tagged blood flow solutions.

## 7.7 Overview of machine learning results

Throughout the rest of this paper, we generally refer to our entire (signal encoder + sPINN) model as the sPINN. We trained sPINN models for each of the four datasets described in Supplementary Discussion 7.1.1 (Supplementary Table 8–9). A summary of model accuracies for the Group 1, Group 2 and PulseDB[100] datasets is given in Extended Data Table 4. Model accuracies for the Graphene-HGCPT[99] dataset can be found in Supplementary Table 11. For the Group 1 and Group 2 experimental data that we collected, blood pressure results for population-wide models are reported in Fig.4b-h, and Supplementary Fig.23,27–28. Results for subject-specific models are reported in Extended Data Fig.5 and Supplementary Fig.23. Additionally, for Group 1 and 2, a comparison of predicted and reference (measured and modeled) peak systolic blood velocity and end diastolic blood velocity is presented in Extended Data Fig.4, and Supplementary Fig.24. For the synthetic dataset generated using PulseDB[100] and our forward model, results of our model are summarized in Extended Data Fig.3 and Supplementary Fig.21. For the Graphene-HGCPT[99] dataset, results for population-wide and subject-specific models are presented in Supplementary Fig.30–31, respectively. Finally, a comparison of our ML results with competing BP studies is reported in Supplementary Table 10 and with respect to baseline traditional ML methods in Supplementary Table 11. These ML results are discussed in Supplementary Discussion 8.3.

# 8 Supplementary Discussion 8. Discussion

## 8.1 Fluid dynamics

### 8.1.1 Modeling

Modeling of arterial fluid dynamics can be broadly categorized into two approaches: the development of analytical solutions to linearized N.S. with some assumptions on the forms of the solutions and the use of numerical methods to solve variants of N.S. In the case of a cylindrical domain with irrotational flow, Poiseuille flow describes fully developed flow and Wormersly solutions describe periodic pulsatile flow in a rigid domain.[79] Additionally, as is detailed in Supplementary Discussion 4 and by Zamir,[79] with assumptions of a thin wall and using an approximate coupling of fluid and wall motion, one can derive periodic pulsatile flow in an elastic domain.

#### Wormersly flow

Gaw et al.[88,113] discuss a modeling framework that uses Wormersly solutions to derive arterial shear stresses for bulk conductivity. In form, the Wormersly solutions resemble the solutions to pulsatile flow in an elastic tube (PFET) detailed in Supplementary Discussion 4. The form of the PFET solutions primarily differ in the presence of nontrivial complex wavenumbers, $K_n$, the elasticity factor, $G_n$, in Equation (9a), the radial fluid velocities in Equation (9b), and the motion of the wall in Equations (9d) and (9e). Quantitatively, they can differ significantly at different axial locations due to the Wormersly solutions having an infinite speed of propagation, i.e., the solutions do not vary axially, while the PFET solutions have a finite speed of propagation in addition to an exponential attenuation of solutions axially.[79] This finite speed of propagation is necessary for us to model the propagation of a BP waveform from the brachial artery to the distal radial artery. Additionally, the Wormersly solutions require as input a Fourier decomposition of the axial BP gradient. This can also be derived from a centerline velocity or flow waveform. However, we are interested in learning the impacts of BP on electrical properties at the wrist so we do not assume to have knowledge of the flow or pressure gradient. The PFET model allows us to directly use a Fourier decomposition of a BP waveform as input to more easily see the impacts of pressure on BioZ.

#### Local flow theory

Another similar modeling framework relies on Ling and Atabek's "local flow theory".[114–116] In these models, additional assumptions are made about the forms of the axial fluid-flow gradient and the partial derivative of the (changing) radius with respect to pressure. As done by Shen et al.,[114,115] this involves an additional data-driven function for the axial fluid-flow gradient and an assumption of a linear relationship between the changes in vessel diameter and changes in BP. They show that with these assumptions, they are better able to match conductivity measurements in pulsatile flow than rigid Wormersly solutions.[114,115] However, similarly to with Wormersly solutions, they require a centerline velocity profile as input to their model. They also require a period of the artery radius. In contrast, our PFET model allows us to compute radius and velocity with only a BP signal as input, and allows us to tie the input of BP directly to the output of electrical quantities of conductivity and resistance.

**Numerical methods**

Several groups also use numerical methods to solve variants of the N.S. equations.[117–126] These models vary in complexity and dimension, namely from three-dimensional simulations on detailed meshes[117] to one-dimensional pressure-flow relations.[118–120,122–124,126] There also exists software for modeling of fluid-solid interactions in cardiovascular applications.[127] Although our model cannot account for patient-specific geometries, our analytical solutions provide structured knowledge on a simple arterial tree that can be interpreted as a patient-generic surrogate model and quickly rerun with new parameter values. Another benefit of our analytical modeling approach is that it provides general model solutions without the need for time-intensive fluid dynamic simulations or the laborious process of image segmentation to create patient-specific arterial models. Additionally, unlike numerical simulations of fluid dynamics –which would not offer a systematic framework for modeling fundamental relationships between fluid properties and electromagnetism parameters– our method facilitates easier sensitivity analyses to determine which parameters most significantly impact our key outputs.

### 8.1.2 Outflow conditions

Two common outflow boundary conditions imposed on arterial fluids modeling are the Windkessel (WK) and the ST.[80,83,84,121,128] In its most simple form, the WK models describe conservation of mass in a vessel by relating inflow, outflow, and stored volume in a vessel.[128] The two-element WK (WK2) model represents stored volume in terms of the time derivative of pressure and the outflow in terms of pressure, introducing the two parameters of arterial compliance and peripheral resistance.[128] This yields a differential equation outlet condition that can be solved in frequency space for the WK2 impedance. In the three and four-element WK models (WK3,WK4), information about a pressure drop in terms of the first few time-derivatives of the flow are provided along with an inertial parameter.[128,129] In each of these models, it is required to parameterize the WK from physiological data. Additionally, without updating parameters, the WK models do not distinguish between the axially-invariant Wormersly solutions and axially-attenuating elastic solutions.

We instead implemented the ST outflow boundary conditions described by Olufsen et al.[83,84] In the ST outlet boundary condition, terminal vessels are continually branched into smaller and smaller vessels until they reach some minimum radius. As was described in Supplementary Discussion 4, the daughter radii are determined by a $\xi$-law and an area ratio. Additionally, due to the effects of attenuation, we chose a smaller length-radius ratio then used by Olufsen et al.[83,84] Finally, in order to enforce arterial stiffening while not breaking our assumption that the wall thickness was sufficiently small compared to the vessel radius, we kept the wall thickness to radius ratio fixed while using a power law to increase Young's modulus as the radii decreased. This modified ST outlet allowed us to utilize our pulsatile elastic fluids solutions throughout the entire outlet and see peaking effects occur at the distal radial artery.

After the blood flow is computed within our arteries of interest, we use the resulting shear stresses to compute conductivity with modeling described by Gaw et al.[88,113] and Hoetink et al.[87] This modeling relies on the Maxwell-Fricke equations and determining the orientation constant by shear-stress induced deformation and orientation of the RBCs.[91] The deformation is determined by conservation of volume and an assumption on the relationship between RBC strain and shear stress.[87,92] The orientation is determined by the steady state of a first-order kinetics model.[88,93,113] With these components, the orientation constant can be determined pointwise by work on conductivity of ellipsoids by Fricke.[91] The pointwise conductivity is then averaged over a region of interest.[87] Note that unlike in Wormersly solutions, our fluids solutions vary axially, so we integrate both radially and axially for our bulk conductivity as is shown in Equation (20).

## 8.2 Bioimpedance

### 8.2.1 Modeling

#### Cylindrical resistor

A common model for the electrical resistance of an artery is a uniform cylindrical resistor.[130–133] Often, this model is used to express an artery's pulsatile resistance as a function of BP driven volumetric changes alone in an expanding cylinder.[130–133] Despite the simplicity, this model contains several assumptions for explaining surface BioZ measurements. First, the model assumes current flows perpendicularly to the artery's cross-section from one end to the other. In practice, this translates into applying electrical current invasively requiring two current electrodes to be placed directly on the artery. In order to satisfy this experimental condition, it would be necessary to have two disk electrodes placed inside the artery blocking the blood flow, and that is not the case in wearable non-invasive applications. Second, for skin-surface measurements, the model neglects intermediate tissues between the surface sensors and the artery. Even though the electrical properties of intermediate tissues might not change within the cardiac cycle, their low-pass frequency behavior will alter the morphology of the recorded signal.[134,135] Third, this model assumes the artery's blood is a homogeneous fluid with constant conductivity. Additionally, this model does not account for blood's electrical changes through pulsatile shear-stress induced cellular dynamics, viscosity, dependence on heart rate, as well as baseline changes due to varying physiological parameters such as hematocrit concentration or plasma as shown in Fig.3 to name a few.[88,113–115,136]

#### Equivalent electrical circuits

Previous studies to model the BioZ behavior of arteries using equivalent electrical circuits rely on describing the electrical behavior of biological tissue using lumped circuit elements.[133,137–142] Each tissue is then represented as its own equivalent electrical circuit, oftentimes a parallel resistor capacitor network to model the tissue and fluid frequency behavior, then connected to other tissues through serial or parallel connections.[133,138–142] These models aim to consider the electrical contribution of intermediate tissues to surface BioZ data. These empirical equivalent circuits are valuable for simplifying the complex nature originating BioZ data into an electrical circuit parameter of interest; however, they represent an oversimplification and lack a direct biophysical relationship describing the interaction between electricity, tissues, and fluids. Alternatively, there is available a biophysics-based framework to model the BioZ of non-homogeneous domains considering series, parallel, and series-parallel-like circuit topologies while still maintaining their biophysical interpretation through their electrical properties.[143] Here, instead, we present an electrodynamic model that provides tractable solutions to Maxwell's equations in a more physiologically realistic three-layered model with an artery, described next.

### Three-layer cylindrical bioimpedance model

We developed our model based on previous geophysics studies that introduced electrodynamic models to describe the apparent surface resistivity of buried electric cables.[95] These initial models included at maximum two layers with the buried cable positioned in the uppermost layer. Our three-layered model builds upon these previous developments by: (1) relocating the conductive cylinder, representing the radial artery, from the uppermost layer to the lowermost layer (muscle tissue); (2) deriving explicit solutions for a three-layer model; and (3) superimposing the electrodynamic model solutions at two surface voltage recording locations to simulate a four-terminal BioZ measurement at the wrist. Our BioZ model translates changes in arterial blood volume and conductivity into a wrist surface resistance, enabling previous models describing the time-varying bulk conductivity of blood to be translated to a wrist surface resistance.[88,115]

This physiological model simulates a wrist BioZ measurement and captures fluid-driven dependencies that extend beyond the simplified cylindrical resistor model commonly used to explain arterial pulsatile resistance (Fig.3b). Our analysis revealed nonlinear relationships between wrist surface resistance and both radial artery radius and blood conductivity, with blood conductivity exerting a greater influence within the tested range (Fig.3c). We used this model to conduct local and global sensitivity analyses, which showed that radial blood pressure and artery radius significantly affect the timing of maximum and minimum resistance values (Fig.3d). By coupling the BioZ and fluid dynamics models, we demonstrated that both fluid and electrical parameters shape the morphology and baseline offset of the resistance signal (Fig.3e). The resulting synthetic signals qualitatively matched the morphology of experimentally recorded data as illustrated in Supplementary Fig.18.

We extended our analytical analysis and performed computational electrodynamic and BioZ simulations to explore the impact of more anatomically realistic subject-specific computational models (Fig.3f-h). We show in Fig.3i and Supplementary Fig.9 our electrode configuration is robust to $\pm 10$ mm changes in electrode position. Finally, our volume impedance density analysis shown in Fig.3m revealed that tissues as deep as 45.7 mm contributed to the overall measurement, confirming the configuration's penetration depth was sufficient to sense the radial artery. To explore the sensitivity of BioZ to with anatomy, previous studies attempted to perform finite element method simulations with simplified anatomical models and fixed parameters.[133,144,145] Compared to the results reported here, these models and results lacked anatomical accuracy and biological variability presented here using the Virtual Population models.[94] Our finite element method simulations extend the results from our analytical model by conducting a comprehensive sensitivity analysis, examining anatomical and physiological sources, experimental factors, isopotential lines, volume resistance and reactance density, penetration depth, and tissue contributions across four computable human phantoms.

### 8.2.2 Cuffless blood pressure monitoring

A summary of previous BioZ work attempting to measure cuffless BP is reported in Supplementary Table 10. Most existing cuffless BP work using BioZ technology has been based on PWA, PTT, or PWV principles, through empirical calibrations based on the Moens–Korteweg and Hughes equations.[137,146,147]

### Whole-body pulse transit time

The only BioZ approach available to cuffless BP estimation is based on PTT. These time-delay based methods rely on detecting PTT or Pulse Arrival Time (PAT), which is related to PWV and BP through Moens-Korteweg and Hughes equations.[148,149] These methods combine whole-body wrist-to-ankle impedance cardiography (ICG) and distal impedance plethysmogram channel (IPG) measuring changes between the knee joint level and the calf, with three-lead electrocardiogram. Then, when the pulse pressure wave enters the aortic arch and the diameter of the aorta changes, the whole-body impedance decreases. By measuring the time difference between the onset of the decrease in impedance in the whole-body ICG signal and, later, the popliteal artery signal as detected by the distal IPG, the PWV can be determined knowing the distance and the time difference between the two recording sites. Alternatively, PTT can also be determined by measuring ECG and PPG waveforms, or two PPG waveforms and detecting again the time delay between these waveforms. The main disadvantage of whole-body PTT is that application outside the research setting for continuous and wearable BP monitoring is questionable due to the complexity and inconvenience of the experimental setup.

### Localized pulse transit time

Previous studies proposed detecting BioZ time-delays recorded over a relatively short distance as a method to infer BP.[146,150] In their work, Ibrahim et al. recorded BioZ using a total of 18 to 48 electrodes placed in the wrist to extract "whole body PTT-like" time-delay differences between different combinations of electrodes. These data were then fed into a black-box ML model to infer SBP and DBP. The authors used a convolutional NN (CNN) autoencoder to reduce the dimension of the BioZ input signal. In a second step, the authors estimated the arterial pulsation in the latent space making a regression to predict systolic/diastolic BP. Huynh et al. calculated PWV by performing localized PTT measurements across a known distance using different BioZ recording locations across the wrist.[151] Hsiao et al. used a multimodal wearable device and ML to predict BP by integrating PPG and BioZ signals, with PTT, and PWA modeling.[152]

While more practical than whole-body assessments, localized PTT methods lack a robust theoretical foundation and are susceptible to confounding physiological variables, raising fundamental concerns about their accuracy.[153–156] A recent review found limited evidence that PWA and PAT approaches encode reliable BP information.[156] As a result, PTT-based systems typically require frequent recalibration, undermining their real clinical utility.[153,156] Moreover, these methods often demand greater hardware complexity, multi-site measurements, additional electrode contacts, and multi-channel acquisition systems–amplifying their vulnerability to motion artifacts and contact impedance variability.

Unlike studies that rely on PTT, here develop an analytical model BioZ model that, when combined with our forward fluid model, provides and end-to-end solution to directly relate BP to wrist surface resistance through modeling the dynamics of radial blood flow with fluctuating immersed RBCs. Since our approach does not rely on PTT methods, we require only four electrodes–the minimum for BioZ measurement. We next discuss materials used in electrode fabrication.

### 8.2.3 Electrodes

Electrodes play a pivotal role in BioZ sensing, as the quality of electrode-skin contact can impact the accuracy of data acquisition. Inadequate electrode contact can introduce measurement artifacts and pose challenges for the validation of wearable BioZ technologies. In terms of materials to serve as electrodes, liquid metal electrodes (LME), such as those based on eutectic gallium-indium, present several advantages over conventional metal electrodes, particularly in wearable applications that demand flexibility.[157–159] Due to their fluidic nature, LME offer enhanced conformability, allowing for improved skin contact during dynamic activities. In contrast to conventional metal electrodes, their inherent stretchability allows them to maintain electrical conductivity even under significant mechanical deformation, such as bending or stretching, making them a good alternative for stretchable devices. Additionally, LME stability to self-heal small deformations or cracks might be beneficial to obtain reliable performance during movement over prolonged use.

Dry metal electrodes and graphene-based electrodes each offer distinct advantages in terms of practicality, usability, biocompatibility, and cost.[159,160] Dry metal electrodes, such as silver or stainless steel, are well-established and widely used due to their low skin-electrode contact impedance at high frequencies, mechanical robustness, and compatibility with existing devices.[161,162] They do not require conductive gels, making them practical for both short-term and long-term applications. However, their rigidity can limit conformability, particularly for dynamic or flexible monitoring applications. In contrast, graphene-based electrodes offer superior flexibility, lightweight properties, and excellent mechanical adaptability.[137,163–165] Graphene is also biocompatible,[160,166] reducing the risk of skin irritation over extended use.[166] Despite these advantages, the relatively recent development of graphene technology may require additional validation before it can be implemented and used in practical applications.[163]

In addition to graphene and LME, several other advanced materials are emerging for biopotential measurement applications. One example is carbon nanotubes (CNTs), which are highly conductive, flexible, and lightweight while offering improved mechanical strength.[159,167] Like graphene, their flexibility allows them to be integrated into stretchable sensors, providing greater adaptability for dynamic environments. These electrodes are currently being explored for use in various applications, including ECG,[168] electroencephalography (EEG),[169] and neural interfaces.[170] However, like LMEs and graphene, the practical usability of CNTs is questionable since this material pose significant challenges to be considered a convenient electrode option for clinical adoption and use in relevant population.

The integration of CNTs, graphene, and LME, into wearable systems presents unique challenges including costly and specialized fabrication techniques.[159,165,167] To ensure functionality and durability, further efforts are needed to develop solutions that mitigate degradation from external environmental factors while facilitating the miniaturization, integration, and interfacing of these materials with electronic components for ultimate patient use. In contrast, metals offer practical advantages as an electrode material, particularly in applications requiring low-cost and readily interfaces with existing electronic components and integrated circuits, eliminating the need for complex integration strategies often required with flexible materials. This compatibility makes it a practical and suitable choice for integrated bioelectronic devices that do not require mechanical deformability. In this regard, our smartwatch device can be comfortably worn on the wrist and its widely accepted form factor holds potential to offer a positive patient experience in terms of usability. This feature could help address patient adherence challenges with medical wearables. Further, to mitigate perceived patient stigma and reduce the likelihood of device abandonment in future clinical studies, we considered aesthetic aspects alongside wearability and functionality when developing our device.

We characterized the quality of the stainless steel electrodes' contact performing electrical imped-ance spectroscopy (EIS) and cyclic voltammetry (Fig.1). The magnitude and phase at 50 kHz were 13,423 ± 3,204 ohms and -15.94 ± 6.03 degrees, respectively. When connected to our wearable smartwatch device, the system exhibits low noise when connected to a resistor-capacitor network simulating low and high contact impedance (Supplementary Fig.1). The resultant noise has a SD of 3.22 mohms and 4.53 mohms considering low and high contact impedances, respectively. Further, our smartwatch electrode configuration may be robust to variability in user-specific wrist placement preferences and artery depth (Fig.3i-k, Extended Data Fig.1, and Supplementary Table 12).

### 8.2.4 Experimental study

#### Study participants

Although monitoring BP and blood flow velocity after exercise may contain real-time insights of hemodynamic deficiencies that otherwise would not be detected, previous studies have only shown proof of concept measuring BP in healthy subjects at rest (Supplementary Table 10).[137,150] These implementations present significant barriers to inpatient or outpatient monitoring, primarily due to insufficient integration, difficulty of use, elevated cost, and limited availability. Our smartwatch solves these challenges with no restriction to the subject's movement. To assess the experimental accuracy of our approach under varying physiological states, we recorded data on $N$ = 75 healthy subjects and $N$ = 86 patients (Methods and Supplementary Discussion 6). Subjects wore the device on the left wrist for continuous recording during the study (Extended Data Table 2 and Supplementary Fig.15).

In our study, healthy participants in Group 1 completed the lab protocol with a male to female ratio of 39:36, age of 25.68±6.28, height of 1.73±0.1 m, weight of 71.27±11.53 kg, and body mass index of 23.7±3.28 kg/m$^2$ (Extended Data Table 1, Supplementary Table 6). Men and women featured different heights and weights ($p$ <0.001), but not age or BMI (Extended Data Table 1). Full statistics for group comparisons are shared in Supplementary Table 12.

Group 2 patients completed the study with a male to female ratio of 51:35, office SBP 128.1±18.13, office DBP 74.51±9.94, and office resting HR 72.75±13.89 (Extended Data Table 3, Supplementary Table 6). Among the HTN cohort, $N$ = 6 presented with Stage 1 HTN, $N$ = 14 stage 2 HTN, and $N$ = 12 with controlled BP.[15] A total of $N$ = 14 patients with diagnoses of HTN were included in the CVD cohort due to the presence of additional CVDs beyond HTN. Of the CVD patients with an HTN diagnosis, $N$ = 3 presented with Stage 1 HTN, $N$ = 4 with Stage 2 HTN, and $N$ = 7 with controlled BP. Among CVD patients without an HTN diagnosis, $N$ = 2 presented with Stage 1 HTN, $N$ = 3 with Stage 2, and $N$ = 3 with normal BP. Additionally, the current or prior CVDs encountered in the study included atrial-fibrillation, $N$ = 7; coronary artery disease, $N$ = 10; myocardial infarction, $N$ = 3; stroke, $N$ = 3; aortic stenosis, $N$ = 1; heart failure, $N$ = 2; aortic dissection, $N$ = 2; and transient ischemic attack, $N$ = 1. Finally, among the other cohort, $N$ = 9 presented with Stage 1 HTN, $N$ = 3 with Stage 2 HTN, and $N$ = 20 with normal BP. HTN stage classification was based on office systolic and diastolic BP measurements collected at the time of study protocol completion, and may not represent an actual clinical HTN diagnosis. It is important to note that Group 2 patients in each cohort may have had additional comorbidities or conditions not specifically screened by the study, including diabetes, $N$ = 6; asthma, $N$ = 4; hypercholesterolemia, $N$ = 5; kidney disease, $N$ = 2; cancer, $N$ = 2; and other conditions which may have influenced the BP and BioZ measurements. Within the cohorts, $N$ = 15, $N$ = 6, and $N$ = 2 patients had comorbidities, among the other, HTN, and CVD cohorts, respectively.

Group 3 completed the study with the following patient-specific information. Patient 3.1, male aged 61, had a diagnosis of non-ischemic cardiomyopathy and HF and was admitted to the hospital follow-ing cardiogenic shock post ventricular tachycardia storm and 10 shocks from his defibrillator. Patient 3.2, male aged 45 was diagnosed with HF at age 21 and developed non-ischemic cardiomyopathy with HCN-4 positive mutation on genetic testing. He was admitted to the hospital following a syncopal episode during ventricular fibrillation which corrected after three shocks from defibrillator then went into ventricular tachycardia requiring a fourth shock. Patient 3.3, a male aged 22, was diagnosed with anthracycline-induced dilated cardiomyopathy and HF at age 16. He was admitted to the hospital following septic shock with concern for right ventricle dysfunction (Supplementary Table 6).

Previous studies evaluating BioZ for cuffless BP estimation only demonstrated proof of concept in a lab setting on relatively small populations of young primarily male healthy subjects (Supplementary Table 10). Although our work is a pilot feasibility study, it represents the largest academic investigation to date of cuffless BP monitoring using BioZ. Critically, it is also the first BioZ-based BP study to collect data in real-world, non-lab settings and to enroll a clinically relevant population, including individuals with HTN and other CVDs. By focusing on high-risk patients, the study addresses a key and often missing translational gap in previous wearable BioZ studies. Including these data might be inherently more complex and challenging to obtain accurate BP results than measurement in healthy individuals.

#### Anthropometric data and ultrasound imaging

Group 1 subjects underwent anthropometric measurements of the left forearm and wrist (Extended Data Table 1 and Supplementary Fig.16). The population featured a wrist circumference of 15.5$\pm$1.62 cm, a forearm length of 25.41$\pm$2.48 cm, and an upper arm length of 32.80$\pm$3.61 cm. We found that men had larger wrist circumference ($p$ <0.001), forearm length ($p$ <0.001), and upper arm length ($p$ <0.005) than women (Supplementary Table 12). We also collected transverse B-mode, longitudinal B-mode, and pulsed Doppler sonograms to measure arterial depth, arterial major diameter, arterial minor diameter, and peak systolic velocity for the radial, ulnar, and brachial arteries (Extended Data Fig.3 and Extended Data Table 2). We found that men had larger major and minor diameters in the radial, ulnar, and brachial arteries ($p$ <0.001) than women. Men also featured deeper ulnar arteries and an increased radial artery peak systolic velocity than women ($p$ <0.05). The radial artery was more superficial ($p$ <0.01), had larger major ($p$ <0.005) and minor diameters ($p$ <0.001), and similar peak systolic velocity compared with the Ulnar artery, suggesting that the radial artery is a better candidate for noninvasive hemodynamic sensing (Supplementary Table 12).

Our ultrasound evaluation collected diameter, depth, and flow at the major arteries typically selected for noninvasive hemodynamic sensing: the brachial, radial, and ulnar arteries (Extended Data Fig.3). Our work was the first to use ultrasound to collect subject-specific anatomical information in conjunction with BioZ. These data were used to assess how incorporating physiological information influences hemodynamic prediction in our Group 1 cohort. Our finite element simulations on human phantoms revealed an intimate relationship between arterial anatomy and the BioZ signal suggesting that the collected anatomical information can be used to inform parameter selection for numerical or analytical anatomical simulations and as physiological information to contextualize ML models.

## 8.3 Machine learning

Here, we discuss our ML methodology (Supplementary Discussion 7) and ML results (Supplementary Table 11). First, we remark that it is generally challenging to compare both our ML methodology and results to the existing literature for the following reasons:

1. Relatively few studies have explored BP prediction using BioZ data, in contrast to the extensive literature based on PPG or combined PPG and ECG signals (Supplementary Table 10). This disparity likely reflects the broader availability and accessibility of commercial devices for PPG measurement and the existence of large and publicly accessible PPG datasets.

2. It is commonplace in cuffless BP studies to report mean error (ME)[137,144,150] instead of mean absolute error (MAE) or root mean square error (RMSE). While ME does reveal whether a model is biased and is part of the AAMI grading criteria, in isolation, due to cancellation of errors, it is not a suitable metric for model evaluation or model comparison.

3. It is standard practice to use an *individual user calibration* step in model training to improve BP prediction results, that either (i) fine tunes a ML model to an individual or (ii) inputs demographic data (e.g., age, sex, body mass index).[156] Studies where models have been calibrated should be interpreted as predicting BP changes relative to the calibration, rather than predicting absolute BP. We view the latter as a more challenging prediction problem.

4. Many BP ML studies suffer from *data leakage*, where information about the BP testing data is leaked (typically unintentionally and subtly) into the training data, superficially improving and hampering generalizability of ML results.[156]

As we describe our results below, we also explain how we address these challenges.

### 8.3.1 Methodology

PINNs are a deep learning technique, receiving significant interest for making predictions involving physical quantities that solve physical governing (partial differential) equations.[104,105] These models have been employed to model fluid motion and, in particular, for blood flow.[111,171,172] PINNs have been extended to neural operators, that learn the mapping from an input function (typically an initial condition or boundary condition) to an output function (typically the solution to a PDE).[110,173] The application of neural operators to a broad set of problems has recently seen the rise of an idea we consider similar to our own: solutions that are parameterized by an input.[172,174–176] Broadly, the advantage of using PINNs and neural operators over traditional ML regression methods for predicting BP is that they involve loss terms that are grounded in physics, ensuring simulated blood flow dynamics are consistent with the physics of fluid flow in a modeled artery. The implications for our sPINN model (Fig.4a) are multifold:

1. We predict the velocity field of blood flow and since we did not measure blood velocity as a vector field with full-field blood flow imaging methods, traditional loss terms cannot be used. However, by enforcing that the velocity satisfy N.S. equations, we are ensuring that the predicted velocity fields are consistent with the measured BP (Extended Data Fig.3 and Supplementary Fig.24).

2. We predict spatially- and temporally- varying BP and velocity fields, offering unprecedented hemodynamic insights beyond current cuffless BP models and opening new avenues for investigating blood flow and its interaction with the arterial wall. Notably, our ability to estimate radial blood velocity at the arterial wall may shed light on how endothelial cells detect and respond to shear stress. In conditions with endothelial dysfunction, such as atherosclerosis or HTN, these shear forces are often altered and contribute to disease progression. Measuring experimentally flow dynamic effects near the wall could enable earlier detection and more precise monitoring of vascular health.

3. Unlike traditional ML regression techniques, our method does not require knowledge of the true distribution of resistance and BP measurements to always make physically grounded BP and velocity predictions. For example, our models typically predict BP waveforms that are smoothly varying and subjectively look like BP waveforms whereas other ML methods can predict qualitatively out-of-sample BP waveforms.

In training our population-wide sPINN model (both with and without metadata), we do not use an individual user calibration step and we do not include demographic data, and therefore we view them as true calibration-free models. Population-wide sPINN models are tested using BP signals from either (i) subjects that the model has previously seen or (ii) randomly chosen test-exclusive subjects that the model has not seen. Although the former test data is from subjects that are represented in both training and test data, we do not consider this to constitute data leakage as it is not intent of this study to perform a formal validation study nor report validation metrics from this model. To explore the generalizability of our population model, we instead rely on separate pilot (and limited) cohort comprising text-exclusive subjects. Our subject-specific sPINN models are trained and tested only on data from a single subject, and more comparable to published models trained with an individual user calibration step. For population-wide sPINN models with metadata, we include physiological data and not demographic data. Unlike demographic information such as age, weight, or BMI which has been shown to correlate with BP alone,[177–179] physiological data described in Supplementary Discussion 7.1.3 serves to contextualize the experimental input BioZ signal for the PINN. In particular, as we have developed our three-layer cylindrical BioZ model (Supplementary Discussion 5 and Fig.3), we established the sources influencing the resistance signal and found the resistance is sensitive to factors outside fluid dynamics (and therefore BP), that are not modeled by the sPINN, such as the depth of the artery or the thickness of intermediate tissues like skin or subcutaneous adipose tissue.

From an ML methods perspective, the closest work to our own is Li et al.,[172] which uses a neural operator (BP-DeepONet) to estimate BP and blood flow rate from ECG and PPG data. In particular, the model penalizes predictions that do not satisfy a one-dimensional reduced N.S. equation with a time-periodic condition and a WK boundary condition. This one-dimensional model includes an assumed constitutive equation which makes the pressure a direct, algebraic function of the area for which a complex, non-functional relationship has been found in human subjects.[180] Our sPINN model is able to infer quantities that cannot be obtained from the one-dimensional model such as shear stress and axial and radial blood flow velocities anywhere in the artery. The derivation of the underlying one-dimensional equations requires a simplifying assumption on the axial velocity profile which may not capture the spatiotemporal variation in flow profile (Supplementary Video 2-8). Notably, common assumptions on the flow profile deviate most significantly with our findings near the arterial wall. Importantly, these differences at the arterial wall between Li et al.[172] and our sPINN may be especially relevant for enabling the assessment of shear stress and radial blood velocity in individuals with endothelial dysfunction, a precursor to various CVDs including atherosclerosis, HTN, diabetes, HF, and stroke (Supplementary Discussion 1).[181–185]

Importantly, cuffless BP studies that rely on PPG data lack a solid theoretical underpinning that causally relates PPG to BP.[51,62,172,186] On the other hand, we have established a theoretical physiological basis causally relating BP to BioZ (Supplementary Discussion 4–5). Since we use BioZ and not ECG+PPG data, we do not directly tabulate their BP prediction errors in Supplementary Table 10, but they are similar to our results; for their HyperBP-DeepONet model, they attain 4.93 and 2.57 mm Hg MAE for SBP and DBP predictions, respectively while our subject-specific sPINN model attains 5.55 and 2.98 mm Hg MAE (Extended Data Fig.5 and discussion below). The train/test data for this model is different from our own; 75/25% of subjects are randomly assigned to training/test datasets, but then the first 15% of data in each testing subject is used for individual fine-tuning (calibration) to improve BP prediction results. Due to this calibration step, we believe it is appropriate to compare these results to our subject-specific sPINN model results.

Other methodological distinctions between Li et al.[172] and our approach that could impact the accuracy of the BP results lie in the experimental data sources: BP-DeepONet was trained and tested exclusively on the MIMIC ECG+PPG database, which contains clinically measured baseline data from ICU patients. In contrast, our study involves direct baseline and dynamic BioZ measurements in ICU, bedside, and lab settings. Accordingly, the quality and variability of these datasets –and the associated difficulty of predicting BP– may differ markedly, reflecting the contrast between the controlled conditions of baseline ICU monitoring and the more dynamic perturbations introduced during experimental testing (e.g., Valsalva, cold pressor test, and deep breathing). Further, BioZ might offer several additional advantages compared to PPG data whose accuracy is lower in subjects with darker skin.[177,187] BioZ is less sensitive to ambient light and skin pigmentation, thus being robust across various light environments and skin tones.[144,152] Here, we also found through our volume impedance distribution analyses that BioZ might offer greater depth of penetration than green light PPG with $\approx$15 times greater depth than that of the radial artery (Fig.3n, Supplementary Fig.9, and Extended Data Fig.4).[188]

Sel et al.[189] uses a NN to estimate BP from BioZ data with a loss function that penalizes inconsistencies between the predictions and Taylor's approximation. We view their approach as a regularization term to help smooth the BP time-signal predictions, however, although the authors refer to their NN as a "PINN" (for example Figure 1 therein), their architecture does not embed the knowledge of any governing physical law in the learning process. In contrast to our sPINN that integrates the N.S. equations (Supplementary Discussion 7.4), their NN does not predict axial and radial blood velocity fields, or model arterial wall movement (Extended Data Fig.3 and Supplementary Fig.19–20). In regard to how the data is processed, our method encodes the BioZ signal using a signal encoder that is updated based on the phase 1 loss function, allowing the network to learn how to best encode the resistance signal to maximize BP predictions. This contrasts previous work, which relies on hand-selected features from the BioZ signal for input to a NN, e.g., Sel et al. (Figure 2 therein).[144,150,189] Although pretraining a signal encoding to learn compact, faithful representations is fairly standard in ML, as far as we know, using a (pretrained) signal encoding as input to a PINN represents a novel aspect of our architecture.

### 8.3.2 Results and discussion

A summary of all model accuracies for our experimental (Group 1 and Group 2) and synthesized (PulseDB[100]) datasets is given in Extended Data Table 4 (accuracies for the Graphene-HGCPT[99] dataset are tabulated in Supplementary Table 11), a comparison of our results with baseline traditional ML methods is given in Supplementary Table 11, and a comparison of our ML results with competing BP studies is given in Supplementary Table 10. Overall, our population-wide model achieves state-of-the art results for predicting BP based on BioZ cuffless studies. This is a promising step forward for cuffless BP prediction as we demonstrate the potential for population-wide and calibration-free general BioZ based inference models.

In what follows, we discuss results for models trained for each of the four datasets described in Supplementary Discussion 7.1.1 (Supplementary Table 8 and Supplementary Table 9). For each model, we report and discuss

1. The coefficient of determination, $r^2$, along with the statistical significance, $p$, obtained from linearly regressing the predicted SBP/DBP values on the ground truth values. We chose to report $r^2$ values instead of concordance correlation coefficient values to facilitate their comparison with previous studies.

2. MAE$\pm$SD for SBP/DBP predictions.

3. ME and limits of agreement (LOA) results of a Bland-Altman analysis for SBP/DBP predictions, the latter defined as $\pm$1.96 SD representing the 95% confidence interval.

4. For the waveform predictions, we report average RMSE, ARMSE $= \frac{1}{N\sqrt{n}} \sum_{i=1}^{N} \|p_i - \hat{p}_i\|_2$, and average MAE, AMAE $= \frac{1}{Nn} \sum_{i=1}^{N} \|p_i - \hat{p}_i\|_1$, where $p_i$ and $\hat{p}_i \in \mathbb{R}^n$, are the reference and predicted BP waveforms, respectively, $n = 100$ is the length of the BP waveform, and $N$ is the number of waveforms in the test set.

We believe that these statistics, considered together, are suitable for model evaluation and model comparison (Fig.4b-h, Extended Data Fig.2, 4, 5, Extended Data Table 4, Supplementary Fig.21, 28, 30, and 31).

### Blood pressure and blood velocity for Group 2 experimental dataset

For our Group 2 experimental dataset (clinic cohort), results for our population-wide model when evaluating all patients together are presented in Fig.4b-c and are further broken down by HTN patients Fig.4d-e, patients with CVD Fig.4f-g, and patients with other conditions Fig.4h-i. Our population-wide sPINN models yielded SBP and DBP regression coefficients $r^2$ = 0.774 ($p$ <0.001) and $r^2$ = 0.807 ($p$ <0.001) when evaluating all patients together, $r^2$ = 0.752 ($p$ <0.001) and $r^2$ = 0.854 ($p$ <0.001) for HTN patients, $r^2$ = 0.859 ($p$ <0.001) and $r^2$ = 0.811 ($p$ <0.001) for those with CVD, and $r^2$ = 0.588 ($p$ <0.001) and $r^2$ = 0.642 ($p$ <0.001) for other patients. The corresponding MAE$\pm$SD for SBP and DBP predictions, respectively, are 7.26 $\pm$ 8.46 and 3.89 $\pm$ 4.63 mm Hg when evaluating all patients together, 8.25$\pm$10.23 and 4.30$\pm$5.45 mm Hg for HTN patients, 6.67$\pm$6.96 and 3.30$\pm$3.24 mm Hg for those with CVD, and 6.67$\pm$7.24 and 3.86$\pm$4.47 mm Hg for other patients. A Bland-Altman analysis demonstrates the model displays a low prediction bias; the ME and LOA, for SBP and DBP, respectively, are -0.46, [-22.30, 21.37] and -0.01, [-11.87, 11.85] mm Hg when evaluating all patients together, 0.75, [-26.46, 24.96] and -0.40, [-13.99, 13.19] mm Hg for HTN patients, 0.52, [-18.35, 19.39] and 0.25, [-9.79, 9.30] mm Hg for those with CVD, and -0.81, [-20.03, 18.42] and 0.20, [-11.37, 11.76] mm Hg for other patients. These results were evaluated on test sets that exclude test-exclusive subjects. The other patients cohort had a higher prevalence of comorbidities (47%) compared to the HTN (19%) and CVD (23%) cohorts, potentially contributing to greater variance in BP and BioZ measurements.

Subject-specific results for Group 2 are presented in Extended Data Fig.5 and, again, are broken down by HTN patients, patients with CVD, and patients with other conditions. Our subject-specific sPINN models yielded SBP and DBP regression coefficients $r^2$ = 0.860 ($p$ <0.001) and $r^2$ = 0.895 ($p$ <0.001) when evaluating all patients together, $r^2$ = 0.861 ($p$ <0.001) and $r^2$ = 0.922 ($p$ <0.001) for HTN patients, $r^2$ = 0.944 ($p$ <0.001) and $r^2$ = 0.913 ($p$ <0.001) for those with CVD, and $r^2$ = 0.662 ($p$ <0.001) and $r^2$ = 0.836 ($p$ <0.001) for other patients. The corresponding MAE$\pm$SD for SBP and DBP predictions, respectively, are 6.25±5.75 and 3.3±2.97 mm Hg when evaluating all patients together, 7.16±6.45 and 3.77±3.38 mm Hg for HTN patients, 5.92±4.91 and 3.07±2.46 mm Hg for those with CVD, and 5.55±5.36 and 2.98±2.76 mm Hg for other patients. A Bland-Altman analysis demonstrates the model displays a low prediction bias; the ME and LOA, for SBP and DBP, respectively, are 0.1, [16.74, -16.54] and -0.25, [8.44, -8.95] mm Hg when evaluating all patients together, -1.01, [-19.79, 17.77] and -0.82, [-10.62, 8.99] mm Hg for HTN patients, 1.03, [-13.91, 15.97] and 0.31, [-7.39, 8] mm Hg for those with CVD, and 0.62, [-14.46, 15.69] and -0.05, [-8.01, 7.91] mm Hg for other patients.

As expected, the subject-specific models outperform the population-wide model on all metrics. However, direct comparison of results is additionally complicated by the difference between datasets used to train and test the subject-specific and population-wide models (Supplementary Table 8). We removed 21 subjects from the Group 2 subject-specific dataset because these subjects had less than 100 BP/resistance pairs in their training dataset, after preprocessing and filtering (Section 7.1). This threshold was necessary to ensure sufficient data for model training subject-specific models. However, these patients were retained in both the training and test sets for the Group 2 population-wide models, where the aggregation of data across subjects mitigated the impact of reduced number of data for certain individuals.

The results of our subject-specific models on Group 2 clinic data (MAE 6.25 mm Hg SBP and 3.30 mm Hg DBP) are generally better than or competitive with results previously published in other studies using BioZ data (Supplementary Table 10), although a fair comparison is difficult to make due to fundamental differences. For example, the results reported by Wang et al.[130] (2.01 mm hg SBP MAE and 2.26 mm Hg DBP MAE) and the follow up study[131] (2.63 mm Hg SBP MAE and 2.66 mm Hg DBP MAE) are better than ours. The results reported by Ibrahim et al.[146] (3.44 mm Hg SBP RMSE and 2.63 mm Hg DBP RMSE), assuming MAE≈RMSE, are also stronger.

However, we believe these results would be significantly degraded if tested considering real-world experimental conditions similar to ours, also including clinically relevant population. For instance, prior studies collected data exclusively from healthy individuals using wet contact electrodes,[130,146] or involved a limited sample size ($N$ = 6) comprising measurements in participants in their twenties.[131] Wet electrodes or ultraconductive materials such as graphene provide superior skin-electrode coupling, resulting in markedly lower contact impedance than stainless steel used in our study (Fig.1).

Our approach represents a fundamental departure from these studies in that we do not rely on wet electrodes or ultraconductive materials to detect low signal-to-noise pulsatile changes at the wrist. We demonstrate unprecedented high-fidelity BioZ signal detection using rigid and untreated stainless steel, embedded directly into the watch back case, allowing effortless use —particularly for older adults— without the need for gels, cables connected to external electronics, electrode replacement over time, and costly integration and manufacturing techniques.

In Supplementary Fig.24 we also present comparisons of our forward fluid modeled velocity and population-wide sPINN predicted velocity. Specifically, the cross-sectional average velocity in the middle axial 50% of the arterial domain at both systole and diastole, for which the regression coefficients are $r^2$ = 0.759 ($p$ <0.001) and $r^2$ = 0.685 ($p$ <0.001), respectively. The corresponding MAE±SD for average velocity at systole and diastole, respectively, are 3.61±4.78 and 2.36±2.76 cm/s. A Bland-Altman analysis shows a low velocity prediction bias; the ME and LOA, for average velocity at systole and diastole, respectively, are -0.13, [11.86, -11.60] and 0.41, [7.48, -6.66] cm/s. These results suggest the sPINN also predicts velocity values in line with our forward fluid model. Despite this modeled/predicted velocity agreement not necessarily being an objective or a requirement, it further highlights the physical consistency of the sPINN fluid predictions.

### Blood pressure and blood velocity for Group 1 experimental dataset

For our Group 1 experimental dataset (healthy lab cohort), results for our population-wide and subject-specific models are presented in Extended Data Fig.4. Our population-wide sPINN models yielded SBP and DBP regression coefficients $r^2$ = 0.575 ($p$ <0.001) and $r^2$ = 0.629 ($p$ <0.001). The corresponding MAE±SD for SBP and DBP predictions, respectively, are 7.24±6.98 and 5.45±5.19 mm Hg. A Bland-Altman analysis demonstrates the model displays a low prediction bias; the ME and LOA, for SBP and DBP, respectively, are 0.10, [-16.54, 16.74] and -0.33, [-15.10, 14.44] mm Hg. These results were evaluated on test sets that exclude test-exclusive subjects. Our subject-specific sPINN models yielded SBP and DBP regression coefficients $r^2$ = 0.626 ($p$ <0.001) and $r^2$ = 0.680 ($p$ <0.001) The corresponding MAE±SD for SBP and DBP predictions, respectively, are 6.62±6.42 and 4.99±4.78 mm Hg. A Bland-Altman analysis demonstrates the model displays a low prediction bias; the ME and LOA, for SBP and DBP, respectively, are -0.22, [17.84, -18.28] and -0.14, [13.40, -13.68] mm Hg.

The sPINN BP prediction metrics for the Group 2 dataset are broadly better than for Group 1; we think this is at least partially explained by discrepancies between the datasets, rather than systemic model issues. Summarily, the Group 1 dataset has a higher variance in its resistance signals than Group 2 and Group 2 has a higher BP variance than Group 1. Specifically, we compute the sample BP and resistance covariance matrices, $\Sigma$, for both the Group 1 and Group 2 training sets. For the resistance data, we subtract the resistance signal average from the each resistance signal prior to covariance computation and compute the variance of the average resistance value separately. Then, for both the BP and resistance Group 1 and Group 2 datasets, we compute the total variance, tr $(\Sigma)$. The BP total variance for Group 2 (31,047) is higher than for Group 1 (20,449). Yet, the (average removed) resistance total variance for Group 1 (0.50) is higher than for Group 2 (0.29), and similarly, the variance of the per-signal average resistance for Group 1 (35.08) is higher than for Group 2 (15.76). The higher resistance variance likely indicates more resistance measurement noise in the Group 1 dataset, and thus, a more challenging prediction problem.

Further, recall that a novel aspect of our work is disentangling the anatomical, physiological, and experimental factors and their influence on wrist surface resistance (Fig.3). The physiological factors with the largest influence (i.e., depth and radius of the radial artery) were measured with ultrasound (Extended Data Fig.2, Extended Data Table 2 and Supplementary Fig.15) and included into the population-wide sPINN model Supplementary Discussion 7.1.3), along with other physiological data. The sPINN model is sensitive to the inclusion/exclusion of this information. Indeed, as we report in Extended Data Table 4, without the inclusion of metadata, our population-wide sPINN model yielded SBP and DBP regression coefficients $r^2$ = 0.428 ($p$ <0.001) and $r^2$ = 0.448 ($p$ <0.001). The corresponding MAE±SD for SBP and DBP predictions, respectively, are 8.59±8.34 and 6.67±6.56 mm Hg. These results are worse across the board, suggesting that the inclusion of physiological parameters is important for contextualizing the input resistance signal in order to make accurate BP predictions. This is unsurprising since BioZ is sensitive to physiological factors such as artery depth and subcutaneous adipose tissue thickness, as explained by our three-layer cylindrical model and sensitivity analyses (Supplementary Discussion 5 and Fig.3); physiological factors that are not accounted for in the constitutive N.S. equations the sPNN enforces.

In Extended Data Fig.4, the sPINN predicted peak systolic velocity (PSV) is compared with ref-erence Doppler brachial flowmetry for Group 1 subjects. Doppler flowmetry imaging was performed over ≈15 s (Supplementary Video 9-11), and, a single representative operator-informed PSV measured value was assigned to each subject using the ultrasound's built-in signal processing tools. The sPINN, on the other hand, yields an arterial blood velocity waveform for every BP period in the test set. Further, the sPINN allows us to specify the velocity anywhere in the modeled arterial domain, whereas Doppler flowmetry measurements do not have the same level of experimental spatial resolution; so for comparison's sake, we consider the sPINN's spatial maximum velocity on a focused slice of the sPINN's arterial domain–constituting the middle 50% of the axial and the bottom 95% of the radial range–at systole to be the most similar value to measured PSV. Importantly, the measured PSV values and the corresponding sPINN predictions were acquired at different time points for the same subject. As a result, direct correlation, Bland–Altman, and histogram error analyses are not appropriate. Instead, we assess agreement in Extended Data Fig.4 by overlaying the distributions of the measured and predicted values, providing both subject-level and population-level comparisons for PSV.

Extended Data Fig.4 showcases the impressive agreement between the sPINN's predicted PSV distribution and measured PSV values. This is evident on both a per-subject basis and for the PSV predicted and measured distributions as a whole. Indeed, for 76% of subjects, the measured PSV values are within the sPINN's min-max predicted range; furthermore, for 38% of subjects, the measured values are even within the sPINN's predicted inner quartile range. We view this experimental validation as entirely novel (to the best of our knowledge) and, further, powerful evidence regarding the sPINN's ability to construct physically relevant arterial fluid dynamics. As a validation procedure, it is especially cogent, considering that the sPINN does not reference any experimental velocity during training, coupled with variability between the experimental and predicted velocity collection procedures. This principled approach stands in marked contrast to prior wearable cuffless studies broadly, which have focused primarily on BP prediction only. Moreover, existing studies often relied on either (i) methods that lack theoretical grounding and depend on empirical techniques such as PWA, or (ii) are susceptible to confounding by unmeasured physiological factors, as seen with PAT, PTT, and arterial dynamics (i.e., wall motion measurements) –ultimately limiting their clinical utility. Given our validating results for BP and velocity, our sPINN offers unprecedented cuffless hemodynamic capabilities and provides physically-explainable BP and radial and axial blood velocity field predictions.

We also compare predicted and corresponding experimentally measured and BP-modeled reference cross-sectional average velocities. The predicted and reference modeled velocities are generated via the sPINN and our forward fluid model, respectively, for the same subsection of the arterial domain as described above for the experimental validation procedure. The results are presented in Supplementary Fig.24. For average velocity at systole and diastole, the regression coefficients are $r^2$ = 0.753 ($p$ <0.001) and $r^2$ = 0.525 ($p$ <0.001). The corresponding MAE±SD for average velocity at systole and diastole, respectively, are 4.74±3.75 and 2.60±1.80 cm/s. A Bland-Altman analysis shows the ME and LOA, for average velocity at systole and diastole, respectively, are -3.44, [6.34, -13.19] and -2.25, [2.13, -6.61] cm/s. The Bland-Altman analysis indicates the sPINN consistently predicts smaller velocities than the reference at both systole and diastole. It is important to note that the sPINN model velocities and the reference velocities do not necessarily need to fully align and these analyses serve merely as a benchmark to further validate the physical relevance of our sPINN predictions. Indeed, we believe these results corroborate the experimental PSV analysis above, reinforcing our conclusions regarding the sPINN's ability to model physically consistent arterial fluid field predictions.

## Generalizability for Group 1 and 2

In Supplementary Fig.28, the results for our population-wide sPINN model are presented for Group 1 and Group 2 with test-exclusive subjects excluded and included (see Supplementary Discussion 7.1.3 for further discussion of the data train/test splitting procedure). For the Group 1 dataset, with test-exclusive subjects *excluded*, our model has SBP and DBP regression coefficients $r^2$ = 0.575 ($p$ <0.001) and $r^2$ = 0.629 ($p$ <0.001). The corresponding MAE±SD for SBP and DBP predictions, respectively, are 7.24±6.98 and 5.45±5.19 mm Hg. A Bland-Altman analysis demonstrates the model displays a low prediction bias; the ME and LOA, for SBP and DBP, respectively, are 0.10, [-16.54, 16.74] and 0.33, [-15.10, 14.44] mm Hg. For the Group 1 dataset, with test-exclusive subjects *included*, our model performs worse; it yields SBP and DBP regression coefficients $r^2$ = 0.357 ($p$ <0.001) and $r^2$ = 0.397 ($p$ <0.001). The corresponding MAE±SD for SBP and DBP predictions, respectively, are 9.12±8.39 and 7.39±6.84 mm Hg. A Bland-Altman analysis demonstrates the model displays a low prediction bias; the ME and LOA, for SBP and DBP, respectively, are 0.06, [-24.23, 24.36] and -2.43, [-21.59, 16.72] mm Hg.

For the Group 2 dataset, with test-exclusive subjects *excluded*, our model has SBP and DBP regression coefficients $r^2$ = 0.774 ($p$ <0.001) and $r^2$ = 0.807 ($p$ <0.001). The corresponding MAE±SD for SBP and DBP predictions, respectively, are 7.26±8.46 and 3.89±4.63 mm Hg. A Bland-Altman analysis demonstrates the model displays a low prediction bias; the ME and LOA, for SBP and DBP, respectively, are -0.46, [-22.30, 21.37] and -0.01, [-11.87, 11.85] mm Hg. For the Group 2 dataset, with test-exclusive patients *included*, our model again performs worse; it yields SBP and DBP regression coefficients $r^2$ = 0.328 ($p$ <0.001) and $r^2$ = 0.502 ($p$ <0.001). The corresponding MAE±SD for SBP and DBP predictions, respectively, are 11.82±12.45 and 6.42±5.98 mm Hg. A Bland-Altman analysis demonstrates the model displays a low prediction bias; the ME and LOA, for SBP and DBP, respectively, are 2.56, [-30.70, 35.83] and -0.13, [-17.32, 17.06] mm Hg.

It is important to note that the amount of test data is split roughly in half between test-exclusive subjects'/patients' data and non-test-exclusive subjects'/patients' test data. Therefore, poor predictions for test-exclusive patients significantly impact the overall model metrics. However, we still conclude that the population-wide sPINN model struggles with generalization for BP prediction on unseen subjects in both the Group 1 and Group 2 datasets. This subject/patient generalization is somewhat expected, given the relatively small subject/patient sample size of our training datasets.

In Supplementary Fig.27, we present results for the population-wide model trained on the Group 2 dataset. We see impressive agreement for BP predictions across patients, both in the example individual BP periods plots and the entire predicted-true BP test distribution plot. The box plots for patient $i$ distribution of RMSE errors, $E_j = \frac{1}{\sqrt{n}}\|\hat{\mathbf{p}}_j - \mathbf{p}_j\|_2$, $j \in [N_i]$ reveal that the population-wide model behaves differently for different patients. The subject-median of the median-RMSE error is 5.01 mm Hg (and 4.75 mm Hg, excluding test-exclusive patients). However, the median-RMSE errors for test-exclusive patients are 9.06 mm Hg (M9), 13.52 mm Hg (M26), 21.74 mm Hg (M61), 21.4 mm Hg (M68), 9.1 mm Hg (M76), and 8.87 mm Hg (M82), which are worse than the subject-median value (4.75 mm Hg). Although we find the sPINN's ability to predict BP across a breath of patients impressive and promising, we again conclude that the sPINN model struggles with generalization for predicting BP on unseen patients.

The results of our population-wide models compare favorably to Hsiao et al.[152] They report that their AMAE (SBP and DBP combined) for their leave-one-subject-out test study is 10.63 mm Hg (see Table S1 therein) and varies greatly by test subject, allowing the authors to conclude that the method does not generalize well to test subjects whose BP distribution is dissimilar to the overall dataset. Our sPINN prediction accuracies *with* test-exclusive patients outperform these accuracies, achieving individual MAE for SBP and DBP predictions, respectively, 9.12 and 7.39 mm Hg for Group 1 and 11.82 and 6.42 mm Hg for Group 2 (Supplementary Fig.28). This is despite the fact that Hsiao et al.: (i), use multi-modal data (two PPG and two BioZ datasets); and (ii), only predict SBP and DBP values and not the entire BP waveform. Hsiao et al. also report validation MAE for their study, which are comparable to our study *without* test-specific subjects. These MAE are excellent: 1.7$\pm$2.2 and 1.2$\pm$1.5 mm Hg for SBP and DBP, respectively, outperforming our model. However, an unusual aspect of reported validation MAE values is that they are $\approx$12% on average better than MAE observed on the training set for every single combination in their leave-one-subject-out study (Table S1 therein). Assuming no data leakage, we find this pattern surprising, as ML models typically perform better on training data, where optimization is directly applied, than on unseen validation data. Also, they present a "residual plot" (Figure 8 therein) instead of the norm of performing a Bland-Altman analyses. Finally, the authors combine SBP and DBP predictions to compute a single $r^2$ value, which can be misleading of true precision. Doing so, it greatly increases regression leverage, potentially inflating the $r^2$ value compared to the single SBP and DBP $r^2$ predictions. For example, they report a value of $r^2$ = 0.94 in comparing their predictions on validation data to reference BP value, and for their predictions on their test+validation data, they report $r^2$ = 0.85. If we combine our SBP and DBP predictions, we obtain $r^2$ = 0.922 ($p$ <0.001) on Group 2 data, which inflates our SBP and DBP values of $r^2$ = 0.774 ($p$ <0.001) and $r^2$ = 0.807 ($p$ <0.001), respectively (Supplementary Fig.28). This inflationary phenomenon can obfuscate actual results when considering lower separate DBP and SBP $r^2$ values; for our Group 2 data *with* test-exclusive patients, we see a combined $r^2$ = 0.809 ($p$ <0.001) compared to $r^2$ = 0.328 ($p$ <0.001) and $r^2$ =0.502 ($p$ <0.001) for SBP and DBP, respectively (Supplementary Fig.28). This $r^2$ inflation is not isolated to our Group 2 data, indeed, for our Group 1 data the combined $r^2$ = 0.890 ($p$ <0.001) and $r^2$ = 0.831 ($p$ <0.001) for *with* and *without* test-exclusive subjects, respectively; whereas, the corresponding SBP $r^2$ = 0.575 ($p$ <0.001) and DBP $r^2$ = 0.629 ($p$ <0.001), *with* test-exclusive subjects, and SBP $r^2$ = 0.357 ($p$ <0.001) and DBP $r^2$ = 0.397 ($p$ <0.001), *without* test-exclusive subjects (Supplementary Fig.28). Their choice to display the $r^2$ values for their predictions in this manner is a departure from the norm of reporting individual SBP and DBP $r^2$ values, further contributing to the present ambiguity in their $r^2$ metrics. Summarily, these omissions limit our ability to assess their model's true performance and generalizability. Finally, we mention that the results reported by Hsiao et al.[152] show BioZ signals that increase during the systole phase. Their measured data contradict both our experimental findings and first-principles modeling results, which consistently show that pulsatile BP increase arterial blood volume (by expanding arterial radius) and blood conductivity, and decrease the net resistance (Fig.3b-c and Supplementary Fig.18). This observation raises the possibility that the authors analyzed conductance rather than resistance–the latter referred in the literature to as BioZ.

### Blood pressure for PulseDB synthetic dataset

For the synthetic dataset generated using PulseDB,[100] our sPINN model achieved high accuracy solving for axial blood velocity in space and time compared to the reference forward fluid dynamic model (Extended Data Fig.3 and Extended Data Table 4). The sPINN likewise accurately evaluated velocity profiles at different time snapshots of a single period, with a maximum RMSE of 1.75 cm/s. The sPINN also excelled at learning the underlying pressure driving pulsatile flow. Four randomly selected BP periods show the network's ability to discriminate between morphologically different experimental BP waveforms. The ML model maintained a high correlation when assessing the SBP ($r^2$ = 0.998, $p$ <0.001), DBP ($r^2$ = 0.974, $p$ <0.001), cross-section average velocity at systole ($r^2$ = 0.994, $p$ <0.001), and cross-section average velocity at diastole ($r^2$ = 0.888, $p$ <0.001). A Bland-Altman analysis demonstrates the model displays a low prediction bias with 0.03, [-1.83, 1.90] mm Hg for SBP and 0.45, [-2.00, 2.90] mm Hg for DBP. The corresponding MAE±SD for SBP and DBP predictions are 0.57±0.76 and 0.85±1.02 mm Hg, respectively.

**Impact on blood pressure for PulseDB synthetic dataset with biological variability**

To study how our sPINN handles biological variability, we used our forward model in a controlled setup to permute seven different biological parameters: (i), radial artery radius; (ii), radial artery wall Young's modulus; (iii), blood hematocrit; (iv), heart rate; (v), radial mean wall shear stress; (vi), subcutaneous adipose fat thickness; and (vii), skin conductivity. We trained the sPINN using the training set generated by sampling all parameters randomly and simultaneously; then, we evaluated the model on the test set generated via isolating and randomly sampling a single parameter and leaving the others fixed (see Supplementary Discussion 7.1.2 for further discussion of data generation and Supplementary Table 8 and 9). The results for each isolated parameter test set are displayed in Extended Data Fig.3 and Supplementary Fig.21 and summarized below.

i Radial artery radius test set: regression coefficients $r^2$ = 0.965 ($p$ <0.001) and $r^2$ = 0.666 ($p$ <0.001) for SBP and DBP; MAE±SD for SBP and DBP predictions are 2.55±2.72 and 3.36±3.12 mm Hg; Bland-Altman analysis shows a ME and LOA of 0.08, [7.39, -7.23] mm Hg for SBP and 0.86, [9.69, -7.97] mm Hg for DBP.

ii Radial mean wall shear stress test set: regression coefficients $r^2$ = 0.971 ($p$ <0.001) and $r^2$ = 0.680 ($p$ <0.001) for SBP and DBP; MAE±SD for SBP and DBP predictions are 2.39±2.45 and 3.31±3.07 mm Hg; Bland-Altman analysis shows a ME and LOA of 0.27, [6.96, -6.42] mm Hg for SBP and 0.97, [9.61, -7.68] mm Hg for DBP.

iii Radial artery wall Young's modulus test set: regression coefficients $r^2$ = 0.969 ($p$ <0.001) and $r^2$ = 0.688 ($p$ <0.001) for SBP and DBP; MAE±SD for SBP and DBP predictions are 2.41±2.52 and 3.27±3.03 mm Hg; Bland-Altman analysis shows a ME and LOA of 0.17, [7.01, -6.66] mm Hg for SBP and 0.94, [9.48, -7.59] mm Hg for DBP.

iv Blood hematocrit test set: regression coefficients $r^2$ = 0.969 ($p$ <0.001) and $r^2$ = 0.693 ($p$ <0.001) for SBP and DBP; MAE±SD for SBP and DBP predictions are 2.42±2.54 and 3.23±3.02 mm Hg; Bland-Altman analysis shows a ME and LOA of 0.14, [7.02, -6.73] mm Hg for SBP and 0.93, [9.40, -7.53] mm Hg for DBP.

v Heart rate test set: regression coefficients $r^2$ = 0.971 ($p$ <0.001) and $r^2$ = 0.688 ($p$ <0.001) for SBP and DBP; MAE±SD for SBP and DBP predictions are 2.37±2.46 and 3.25±3.04 mm Hg; Bland-Altman analysis shows a ME and LOA of 0.32, [6.98, -6.34] mm Hg for SBP and 0.92, [9.45, -7.61] mm Hg for DBP.

vi Skin conductivity test set: regression coefficients $r^2$ = 0.972 ($p$ <0.001) and $r^2$ = 0.697 ($p$ <0.001) for SBP and DBP; MAE±SD for SBP and DBP predictions are 2.32±2.40 and 3.23±3.01 mm

Hg; Bland-Altman analysis shows a ME and LOA of 0.36, [6.87, -6.17] mm Hg for SBP and 1.00, [9.42, -7.42] mm Hg for DBP.

vii Subcutaneous adipose fat thickness test set: regression coefficients $r^2$ = 0.966 ($p$ <0.001) and $r^2$ = 0.691 ($p$ <0.001) for SBP and DBP; MAE±SD for SBP and DBP predictions are 2.53±2.68 and 3.24±3.01 mm Hg; Bland-Altman analysis shows a ME and LOA of 0.05, [7.28, -7.17] mm Hg for SBP and 0.88, [9.36, -7.61] mm Hg for DBP.

We first note, across all of the biological parameters, we see excellent agreement for SBP predictions. However, for DBP predictions the sPINN consistently performs worse across all the parameters. The consistency of the lower prediction accuracy of DBP (compared to SBP) across all seven biological parameters suggests that, under variable biological parameters, the resistance waveforms generated from our BioZ forward model preserve BP information at systole better than at diastole. This results in a more challenging DBP prediction problem (compared to SBP) for the sPINN, independent of the particular biologically variable parameter. To investigate this further, we computed the sample covariance between the BioZ waveform and both the DBP and the SBP by finding the sample covariance between each BioZ waveform index and the DBP (or SBP). Then, we consider the maximum absolute variability between the BioZ waveform and the DBP (or SBP); simply, at what index (point in time), does the DBP (or SBP) vary the most with the BioZ and what is the sample covariance at that time. The maximum absolute covariance between BioZ and DBP is 0.21 and between BioZ and SBP is 2.21; the maximum covariation between BioZ and SBP is ≈10 times the maximum covariation between BioZ and DBP. This is further evidence that SBP and our synthetically generated BioZ waveforms have a stronger (linear) relationship than DBP and the same BioZ waveforms and, in essence, reinforces our conclusion that DBP is more challenging to predict than SBP, using these synthetic BioZ signals.

A potential modeling explanation for this DBP prediction challenge relates to the computation of conductivity around diastole (Supplementary Discussion 4). The RBCs orient and deform as a function of the magnitude of the shear stress in the blood. Around diastole, the shear stress is likely close to zero and possibly changing sign multiple times. It is possible that information about the diastolic BP is lost during this process. Fig.3b,c,e illustrate the flattening of the resistance waveform away from systole. After subtracting the means off of the PulseDB biological variability training dataset, systolic and diastolic synthetic BioZ variances were computed by identifying the minimum and maximum of the mean waveform respectively. A diastolic variance of 2.33 mohms and systolic variance of 18.64 mohms were found, demonstrating a ≈9 times larger variability for BioZ at systole and diastole.

We consider the degradation in DBP prediction performance to be likely a modeling and data generation limitation and not a systemic sPINN issue. As such, it is notable that the sPINN's BP predictions are negligibly variable across the seven biological parameters. This suggests that the sPINN model can indeed handle biological variability influenced input resistance signals and still yield reasonable BP predictions. For posterity, the results aggregated across all of the biologically variable parameters (i.e. all of the biological variability test datasets) are tabulated in Extended Data Table 4.

### Blood pressure for Graphene-HGCPT experimental dataset

For the Graphene-HGCPT[99] dataset, results for population-wide and subject-specific models are presented in Supplementary Fig.30 and Supplementary Fig.31, respectively. Since there are only seven subjects in this dataset (Supplementary Table 8), we did not exclude any subjects from this dataset for our population-wide model; the training and test datasets for both the subject-specific and population-wide models are the same.

The population-wide model shows good agreement between true and predicted BPs for four randomly selected BP periods from the test set. The model achieves correlations of $r^2$ = 0.699 ($p$ <0.001) for SBP and $r^2$ = 0.734 ($p$ <0.001) for DBP. A Bland-Altman analysis demonstrates the model displays a low prediction bias with -0.43, [-20.24, 19.34] mm Hg for SBP and -0.26, [-15.18, 14.67] mm Hg for DBP. The corresponding absolute prediction errors for SBP and DBP are 6.87±7.42 and 5.19±5.58 mm Hg, respectively. When comparing the entire predicted and true waveforms, we see an AMAE of 5.99 mm Hg and an ARMSE of 6.43 mm Hg. The subject-specific model achieves correlations of $r^2$ = 0.770 ($p$ <0.001) for SBP and $r^2$ = 0.828 ($p$ <0.001) for DBP. A Bland-Altman analysis demonstrates the model displays a low prediction bias with -0.96, [-18.29, 16.37] mm Hg for SBP and -0.71, [-12.64, 11.23] mm Hg for DBP. The corresponding MAE±SD for SBP and DBP predictions are 6.52±6.05 and 4.51±4.15 mm Hg, respectively. For waveform comparisons, we see an AMAE of 5.42 mm Hg and an ARMSE of 5.83 mm Hg.

Compared with the results in Sel et al.,[189] our population-wide (and subject-specific) model obtains both higher SBP $r^2$ = 0.699 (ours) vs $r^2$ = 0.66 (theirs) and DBP $r^2$ = 0.734 (ours) vs $r^2$ = 0.53 (theirs) on the Graphene-HGCPT[99] dataset. It is important to note that a direct comparison of results is challenging as we do not use their "minimal training" criterion for defining our training set and we preprocess and filter the data differently (Supplementary Discussion 7.1.3). Additionally, we do not compare with their leave-one-out model training methodology, even though it would be most similar to our population-wide model, for the following reasons:

1. Their model is trained using two datasets which are not publicly available, in addition to the Graphene-HGCPT dataset.
2. They only report ME and SD, a performance metric that is also reported in other cuffless BP literature (Supplementary Table 10). As noted earlier in this Section, we strongly prefer MAE over ME (or *absolute* ME) when assessing ML model accuracy, as (*absolute*) ME can yield deceptively low values and obscure true predictive performance.
3. They calibrate their models using subject-specific BP information from the held out subject's data before model evaluation.
4. We do not hold out single subjects during our own training, due to the limited size of the dataset.

Our population-wide sPINN model performs better on our Group 2 experimental dataset for SBP ($r^2$ = 0.774, $p$ <0.001) and for DBP ($r^2$ = 0.807, $p$ <0.001) compared to our population-wide sPINN model model on the Graphene-HGCPT[99] dataset ($r^2$ = 0.699, $p$ <0.001 for SBP, and $r^2$ = 0.734, $p$ <0.001 for DBP), despite our experimental dataset being larger and having been collected with stainless steel electrodes instead of graphene, the former a more challenging material for BioZ recordings (Fig.1). We believe that this is because our smartwatch is capable of measuring with high precision low impedance changes (Supplementary Fig.1). Furthermore, BP can be regarded as a DC coupled signal, with its baseline encoding mean arterial pressure. Accordingly, we input the AC-coupled BioZ waveform and its DC (mean) component into our sPINN, rather than discarding the latter via band-pass filtering as in the Graphene-HGCPT dataset (Supplementary Fig.17). This lost resistance information likely explains why population-wide models (theirs and ours) using this dataset require calibration, in their case, and limit comparative performance, in our case.

### Comparison of sPINN model results against baseline traditional machine learning methods

In Supplementary Table 11, we provide a benchmark comparison of our sPINN model results with traditional ML methods. In part, we compare against traditional ML methods because there have been few previous studies for predicting BP based on BioZ data (Supplementary Table 10), so it is helpful to include an evaluation of the performance of other methods to establish expectations in terms of predictive accuracy for BioZ data. This is especially true for our own datasets, as they have not been analyzed in any other publication. We also emphasize that this comparison with other ML methods is meant to establish a set of simple baselines to contextualize the performance of our developed sPINN model. Our goal is not to explore a breath of ML architectures, as the sPINN model is implemented with a relatively small NN (number of tuneable parameters) by modern standards and is instead largely defined by its unique adherence to the physics of the problem. In comparing the sPINN to other ML methods, we are not engaging in a form of data leakage[156] or "P hacking",[190] where we train many ML models and either prominently or only report the test results for the best performing ML models.

We implemented and evaluated three methods in addition to our sPINN. In each case below, the model predicts the entire BP waveform (not just SBP/DBP) so that the results are a fair benchmark with the sPINN model. The standard ML methods evaluated are

1. Simple linear regression (LR).
2. Multilayer perceptron (MLP) —a NN with fully connected layers—, implemented in sklearn. We use five hidden layers, each with 100 neurons, so that the total number of tunable parameters is similar to the sPINN (excluding the signal encoder). We compare with a NN of comparable size to ensure the sPINN can solve the BioZ to BP regression problem with comparable performance, despite having a very different loss structure and therefore less direct learning incentive for regression.
3. Signal encoder and AF BP prediction model that is trained in phase 1 (Fig.4a and Supplementary Discussion 7.3.1).

We now discuss the results in Supplementary Table 11. A first general observation is that all models are better at predicting DBP than SBP. This might be attributed to the fact that DBP is generally more predictable than SBP due to its lower variability and more consistent temporal dynamics.

For the PulseDB[100] synthetic dataset, we see the sPINN outperforms both MLP and LR. However, both the MLP and LR methods also have strong results on this dataset. This shows that the regression problem for predicting BP from synthesized resistance is oversimplified; this dataset consists of resistance waveforms generated using a deterministic map from BP waveforms.

The subject-specific prediction task is generally easier than the population-wide prediction task. For the subject-specific models for the Group 1, Group 2, and Graphene-HGCPT[99] datasets, we find that the sPINN and MLP are commensurate in performance, and always outperform LR. However, on population-wide Group 1 and Group 2 models, the results for sPINN are better than MLP/LR. These results suggest that our physics-based model may provide better generalization towards calibration-free models than traditional ML methods. Specifically, we believe this is the result of the sPINN's ability to capture a broader distribution of physically relevant BP without necessarily having to see like-signals in the training set, via the governing physical loss terms. Our sPINN approach therefore may enable the training of more generalizable cuffless BP models with less reliance on experimental (BioZ-BP pairs) data.

We also observe that predictions for the Group 1 dataset are generally worse than those for Group 2. This likely reflects differences in the quality of the data. The Group 2 dataset consists exclusively of baseline measurements obtained in a clinical setting, and includes elevated SBP values in some patients with uncontrolled HTN, whereas the Group 1 dataset includes changes in vascular tone and sympathetic nervous system activity induced by the Valsalva maneuver, deep breathing, cold pressor, and handgrip test, intentionally conducted to induce BP fluctuations. These BP dynamics almost certainly make Group 1 a more challenging dataset compared to Group 2. Qualitatively (and quantitatively), our Group 1 data also had more measurement noise and electrode motion artifacts, as compared to the Group 2 data collection, further contributing to the problem complexity difference between datasets.

Comparing the AF and sPINN models indicates how the sPINN's PINN-style learning techniques impact model performance relative to a simple NN with standard regression loss structure; given that both use a learned encoded signal resistance signal representation (Supplementary Discussion 7.3.1). Additionally, we wanted to explicitly analyze BP prediction changes between Phase 1 training and Phases 2 and 3 training. For the population-wide Group 2 and Graphene-HGCPT[99] datasets, we observe small performance gains and for the population-wide Group 1 dataset, we see a small performance loss due to the difference in learning incentives (standard regression vs physics), between Phase 1 and Phases 2 and 3. Note that AF model comparisons are only presented for population data, where Phase 1 training was performed.

We have reported multiple metrics for all of our ML model results including $r^2$, MAE, SD, ME, and LOA when comparing SBP/DBP, as well as ARMSE and AMAE when assessing waveform accuracy. Supplementary Table 11 highlights the importance of providing a comprehensive set of model performance metrics. In particular, the MAE can misconstrue model BP predictions as reasonable when considered in isolation, but when considered in conjunction with $r^2$ the model can display little predictive power. For example, the population-wide (no metadata) MLP model on the Group 1 dataset has $r^2$ = 0.087 but a relatively misleading MAE of 9.16±7.01 for DBP prediction. A similar, but less extreme, observation is evident for the Group 2 population MLP model, where the model has a decent MAE (7.93±6.61) for DBP prediction, but an $r^2$ = 0.409. Another misleading example is Kireev et al,[137] where the authors reported *absolute* ME values of 0.2±5.8 and 0.2±4.5 mm Hg for SBP and DBP, respectively. Although the authors did not follow the IEEE 1708-2014 standard in terms of the validation protocol or sample size, they assigned their model an 'A' grade based on these values rather than mean absolute difference (MAD) values, which are equivalent to our definition of MAE, as specified by the standard. The authors did not report MAD or MAE values, however, their reported RMSE values range from 5 to 7 mm Hg. The substantial discrepancy between *absolute* ME and RMSE values suggests also large MAD/MAE values —likely inconsistent with an 'A' grade under IEEE 1708-2014 standard validation protocol.

# 9 Supplementary Discussion 9. Limitations and future directions

## 9.1 Fluid dynamics

There are several assumptions that must be made to derive the analytical forms of our fluids equations. These include that the artery radius is sufficiently small compared to the wavelength of the propagating pressure waves; the artery wall is sufficiently thin compared to the artery radius; the equations of motion of the wall, dictated by the Young's modulus, is dictated by linear equations; and the artery radius cannot continuously taper as it travels further distally from the aorta. Some of these assumptions carry further validity than others. In particular, the long-wave approximation holds in the arteries of the arms as the wavelength is typically on the order of meters while the corresponding radii are on the orders of millimeters.[118] On the other hand, the artery wall thickness is on the order of a tenth of the artery radii,[191] the arterial wall displacements are known to be nonlinear,[192] and it is known that the brachial and radial arteries taper in radius as one travels further distally.[193] Introducing nonlinearities in the artery wall, or the fluids solutions, require nonlinear PDE solvers that may be more expensive to compute than the current forms of our solutions and reduce analytical interpretability. Future research will explore introducing nonlinearities such as those described above to explore models that can include such finer details.

Additionally, the following computation of bulk conductivity of the artery requires several assumptions to be made about the immersed RBCs. For simplicity, the RBCs are geometrically modeled as elastic ellipsoids, as opposed to more accurate biconcave disks,[194] which deform and orient depending on the surrounding flow.[87,88,113] The model of ellipsoidal RBC deformation makes very few modeling assumptions but another avenue of research could study how this deformation compares to that of biconcave disks. Their orientation is determined by assuming some convex combination of $a$-aligned or $b$-aligned RBCs. The convex combination is determined by a $f$-function defined by Bitbol and Leterrier which is the steady state of a first-order kinetics model.[195] This method does not account for RBC movement and collision that can be simulated in particle-based methods such as the immersed boundary method.[196,197] This model requires experimentally difficult-to-obtain parameters that can be the source of future research. Future research may also study how this model of orientation differs from that of biconcave disks. In Supplementary Discussion 8.3, we discussed how the resistance waveform at diastole is 9x less sensitive to biological variability than at systole and that this is likely due to these limitations associated with the synthetic data generation. An improved model and updated parameter range values could further improve our ability to perform a biological variability study especially during diastole.

Finally, one difficulty in applying this model is in specifying its large number of parameters (Supplementary Table 1). For any given individual, many of the parameters are difficult or currently impossible to establish non-invasively in vivo. Therefore, we rely on population average values reported in the literature. Through more physiological studies we can establish more accurate population average values and variances across individuals (Supplementary Discussion 9.3). This information would allow for more accurate patient-specific simulation and more thorough uncertainty quantification of models similar to the one described here.

## 9.2 Bioimpedance

The cylindrical BioZ model is based on several assumptions regarding forearm anatomy. First, it assumes that the artery depth is sufficiently larger than the arterial radius. Second, although the model provides a close approximation to human anatomy, it neglects the finite and curved boundaries of the forearm. Third, it assumes that tissues are organized in isotropic, layered structures with a single cylindrical artery embedded at a fixed depth. Fourth, the model omits anatomical features such as the ulnar artery and major veins. Another limitation of the model derivation is the assumption of direct current conditions. While the solution remains approximately valid at low frequency alternating current frequencies, it does not explicitly include the effects of permittivity, and thus cannot account for the reactive signal of the measured wrist BioZ. Future theoretical and computational work will explore improving the BioZ model by accommodating a domain with curvature to more accurately resemble the anatomy of the wrist, as well as to incorporate key additional physiological structures including the contribution of tissue electrical anisotropy and permittivity.

Additionally, the electrostatic finite element simulations were completed with some assumptions. First, the computational models utilized a database of tissue electrical properties. Similarly to our fluid dynamic model parameters, these values may not accurately reflect in vivo human values.[198] Second, the simulations did not account for contact impedance present during experimental measurements. Third, all tissues were assumed with isotropic conductivity and permittivity properties. Finally, the Fats and Glenn human phantoms lacked complete vasculature with the radial and ulnar arteries terminating at mid-forearm and were excluded from the blood conductivity sweep simulations but included in the VID and tissue contribution analyses (Supplementary Table 4). As such, the specific arterial resistance and reactance contribution might had been underestimated (Fig.3n, Extended Data Fig.1, Supplementary Fig.10–12 and Supplementary Table 5).[94,199]

## 9.3 Experimental data collection

Although our experimental protocol generated the largest dataset of simultaneous BP and BioZ resistance signals reported that we are aware of, both the protocol and subsequent analysis have limitations. First, we did not follow validation protocols recommended by IEEE 1708-2014 or the European Society of Hypertension.[200,201] Rather, we intentionally designed our experimental protocol as a pilot study to evaluate the feasibility of our approach in a cohort comprising both healthy subjects and patients with HTN, CVD, and other conditions. Second, we did not assess inter- nor intra-session repeatability with multiple operators which would had been necessary to study the reliability, reproducibility, and generalizability of the experimental results. Third, while our sPINN model provides continuous estimates of BP and blood velocity, the primary error metrics were computed at systole and diastole. Fourth, we used two different cuff-based reference devices for the lab and clinical cohorts, with the NOVA system measuring brachial artery pressure and the Caretaker system estimating central aortic pressure. These systematic differences in BP reference measurements necessitated separate analysis of the two datasets and may have affected model training and evaluation. Furthermore, since BioZ signals were collected at the wrist, the use of central BP estimates may have impacted accuracy due to the greater anatomical distance from the BioZ measurement site. Fifth, we did not collect simultaneous BioZ and blood velocity data, warranting further work to assess the sPINN's blood velocity waveform prediction accuracy. Also, we have insufficient gold standard A-line BP measurements for sPINN training and test thus limiting a direct comparison. These data are now being collected as part of an ongoing study and the results will be shared in future works. Further, although physiological data were collected during exercise for healthy participants in Group 1, these data were not analyzed in the present study. Future work will investigate whether such kind of data could reveal cardiovascular deficiencies that may remain undetected at rest or post-physical activity. Finally, another limitation of our experimental protocol is the distribution of BP within the HTN patient group. Specifically, 12 out of 32 patients had controlled BP. As a result, the distribution of SBP predictions in our testing dataset does not consistently exceed the threshold required to be classified as hypertensive (Fig.4b-e and Extended Data Fig.5). Future work will focus on collecting data on a larger population of uncontrolled Stage 1 and 2 hypertensive patients.

## 9.4 Machine learning

Although our subject-specific and population-wide sPINN (Supplementary Discussion 7) showcase potential for cuffless BP and blood velocity monitoring (Supplementary Discussion 8.3), there are several limitations and extensions that could further improve results.

For BP prediction, it is well-known that generalization for BP prediction models is a challenge and as a result, the common approach adopted in many studies is to use individual calibration, also known as fine-tuning or personalization, to tailor a model to an individual.[156] We demonstrate generalization performance for a population-wide model trained on BioZ data without individual-specific calibration (excluding physiological information or metadata). While this opens the door to (re-)calibration-free models, it also presents a more challenging prediction and future validation problem. In this context, a key limitation of our trained sPINN model is its reduced generalization to entirely unseen, test-exclusive subjects (Supplementary Fig.28 and Supplementary Fig.27). This is unsurprising because, although we have collected a relatively large BP-BioZ dataset, especially when compared to previous studies (Supplementary Table 10), this is still a comparably small number of individuals relative to the complexity of the ML task of predicting BP. The dataset size generally expected to train other modern ML models in order to capture population information is significantly larger than our own.[202–205]

Another challenge affecting model performance and generalizability is that we deliberately avoided moistening the skin underneath the electrodes with saline solution to improve electrode contact impedance, in our clinical cohort (Group 2), as such a procedure would diminish the significance of our results and the real use of our technology. Rather, electrodes were applied directly to unprepared skin, and the resulting data was preprocessed with a pipeline to reflect a more relevant and challenging experimental scenario (Supplementary Discussion 6, Supplementary Discussion 7.1.3, and Supplementary Fig.17). Furthermore, patients received instruction from the operator on how to wear and position the smartwatch. Consequently, our experimental data reflect natural variability in both the applied pressure and positioning of the BioZ sensors relative to the radial artery. We anticipate that learning the mapping from these experimentally noisy and more variable resistance signals to underlying BP may require larger datasets to enable robust performance and further, generalization.

In the Group 1 dataset we do not isolate any of the data measured during/after interventions (cold pressor test, Valsalva, etc.) nor post exercise in the test or training sets. This modeling choice likely adds to the BP and resistance variance in both the training and test sets for the Group 1 dataset. Further, by not isolating these segments in the test set, we are unable to quantify how they specifically impact our sPINN performance metrics. However, by including them in the training (and test) sets our model likely has the ability to predict (and showcase) a more diverse range of BP signals from a more dynamic set of resistance signals.

The sPINN training relies on the forward fluid model to generate reference pressure and axial velocity fields. This propagates the assumptions from the forward fluid model (Supplementary Discussion 4) into the sPINN pressure and velocity predictions; specifically, this includes a single elastic artery with no-reflection outflow boundary conditions and the limitations discussed in Supplementary Discussion 9.1. Additionally, it likely constrains the sPINN fluid predictions to mimic those of the forward fluids model. The sPINN allows for the inclusion of nonlinear terms when enforcing the N.S. equations, yet by relying so heavily on the forward fluid model we may be preventing the sPINN from expressing these nonlinear effects in its solutions. While it is not clear that this is exclusively a limitation, we would like to extend the sPINN model so that it is more independent from the forward fluids model during training. Further, our wall motion method does not include the coupled axial wall displacement term like in the forward fluids modeling (Supplementary Discussion 4). Although the equation we use to model radial displacement is standard for modeling linear elastic thin wall radial displacement in a tube, our choice to do so was motivated by practical constraints. We were limited in our ability to construct a method to solve for axial displacement in a manner similar to radial displacement as in Supplementary Discussion 7.5.3.

In order to ensure that the enforced N.S. equation residuals in Supplementary Discussion 7.5.2 are finite near the center of the tube, we multiply each residual term by the square of the value of the radial collocation point. This is necessary for computation, but allows for the radial velocity at the center of the tube to be non-zero (Supplementary Fig.20) and therefore violate the N.S. incompressibility condition (Equation (40c)). This could likely be remedied by specifying an additional loss term for the radial velocity at the center of the tube to penalize nonzero radial velocity predictions. Our sPINN model only incorporates residuals from the N.S. equations to enforce fluid dynamics constraints, but does not currently integrate Maxwell's electromagnetism equations to constrain the resistance signal input. Incorporating electromagnetic priors represents a promising future direction to further enhance the physical fidelity of our sPINN framework.

Another limitation of our current study is that we utilized only the resistance signal when tagging BP periods, despite resistance and reactance signals being measured and available from the smartwatch. This decision was driven by the fact that our current forward model does not accommodate the permittivity as a parameter and, therefore, the causal source of the reactance signal. However, future work that establishes the biophysical relationship underpinning both real and imaginary parts of the BioZ signal could enable the incorporation of both experimental resistance and reactance signals. This may enhance the signal tagging aspect of the sPINN and lead to improved model performance, allowing for fairer BP accuracy comparison with other studies that have used more than one source of experimental physiological data, such as combined ECG and PPG.[62,172,186,206]

Training time is a significant challenge (Supplementary Table 11). Training the sPINN model during phase 2 or 3, on either the Group 1 or the Group 2 dataset requires more than 72 hours. The current NN architecture we implement for the sPINN is small compared to other modern ML methods, so most of the training time speed-ups are likely to come from larger batch sizes and optimizing the batch data generation process (Supplementary Discussion 7.6.2).

We did not utilize transfer learning techniques,[207] however, we think that they could be used to learn better representations of resistance signals, improve training times, and supplement data scarcity, thus potentially helping the sPINN model generalize to unseen subjects. For example, forward-modeled resistance closely approximates measured resistance waveform morphology (Supplementary Fig.18) and with further magnitude investigation and alignment, first training the sPINN with synthetic data (i.e. transfer learning) could be implemented to accelerate training times and enhance performance on experimental datasets. Alternatively, additional resistance data without synchronized (or measured) BP signals could be collected and used to better understand and encode the distribution of resistance signals.

Also, while we explored a range of architectures and network sizes for both the signal encoder and the fully connected components of sPINN, a more extensive hyperparameter and architecture search could further improve performance. Our sPINN model (not including the signal encoder) has a relatively standard parameter size compared to other PINN models. However, our model generalizes the ideas behind PINNs to regression and thus it is unclear if a larger number of parameters would improve performance. Subsequently, changes to the signal encoder, such as, utilizing long short term memory (LSTM), recurrent neural network (RNN), or transformer architectures, could improve the feature extraction regarding the temporal dynamics of the input signal.

Although it is common during PINN training to begin with a first-order optimizer (e.g., SGD or Adam) and later switch to L-BFGS to better navigate the complex PDE loss landscape,[208] we did not adopt this strategy. Future work may benefit from incorporating this two-phase optimizer approach to potentially enhance BP and blood velocity solutions.

To ensure clinical trust and facilitate adoption by physicians, it is crucial that models indicate when their predictions may be unreliable. A promising future direction is the development of a risk estimator that quantifies uncertainty or confidence alongside BP and blood velocity outputs. This could be implemented by training models to predict error bounds or by estimating predictive variance during inference.

Finally, in our implementation, the sPINN framework operates deterministically. However, we are inputting experimental BioZ data that inherently contains stochastic noise and potential measurement artifacts. While our current preprocessing pipeline applies deterministic smoothing to these inputs, this process may inadvertently suppress physiologically meaningful features–particularly those related to BP dynamics. Moreover, the BP signals we treat as ground truth are themselves not definitive representations of the true BP waveform; rather, they are indirect and potentially incomplete proxies. Without a means to perfectly measure and input the true BP and BioZ signals into the sPINN, we cannot fully assess what information may be lost or distorted in our models.

In conclusion, despite these limitations, our study demonstrates a viable path toward cuffless, calibration-free blood pressure monitoring. Addressing the above limitations will be the focus of future work as we refine this technology for real-world application.

# 10 Supplementary Discussion 10. Statistical analyses

Data were analyzed using Prism 10 (Dotmatics, Boston, MA). All datasets were evaluated for normality using the Shapiro-Wilks test. Unpaired T-tests, Mann-Whitney tests, and Welch's $t$-tests were performed between male and female populations and between arteries (Extended Data Fig.3). Additionally, to evaluate the dependence of anatomical and experimental factors on wrist surface resistance, we performed linear regression analyses and assigned $p$-values using an $F$-test with the null hypothesis that the slope was zero and correlation values assigned using the mean value across human phantoms when applicable (Fig.3 and Extended Data Fig.1-2). Statistical significance was evaluated to assess the effect of electrode position and arterial anatomy using 1-way ANOVA with Dunnet's multiple comparison tests (Fig.3). Group 1 vital data were scaled with respect to the mean value of that subgroup's subjects baseline vitals to determine if the interventions (i.e., treadmill recovery, bike recovery, cold pressor, Valsalva, hand grip, and deep breathing) created changes in patient vitals (Extended Data Fig.4). Finally, the one sample $t$-test or Wilcoxon signed rank test were used to assess if the scaled data was significantly different from non-zero. The statistical results are reported in a separate Supplementary Table 12.

# Supplementary Figures

## 11 Supplementary Fig.1. Smartwatch noise characterization

The measurement setup consists of four identical **a**, low, and **b**, high contact impedances connected to 49.9 and 51 ohmsresistor terminals, respectively. The electrical current frequency was 51.2 kHz, and the amplitude was 100 $\mu A_{rms}$. SD, standard deviation. Scale bars, 1 s.

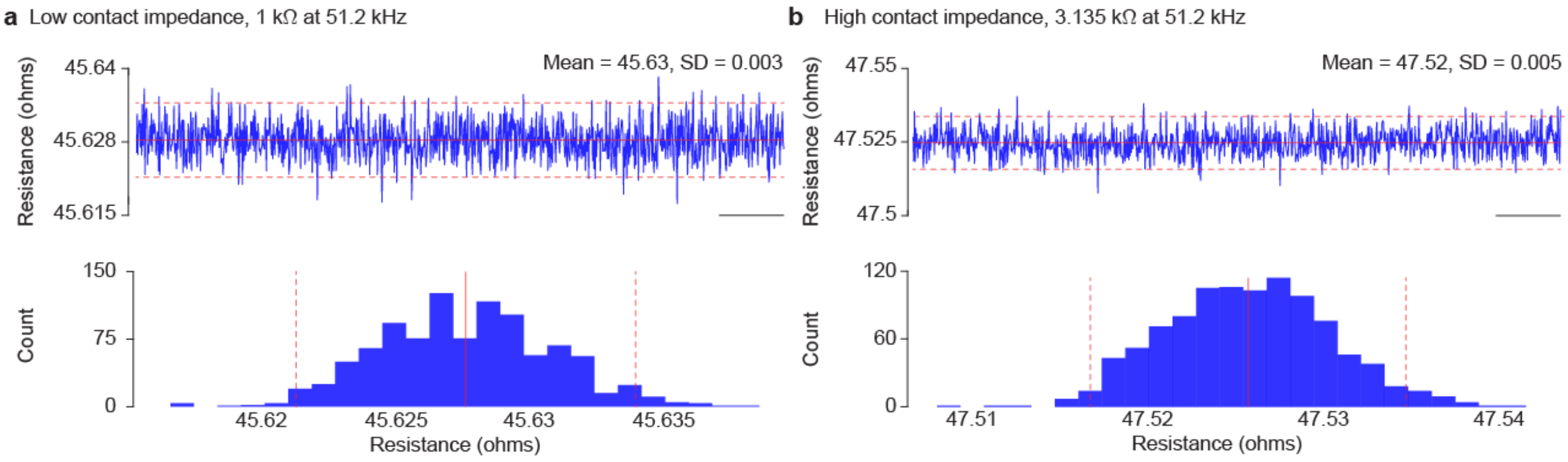

# 12 Supplementary Fig.2. Three-layer cylindrical bioimpedance model of the wrist

Illustration of the electrodynamic model developed for a tetrapolar resistance measurement over the skin. Parameters: $\sigma_1$, skin conductivity; $\sigma_2$, Subcutaneous adipose tissue (SAT) conductivity; $\sigma_3$, skeletal muscle conductivity; $I_-$, current collection electrode; $V_-$, negative voltage sensing electrode; $V_+$, positive voltage sensing electrode; $I_+$, current injection electrode; $d$, artery offset; $a$, artery radius; $h$, artery depth; $\sigma_a$, artery conductivity; $s_1$, depth of skin layer; $s_2$, depth of SAT layer; $e_1$, gap between current and voltage electrodes; $e_2$, gap between voltage electrodes; $d_1$, thickness of skin layer; $d_2$, thickness of SAT layer.

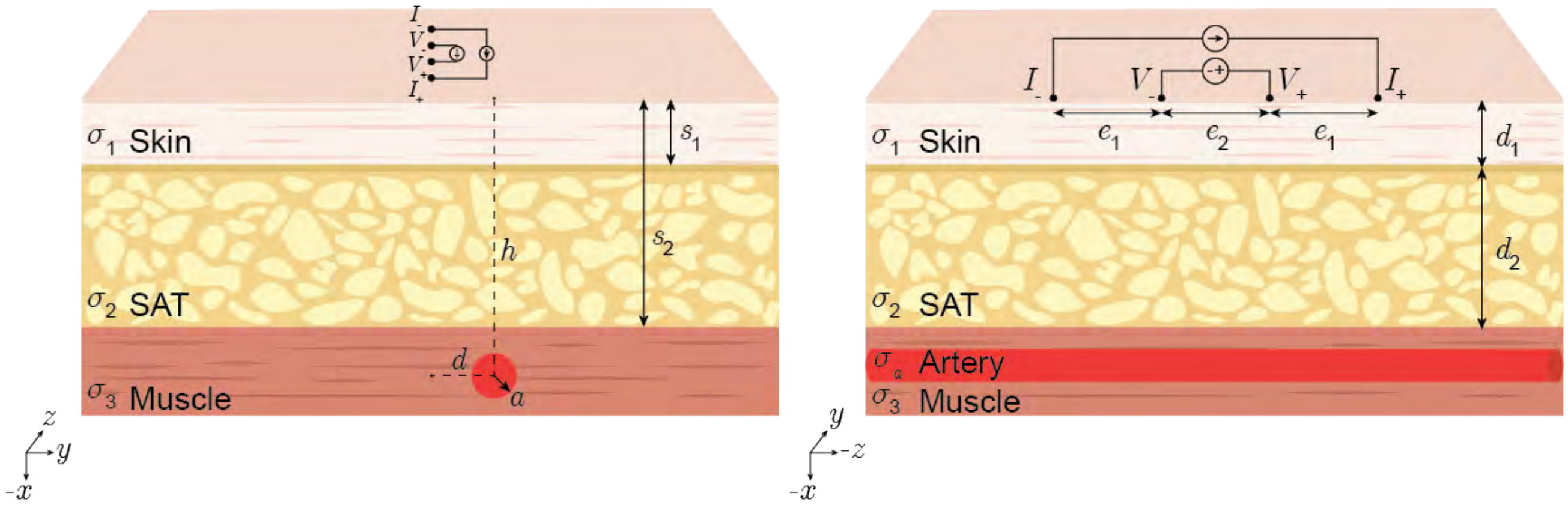

## 13 Supplementary Fig.3. Validation of the three-layer cylindrical bioimpedance model of the wrist

**a, i.** Three-layered domain with skin, subcutaneous adipose tissue (SAT), muscle and a cylindrical artery with a single point current source. The half-space was represented using infinite-element domains. **ii.** The domain was adaptively meshed using at least 1M tetradhedrons. **iii.** As expected, the error is larger near the current singularity at 0 mm where the potential is not defined and it drops to $<$ 1.3 mV beyond $\pm$10 mm. **b, i.** COMSOL domain with a four-terminal bioimpedance measurement configuration. **ii.** The relative error between COMSOL and the model was less than 1% over 21 linearly-spaced blood conductivity values between 0.5 and 0.9 S/m. **iii.** The relative error between COMSOL and the model was less than 1% for 21 linearly-spaced values electrode gaps between 1 and 12 mm. The excellent agreement between our model and COMSOL confirms the accuracy of our analytical model. Scale bars, 1 cm.

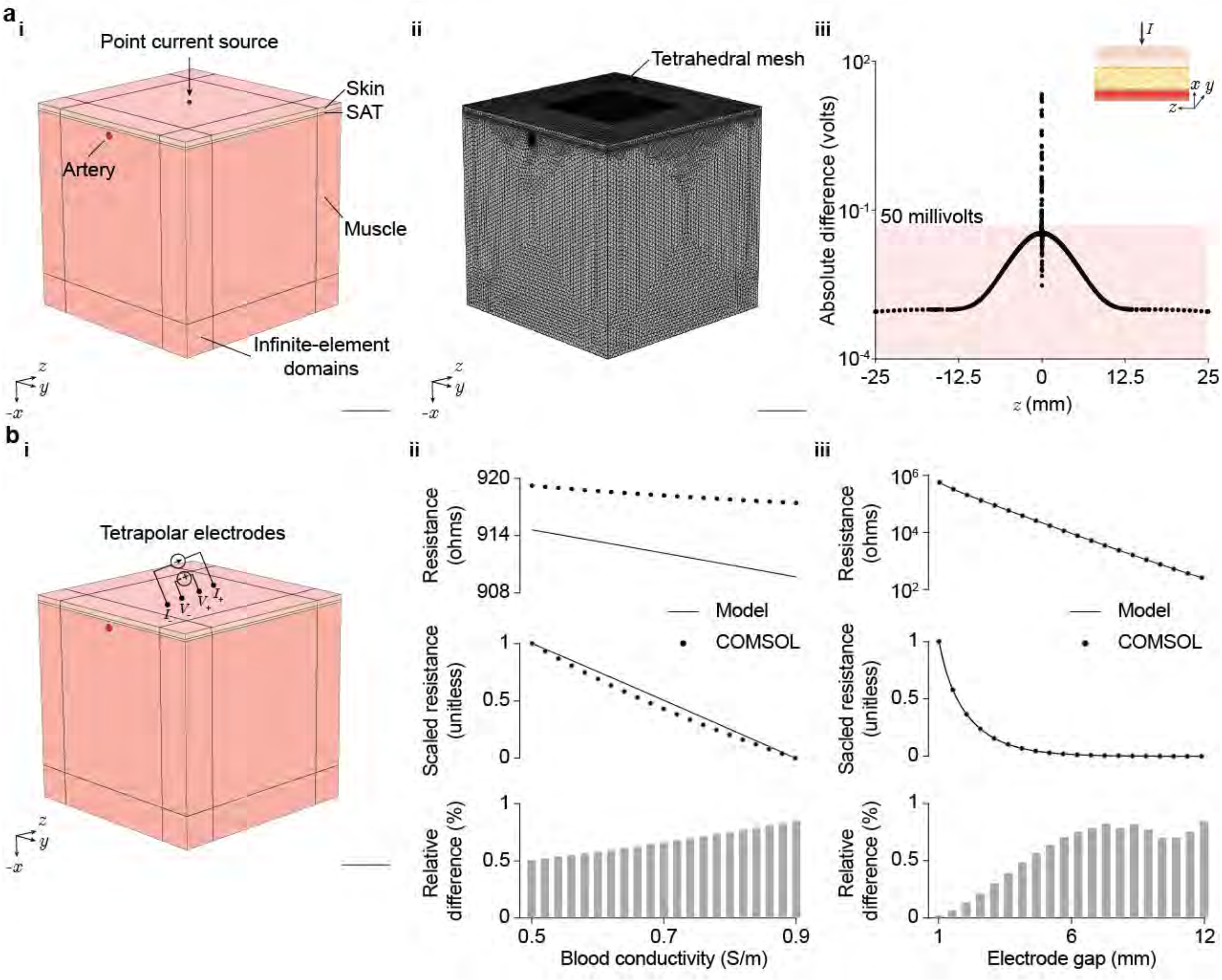

# 14 Supplementary Fig.4. Extended resistance local sensitivity analyses

Extended local sensitivity analyses for: **i**. Mean arterial pressure, **ii**. Heart rate, **iii**. Blood viscosity, **iv**. Radial artery wall thickness to radius ratio, and **v**. Undeformed red blood cell axis ratio. Scale bars, one quarter period. Parameters are sampled at their minimum, mean, and maximum values. BPM, beats per minute.

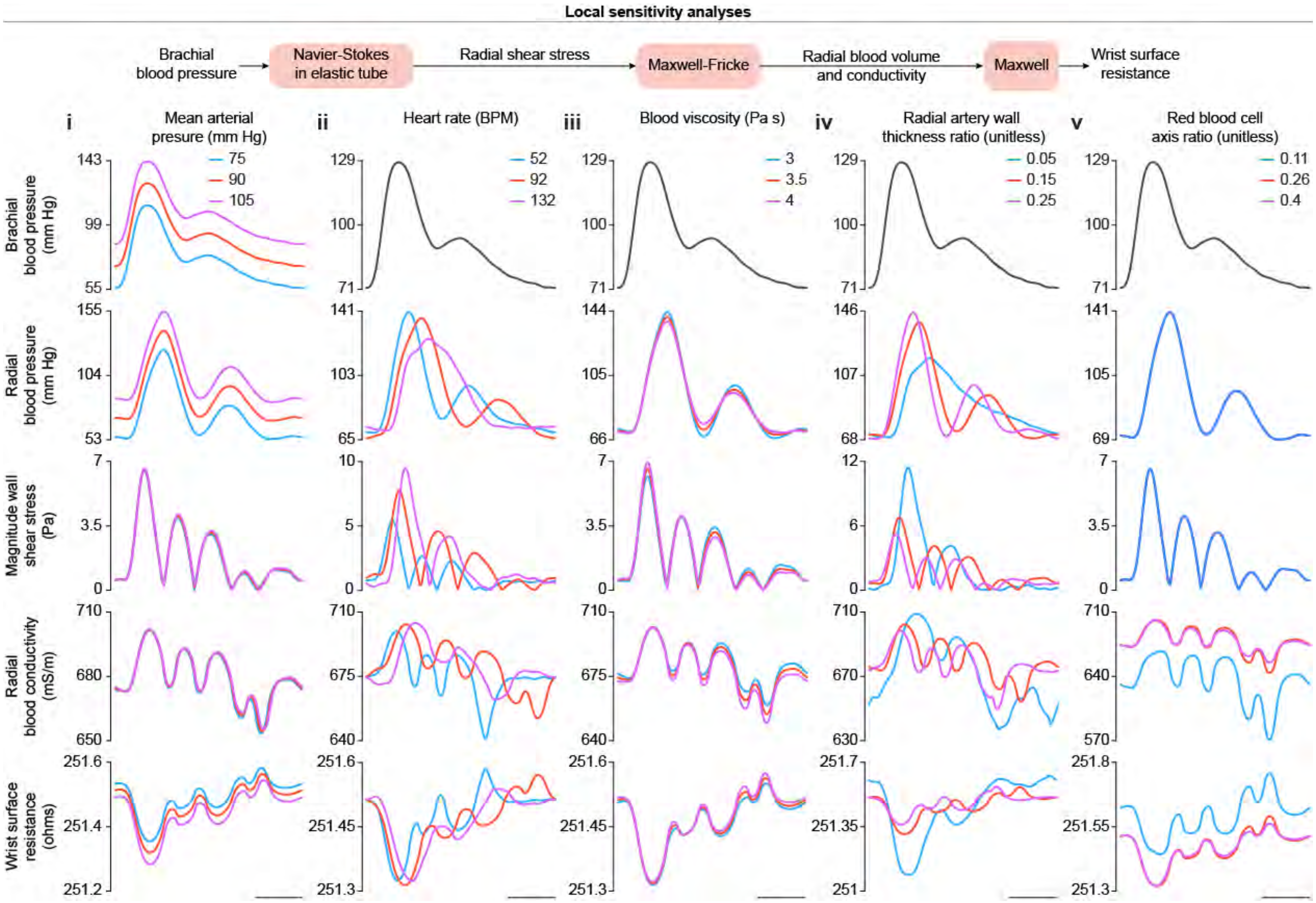

# 15 Supplementary Fig.5. Isopotential lines on Ella phantom with smartwatch circular electrode array

Isopotential lines (V) for different electrode configurations. Simulated baseline impedances at 50 kHz for each configuration **a**, 94 - 83$i$ ohms. **b**, 78 - 92$i$ ohms. **c**, 96 - 82$i$ ohms. **d**, 83 - 96$i$ ohms. **e**, 15 + 11$i$ ohms. **f**, 17 + 8$i$ ohms. $i$ represents the imaginary number, $\sqrt{-1}$. Red and blue electrodes indicate current injection and voltage sensing, respectively. Scale bars, 1 cm.

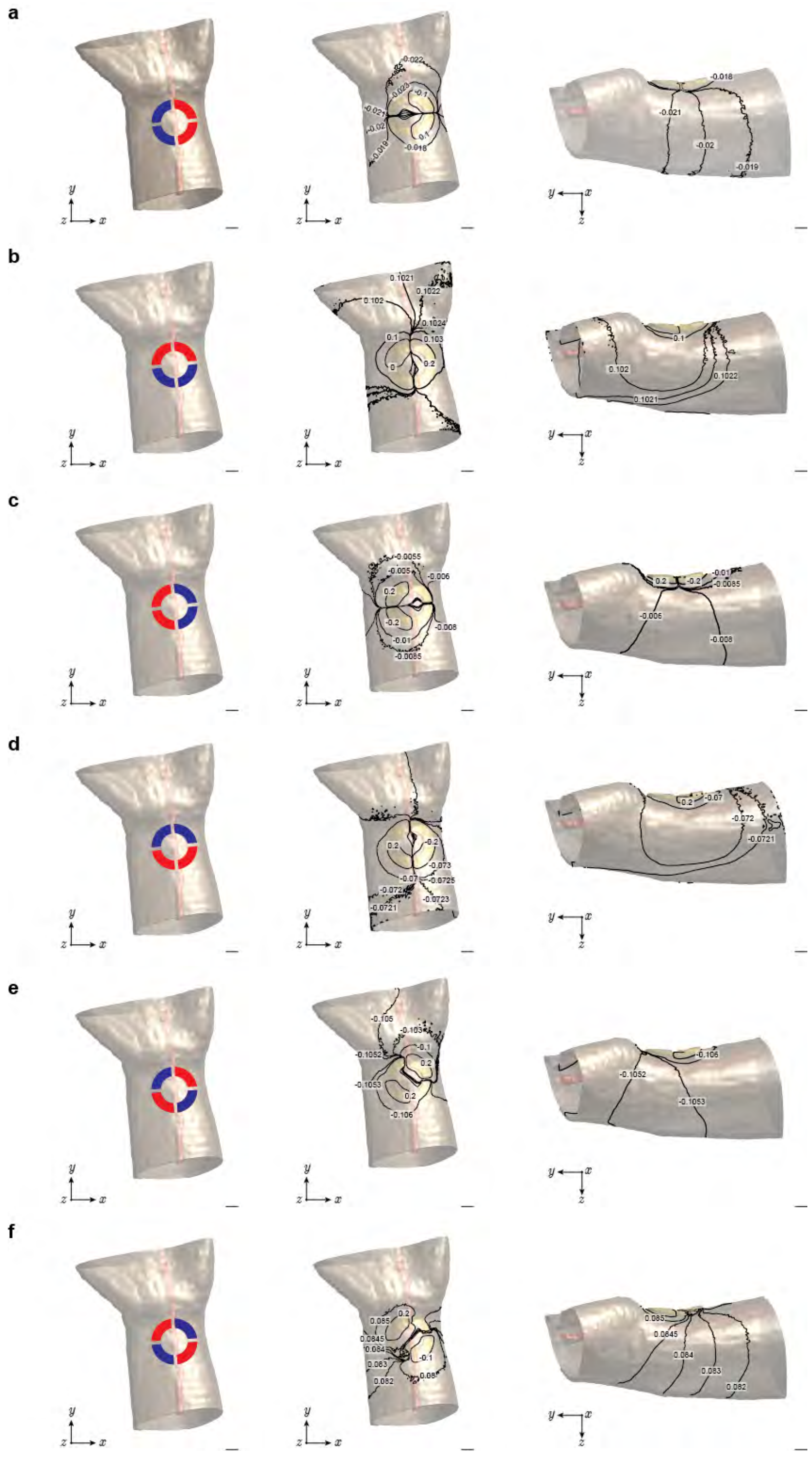

# 16 Supplementary Fig.6. Isopotential lines on Fats phantom with smartwatch circular electrode array

Isopotential lines (V) for different electrode configurations. Simulated baseline impedances at 50 kHz for each configuration **a**, 211-357$i$ ohms. **b**, 153-296$i$ ohms. **c**, 207-348$i$ ohms. **d**, 152-292$i$ ohms. **e**, 56-62$i$ ohms. **f**, 56-61$i$ ohms. $i$ represents the imaginary number, $\sqrt{-1}$. Red and blue electrodes indicate current injection and voltage sensing, respectively. Scale bars, 1 cm.

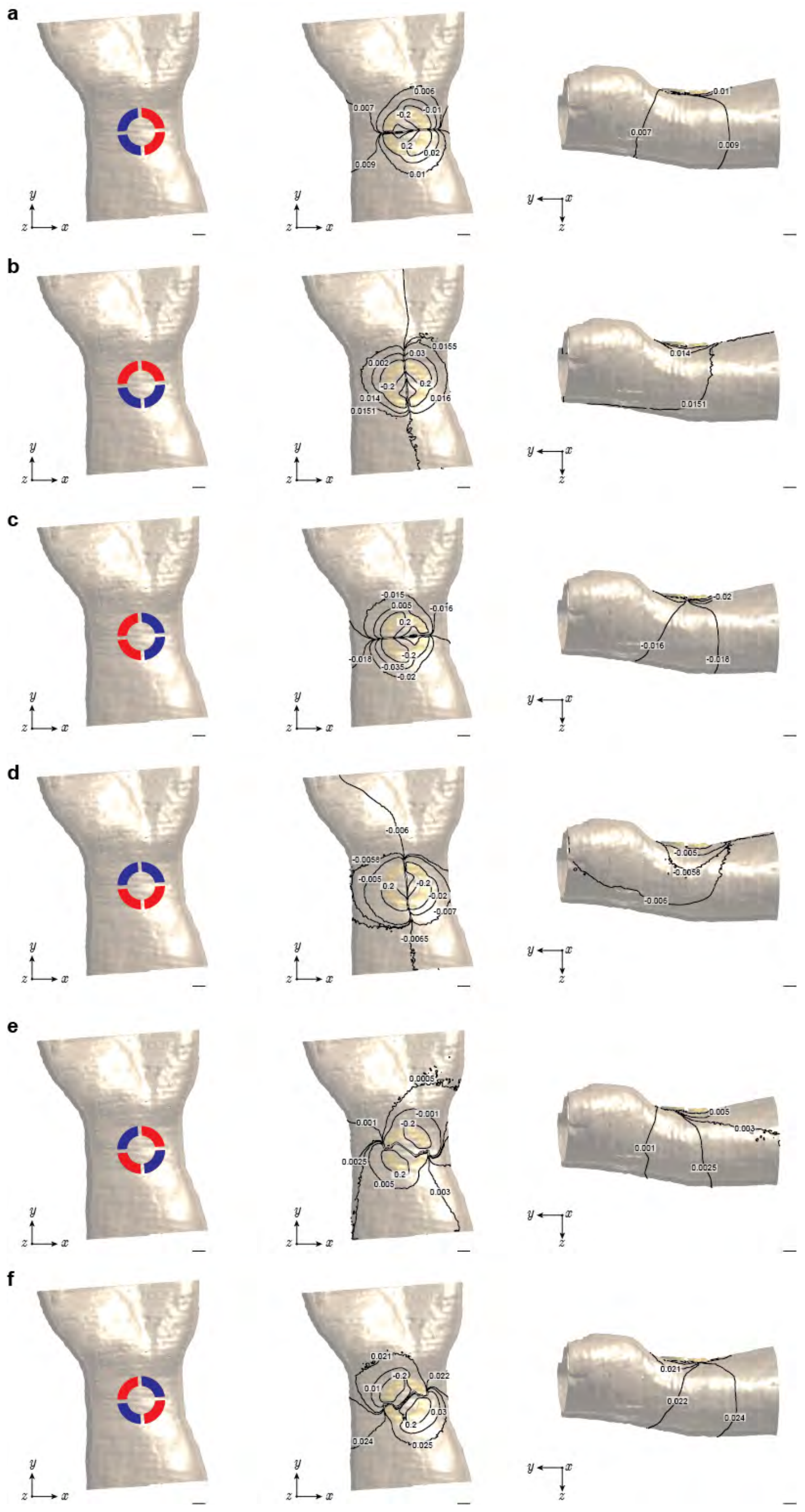

# 17 Supplementary Fig.7. Isopotential lines on Glenn phantom with smartwatch circular electrode array

Isopotential lines (V) for different electrode configurations. Simulated baseline impedances at 50 kHz for each configuration **a**, 131 - 284$i$ ohms. **b**, 144 - 388$i$ ohms. **c**, 147 - 310$i$ ohms. **d**, 146 - 393$i$ ohms. **e**, 6 - 87$i$ ohms, **f**, 6 - 90$i$ ohms. $i$ represents the imaginary number, $\sqrt{-1}$. Red and blue electrodes indicate current injection and voltage sensing, respectively. Scale bars, 1 cm.

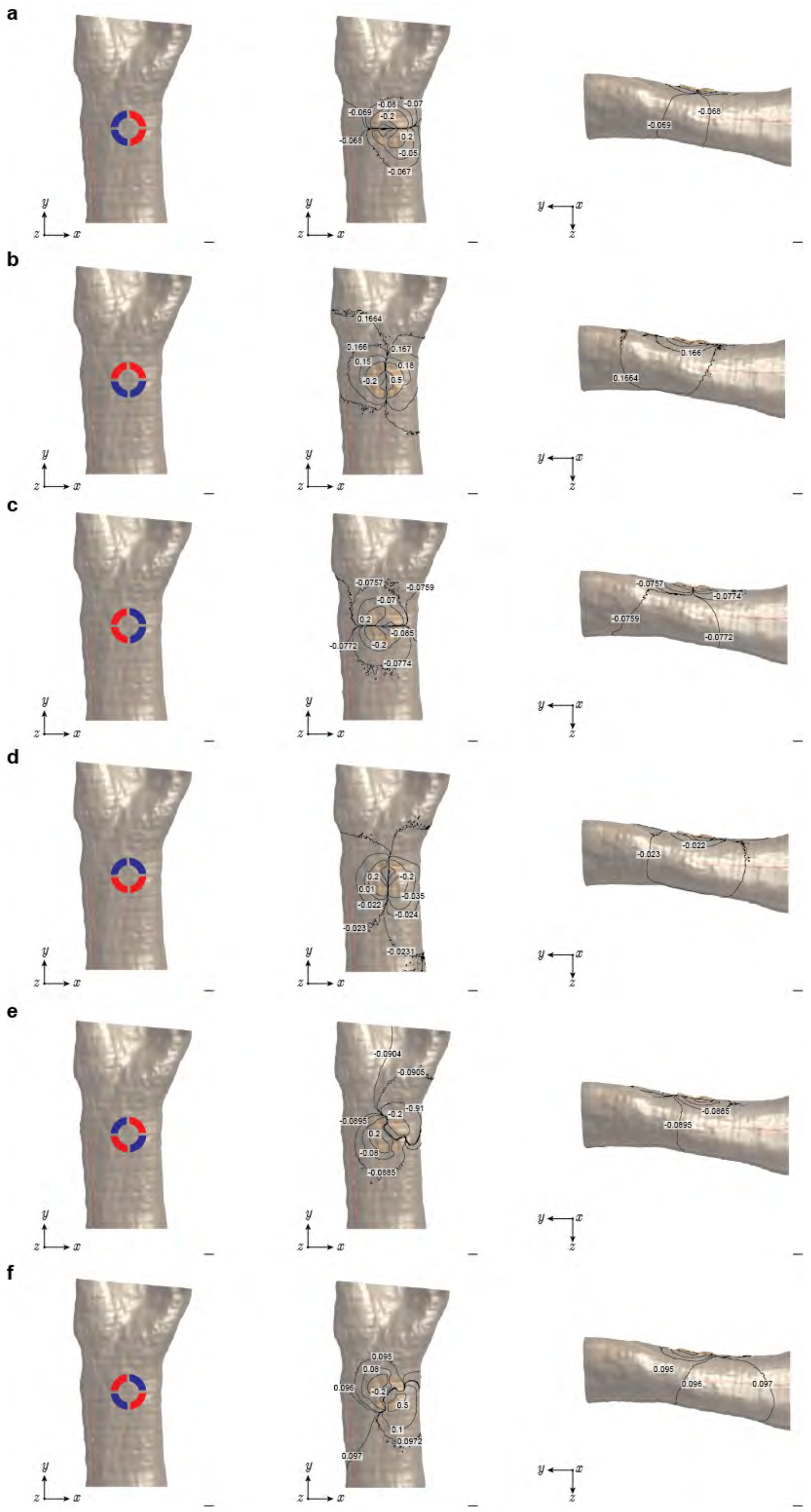

# 18 Supplementary Fig.8. Isopotential lines on morphed Ella phantom with smartwatch circular electrode array

Isopotential lines (V) for different electrode configurations. Simulated baseline impedances at 50 kHz for each configuration **a**, 89-75$i$ ohms. **b**, 80- 99$i$ ohms. **c**, 94 - 78$i$ ohms. **d**, 84 - 101$i$ ohms. **e**, 10 + 23$i$ ohms. **f**, 11 + 24$i$ ohms. $i$ represents the imaginary number, $\sqrt{-1}$. Red and blue electrodes indicate current injection and voltage sensing, respectively. Scale bars, 1 cm.

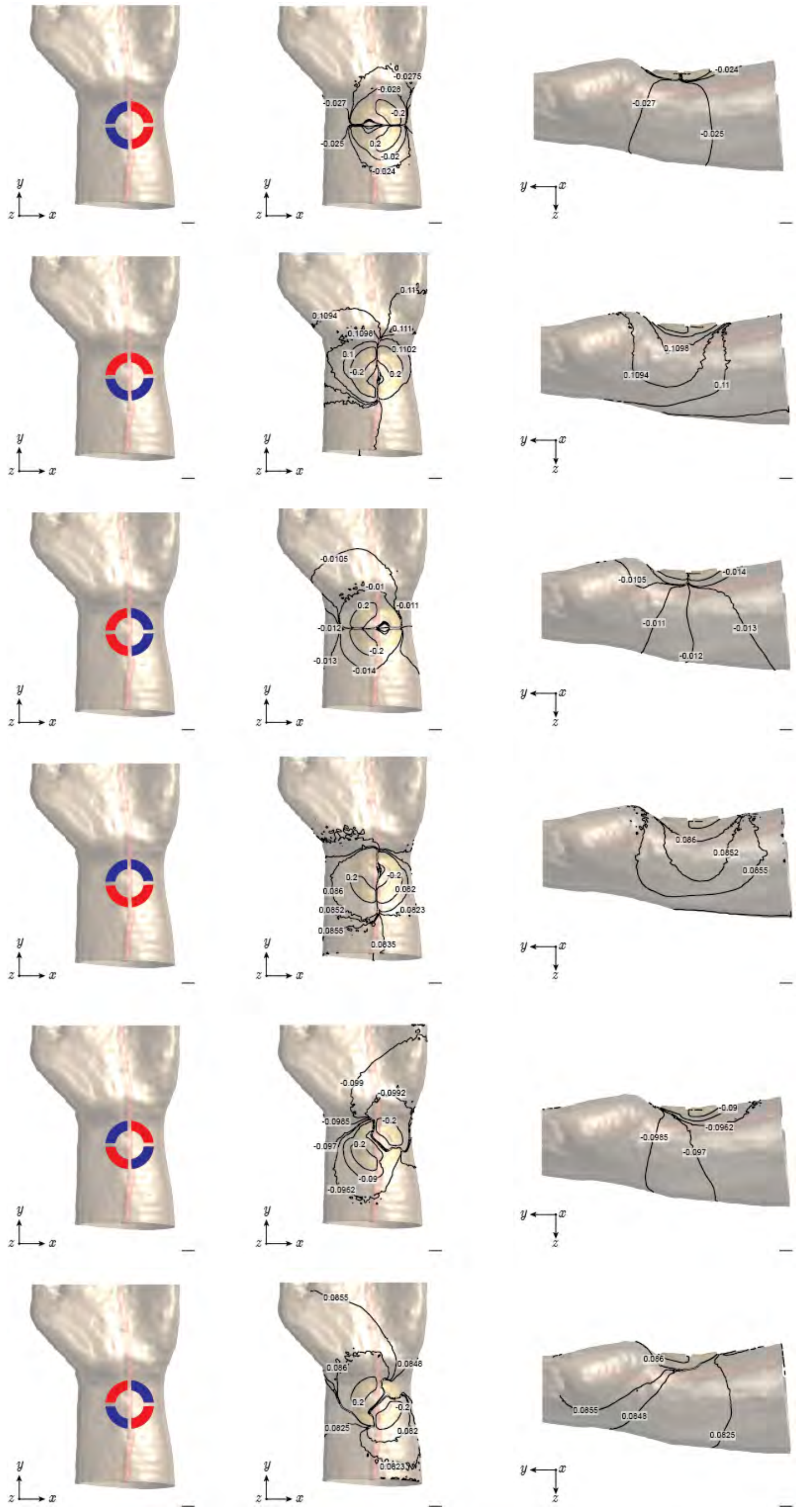

# 19 Supplementary Fig.9. Relative change in resistance and reactance for multiple electrode positions

**a**, Resistance. **b**, Reactance. The percent change is calculated with respect to the nominal position.

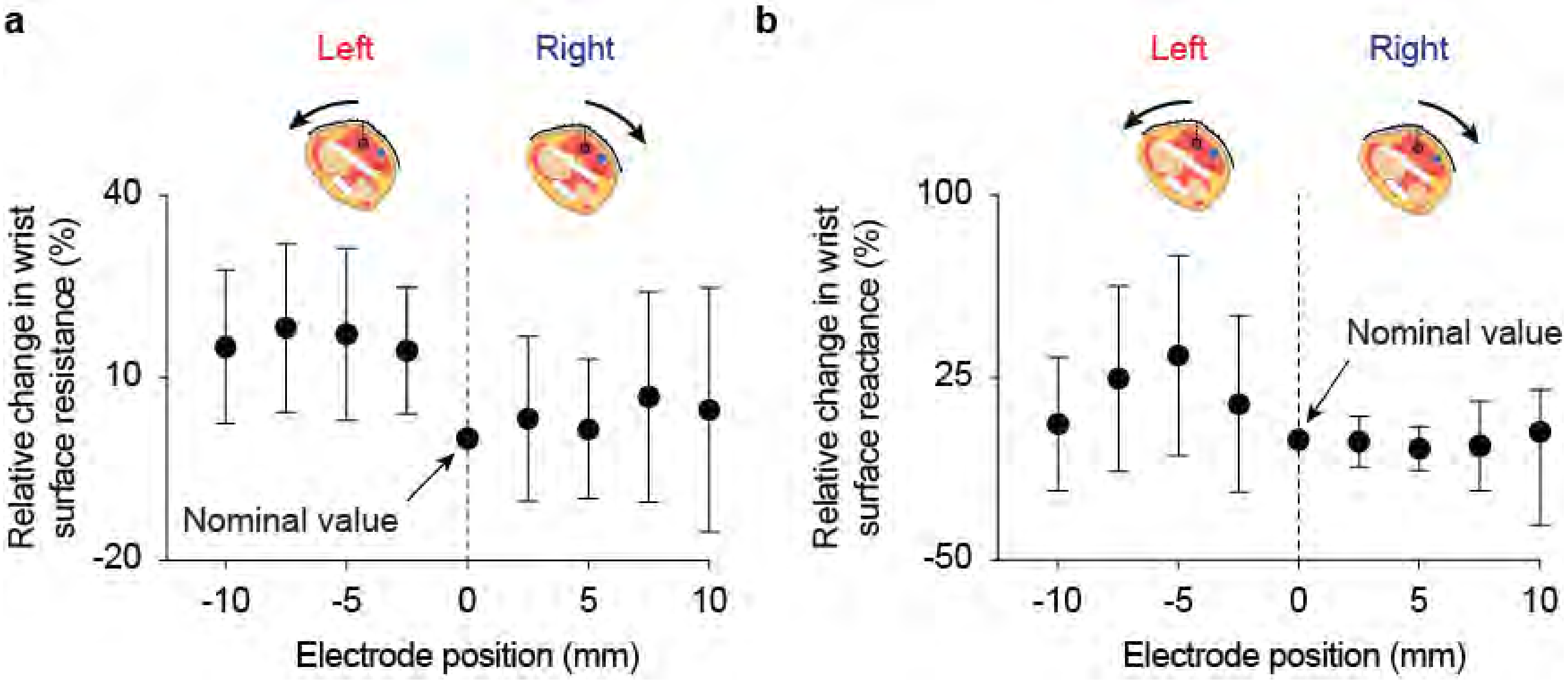

# 20 Supplementary Fig.10. Volume resistance density distribution

The volumes visualize the most sensitive voxels that sum to 95% of the total resistance. **a**, Ella model featured a 49x59x45 mm$^3$ volume. **b**, Morphed Ella model featured a 49x61x44 mm$^3$ volume. **c**, Fats model featured a 60x54x45 mm$^3$ volume. **d**, Glenn model featured a 62x58x49 mm$^3$ volume. VRD, volume resistance distribution.

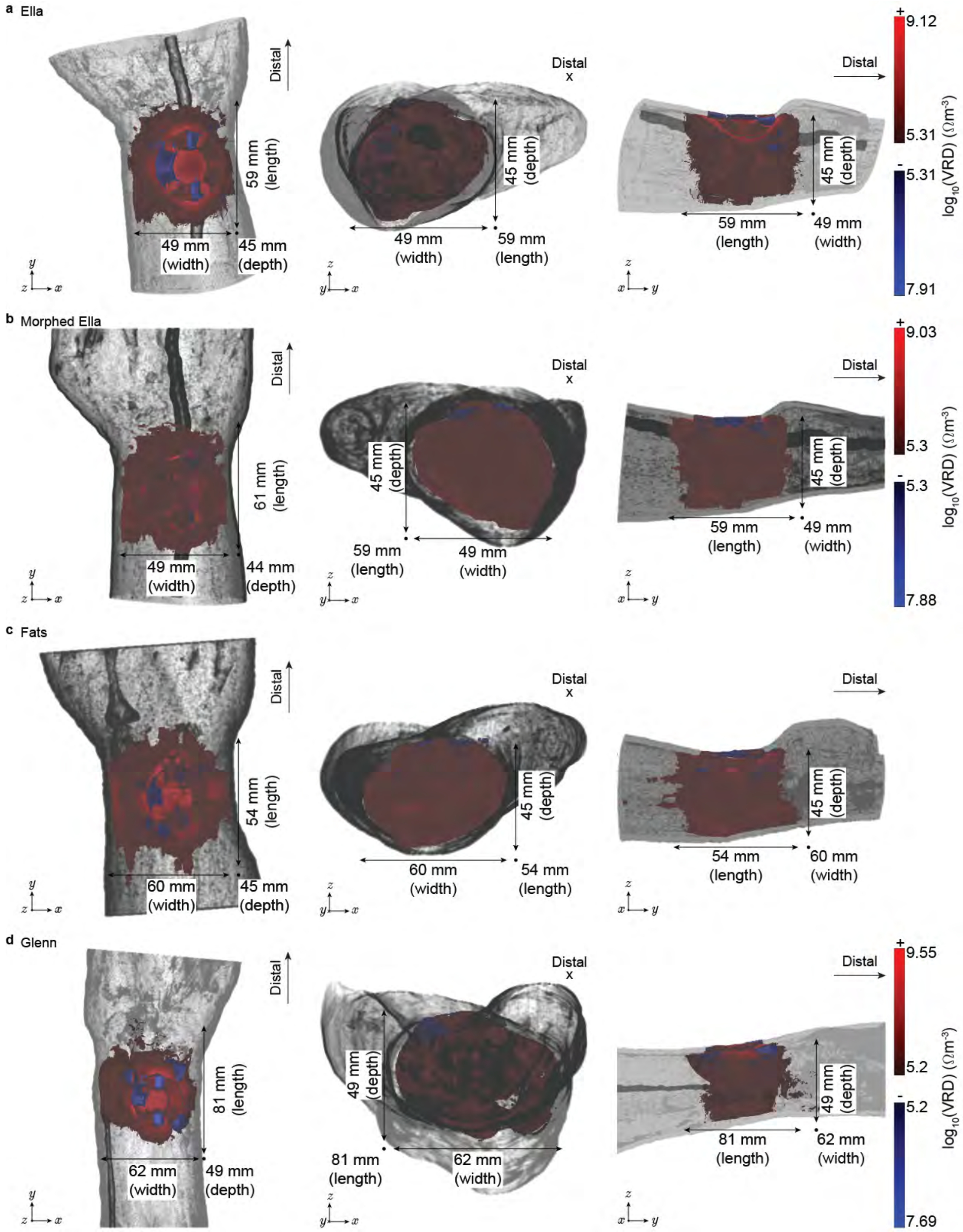

# 21 Supplementary Fig.11. Volume reactance density distribution

The volumes visualize the most sensitive voxels that sum to 95% of the total reactance. **a**, Ella model featured a 47x53x44 mm$^3$ volume. **b**, Morphed Ella model featured a 47x54x44 mm$^3$ volume. **c**, Fats model featured a 51x53x46 mm$^3$ volume. **d**, Glenn model featured a 51x58x46 mm$^3$ volume. VXD, volume reactance distribution.

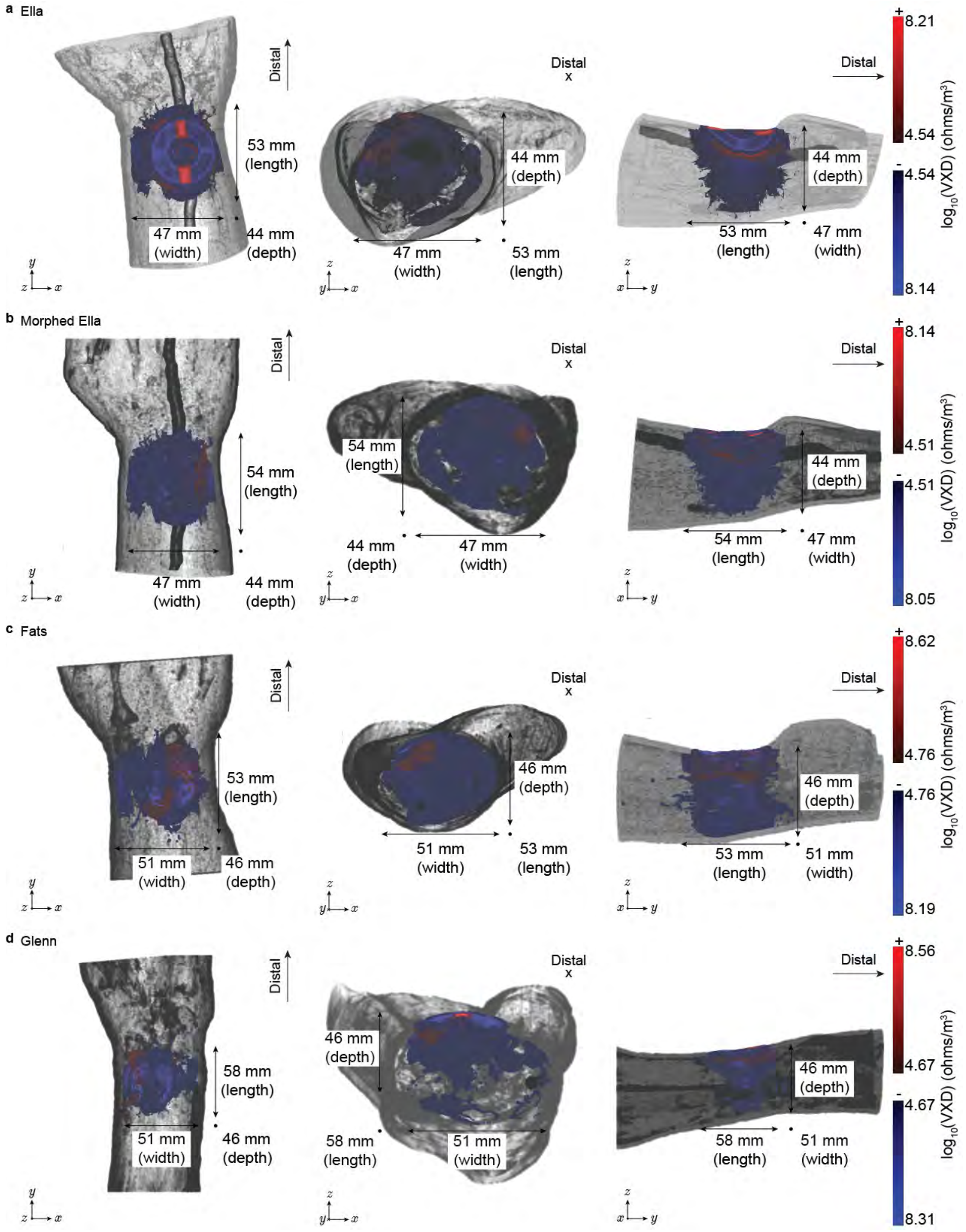

# 22 Supplementary Fig.12. Resistance and reactance contribution of each tissue for individual human phantoms

Tissue resistance and reactance contributions across four human phantoms with circular electrodes and adjacent current injection at 50 kHz. **a**, Resistance and **b**, Reactance contributions. Red and blue electrodes indicate current injection and voltage sensing, respectively. Scale bars, 1 cm. SAT, subcutaneous adipose tissue.

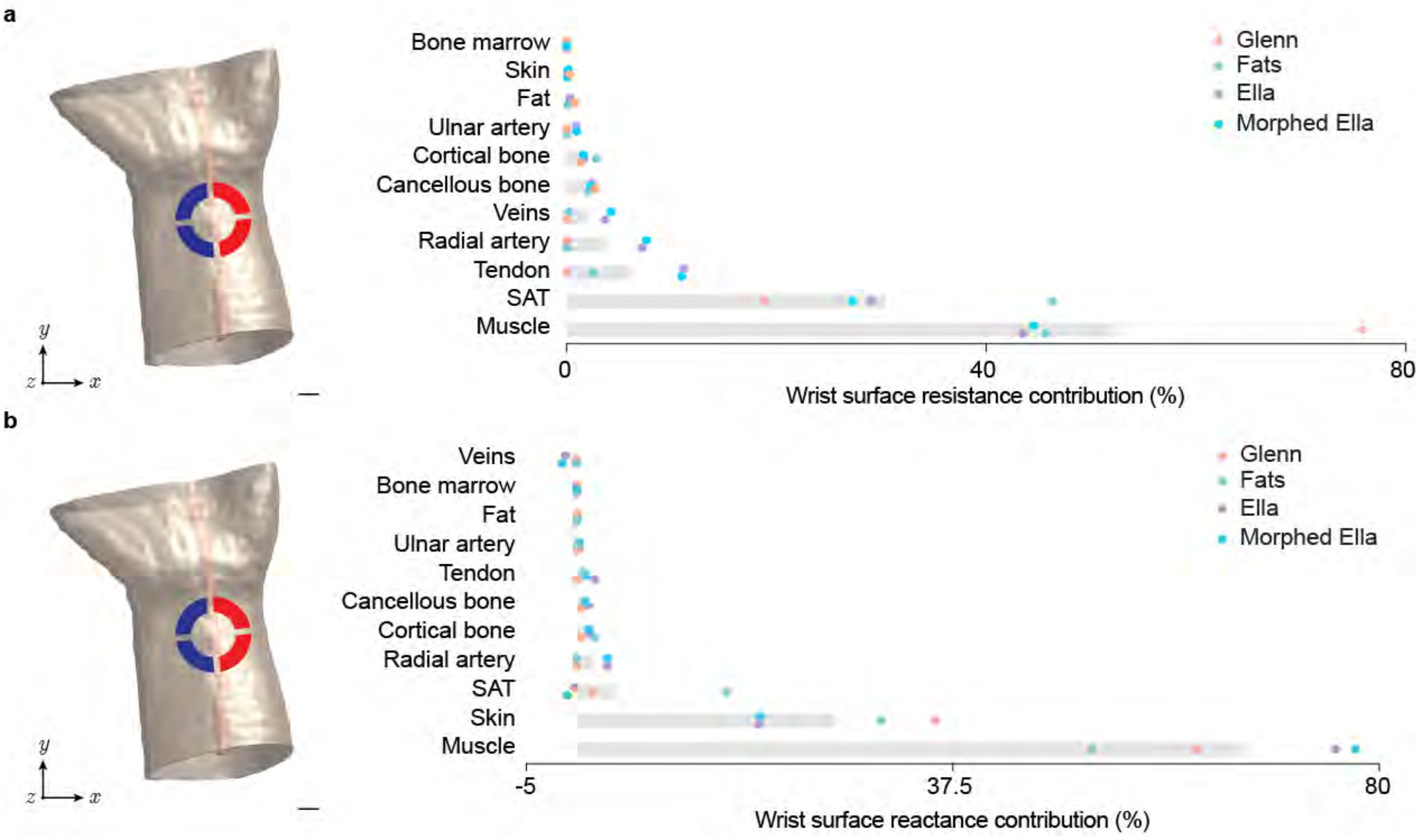

# 23 Supplementary Fig.13. Isopotential lines on Ella phantom with a linear electrode array

Isopotential lines (V) for a linear electrode array with longitudinal (**a**) and transverse (**b**) orientations. The simulations featured a baseline impedance of 201-111$i$ and 102-61$i$ ohms at 50 kHz for the linear and transverse configurations, respectively. $i$ represents the imaginary number, $\sqrt{-1}$. Red and blue electrodes indicate current injection and voltage sensing, respectively. Scale bars, 1 cm.

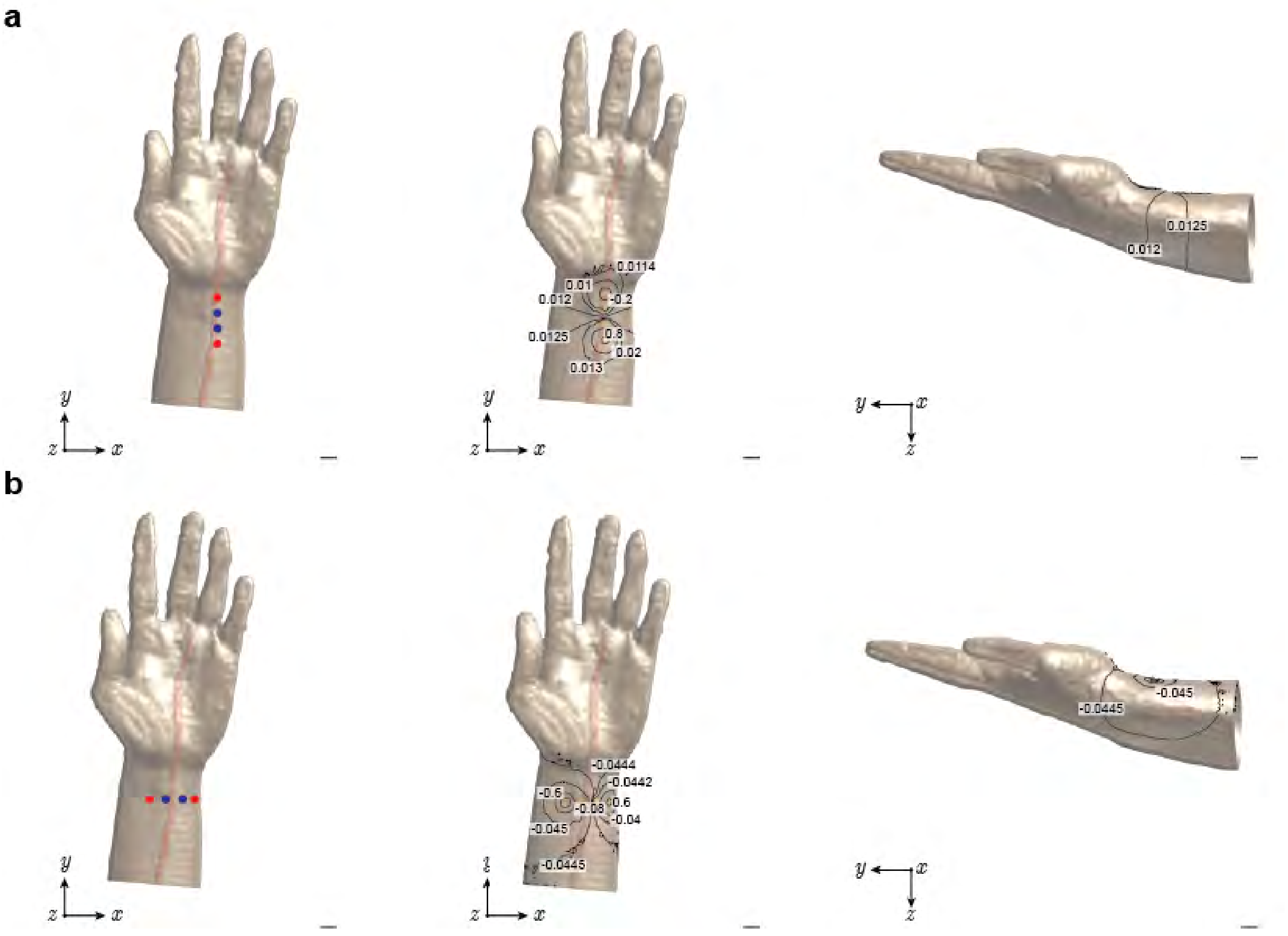

# 24 Supplementary Fig.14. Volume and tissue resistance contribution for Ella phantom with four disc electrodes

Electrodynamic simulations for a linearly aligned longitudinal and transverse electrode array configuration on the Ella computable phantom with 5 mm diameter electrodes and 10 mm center-to-center spacing. **a**, The linear electrode configuration featured a 30 (width) x 46 (length) x 27 (depth) $mm^3$ volume contributing to 99% of the measured resistance and the radial artery contributed 7.95% to the wrist surface resistance. **b**, The transverse electrode configuration featured a 32 (width) x 35 (length) x 26 (depth) $mm^3$ volume contributing to 99% of the measured resistance and the radial artery contributed 5.45% to the wrist surface resistance. Scale bars, 1 cm. SAT, subcutaneous adipose tissue; VRD, volume resistance distribution.

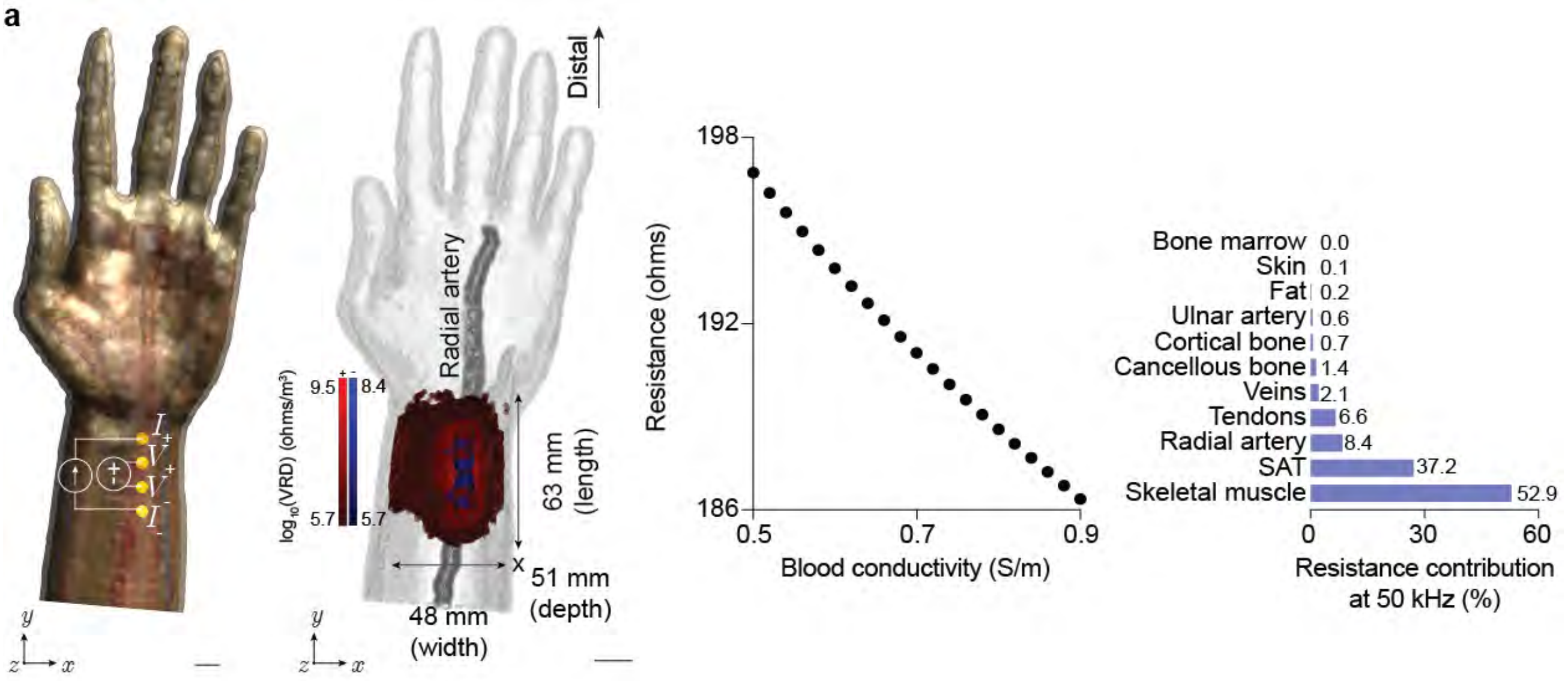

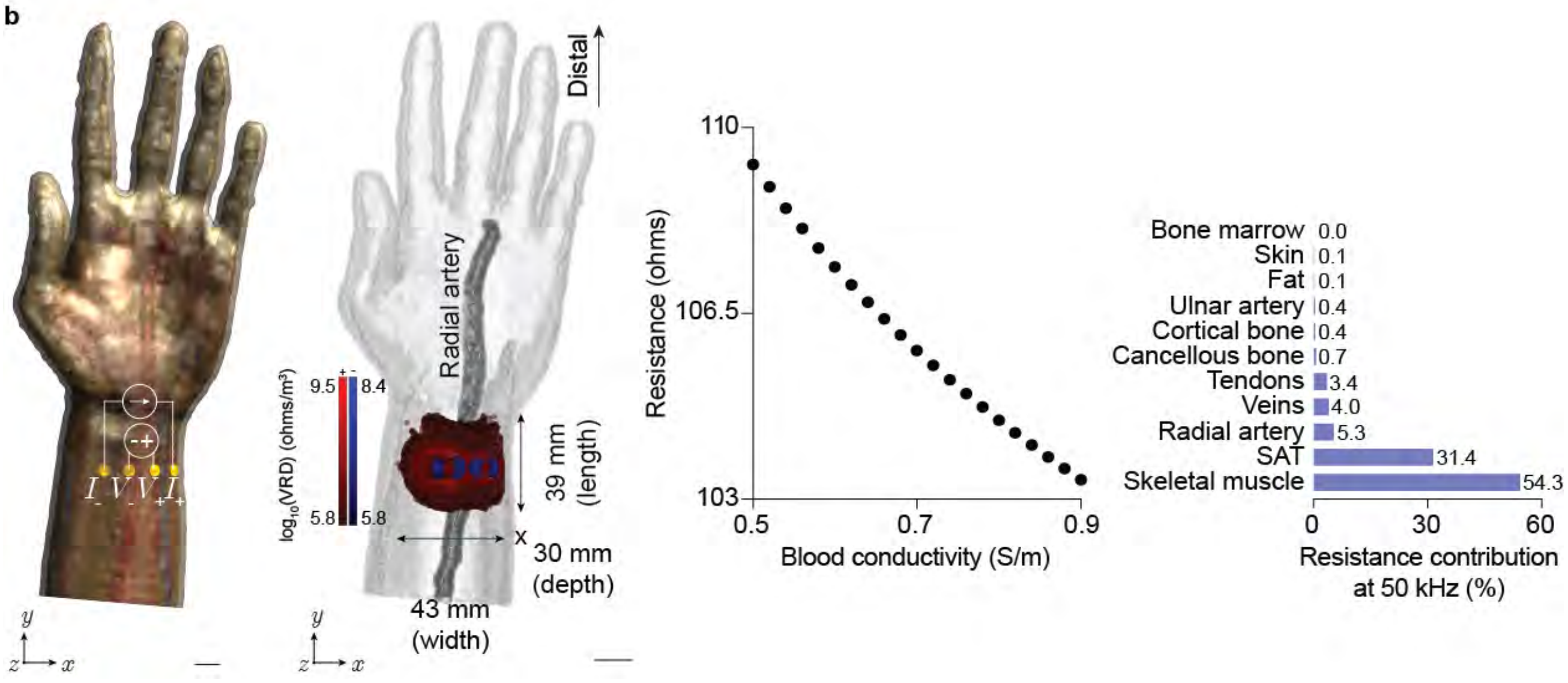

## 25 Supplementary Fig.15. Enrollment flowchart and study protocols for Group 1

Flow diagrams of participant enrollment for the healthy lab study and corresponding study protocols. The lab study was divided into three sub-cohorts each performing a different protocol. A total of $N$ = 76 healthy subjects were enrolled and $N$ = 75 completed the study in the lab setting. **a, i.** Enrollment flowchart for Group 1a. **ii.** Study protocol for Group 1a. **b, i.** Enrollment flowchart for Group 1b. **ii.** Study protocol for Group 1b. **c, i.**, Enrollment flowchart for Group 1c. **ii.** Study protocol for Group 1c. Enrollment and study protocol details are described in Supplementary Discussion 6.

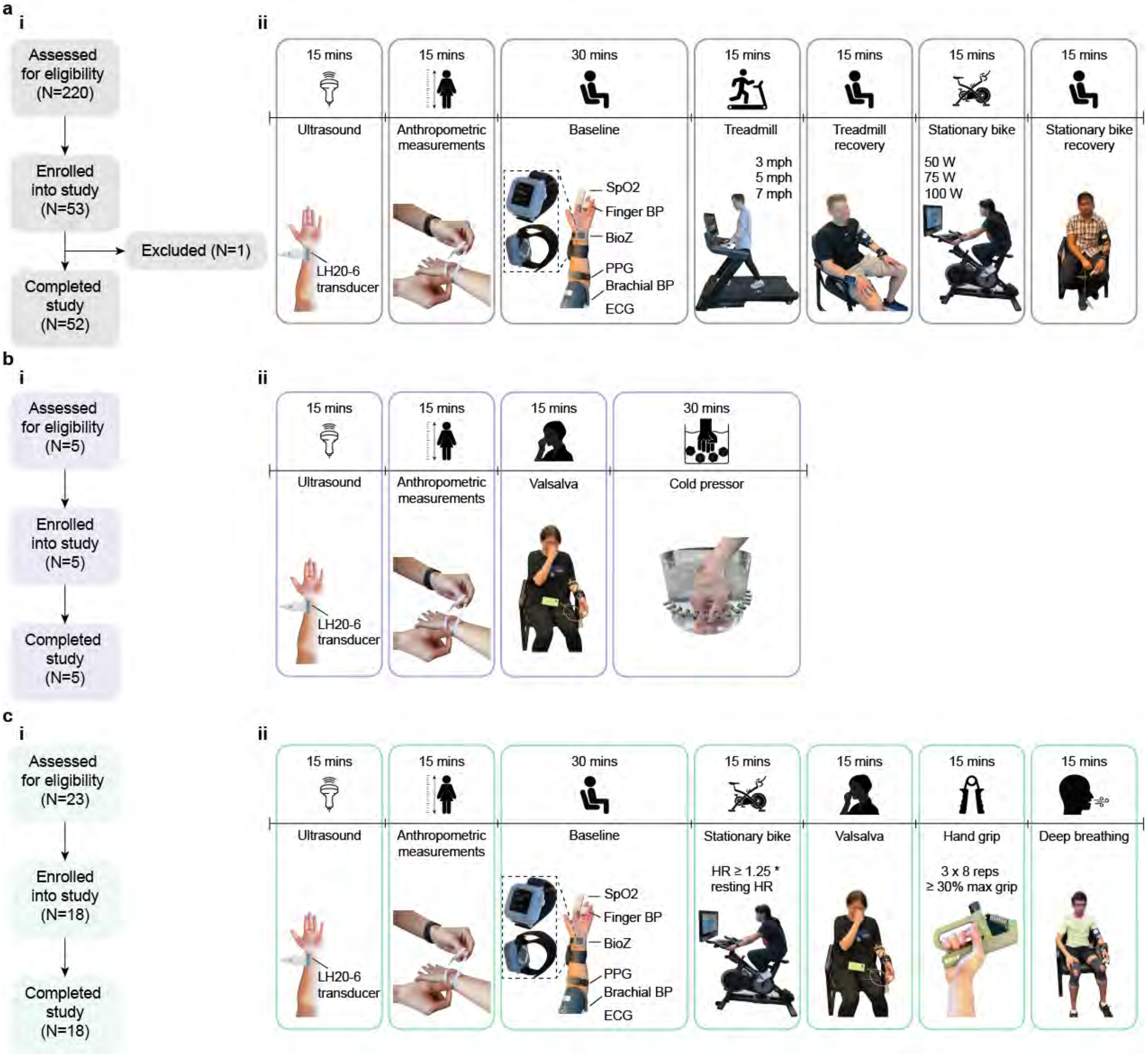

# 26 Supplementary Fig.16. Demographic and anthropometric data for Group 1

**a**, Age. **b**, Height. **c**, Weight. **d**, Body mass index (BMI). **e**, Wrist circumference. **f**, Forearm length. **g**, Upper arm length. Data shown for male, female, and male and female combined. Statistical tests results in Supplementary Table 12.

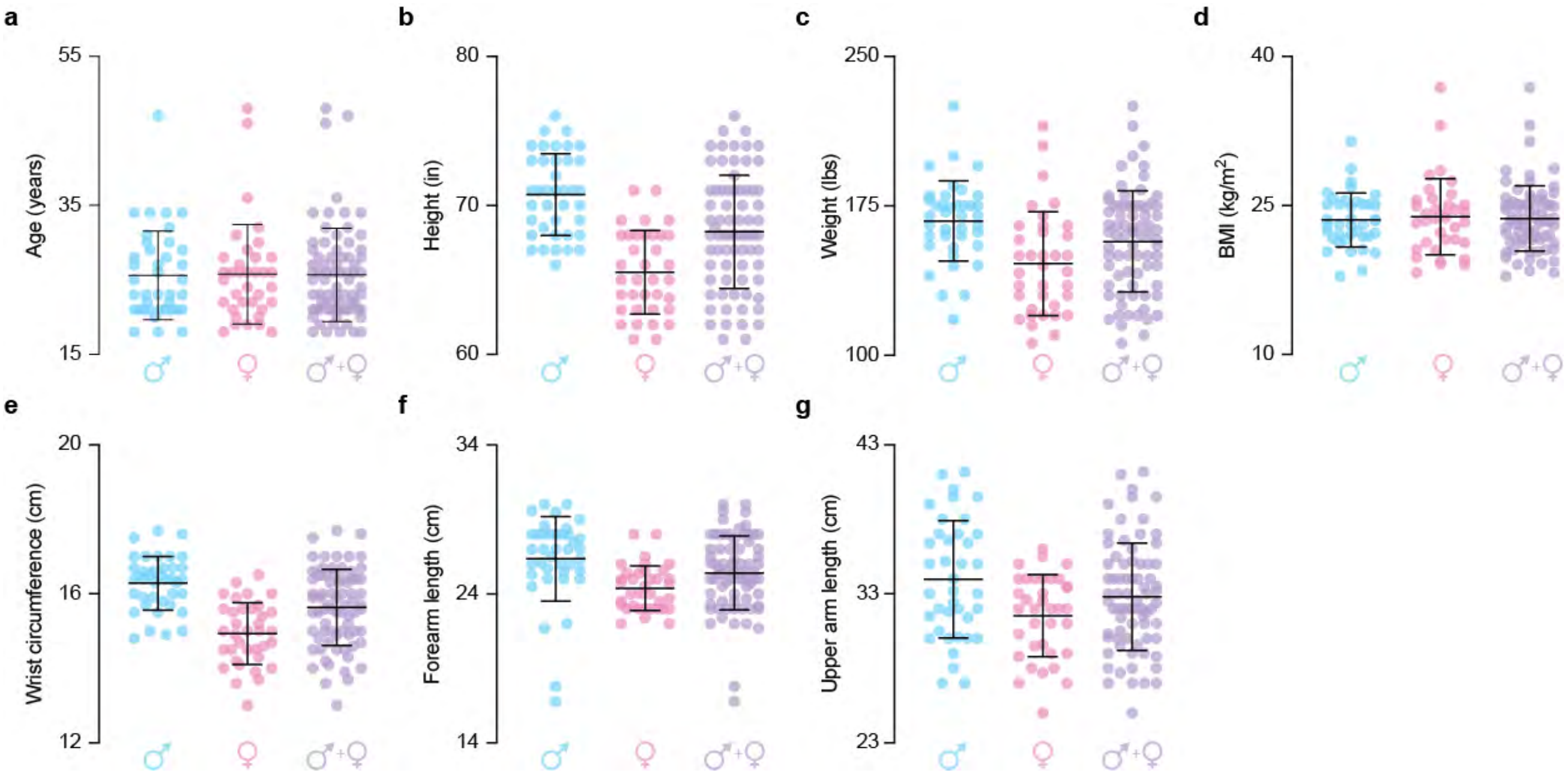

# 27 Supplementary Fig.17. Data preprocessing pipeline for training and testing a signal-tagged physics-informed neural network model

The experimentally collected data underwent a complete processing pipelines including filtering, synchronization, segmenting, and signal quality assessment to prepare the data for machine learning. Abbreviations: BioZ, bioimpedance; BP, blood pressure; UC, uncalibrated; LPF, low pass filter; IBI, inter beat interval; PCA, principle component analysis; LRR, low rank reconstruction.

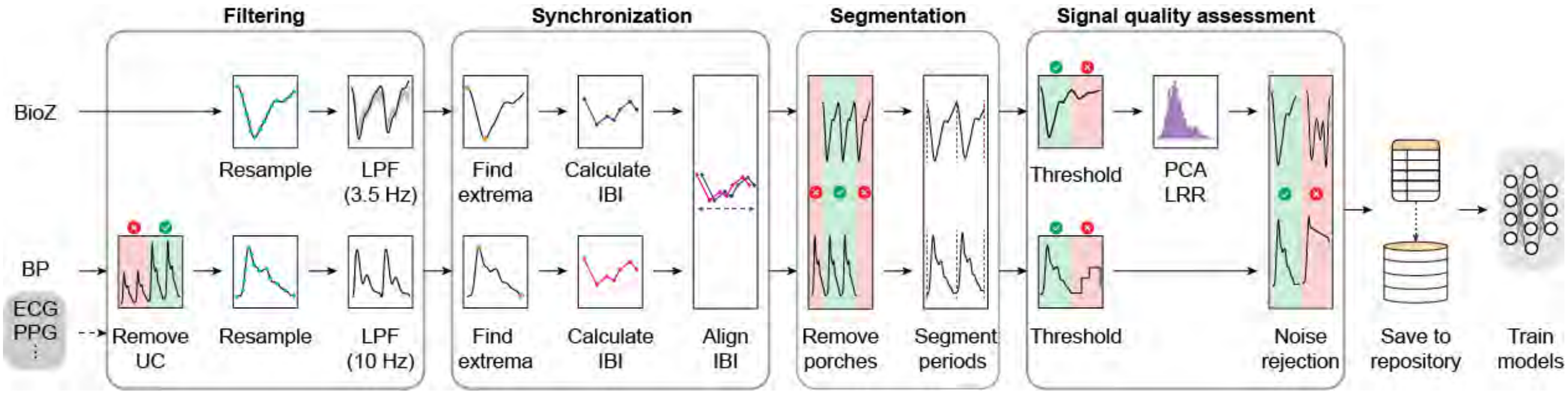

# 28 Supplementary Fig.18. Comparison of synthetically generated resistance signals from the multiscale modeling framework with experimentally measured resistance signals

**a**, Five randomly selected synthetic and corresponding experimental resistance periods. **b**, One thousand overlaid synthetic and corresponding resistance periods, and mean resistance period shown in dotted red line. All synthetic data was generated using the experimentally collected blood pressure corresponding to the shown resistance data. Scale bars, one-quarter period. Experimental resistance data underwent the full processing pipeline while the synthetic was only low pass filtered. Each period was individually rescaled using min-max normalization for easier visual comparison.

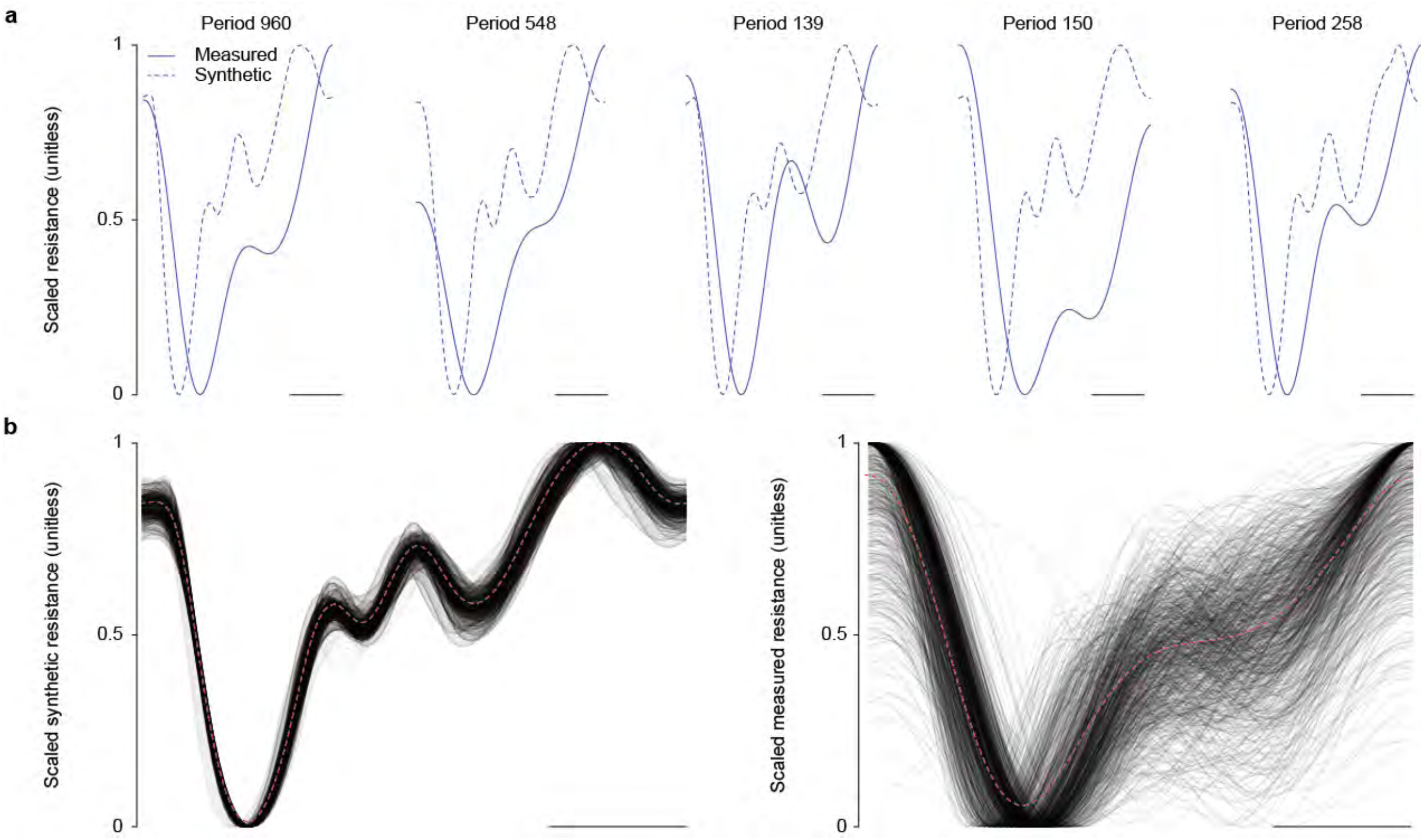

# 29 Supplementary Fig.19. Collocation point sampling procedure for training a signal-tagged physics-informed neural network

**a**, Illustration of collocation point sampling from the dynamic space-time domain representing an artery (Section 7.6.2). Each sampled point, $(r, z, t)$, is used to enforce a particular type of loss, $\mathcal{L}_i$ for $i \in \{\text{PDE}, \text{BC}, \text{IC}, \text{data}, \text{stitch}\}$ during signal-tagged physics-informed neural network training (Supplementary Discussion 7 and Equation (42)). "Supp" points show supplementary sampled points that are treated as "data" points in the loss function. **b**, Initial collocation points are sampled from a reference domain and boundary Equation (45). **c**, Example of the collocation point sampling procedure. **i.** Sampled pressure $\mathbf{p}_k$ and resistance signal $\tilde{\mathbf{r}}_k$. **ii.** Sampled collocation points $(r, z, t)$ for the different loss types (PDE, BC, IC, data, stitch). PDE, partial differential equation; BC, boundary condition; IC, initial condition. Scale bars, one-quarter period.

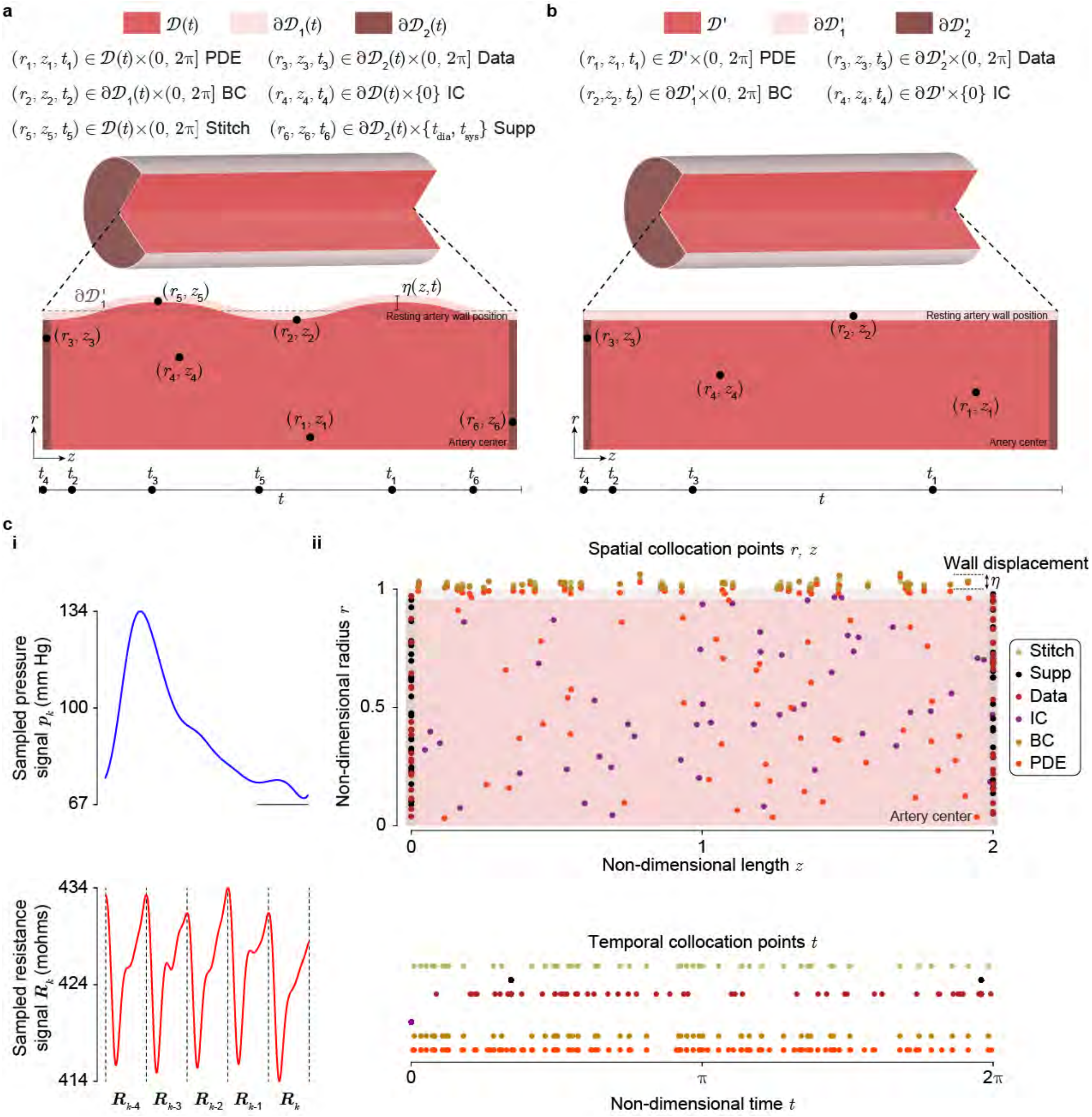
a
(r1, z1, t1) ∈ D(t)×(0, 2π] PDE
(r3, z3, t3) ∈ ∂D2(t)×(0, 2π] Data
(r2, z2, t2) ∈ ∂D1(t)×(0, 2π] BC
(r4, z4, t4) ∈ ∂D(t)×{0} IC
(r5, z5, t5) ∈ D(t)×(0, 2π] Stitch
(r6, z6, t6) ∈ ∂D2(t)×{tdia, tsys} Supp
η(z,t)
Resting artery wall position
Artery center
b
(r1, z1, t1) ∈ D'×(0, 2π] PDE
(r3, z3, t3) ∈ ∂D'2×(0, 2π] Data
(r2, z2, t2) ∈ ∂D'1×(0, 2π] BC
(r4, z4, t4) ∈ ∂D'×{0} IC
Resting artery wall position
Artery center
c
i
Sampled pressure signal pk (mm Hg)
134
100
67
Sampled resistance signal Rk (mohms)
434
424
414
Rk-4 Rk-3 Rk-2 Rk-1 Rk
ii
Spatial collocation points r, z
Wall displacement
Non-dimensional radius r
Non-dimensional length z
Artery center
Stitch
Supp
Data
IC
BC
PDE
Temporal collocation points t
Non-dimensional time t
0
π
2π

# 30 Supplementary Fig.20. Elastic wall motion method for signal-tagged physics-informed neural network

**a**, Diagram of the wall motion method described in Section 7.5.3. **b**, Representative plots showing the convergence of the wall motion method iterates; we plot iteration $j$ vs. the percent relative error at iteration $j$, namely

$$\frac{\sqrt{\sum_{i=1}^{100}\left(\eta^{(j)}\left(z=\frac{1}{2}\frac{L_{\mathcal{D}}}{\bar{a}},t_i\right)-\eta^{(100)}\left(z=\frac{1}{2}\frac{L_{\mathcal{D}}}{\bar{a}},t_i\right)\right)^2}}{\sqrt{\sum_{i=1}^{100}\left(\eta^{(100)}\left(z=\frac{1}{2}\frac{L_{\mathcal{D}}}{\bar{a}},t_i\right)\right)^2}}\cdot 100\%$$

for $j=1,\ldots,10$. Here, the reference solution is the 100-th iteration, $\eta^{(100)}$. The inset plots are the signal-tagged physics-informed neural network (sPINN) modeled BPs at $r=1$, $z=\frac{1}{2}\frac{L_{\mathcal{D}}}{\bar{a}}$, and $t\in[0,2\pi]$. **c**, We plot the $j=1,\ldots,5$ first iterations of the wall position, $\eta^{(j)}\left(z=\frac{1}{2}\frac{L_{\mathcal{D}}}{\bar{a}},t\right)$, $t\in[0,2\pi]$ for the elastic wall motion method. Radial location vs. time heatmap plots of **d**, Blood pressure, **e**, Axial velocity, and **f**, Radial velocity predictions from the sPINN model for two different example input resistance sequences. Scale bars, one-quarter period.

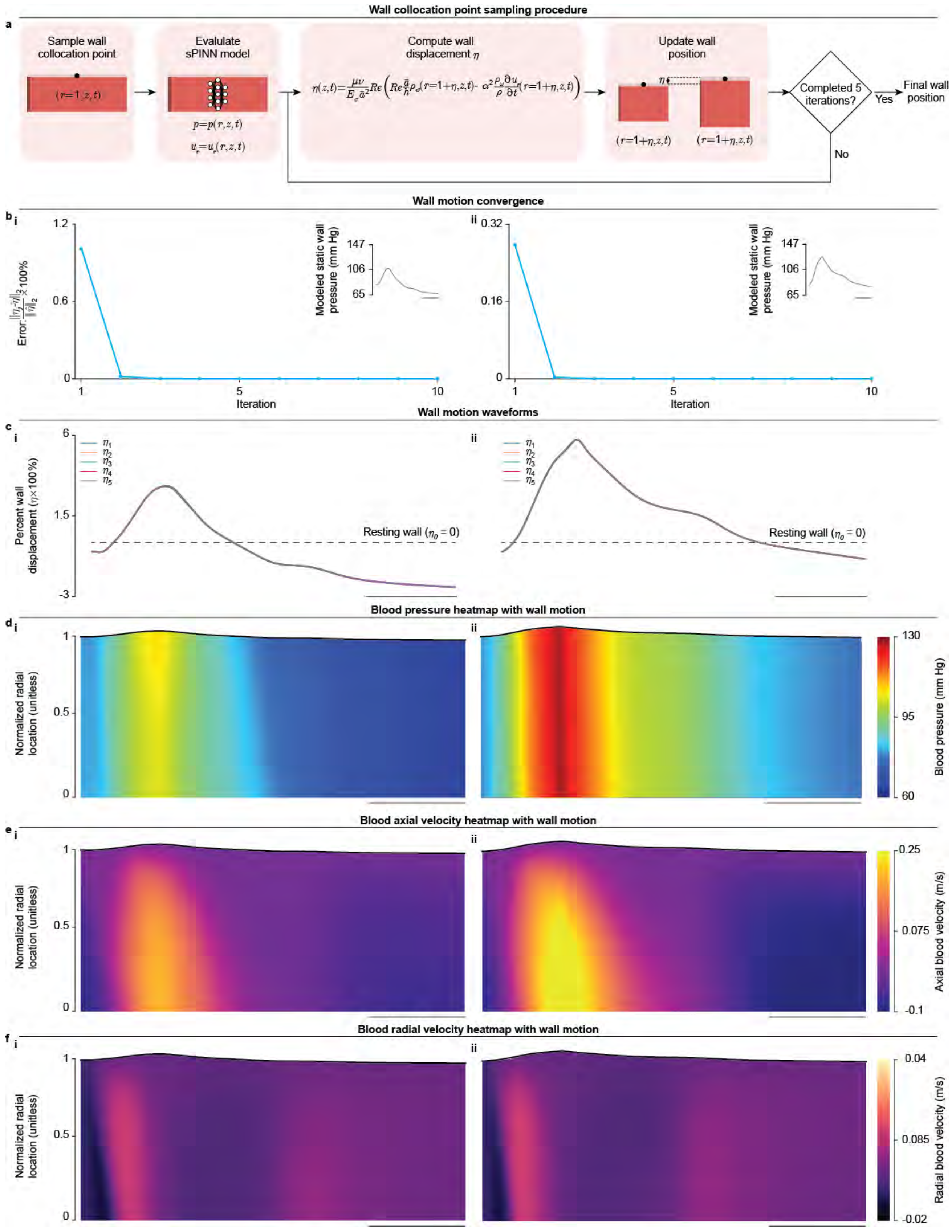
Wall collocation point sampling procedure
a
Sample wall collocation point
Evalulate sPINN model
Compute wall displacement η
Update wall position
Completed 5 iterations?
Yes
Final wall position
No
Wall motion convergence
b
i
ii
Modeled static wall pressure (mm Hg)
Iteration
Wall motion waveforms
c
Percent wall displacement (η×100%)
Resting wall (η0 = 0)
Blood pressure heatmap with wall motion
d
Normalized radial location (unitless)
Blood pressure (mm Hg)
Blood axial velocity heatmap with wall motion
e
Axial blood velocity (m/s)
Blood radial velocity heatmap with wall motion
f
Radial blood velocity (m/s)

# 31 Supplementary Fig.21. Bland-Altman plots and error histograms for results of signal-tagged physics-informed neural network on data generated with PulseDB synthetic dataset including biological variability

Signal-tagged physics-informed neural network predicted systolic and diastolic blood pressure for biological variability parameters. **a**, Bland-Altman plots. **b**, Error histograms. $x$- and $y$- axes are consistent across columns. AE, absolute error; DBP, diastolic blood pressure; SBP, systolic blood pressure; SD, standard deviation; SAT, subcutaneous adipose tissue.

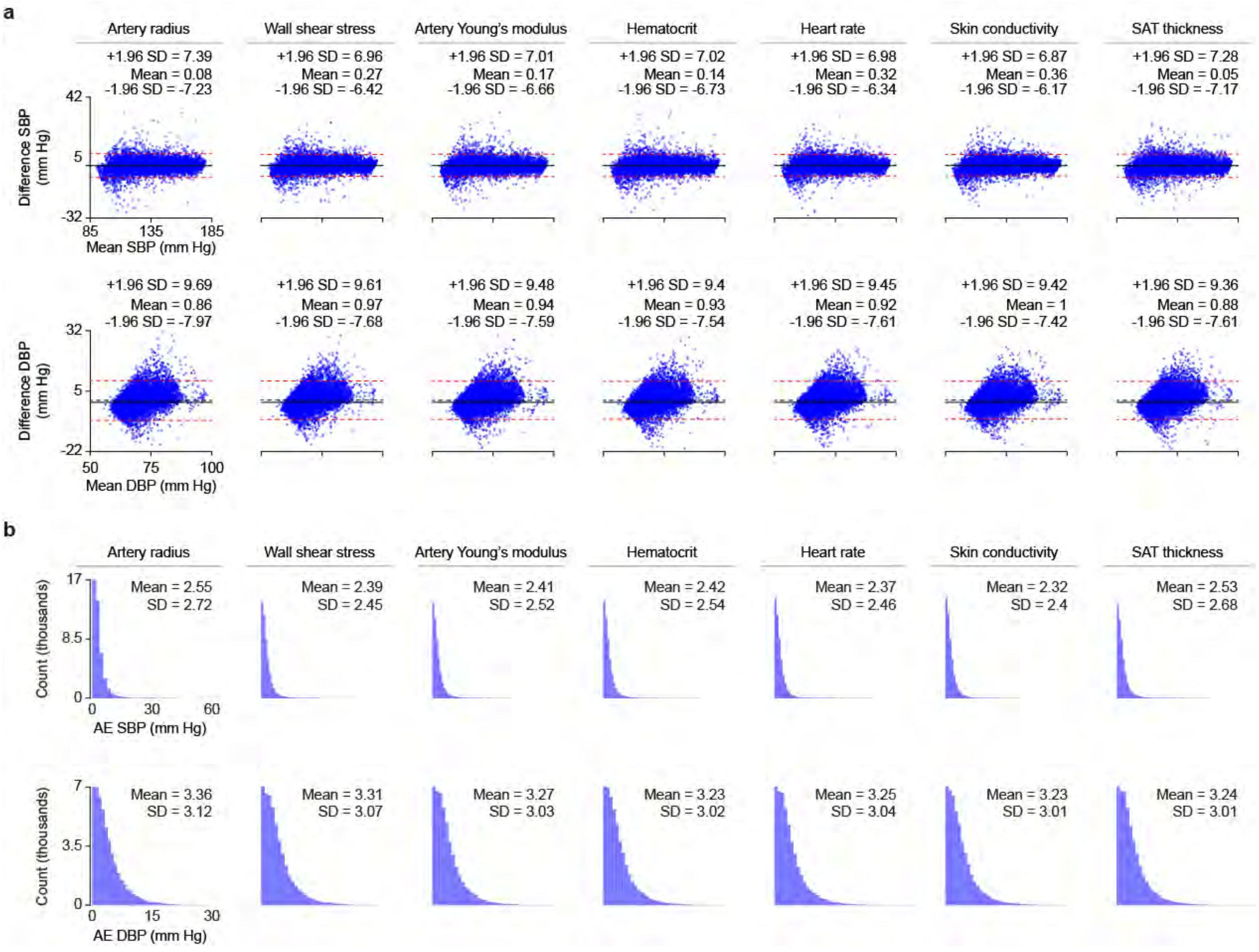

## 32 Supplementary Fig.22. Synchronized physiological signals and heart rate during exercise protocol for Group 1

**a**, Synchronized physiological signals recorded during the Group 1 protocols. **b**, Heart rate during the Group 1a treadmill and stationary bike exercise trials. Scale bar, 1 s. Mean and standard deviation shown.

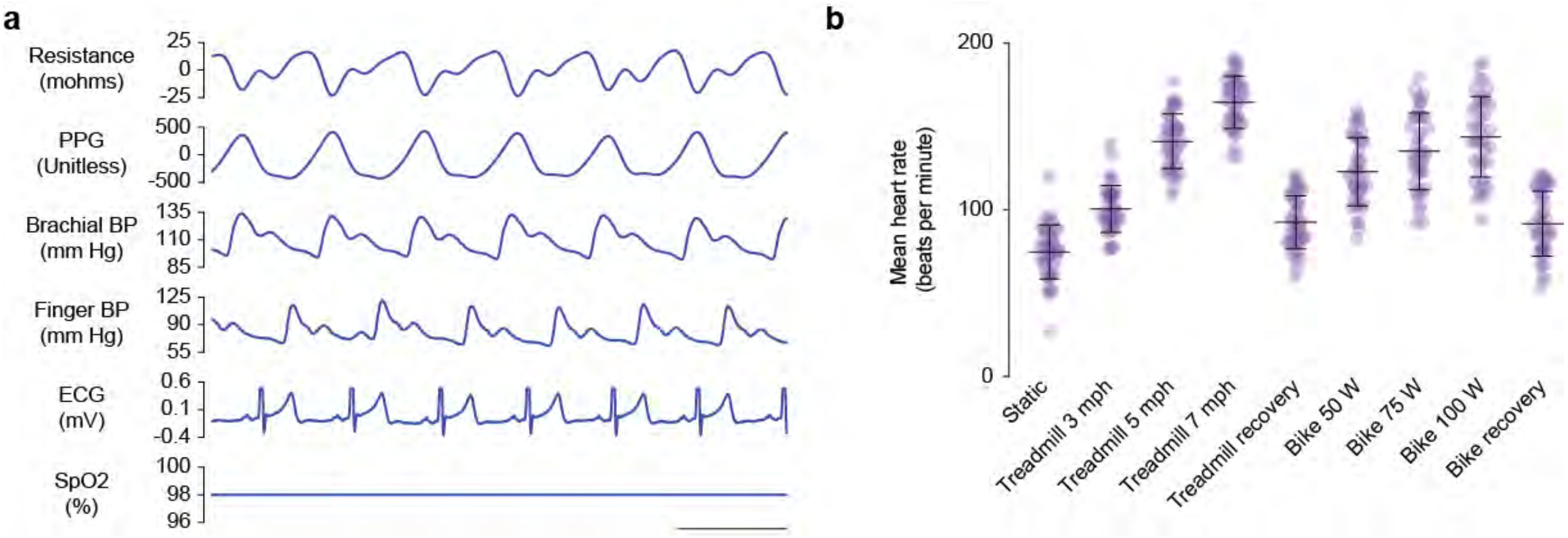

# 33 Supplementary Fig.23. Signal-tagged physics-informed neural network blood pressure results for Group 1

Signal-tagged physics-informed neural network (sPINN) predictions for **a,b**, Group 1 population-wide model and **c,d**, Group 1 subject-specific model. **i.** Correlation plots; **ii.** Bland-Altman plots; **iii.** Error histograms. SBP, systolic blood pressure; DBP, diastolic blood pressure; SD, standard deviation; AE, absolute error. All results are shown without text-exclusive subjects.

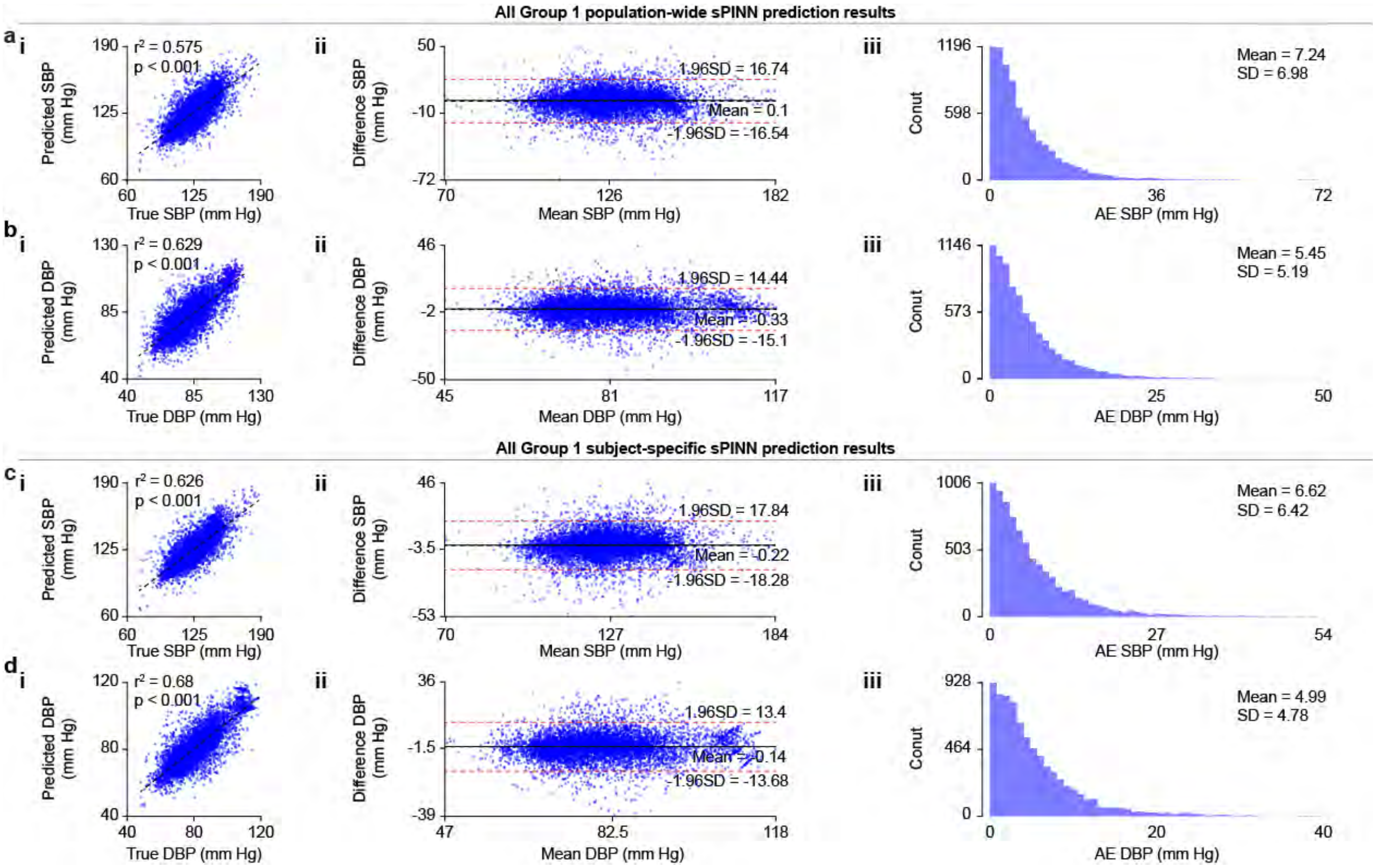

# 34 Supplementary Fig.24. Signal-tagged physics-informed neural network blood velocity results against fluid model for Group 1 and 2

Population-wide signal-tagged physics-informed neural network (sPINN) predicted cross sectional averaged blood velocity versus fluid forward blood pressure-modeled cross sectional average blood velocity for **a,b**, Group 1 and **c,d**, Group 2. **a,c**, Peak systolic velocity (PSV). **b,d**, End diastolic velocity (EDV). **i.** Correlation plots; **ii.** Bland-Altman plots; **iii.** Error histograms. AE, absolute error; SD, standard deviation.

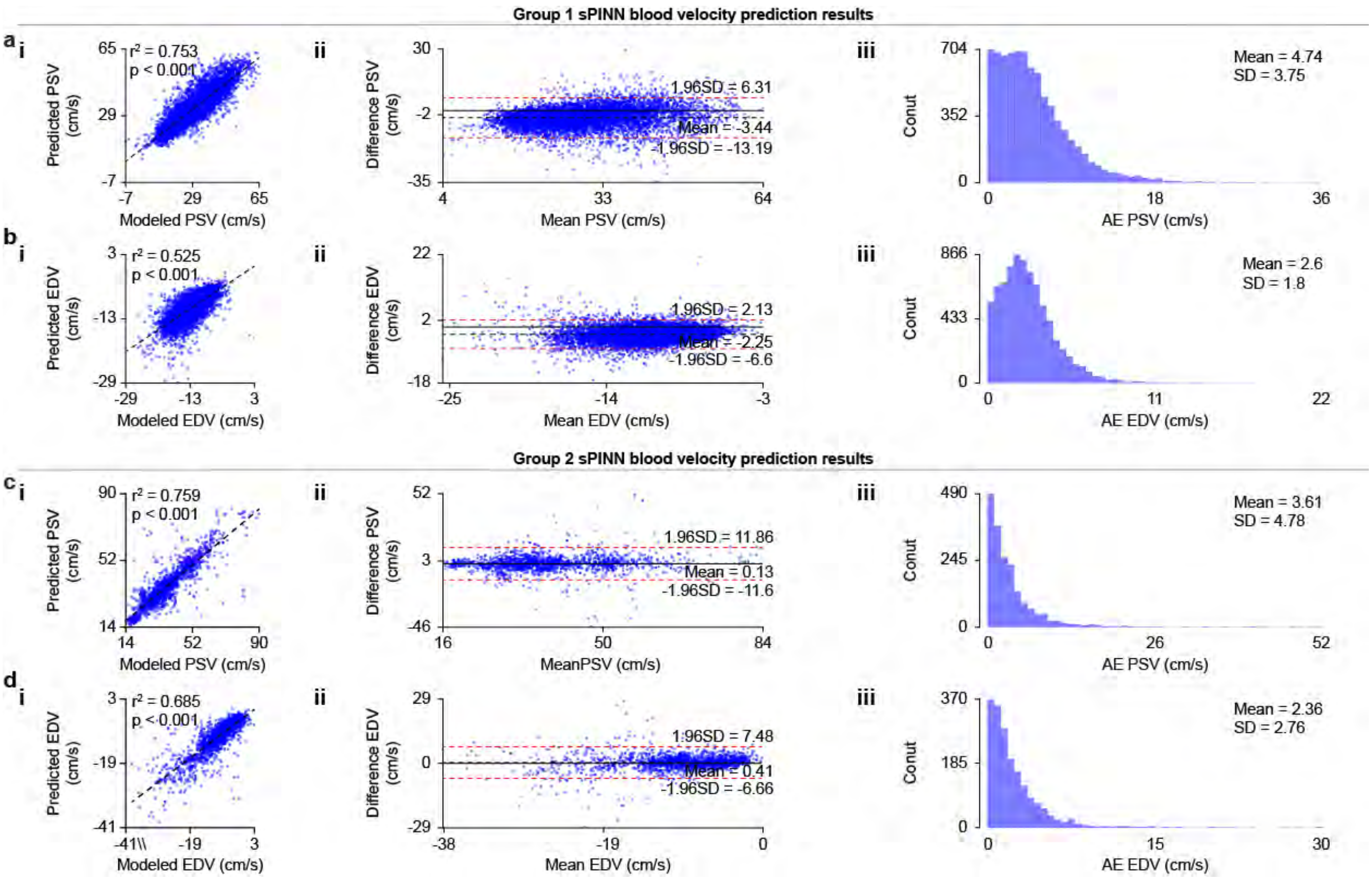

# 35 Supplementary Fig.25. Experimental relationship between blood pressure and smartwatch bioimpedance signals

**a**, Blood pressure (BP) and resistance tracings. **b**, BP versus resistance curves showing nonlinear relationship between pressure and resistance featuring hysteresis. **i**. Hypertensive patient; **ii**. Cardio-vascular disease patient; and **iii**. Other patient. Scale bars, 1 s.

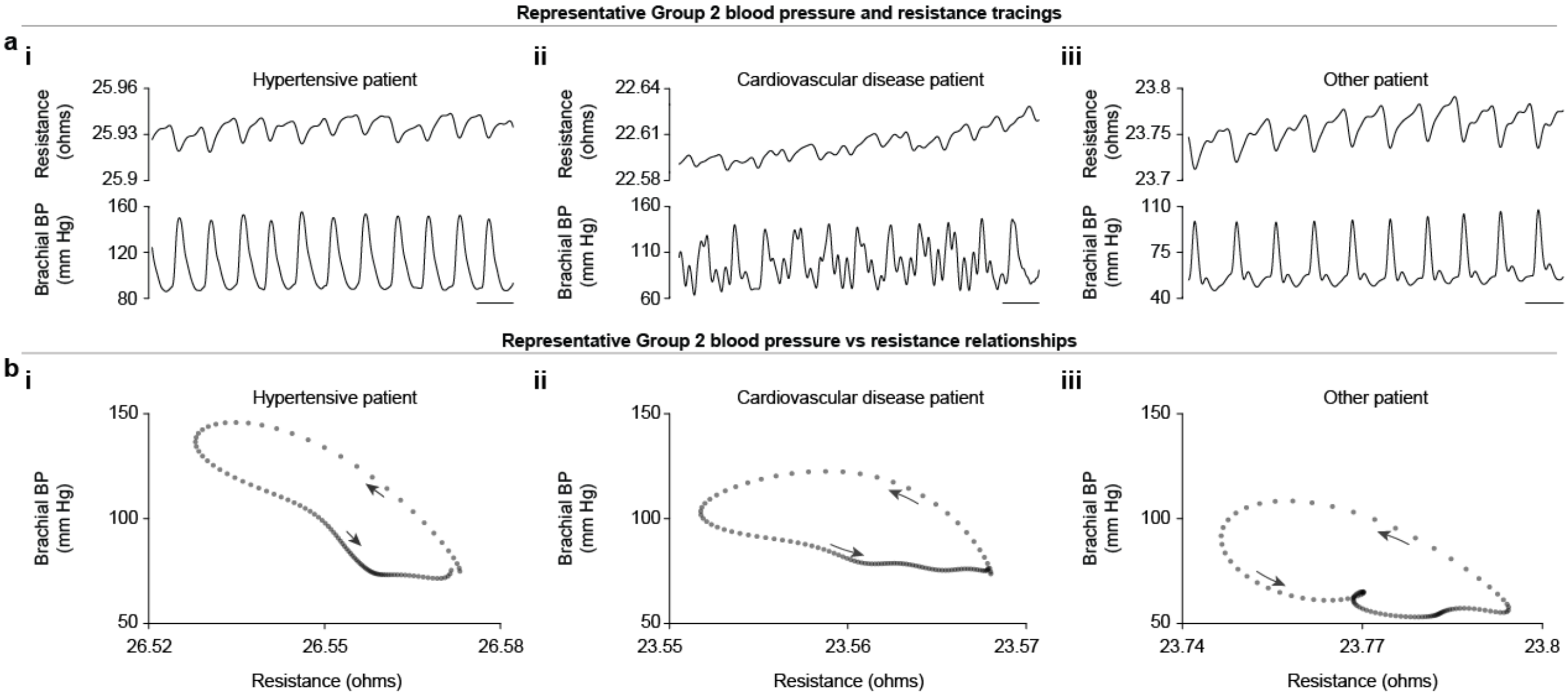

# 36 Supplementary Fig.26. Enrollment flowchart and study protocol for Group 2

**a**, Enrollment flow chart. **b**, Study protocol. **c**, Age histogram. **d**, Histogram of office systolic and diastolic blood pressures.

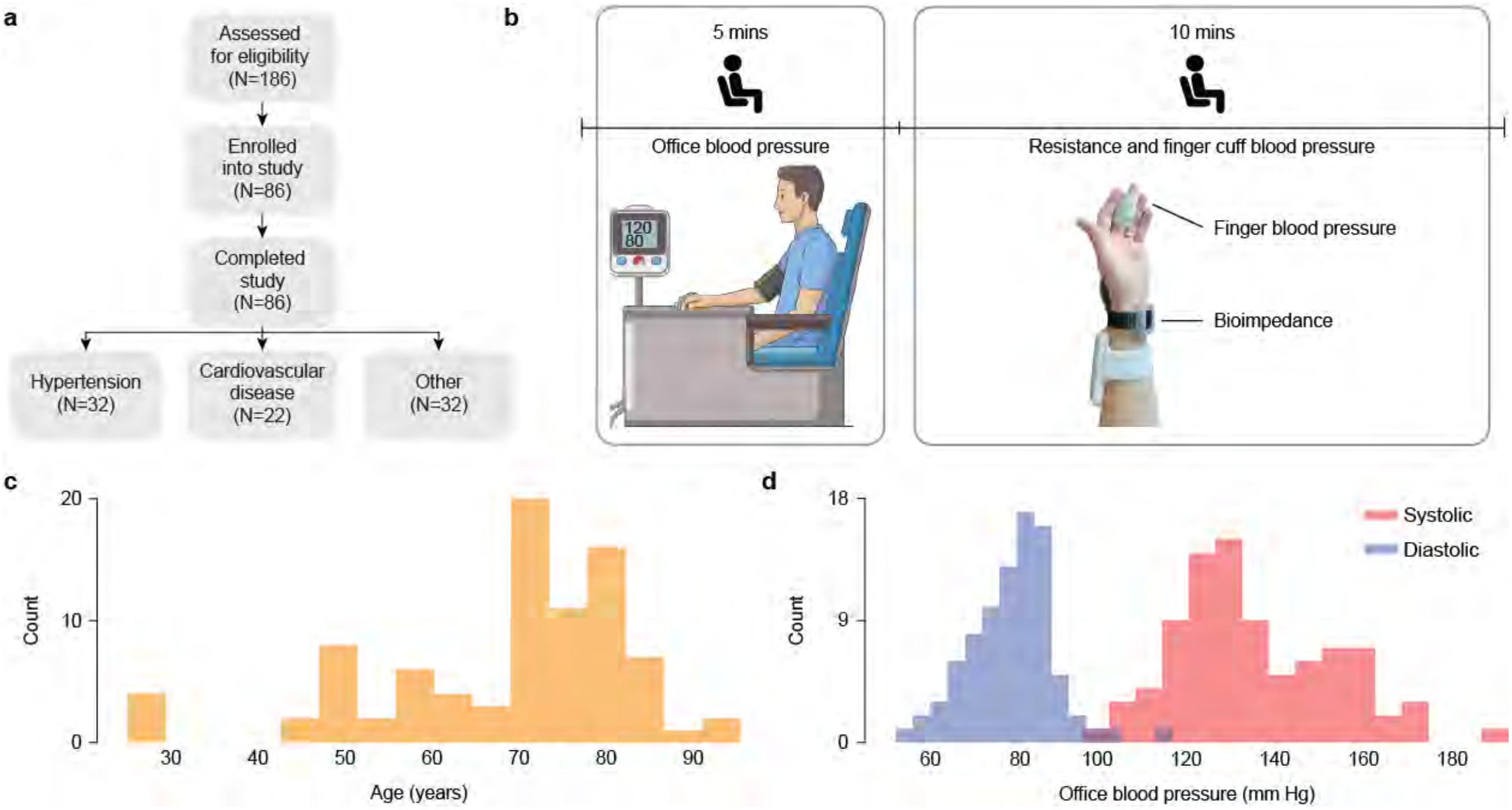

# 37 Supplementary Fig.27. Individual subject blood pressure results of a population-wide signal-tagged physics-informed neural network for Group 2

**a**, Measured and population-wide signal-tagged physics-informed neural network (sPINN) predicted blood pressure waveforms for all the patients in Group 2 dataset. Waveforms are colored by subject. **b**, Box plot of the root mean square error (RMSE) of the sPINN predicted waveforms versus measured waveforms for each of the $N$ = 75 subjects in Group 2. **c**, Randomly sampled blood pressure sPINN predictions from $N$ = 6 patients in Group 2. Scale bars, one-quarter period.

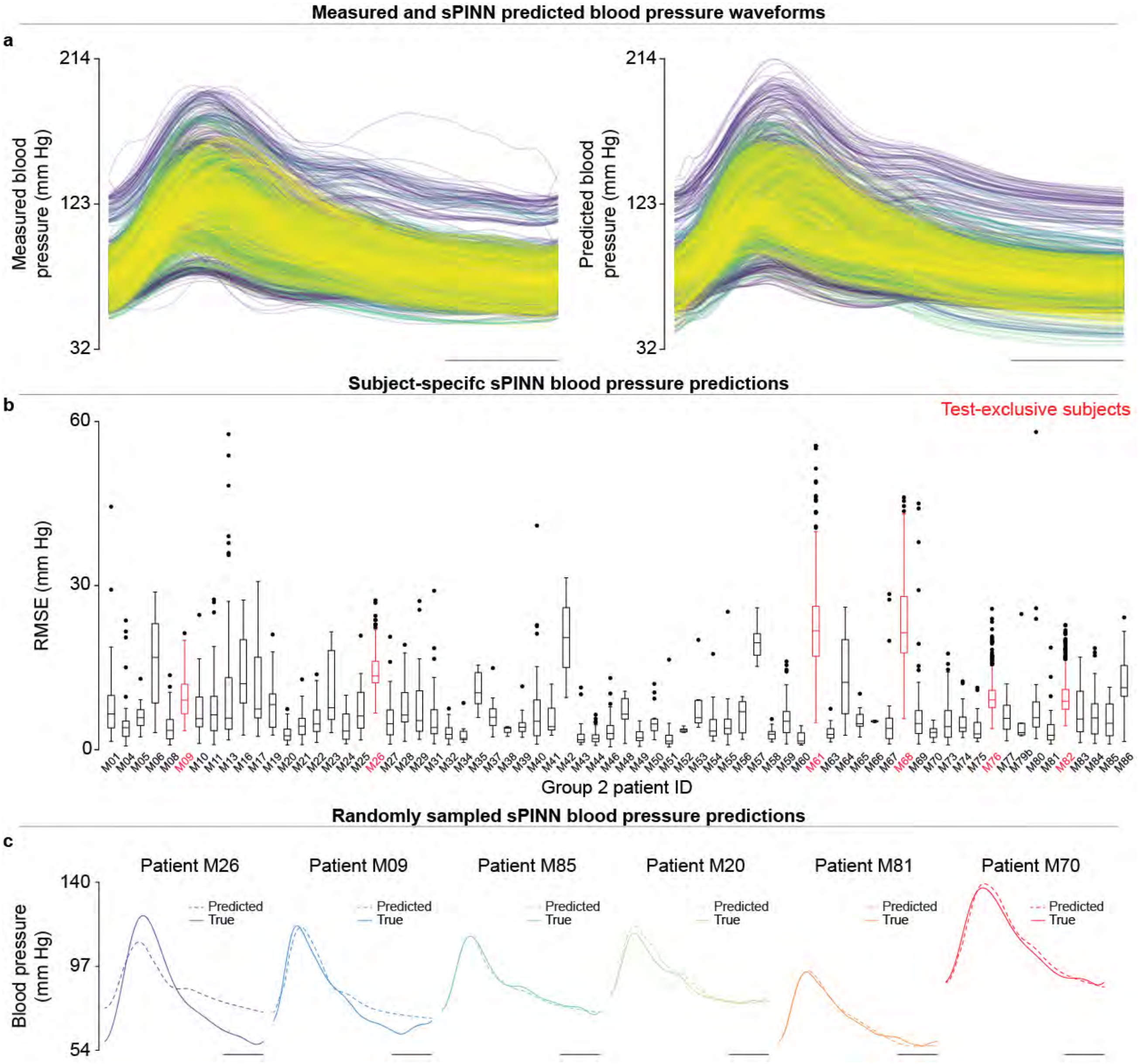

# 38 Supplementary Fig.28. Population-wide signal-tagged physics-informed neural network generalizability blood pressure results for Group 1 and 2

**a**, **b**, Group 1 test-exclusive subjects *excluded*. **c**, **d**, Group 1 test-exclusive subjects *included*. **e**, **f**, Group 2 test-exclusive patients *excluded*. **g**, **h**, Group 2 test-exclusive patients *included*. For **a-h**, **i**. Correlation plots. **ii**. Bland-Altman plots. **iii**. Absolute error histograms. SBP, Systolic blood pressure; DBP, diastolic blood pressure; AE, absolute error; SD, standard deviation.

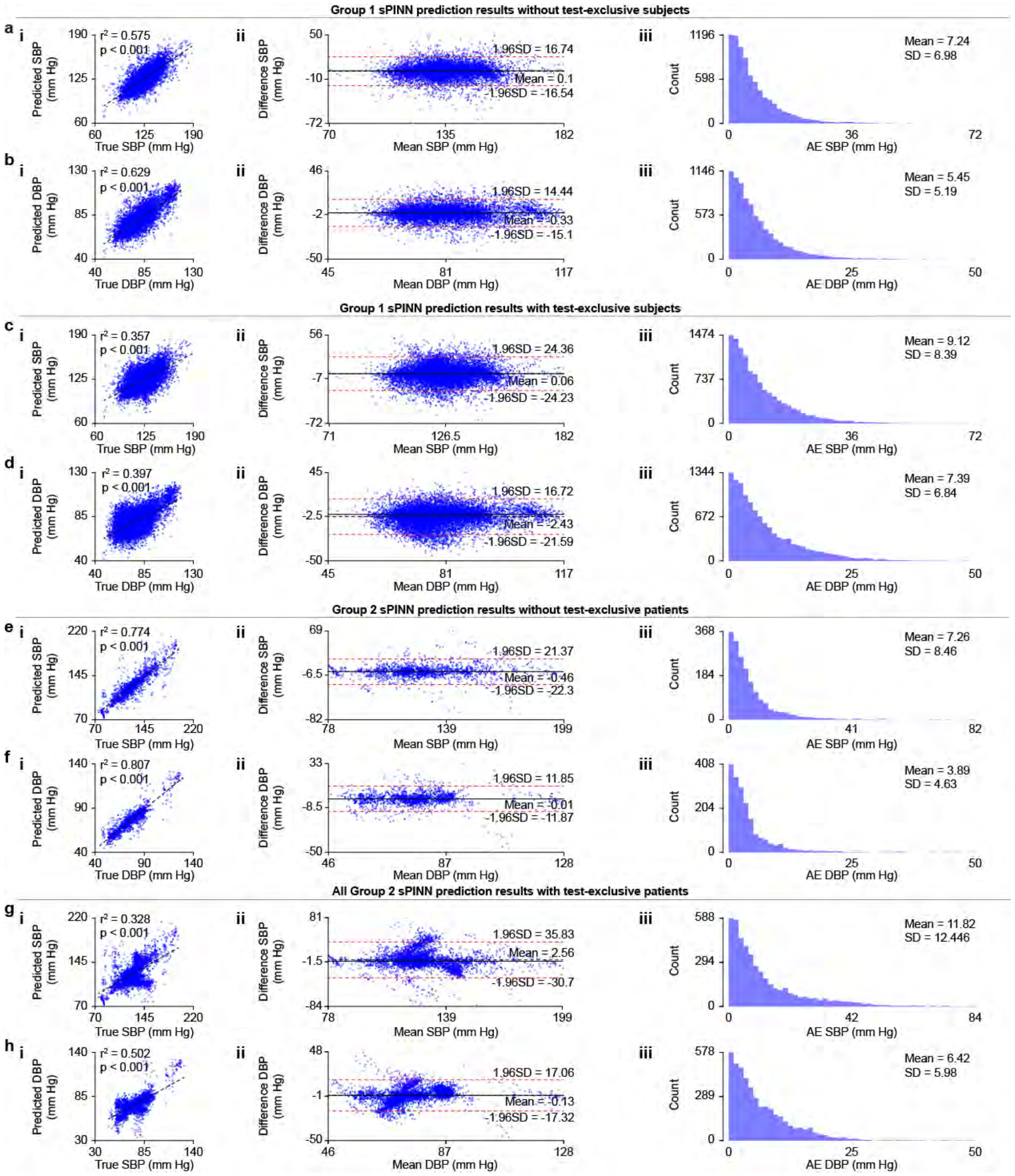

# 39 Supplementary Fig.29. Physiological recordings at the intensive care unit for Group 3

**a**, Illustration of data collection in the intensive care unit (ICU). **b**, Pre-operative ICU vitals in an end-stage heart failure patient with cardiopulmonary bypass. **c**, ICU vitals from left ventricle assist device implant patient seven months post-operative admitted for mixed distributive shock. **d**, Post-operative ICU vitals from left ventricle assist device implant patient. Scale bars, 1 s. BP, blood pressure.

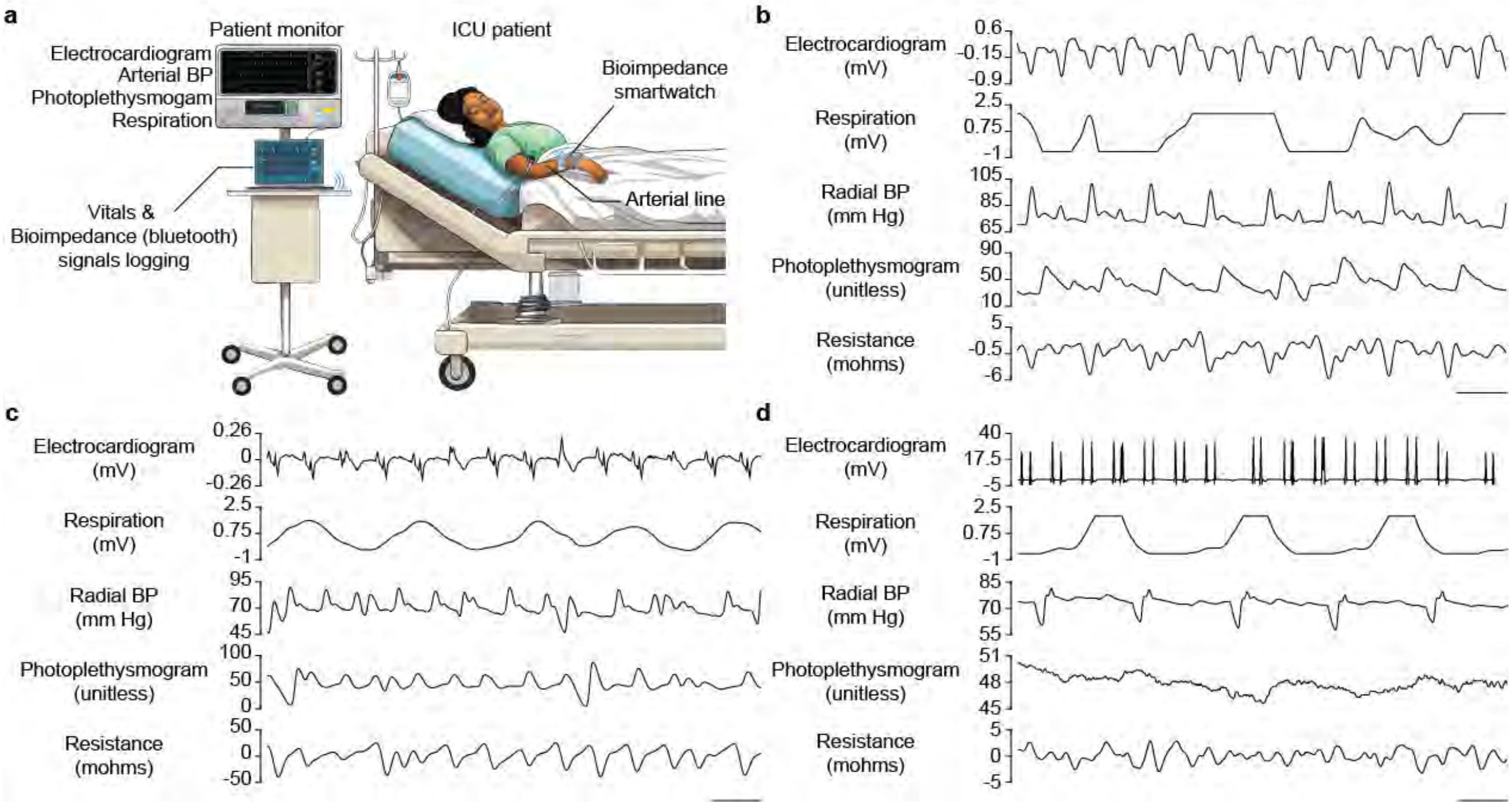

# 40 Supplementary Fig.30. Population-wide signal-tagged physics-informed neural network results on Graphene-HGCPT dataset

Radial blood pressure (BP) predictions from the Graphene-HGCPT dataset[99] using a population-wide signal-tagged physics-informed neural network. **a**, Representative examples of four continuous BP periods predictions. **b**, Correlation plots. **c**, Bland-Altman plots. **d**, Absolute error histograms. Scale bars, one-quarter period. SBP, systolic blood pressure; DBP, diastolic blood pressure; AE, absolute error; SD, standard deviation.

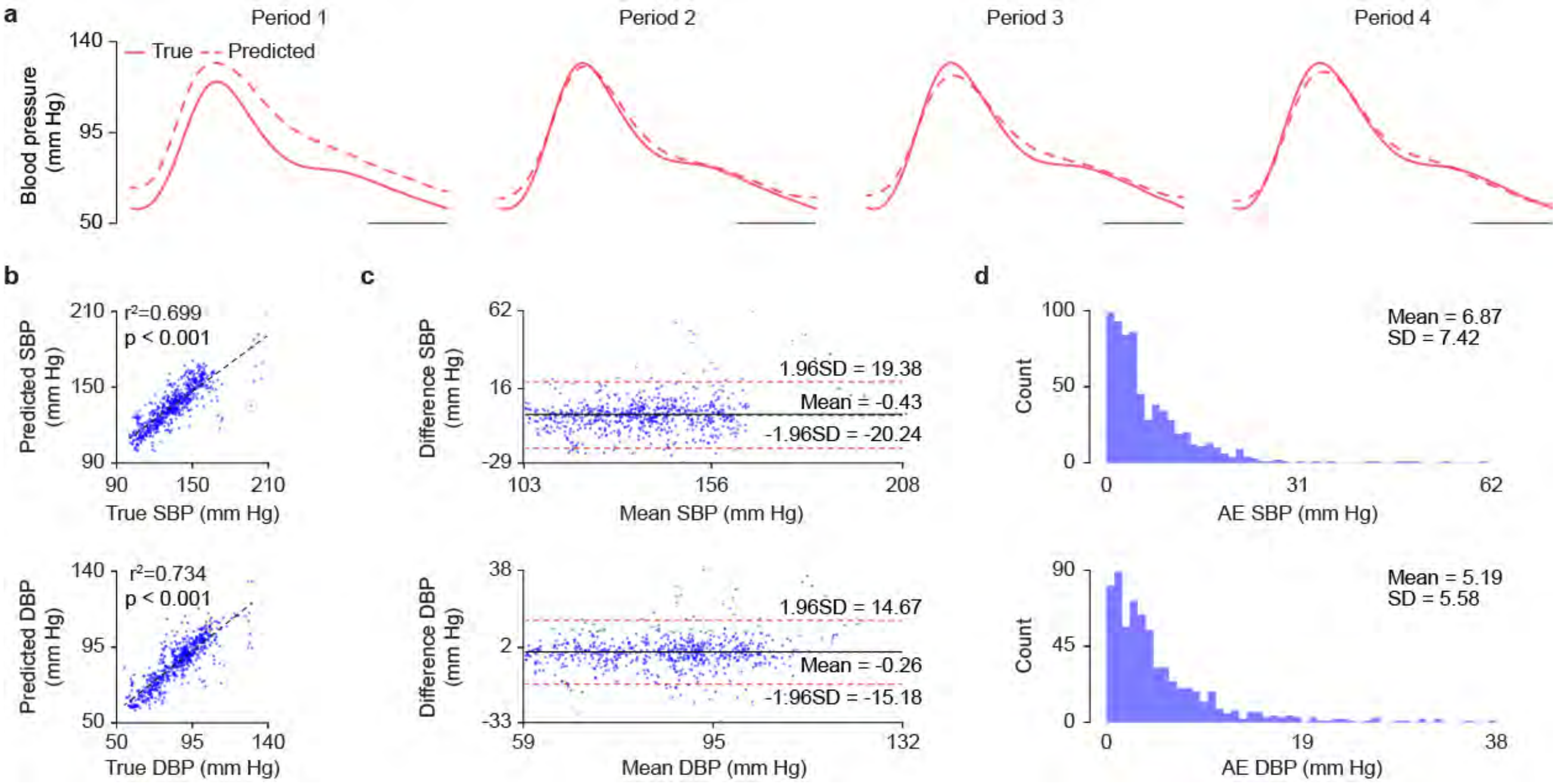

# 41 Supplementary Fig.31. Subject-specific signal-tagged physics-informed neural network results on Graphene-HGCPT dataset

Radial blood pressure (BP) predictions from the Graphene-HGCPT dataset[99] using a subject-specific signal-tagged physics-informed neural network. **a**, Representative examples of four continuous BP periods predictions. **b**, Correlation plots. **c**, Bland-Altman plots. **d**, Absolute error histograms. Scale bars, one-quarter period. SBP, systolic blood pressure; DBP, diastolic blood pressure; AE, absolute error; SD, standard deviation.

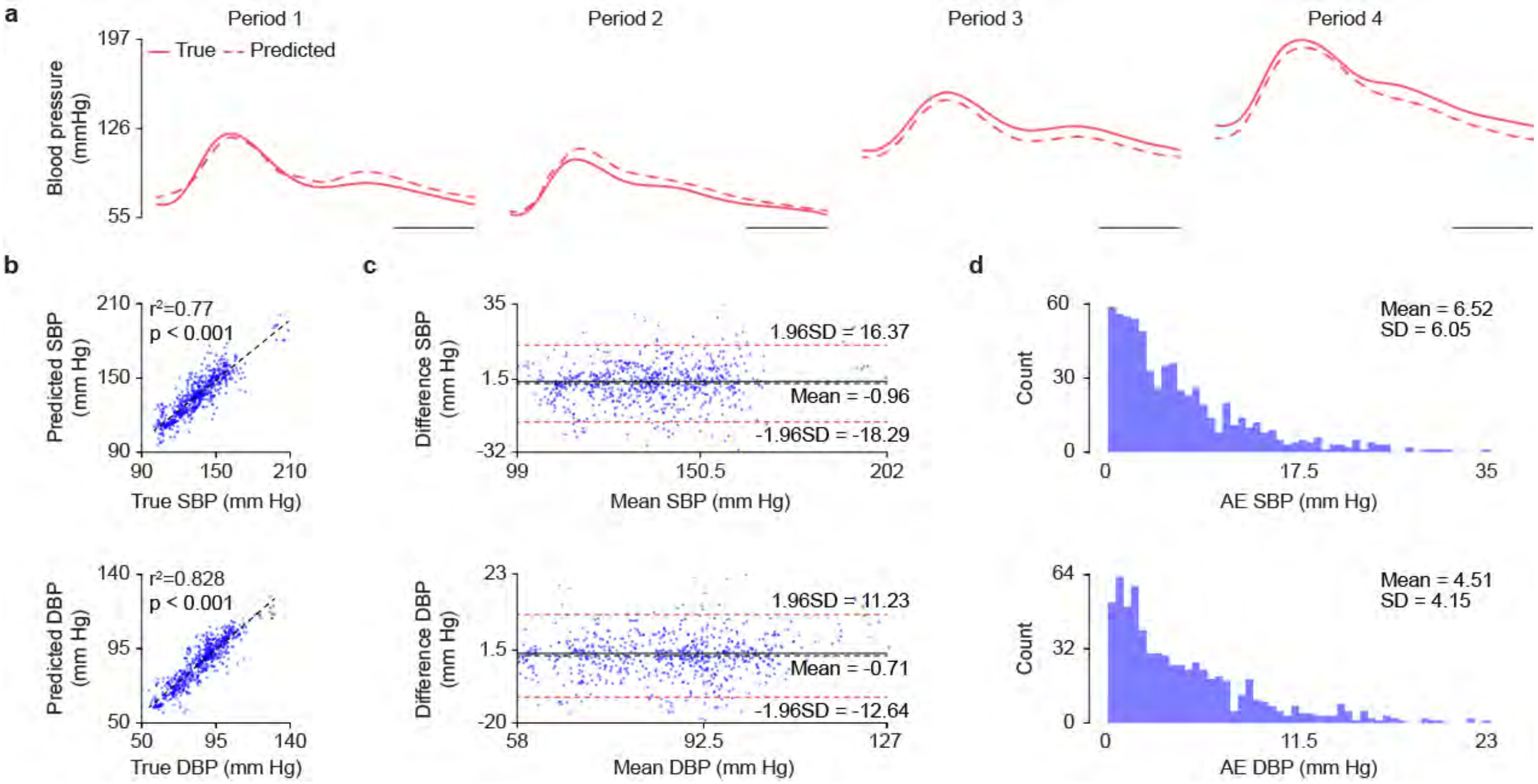

# 42 Supplementary Fig.32. Relationship between arterial wall compliance and artery radius

**a,** Linear and **b**, Logarithmic relationship between inverse arterial wall compliance and artery radius.

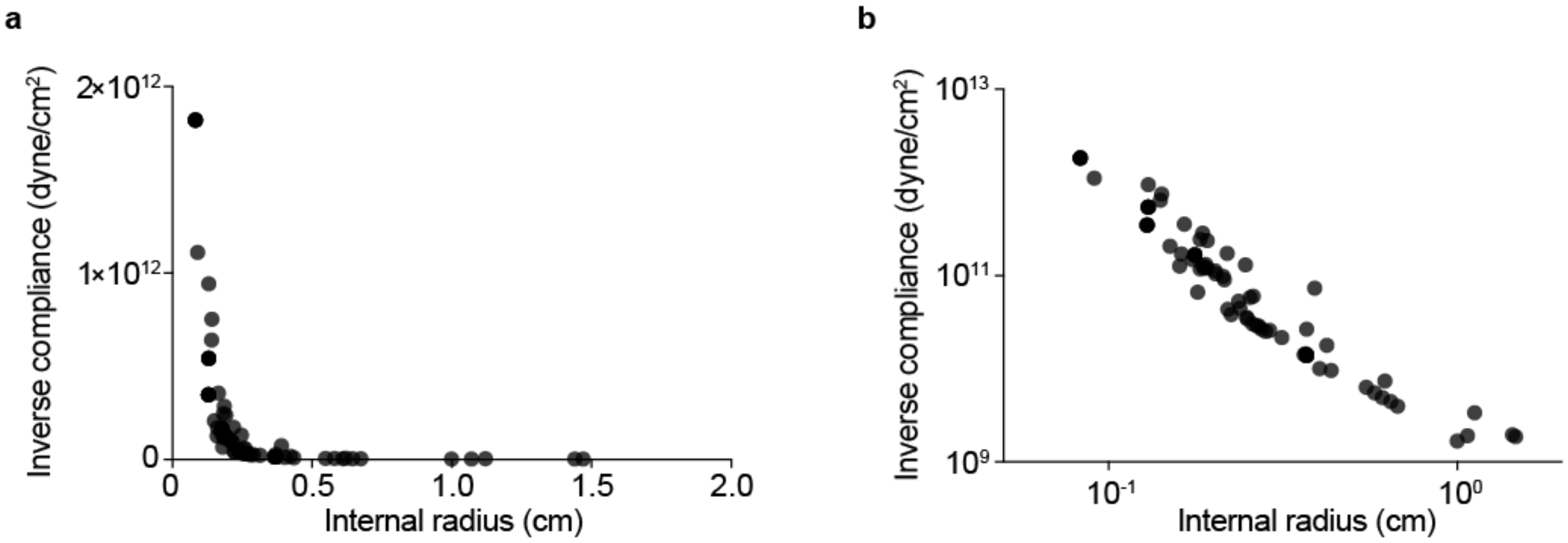

# 43 Supplementary Fig.33. Radial blood pressure results from synthetic wall motion data calibrated with a linear function

We utilized our linearized radial Navier-Stokes model along with experimental radial blood pressure (BP) waveforms from PulseDB database to explore the accuracy of predicting BP from synthetic radial wall motion using an empirical linear relationship. First, a single BP-wall motion period was selected to calibrate systolic (SBP) and diastolic (DBP) BP using a linear fit (**a**). Subsequently, we applied a linear least squares fitting approach to a dataset of $N$ = 500 experimental radial BP periods (**b**). **a, i.** Experimental radial BP waveform used to calibrate the linear model (top) and the prediction error (bottom). Since the linear fit is calibrated at SBP and DBP, the prediction error is 0 mm Hg at these two time instants. **ii-v.** Four representative radial BP waveforms and the prediction error using a single-BP period linear SBP/DBP calibration. **b,** Calibration extension using a training radial BP dataset of $N$ = 500 periods used to fit a least-squares regression line. **ii.** Test radial BP dataset used to predict radial BP prediction errors. **iii.** Train set radial BP versus artery diameter scatter plot with least squares regression line. **iv.** Test radial BP error distributions at different artery diameter values, demonstrating a linear fit to SBP and DBP based on measuring artery diameter only is unable to predict continuous BP. **v.** Radial BP differences at systole and diastole. Scale bars, one-quarter period.

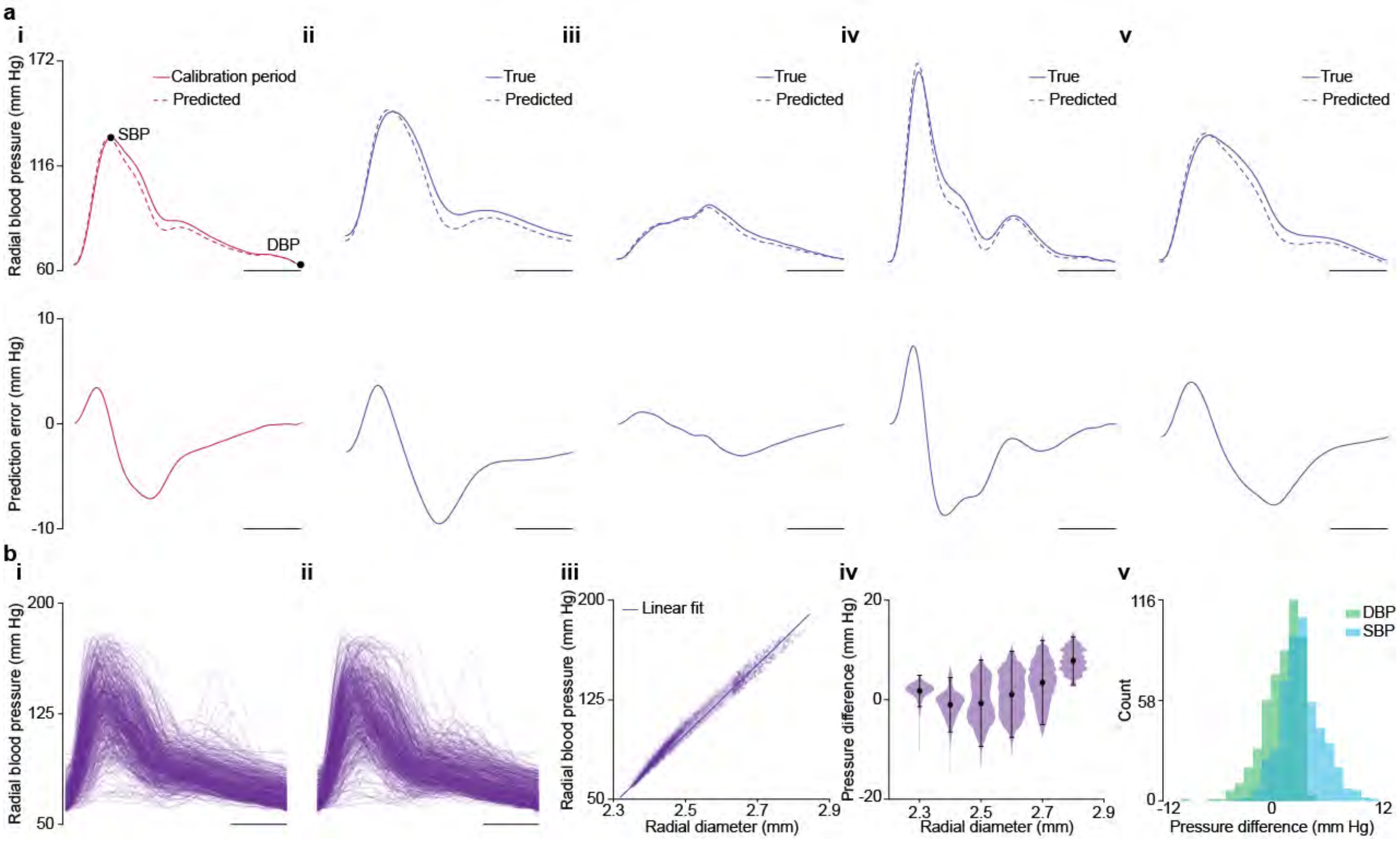

## 44 Supplementary Fig.34. Brachial blood pressure results from experimental wall motion data measured with ultrasound and calibrated with an exponential function

Error metrics from experimental brachial blood pressure (BP) predictions reported in Zhou et al. (Supplementary Fig.20 therein)[72] were used to evaluate the accuracy of predicting continuous BP from measuring brachial wall motion with ultrasound and calibrating the data with an empirical exponential function (Supplementary Discussion 3). **a**, Correlation plot on systolic blood pressure (SBP, red) and diastolic blood pressure (DBP, blue). **b**, Bland-Altman plot for SBP and DBP. **c**, SBP error histogram. **d**, DBP error histogram. **e**, Correlation plot on the BP waveform. **f**, Bland altman plot for the BP waveform. **g**, Absolute error histogram for BP waveform. **h**, Root mean square error (RMSE) for predicted waveform. RMSE, root mean squared error; AE, absolute error; SD, standard deviation.

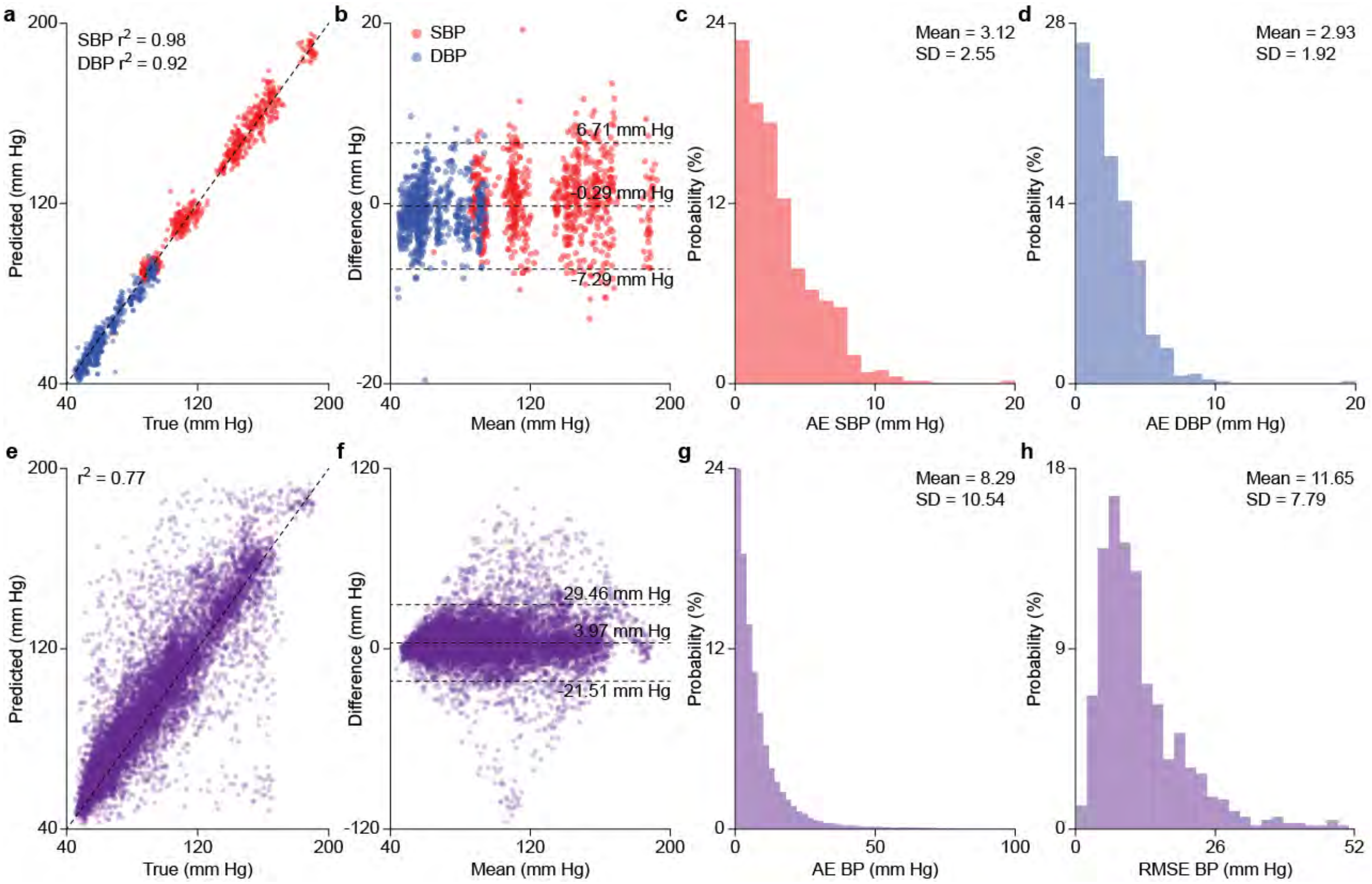

# Supplementary tables

## 45 Supplementary Table 1. List of Navier-Stokes, Maxwell-Fricke, and Maxwell modeling parameters

*Tissue electrical conductivity ranges are reported from 1 kHz to 1 MHz, while the nominal value was selected at 50 kHz. RBC, red blood cell; BP, blood pressure; ST, structured tree.

| Parameters (units) | Nominal value | Minimum | Maximum | References |
|---|---|---|---|---|
| **Scaling parameters** | | | | |
| Radial length scaling $L_{\text{rs}}$ | 1 | 0.875 | 1.125 | |
| Axial length scaling $L_{\text{as}}$ | 1 | 0.875 | 1.125 | |
| **Fluid parameters** | | | | |
| $2\pi$ over period (1/s) | $2\pi$(81.8/60) | $2\pi$(132.1/60) | $2\pi$(52.5/60) | Avram et al.[209] |
| BP Fourier coefficients | - | - | - | |
| Young's Modulus of radial wall (MPa) | 0.45 | 0.2 | 1.0 | van de Vosse et al.[210] |
| Radial wall thickness ratio $h/\hat{a}$ | 0.197 / 1.22 | 0.05 | 0.25 | Mourad et al.[191] |
| Blood density (kg/m$^3$) | 1,050 | 1,025 | 1,060 | IT'IS[211] |
| Wall density (kg/m$^3$) | 1,102 | 1,056 | 1,147 | IT'IS[211] |
| Blood viscosity (kg/m/s) | $3.5\cdot10^{-3}$ | $3\cdot10^{-3}$ | $4\cdot10^{-3}$ | De Forest et al., Ch 3.1[212] |
| Average radial radius $\bar{a}$ (mm) | $1.22\cdot L_{\text{rs}}$ | $0.62\cdot L_{\text{rs}}$ | $1.82\cdot L_{\text{rs}}$ | Kotowycz et al.[213] |
| Average mean BP (mm Hg), $\hat{B}_0$ | 85 | - | - | |
| Poisson's ratio | 0.5 | - | - | Zamir[79] |
| Axial length measured $L$ (cm) | 4.35 | - | - | |
| Mean wall shear stress (Pa) | 0.82 | 0.305 | 1.335 | Girerd et al.[214] |
| **Branching parameters** | | | | |
| Radial length $L_{\text{r}}$ (cm) | $23.5\cdot L_{\text{as}}$ | - | - | Stergiopulos et al.[193] |
| Radial length to wrist/watch | $L_{\text{r}} - L$ | - | - | |
| Brachial radius (mm) | $2.15\cdot L_{\text{rs}}$ | $1.33\cdot L_{\text{rs}}$ | $2.97\cdot L_{\text{rs}}$ | Dibble et al.[215] |
| Brachial wall thickness ratio | (0.29 / 2.15)$(h/\hat{a})$/(0.197/1.22) | - | - | Iwamoto et al.[216] |
| Brachial length of interest (cm) | $25.09/4\cdot L_{\text{as}}$ | - | - | |
| Brachial wall shear stress (MPa) | 0.42 | 0.14 | 0.70 | Dammers et al.[217] |
| Brachial Young's modulus (MPa) | 0.45 | 0.2 | 1.0 | van de Vosse et al.[210] |
| Ulnar A. radius $a_{\text{u}}$ (mm) | $1.07\cdot L_{\text{rs}}$ | $0.54\cdot L_{\text{rs}}$ | $1.60\cdot L_{\text{rs}}$ | Kotowycz et al.[213] |
| Ulnar A. wall thickness ratio | $h/\hat{a}$ | - | - | |
| Ulnar A. length (cm) | $6.7\cdot L_{\text{as}}$ | - | - | Stergiopulos et al.[193] |
| Ulnar I. radius (mm) | $0.91\cdot\hat{a}/1.5$ | 0.9 | 1.05 | Reymond et al.[118] |
| Ulnar I. wall thickness ratio | $h/\hat{a}$ | - | - | |
| Ulnar I. length (cm) | $7.9\cdot L_{\text{as}}$ | - | - | Stergiopulos et al.[193] |
| Ulnar B. radius (mm) | $1.725\cdot\hat{a}/1.5$ | - | - | |
| Ulnar B. radius | $a_{\text{u}}$ | - | - | |
| Ulnar B. wall thickness ratio | $h/\hat{a}$ | - | - | |
| Ulnar B. length (cm) | $17.1\cdot L_{\text{as}}$ | - | - | Stergiopulos et al.[193] |
| **Blood parameters** | | | | |
| Hematocrit concentration | 0.45 | 0.35 | 0.55 | De Forest et al., Ch 3.1[212] |
| Plasma conductivity (S/m) | 1.41 | 1.11 | 1.59 | IT'IS[211] |
| Long RBC axis length ($\mu$m) | 4 | - | - | Hoetink et al.[87] |
| RBC axis ratio | 0.38 | 0.24 | 0.52 | Bitbol et al.[195] |
| Membrane shear modulus (N/m) | $1.5\cdot10^{-5}$ | $1\cdot10^{-5}$ | $1.5\cdot10^{-5}$ | Hoetink et al.[87] |
| Bitbol orientation parameter | 3 | - | - | Gaw[218] |
| **ST parameters** | | | | |
| ST radius law $\xi$ | 2.76 | 2.33 | 3.0 | Olufsen et al.[84] |
| ST area ratio | $\max$(1.16, $f(\xi)$) | - | - | Olufsen et al.[84] |
| ST radius-length multiplier | $25\cdot(L_{\text{as}}/L_{\text{rs}})$ | $15\cdot(L_{\text{as}}/L_{\text{rs}})$ | $35\cdot(L_{\text{as}}/L_{\text{rs}})$ | Olufsen et al.[84] |
| Minimum ST radius (µm) | 30 | | | Olufsen et al.[84] |
| Stiffness power law | -2.25 | - | - | |
| **Electromagnetic parameters** | | | | |
| Radial blood conductivity (S/m) | 0.7* | 0.7 | 0.822 | Hasgall et al.[219] |
| Skin conductivity (S/m) | $2.73\cdot10^{-4}$* | $2\cdot10^{-4}$ | $1.32\cdot10^{-2}$ | Hasgall et al.[219] |
| Sub. Adipose conductivity (S/m) | $4.327\cdot10^{-2}$* | $3.77\cdot10^{-2}$ | $4.41\cdot10^{-2}$ | Hasgall et al.[219] |
| Muscle conductivity (S/m) | $3.518\cdot10^{-1}$* | $2.02\cdot10^{-1}$ | $5.03\cdot10^{-1}$ | Hasgall et al.[219] |
| Skin thickness $d_1$ (mm) | 2 | 1.5 | 2.5 | Gosselin et al.,[94] Christ et al.[199] |
| Sub. Adipose thickness $d_2$ (mm) | 0.5 | 0.25 | 1 | Gosselin et al.,[94] Christ et al.[199] |
| Centerline artery depth $h$ (mm) | 5.72 | - | - | Gosselin et al.,[94] Christ et al.[199] |
| Sub. Adipose depth (mm) | 2.5 | $0.5+d_1$ | $2+d_1$ | Gosselin et al.,[94] Christ et al.[199] |
| Radial artery depth (mm) | 4.5 | 3 | 7.5 | Gosselin et al.,[94] Christ et al.[199] |
| Radial artery offset $d$ (mm) | 0 | -10 | 10 | |
| Current frequency (kHz) | 50 | - | - | |
| Current electrodes gap (cm) | 1 | - | - | |
| Voltage electrodes gap (cm) | 1 | - | - | |

# 46 Supplementary Table 2. List of Sobol indices for blood conductivity global sensitivity analysis

First and total order Sobol indices and uncertainty measures for each blood conductivity model parameter.[220]

| Parameters | Mean conductivity | | | |
|---|---|---|---|---|
| | First order sobol index | First order confidence | Total order sobol index | Total order confidence |
| Heart rate | 1.424E-04 | 1.553E-04 | 3.103E-04 | 5.526E-06 |
| Young's modulus | 9.734E-04 | 3.044E-04 | 1.222E-03 | 1.521E-05 |
| Wall thickness ratio | 1.046E-03 | 2.992E-04 | 1.214E-03 | 1.511E-05 |
| Blood density | 1.156E-06 | 3.810E-06 | 1.935E-07 | 4.836E-09 |
| Artery wall density | 6.063E-08 | 1.044E-07 | 1.490E-10 | 7.247E-12 |
| Blood viscosity | 1.570E-05 | 6.622E-05 | 5.795E-05 | 1.885E-06 |
| Radial artery radius | 2.129E-04 | 1.479E-04 | 2.776E-04 | 3.770E-06 |
| Radial mean shear stress | 3.662E-04 | 2.376E-04 | 7.334E-04 | 1.377E-05 |
| Hematocrit | 5.725E-01 | 5.848E-03 | 5.787E-01 | 4.817E-03 |
| Plasma conductivity | 4.156E-01 | 5.198E-03 | 4.214E-01 | 3.681E-03 |
| Axis length ratio | 1.565E-04 | 1.547E-04 | 2.973E-04 | 3.659E-06 |
| Membrane shear modulus | -1.380E-05 | 4.752E-05 | 2.946E-05 | 1.070E-06 |
| Radial blood pressure | 3.014E-03 | 5.007E-04 | 3.357E-03 | 3.763E-05 |
| | **Peak-to-peak conductivity** | | | |
| | First order Sobol index | First order confidence | Total order Sobol index | Total order confidence |
| Heart rate | 4.072E-02 | 3.605E-03 | 1.674E-01 | 2.998E-03 |
| Young's modulus | 2.730E-02 | 4.092E-03 | 2.212E-01 | 4.038E-03 |
| Wall thickness ratio | 2.433E-02 | 3.927E-03 | 2.190E-01 | 3.966E-03 |
| Blood density | -7.910E-05 | 1.846E-04 | 4.809E-04 | 2.496E-05 |
| Artery wall density | 3.113E-06 | 6.546E-06 | 5.468E-07 | 3.706E-08 |
| Blood viscosity | 9.166E-06 | 1.990E-03 | 4.947E-02 | 1.294E-03 |
| Radial artery radius | 5.464E-02 | 4.042E-03 | 2.086E-01 | 3.227E-03 |
| Radial mean shear stress | 4.967E-02 | 4.897E-03 | 3.119E-01 | 4.965E-03 |
| Hematocrit | 2.091E-03 | 4.208E-04 | 2.505E-03 | 3.739E-05 |
| Plasma conductivity | 9.031E-02 | 2.635E-03 | 1.005E-01 | 1.110E-03 |
| Axis length ratio | 1.836E-02 | 3.083E-03 | 1.050E-01 | 1.841E-03 |
| Membrane shear modulus | -1.684E-04 | 9.369E-04 | 1.122E-02 | 6.621E-04 |
| Radial blood pressure | 2.460E-01 | 6.891E-03 | 6.518E-01 | 7.492E-03 |
| | **Time of minimum resistance** | | | |
| | First order Sobol index | First order confidence | Total order Sobol index | Total order confidence |
| Heart rate | -6.917E-04 | 4.637E-03 | 2.817E-01 | 5.351E-03 |
| Young's modulus | 3.634E-03 | 4.748E-03 | 3.144E-01 | 5.779E-03 |
| Wall thickness ratio | 5.291E-03 | 4.951E-03 | 3.161E-01 | 6.120E-03 |
| Blood density | -1.608E-04 | 9.533E-04 | 1.272E-02 | 1.245E-03 |
| Artery wall density | -7.990E-05 | 2.630E-04 | 8.307E-04 | 3.716E-04 |
| Blood viscosity | 1.650E-03 | 3.344E-03 | 1.483E-01 | 4.038E-03 |
| Radial artery radius | 1.303E-03 | 5.916E-03 | 4.355E-01 | 6.418E-03 |
| Radial mean shear stress | 4.806E-02 | 5.460E-03 | 4.103E-01 | 6.088E-03 |
| Hematocrit | -1.396E-04 | 3.631E-04 | 1.991E-03 | 5.430E-04 |
| Plasma conductivity | -3.630E-05 | 6.975E-05 | 1.034E-04 | 1.986E-04 |
| Axis length ratio | 2.903E-03 | 2.884E-03 | 1.085E-01 | 4.195E-03 |
| Membrane shear modulus | 1.152E-03 | 2.234E-03 | 6.226E-02 | 3.421E-03 |
| Radial blood pressure | 2.359E-01 | 8.715E-03 | 9.030E-01 | 7.469E-03 |
| | **Time of maximum resistance** | | | |
| | First order Sobol index | First order confidence | Total order Sobol index | Total order confidence |
| Heart rate | 7.988E-02 | 5.159E-03 | 3.638E-01 | 5.022E-03 |
| Young's modulus | 4.392E-02 | 4.526E-03 | 2.768E-01 | 4.489E-03 |
| Wall thickness ratio | 4.317E-02 | 4.318E-03 | 2.756E-01 | 4.525E-03 |
| Blood density | 1.412E-04 | 1.461E-03 | 2.606E-02 | 1.459E-03 |
| Artery wall density | -7.206E-04 | 1.285E-03 | 2.039E-02 | 1.316E-03 |
| Blood viscosity | 2.197E-03 | 2.469E-03 | 8.186E-02 | 2.547E-03 |
| Radial artery radius | 9.542E-02 | 5.348E-03 | 3.919E-01 | 5.208E-03 |
| Radial mean shear stress | 2.462E-03 | 2.276E-03 | 6.989E-02 | 2.382E-03 |
| Hematocrit | -7.267E-04 | 1.237E-03 | 2.074E-02 | 1.324E-03 |
| Plasma conductivity | 1.854E-04 | 1.270E-03 | 1.918E-02 | 1.284E-03 |
| Axis length ratio | 5.150E-02 | 4.604E-03 | 3.040E-01 | 4.461E-03 |
| Membrane shear modulus | 2.152E-02 | 3.925E-03 | 1.970E-01 | 4.039E-03 |
| Radial blood pressure | 2.235E-01 | 6.729E-03 | 6.280E-01 | 5.860E-03 |

# 47 Supplementary Table 3. List of Sobol indices for resistance global sensitivity analysis

First and total order Sobol indices and uncertainty measures for each electrodynamic model parameter.[220] SAT, subcutaneous adipose tissue.

| Parameters | First order Sobol index | First order confidence | Total order Sobol index | Total order confidence |
|---|---|---|---|---|
| **Mean resistance** | | | | |
| Skin thickness | 2.756E-04 | 6.072E-04 | 4.849E-03 | 7.429E-05 |
| SAT thickness | 5.721E-03 | 1.504E-03 | 2.883E-02 | 6.984E-04 |
| Electrode offset | 3.817E-01 | 5.983E-03 | 4.973E-01 | 6.245E-03 |
| Radial artery radius | 4.926E-01 | 6.770E-03 | 5.875E-01 | 5.634E-03 |
| Radial artery blood pressure | 2.370E-03 | 5.134E-04 | 3.601E-03 | 5.315E-05 |
| **Peak-to-peak resistance** | | | | |
| Skin thickness | 1.411E-03 | 9.927E-04 | 1.750E-02 | 1.944E-02 |
| SAT thickness | 2.510E-03 | 1.347E-03 | 3.189E-02 | 2.148E-02 |
| Electrode offset | 2.114E-01 | 7.230E-03 | 3.122E-01 | 1.834E-02 |
| Radial artery radius | 5.394E-01 | 1.647E-02 | 6.542E-01 | 2.567E-02 |
| Radial artery blood pressure | 1.008E-01 | 4.448E-03 | 1.729E-01 | 1.861E-02 |
| **Time of minimum resistance** | | | | |
| Skin thickness | 4.395E-03 | 3.570E-03 | 1.460E-01 | 1.169E-02 |
| SAT thickness | 3.688E-04 | 4.293E-03 | 2.326E-01 | 1.364E-02 |
| Electrode offset | 5.895E-03 | 5.932E-03 | 2.551E-01 | 1.294E-02 |
| Radial artery radius | 2.767E-03 | 4.141E-03 | 2.395E-01 | 1.364E-02 |
| Radial artery blood pressure | 7.218E-01 | 1.769E-02 | 9.707E-01 | 1.753E-02 |
| **Time of maximum resistance** | | | | |
| Skin thickness | -4.436E-04 | 2.390E-03 | 8.017E-02 | 3.229E-03 |
| SAT thickness | 8.487E-04 | 2.869E-03 | 1.086E-01 | 3.486E-03 |
| Electrode offset | 2.252E-02 | 4.241E-03 | 2.576E-01 | 4.533E-03 |
| Radial artery radius | 2.759E-02 | 5.993E-03 | 4.589E-01 | 6.265E-03 |
| Radial artery blood pressure | 4.587E-01 | 8.894E-03 | 9.289E-01 | 7.105E-03 |

# 48 Supplementary Table 4. Attributes of the Virtual Population computable human models and their use

These models capture a wide range of anatomical variability due to their difference in age, gender, height, and weight. Fats and Glenn models were excluded from the radial blood conductivity sweep analyses because they feature incomplete vasculature at the forearm. ViP, virtual population; BMI, body mass index.

| ViP Model | Sex | Age | Height (m) | Weight (kg) | BMI ($kg/m^2$) | Frequency sweep | Blood conductivity sweep | Electrode position sweep | Volume impedance density |
|---|---|---|---|---|---|---|---|---|---|
| Ella | Female | 26 | 1.63 | 57.3 | 21.6 | Yes | Yes | Yes | Yes |
| Morphed Ella | Female | 26 | 1.63 | 79.7 | 30 | Yes | Yes | Yes | Yes |
| Glenn | Male | 84 | 1.73 | 61.1 | 20.4 | Yes | No | Yes | Yes |
| Fats | Male | 37 | 1.82 | 96 | 29 | Yes | No | Yes | Yes |

## 49 Supplementary Table 5. List of tissue contributions to resistance and reactance

Tissue breakdown of resistance and reactance percent contribution for each computable human model. SAT, subcutaneous adipose tissue.

| Tissues | Resistance contribution (%) | Reactance contribution (%) |
|---|---|---|
| | **Ella** | |
| SAT | 29.01 | -0.25 |
| Radial artery | 7.23 | 3.10 |
| Ulnar artery | 0.93 | 0.26 |
| Cancellous bone | 2.43 | 1.04 |
| Cortical bone | 1.67 | 1.33 |
| Fat | 0.32 | 0.07 |
| Bone marrow | 0 | 0 |
| Skeletal muscle | 43.51 | 75.65 |
| Skin | 0.08 | 18.10 |
| Tendon | 11.14 | 1.81 |
| Veins | 3.69 | -1.11 |
| All tissues | 100 | 100 |
| | **Morphed Ella** | |
| SAT | 27.25 | -0.93 |
| Radial artery | 7.67 | 3.07 |
| Ulnar artery | 1.00 | 0.26 |
| Cancellous bone | 2.28 | 0.84 |
| Cortical bone | 1.62 | 1.21 |
| Fat | 0.35 | 0.05 |
| Bone marrow | 0 | 0 |
| Skeletal muscle | 44.55 | 77.61 |
| Skin | 0.08 | 18.35 |
| Tendon | 10.98 | 0.98 |
| Veins | 4.24 | -1.44 |
| All tissues | 100 | 100 |
| | **Fats** | |
| SAT | 46.29 | 14.93 |
| Radial artery | 0 | 0 |
| Ulnar artery | 0 | 0 |
| Cancellous bone | 2.15 | 0.83 |
| Cortical bone | 2.85 | 1.81 |
| Fat | 0.10 | 0.01 |
| Bone marrow | 0.01 | 0.01 |
| Skeletal muscle | 45.65 | 51.40 |
| Skin | 0.18 | 30.34 |
| Tendon | 2.56 | 0.64 |
| Veins | 0.21 | 0.03 |
| All tissues | 100 | 100 |
| | **Glenn** | |
| SAT | 18.86 | 1.56 |
| Radial artery | 0.04 | 0.01 |
| Ulnar artery | 0 | 0 |
| Cancellous bone | 2.74 | 0.46 |
| Cortical bone | 1.28 | 0.40 |
| Fat | 0.86 | 0.03 |
| Bone marrow | 0 | 0 |
| Skeletal muscle | 75.88 | 61.76 |
| Skin | 0.30 | 35.78 |
| Tendon | 0.02 | 0 |
| Veins | 0.02 | 0 |
| All tissues | 100 | 100 |

# 50 Supplementary Table 6. Summary of experimental data collection for Group 1, 2, and 3

Enrolled study participants along with health status and study trials performed. ICU, intensive care unit; LVAD, left ventricular assist device.

| Dataset | Setting | Number of subjects enrolled | Number of subjects excluded | Health condition | Trials |
|---|---|---|---|---|---|
| Group 1a | Lab | 53 | 1 | Healthy | Baseline, treadmill, stationary bike |
| Group 1b | Lab | 5 | 0 | Healthy | Cold pressor, Valsalva |
| Group 1c | Lab | 18 | 0 | Healthy | Baseline, stationary bike, Valsalva, hand grip, deep breathing |
| Group 2 | Clinic | 86 | 0 | Hypertensive, cardiovascular disease, other | Baseline |
| Group 3 | ICU | 3 | 0 | Stage IV heart failure | Pre-/Post-LVAD implantation |

# 51 Supplementary Table 7. Average vitals for Group 1

Summary of vital signs recorded on the healthy Group 1 dataset. $N$, number of subjects; BPM, beats per minute; DBP, diastolic blood pressure; SBP, systolic blood pressure; SD, standard deviation; R, resistance.

| **Group 1** | | $N$ | **Heart rate (BPM)** | | **DBP (mm Hg)** | | **SBP (mm Hg)** | | **Baseline R (ohms)** | | **Peak-to-peak R (mohms)** | |
|---|---|---|---|---|---|---|---|---|---|---|---|---|
| Cohort | Trial | | Mean | SD | Mean | SD | Mean | SD | Mean | SD | Mean | SD |
| a | Baseline | 43 | 78.08 | 11.25 | 82.57 | 14.64 | 129.3 | 14.96 | 24.79 | 5.32 | 31.86 | 18.56 |
| | Treadmill recovery | 28 | 85.28 | 12.29 | 82.98 | 10.75 | 124.72 | 12.23 | 25.5 | 3.9 | 31.1 | 17.68 |
| | Bike recovery | 28 | 84.22 | 14.84 | 78.25 | 10.9 | 124.97 | 14.66 | 25.13 | 3.96 | 29.2 | 20.85 |
| b | Baseline | 4 | 73.88 | 13.07 | 72.81 | 7.01 | 120.78 | 10.51 | 27.39 | 1.5 | 31.45 | 8.13 |
| | Cold pressor | 4 | 75.34 | 12.74 | 75.8 | 8.3 | 123.8 | 12.37 | 27.45 | 1.49 | 32.71 | 9.05 |
| | Valsalva | 5 | 77.34 | 11.09 | 66.94 | 9.3 | 112.01 | 11.81 | 27.98 | 1.77 | 30.4 | 9.53 |
| c | Baseline | 18 | 76.62 | 11.1 | 78.26 | 8.64 | 125.24 | 14.11 | 28.49 | 4.98 | 33.81 | 14.08 |
| | Valsalva | 16 | 87.03 | 10.35 | 81.15 | 9.26 | 127.73 | 14.35 | 28.03 | 4.17 | 32.27 | 15.4 |
| | Hand grip | 15 | 87.19 | 10.06 | 78.97 | 9.16 | 124.1 | 11.91 | 28.04 | 4.52 | 38.08 | 18.77 |
| | Deep breathing | 15 | 80.17 | 11.01 | 74.4 | 9.14 | 116.06 | 13.93 | 28.23 | 4.52 | 33.42 | 15.54 |
| All | Baseline | 61 | 77.46 | 11.21 | 80.73 | 12.62 | 127.57 | 14.74 | 26.37 | 5.49 | 32.69 | 16.82 |
| | Valsalva | 21 | 85.23 | 11.14 | 78.51 | 10.79 | 124.81 | 15.2 | 28.02 | 3.84 | 31.92 | 14.51 |
| | All | 69 | 79.8 | 12.12 | 79.92 | 11.92 | 125.91 | 14.62 | 26.58 | 5.05 | 32.57 | 16.96 |

# 52 Supplementary Table 8. Datasets analyzed using signal-tagged physics-informed neural network models

Summary statistics for the four base and derived datasets described in Supplementary Discussion 7.1.1. The parameter $m$ refers to the number of consecutive resistance periods (heartbeats). †: We randomly sampled 100k BP waveforms from the PulseDB dataset subject to the criteria described in Supplementary Discussion 7.1.2. P, population-wide; PM, population-wide + metadata; S, subject-specific.

| Dataset | Model | Subjects | Test-exclusive subjects | $m$ | Raw segments | Training segments | Test segments |
|---|---|---|---|---|---|---|---|
| Group 1 | P & PM | 73 | 4 | 5 | 223,156 | 72,358 | 11,686 |
| | S | 63 | 0 | 5 | 223,156 | 70,977 | 7,750 |
| Group 2 | P | 75 | 6 | 5 | 45,568 | 26,276 | 4,380 |
| | S | 48 | 0 | 5 | 45,568 | 23,207 | 1,732 |
| PulseDB[100] synthetic | P | 2,265 | 0 | 1 | 100,000† | 90,000 | 10,000 |
| PulseDB[100] synthetic with biological variability | PM | 2,265 | 0 | 1 | 500,000 | 500,000 | 350,000 |
| Graphene-HGCPT[99] | P & S | 7 | 0 | 5 | 24,233 | 8,267 | 661 |

# 53 Supplementary Table 9. Signal-tagged physics-informed neural network model specifications and training details

Summary of model specifications and training details for the signal-tagged physics-informed neural network models trained (Supplementary Discussion 7.6). Learning rate scheduler refers to the pytorch learning rate scheduler. Total epochs is an upper bound for the number of epochs trained. Training time is an estimate of the time the model was trained for. G-HGCPT, Graphene-HGCPT; P, population-wide; PM, population-wide + metadata; S, subject-specific; NA, not applicable; RLROP, ReduceLROnPlateau; CALR, CosineAnnealingLR.

| Dataset | Model | Training phases | Starting learning rate | Learning rate scheduler | Total epochs | Phase 2 epochs | Phase 3 epochs | Training time (hr) | Total model parameters |
|---|---|---|---|---|---|---|---|---|---|
| Group 1 | PM | 1, 2, 3 | $3 \cdot 10^{-4}$ | RLROP | 1,200 | 350 | 850 | 280 | 1,020,986 |
| | P | 1, 2, 3 | $3 \cdot 10^{-4}$ | RLROP | 1,200 | 350 | 850 | 280 | 1,016,486 |
| | S | 3 | $3 \cdot 10^{-4}$ | CALR | 1,500 | NA | 1,500 | 2,500 | 611,086 |
| Group 2 | P | 1, 2, 3 | $3 \cdot 10^{-4}$ | RLROP | 1,500 | 250 | 1,750 | 150 | 611,086 |
| | S | 3 | $3 \cdot 10^{-4}$ | CALR | 1,500 | NA | 1,500 | 300 | 611,086 |
| PulseDB[100] synthetic | P | 3 | $3 \cdot 10^{-4}$ | RLROP | 350 | NA | 350 | 72 | 376,486 |
| PulseDB[100] synthetic with biological variability | PM | 3 | $3 \cdot 10^{-4}$ | CALR | 70 | 50 | 70 | 110 | 379,486 |
| G-HGCPT[99] | P | 1, 2, 3 | $3 \cdot 10^{-4}$ | RLROP | 2,350 | 250 | 2,100 | 72 | 611,086 |
| | S | 3 | $3 \cdot 10^{-4}$ | CALR | 2,500 | NA | 2,500 | 300 | 611,086 |

# 54 Supplementary Table 10. Bioimpedance cuffless blood pressure studies

Benchmark comparison of competing blood pressure studies using bioimpedance data. The results for our work are reported on the Group 1 and 2 test datasets excluding test-exclusive subjects and the Group 2 subject-specific dataset (Supplementary Table 8). Here, the sample size refers to the size of the study and not necessarily the number of subjects in the training/test dataset. All values reported for our signal-tagged physics-informed neural network (sPINN) are mean absolute error (MAE). BP, blood pressure; PWA, pulse wave analysis; PWV, pulse wave velocity; PTT, pulse transit time; HTN, hypertension; CVD, cardiovascular disease; DBP, diastolic blood pressure; SBP, systolic blood pressure; ME, mean error; AME, absolute mean error; RMSE, root mean squared error; AMAE, average MAE; SD, standard deviation; ICU, intensive care unit; HF, heart failure. S, subject-specific; P, population-wide.

| Reference | BP prediction model | Electrode material | Sample size (male:female) | Ages | Condition | Setting | Calibration free | SBP error±SD (mm Hg) | DBP error±SD (mm Hg) |
|---|---|---|---|---|---|---|---|---|---|
| Kireev et al.[137] | PWA | Graphene | 7 (Unknown) | mid 20s | Healthy | Lab | No | 0.2±5.8 (AME) | 0.2±4.5 (AME) |
| Sel et al.[144] | PWA | Silver | 10 (10:0) | mid 20s | Healthy | Lab | No | 0.11±5.27 (ME) | 0.11±3.87 (ME) |
| Ibrahim et al.[150] | Adaboost regression | Silver | 4 (Unknown) | 20-25 | Healthy | Lab | No | 0.2±6.5 (ME) | 0.5±5.0 (ME) |
| Rachim et al.[221] | PTT | Silver-plated polyester | 10 (6:4) | 25±4 | Healthy | Lab | No | 6.86±1.65 (RMSE) | 6.67±1.75 (RMSE) |
| Wang et al.[130] | Bramwell-Hill-based impedance plethysmography | Ag/AgCl | 30 (20:10) | 27±8 | Healthy | Lab | No | 2.01±1.4 (MAE) | 2.26±1.43 (MAE) |
| Huynh et al.[151] | Linear regression PWV | Copper foil | 15 (9:6) | 30±5 | Healthy | Lab | No | 7.47±2.15 (RMSE) | 5.17±1.81 (RMSE) |
| Buxi et al.[222] | Linear regression PTT or PAT | Ag/AgCl | 6 (6:0) | Unknown | Healthy | Lab | No | Unknown | Unknown |
| Ibrahim et al.[146] | PWA | Ag/AgCl | 10 (7:3) | 18-30 | Healthy | Lab | No | 3.44±0.84 (RMSE) | 2.63±0.58 (RMSE) |
| Wang et al.[131] | PWA | Silver-plated polyester | 6 (3:3) | 24±1 | Healthy | Lab | No | 2.63±2.58 (MAE) | 2.66±2.52 (MAE) |
| Hsiao et al.[152] | PTT+PWA 2xBioZ + 2xPPG | Tin hot air solder leveled copper | 8 (5:3) | 20-33 | Healthy | Lab | Yes | 10.63±5.36 (AMAE) | |
| Our work | sPINN | Stainless steel | 75 (39:36) | 18-48 | Healthy | Lab | P: Yes<br>S: No | 7.24±6.98<br>6.62±6.42 | 5.45±5.19<br>4.99±4.78 |
| | | | 76 (41:35) | 24-93 | All | Clinic | P: Yes<br>S: No | 7.26±8.46<br>6.25±5.75 | 3.89±4.63<br>3.30±2.97 |
| | | | 32 (23:9) | 28-88 | HTN | Clinic | P: Yes<br>S: No | 8.25±10.23<br>7.16±6.45 | 4.30±5.45<br>3.77±3.38 |
| | | | 22 (15:7) | 58-93 | CVD | Clinic | P: Yes<br>S: No | 6.67±6.96<br>5.92±4.91 | 3.30±3.24<br>3.07±2.46 |
| | | | 32 (13:19) | 24-85 | Other | Clinic | P: Yes<br>S: No | 6.67±7.24<br>5.55±5.36 | 3.86±4.47<br>2.98±2.76 |
| | | | 3 (3:0) | 21-61 | HF | ICU | – | – | – |

# 55 Supplementary Table 11. Benchmark comparison to traditional machine learning methods

Benchmark comparison of signal-tagged physics-informed neural network (sPINN) blood pressure prediction accuracies against conventional machine learning methods. Metrics are presented for the relevant test set without test-exclusive subjects. For each dataset, the best value for each performance metric is italicized. G-HGCPT, Graphene-HGCPT; MLP, multilayer perceptron; DBP, diastolic blood pressure; SBP, systolic blood pressure; MAE, mean absolute error; SD, standard deviation; S, subject-specific; P, population-wide; AF, adaptive filter; NA, not applicable.

| Dataset | Model type | Linear regression | | | | Sklearn MLP | | | | Signal encoder and AF | | | | sPINN | | | |
|---|---|---|---|---|---|---|---|---|---|---|---|---|---|---|---|---|---|
| | | SBP | | DBP | | SBP | | DBP | | SBP | | DBP | | SBP | | DBP | |
| | | MAE±SD (mm Hg) | $r^2$ | MAE±SD (mm Hg) | $r^2$ | MAE±SD (mm Hg) | $r^2$ | MAE±SD (mm Hg) | $r^2$ | MAE±SD (mm Hg) | $r^2$ | MAE±SD (mm Hg) | $r^2$ | MAE±SD (mm Hg) | $r^2$ | MAE±SD (mm Hg) | $r^2$ |
| Group 1 | PM | 11.14±8.45 | 0.089 | 10.89±7.97 | 0.014 | 7.58±6.75 | 0.512 | 5.78±4.71 | 0.619 | *6.86*±6.37 | *0.587* | *5.29*±4.63 | *0.659* | 7.24±6.98 | 0.575 | 5.45±5.19 | 0.629 |
| | P | 10.70±8.50 | 0.117 | 8.88±7.15 | 0.077 | 10.46±8.54 | 0.134 | 9.16±7.01 | 0.087 | *8.11*±7.28 | *0.448* | *6.49*±5.79 | *0.477* | 8.59±8.34 | 0.428 | 6.67±6.56 | 0.448 |
| | S | 10.30±13.08 | 0.276 | 8.26±11.56 | 0.291 | 7.26±6.98 | 0.554 | 5.52±5.11 | 0.617 | NA | NA | NA | NA | *6.62*±6.42 | *0.626* | *4.99*±4.78 | *0.680* |
| Group 2 | P | 16.32±13.62 | 0.128 | 9.02±7.34 | 0.241 | 14.02±12.28 | 0.334 | 7.93±6.61 | 0.409 | 9.22±8.50 | 0.697 | 5.79±4.71 | 0.687 | *7.26*±8.46 | *0.774* | *3.89*±4.63 | *0.807* |
| | S | 36.71±310.68 | 0.003 | 37.68±329.31 | 0.013 | *6.16*±6.21 | 0.857 | *3.17*±3.16 | 0.894 | NA | NA | NA | NA | 6.25±5.75 | *0.872* | 3.30±2.97 | *0.898* |
| PulseDB[100] synthetic | P | 3.14±3.28 | 0.949 | 3.03±2.91 | 0.708 | 1.42±1.46 | 0.989 | 1.32±1.35 | 0.942 | NA | NA | NA | NA | *0.57*±0.76 | *0.998* | *0.85*±1.02 | *0.974* |
| G-HGCPT[99] | P | 12.13±10.75 | 0.228 | 10.47±8.02 | 0.208 | 7.81±7.03 | 0.682 | 5.69±4.98 | *0.734* | 9.20±8.17 | 0.555 | 6.52±5.50 | 0.660 | *6.87*±7.42 | *0.699* | *5.19*±5.58 | *0.734* |
| | S | 8.78±8.83 | 0.576 | 7.19±6.26 | 0.610 | 7.61±6.56 | 0.715 | 5.26±4.77 | 0.770 | NA | NA | NA | NA | *6.52*±6.05 | *0.770* | *4.51*±4.15 | *0.828* |

# 56 Supplementary Table 12. Samples sizes and statistics analyses

Additional Supplementary Table file related to Fig.3, Extended Data Fig.1,2, 4, and Supplementary Fig.16.

# Supplementary Videos

### Instructions to view Supplementary Videos in PDF

To view the embedded Supplementary Videos, Adobe Acrobat Reader is required. By default, multimedia content is disabled in Adobe Reader. To enable video playback:

1. Open the Supplementary Information PDF in Adobe Reader.
2. When prompted with a security warning, select "*Trust this document always*" under the Options button.

## 57 Supplementary Video 1. Traveling blood pressure waveform

Peaking of the blood pressure waveform from the proximal brachial artery, to the brachial-radial-ulnar branch, to the distal radial artery at the wrist.

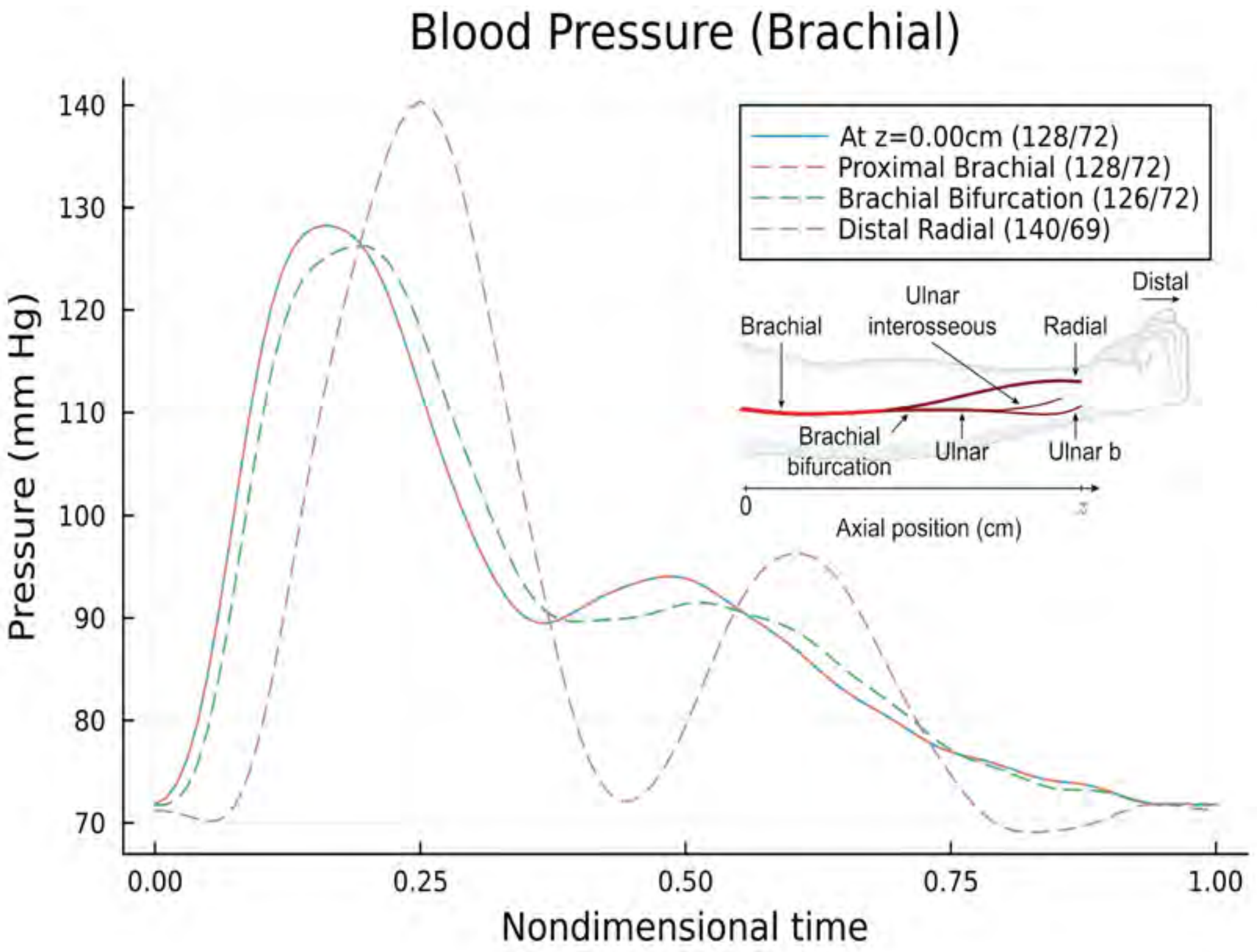

# 58 Supplementary Video 2. Brachial blood pressure, velocity, and wall motion displacement

Axial blood velocity along with wall movement in the brachial artery alongside the blood pressure waveform at the proximal point of the brachial artery.

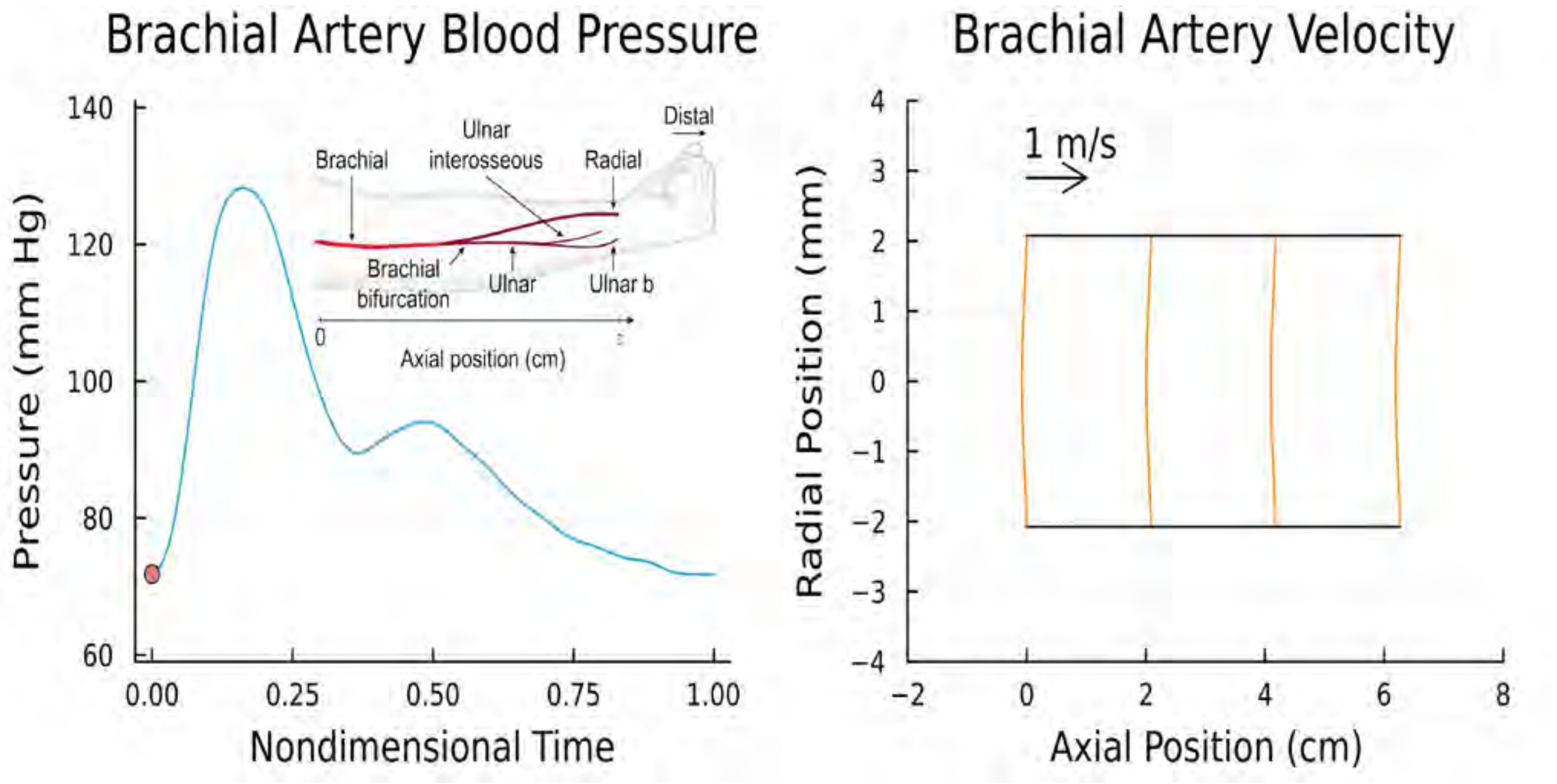

# 59 Supplementary Video 3. Radial blood pressure, velocity, and wall motion displacement

Axial blood velocity along with wall movement in the radial artery alongside the blood pressure waveform at the distal point of the radial artery.

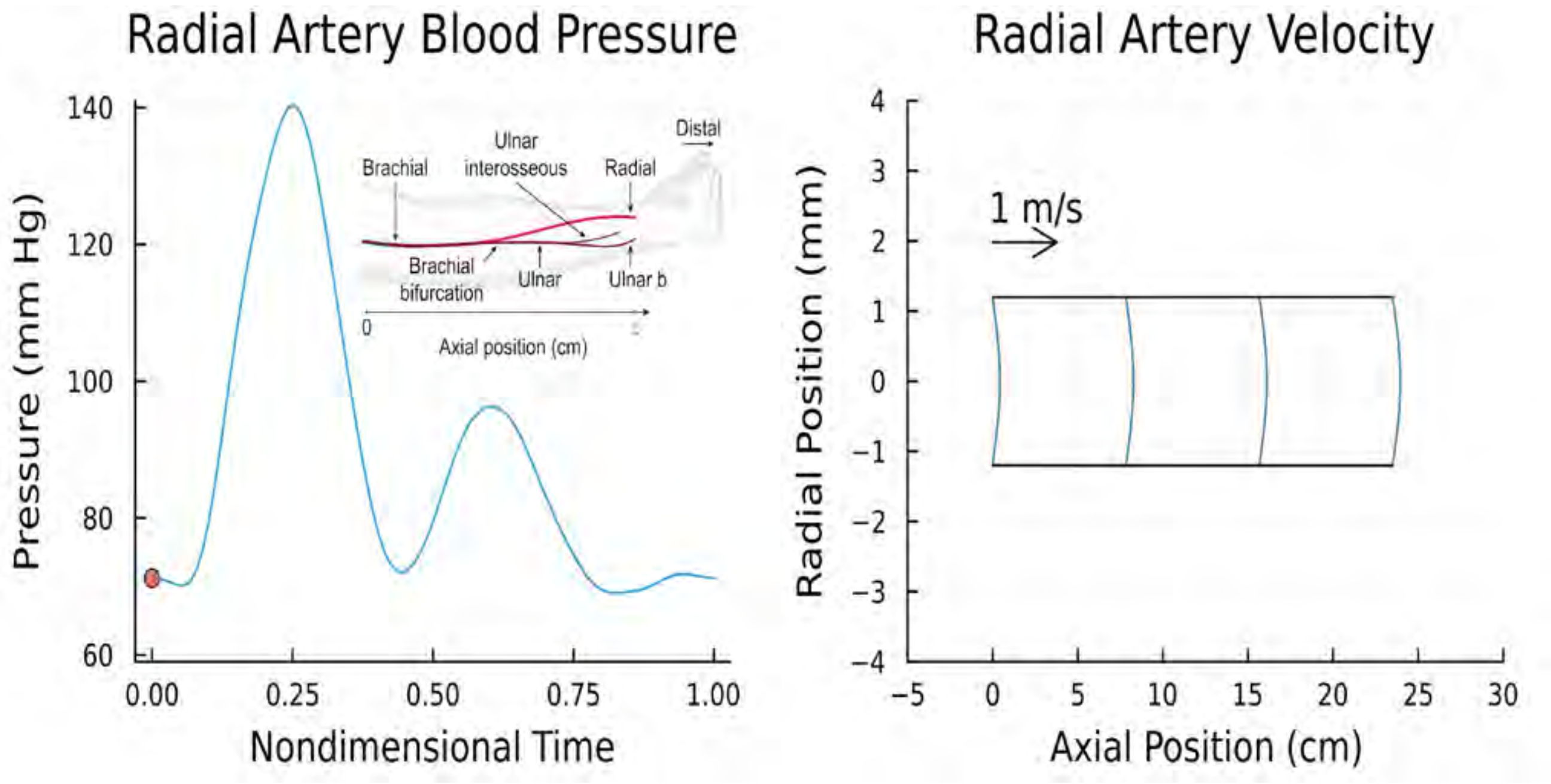

## 60 Supplementary Video 4. Ulnar blood pressure, velocity, and wall motion displacement

Axial blood velocity along with wall movement in the ulnar artery alongside the blood pressure waveform at the proximal point of the ulnar artery. The Poiseuille solution is determined by conservation of flow between the brachial, radial, and ulnar arteries.

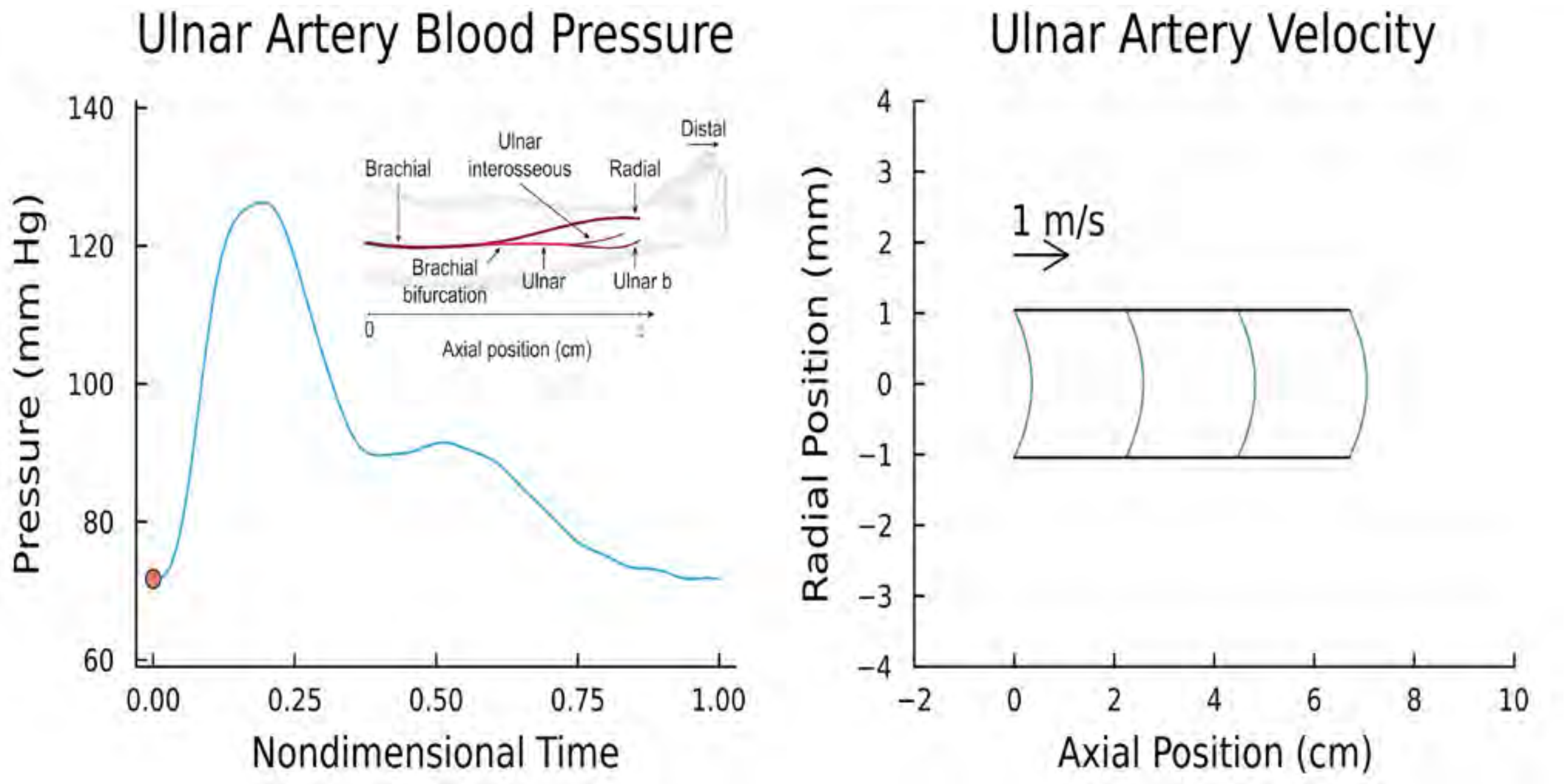

# 61 Supplementary Video 5. Ulnar b blood pressure, velocity, and wall motion displacement

Axial blood velocity along with wall movement in the ulnar b. artery alongside the blood pressure waveform at the proximal point of the ulnar b. artery. The Poiseuille solution is determined by conservation of flow between the brachial, radial, and ulnar arteries and between the ulnar, ulnar interosseus, and ulnar b. arteries, and by assuming that the flow in the ulnar and ulnar b. arteries are equal.

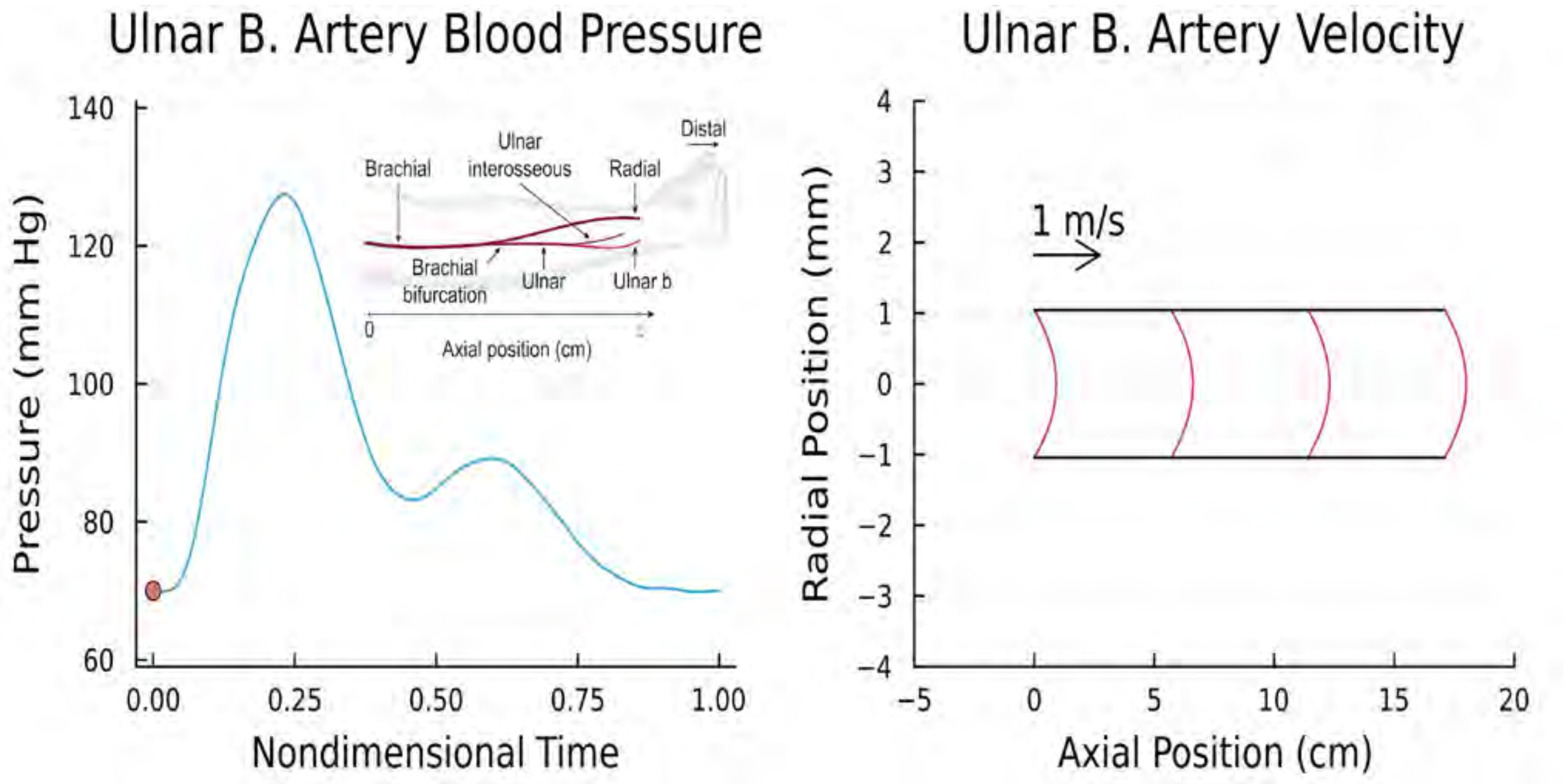

## 62 Supplementary Video 6. Ulnar i blood pressure, velocity, and wall motion displacement

Axial blood velocity along with wall movement in the ulnar interosseus artery alongside the blood pressure waveform at the proximal point of the ulnar interosseus artery. The Poiseuille solution is determined by conservation of flow between the brachial, radial, and ulnar arteries and between the ulnar, ulnar interosseus, and ulnar b. arteries, and by assuming that the flow in the ulnar and ulnar b. arteries are equal.

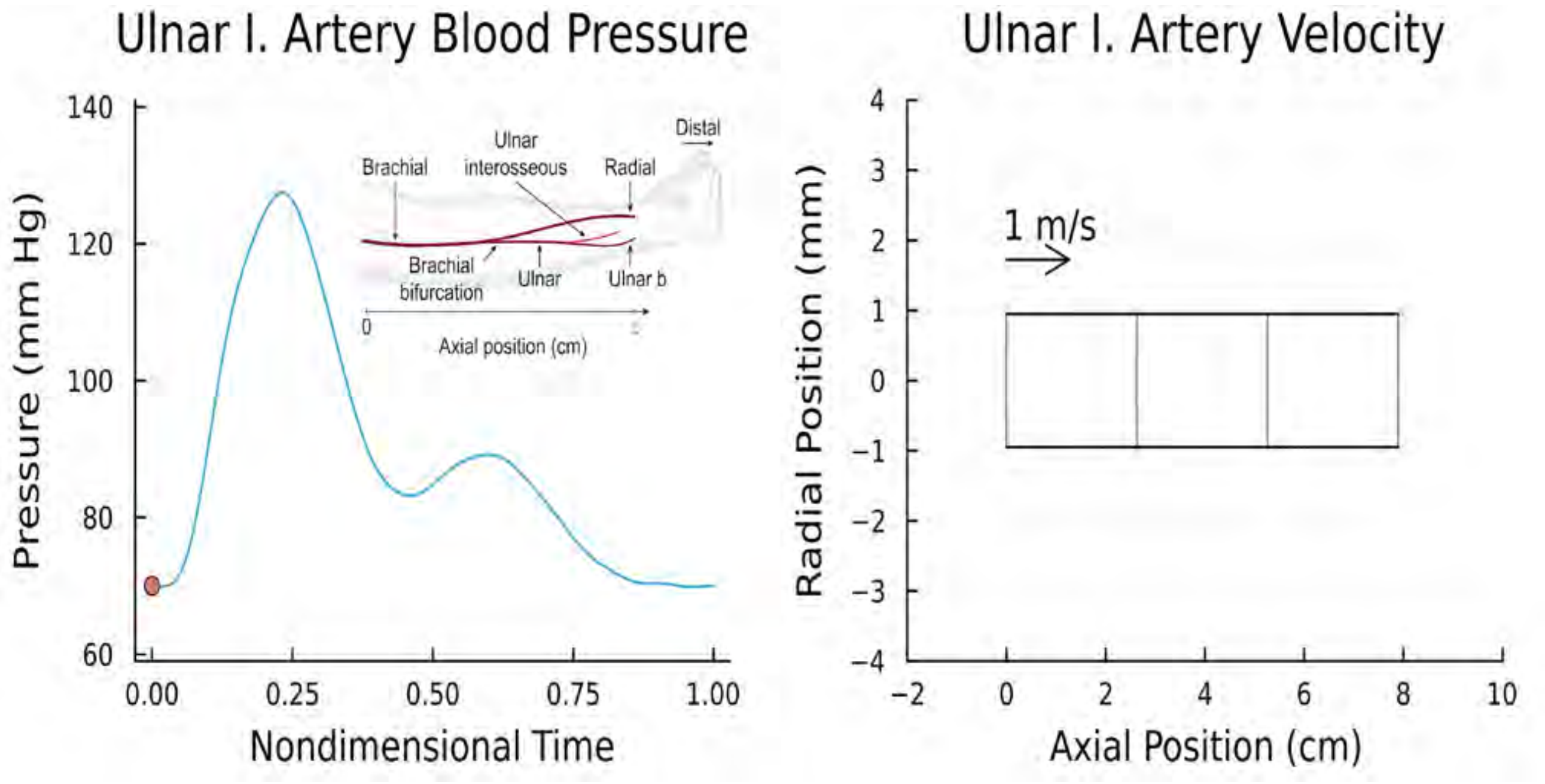

## 63 Supplementary Video 7. Radial shear stress

Filled contour animation of the radial artery axial shear stresses alongside the blood pressure waveform at the distal radial artery.

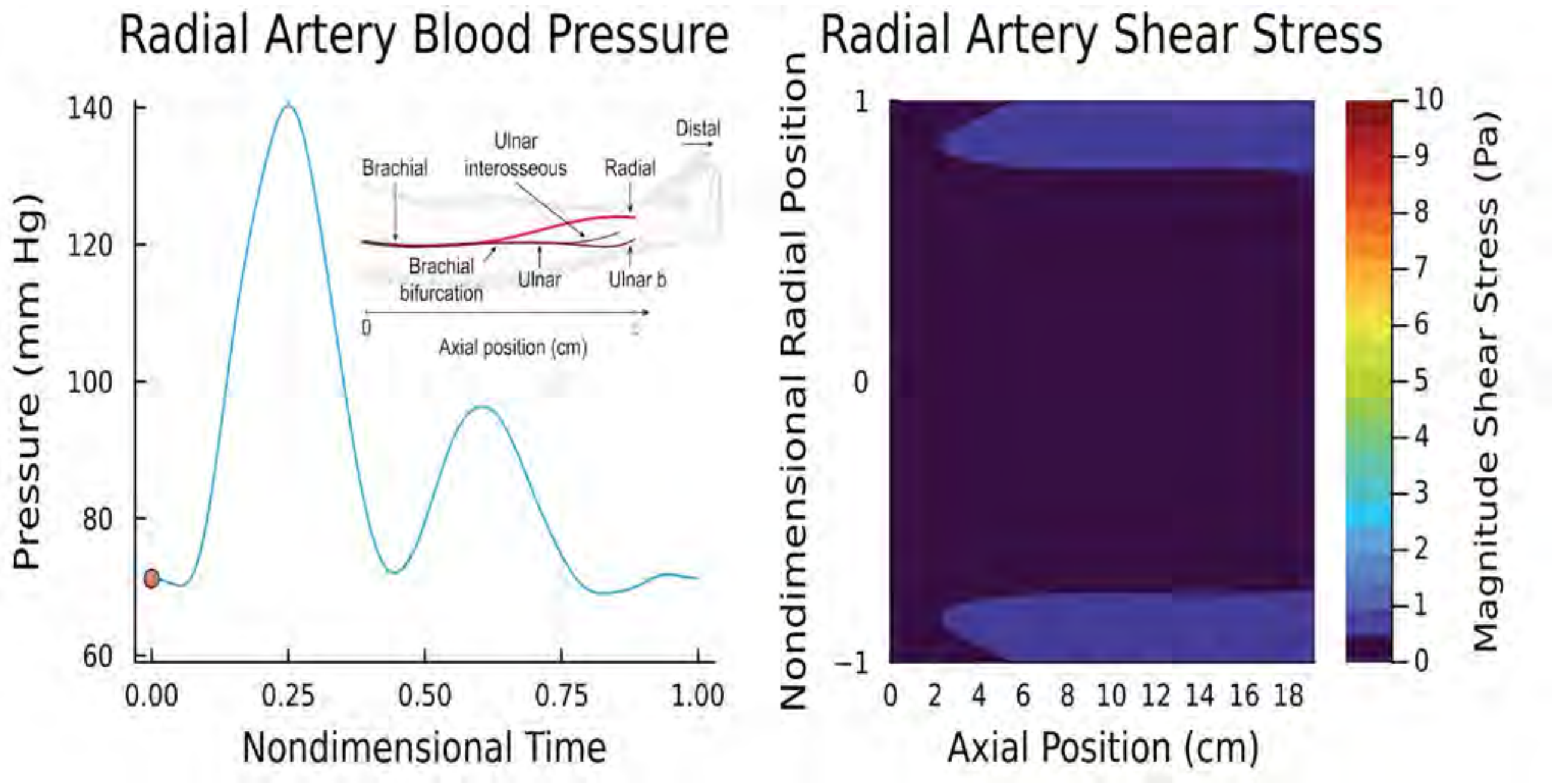

## 64 Supplementary Video 8. Radial quiver velocity plot

Quiver animation displaying the, separately rescaled, axial and radial blood flow velocities in the radial artery.

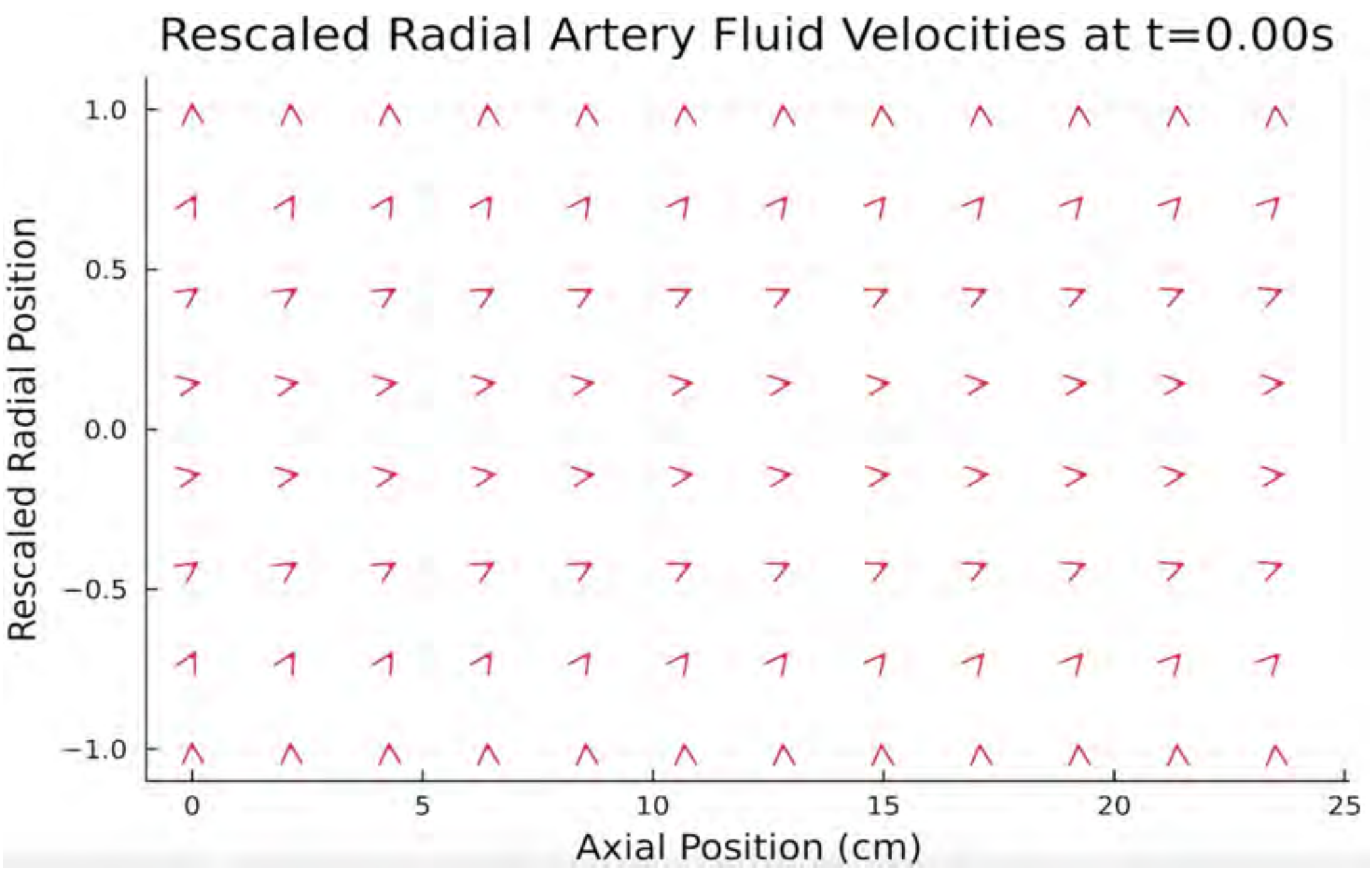

# 65 Supplementary Video 9. Doppler flowmetry of the brachial artery

Doppler flowmetry of the left brachial artery showing a peak systolic velocity of 78 cm/s.

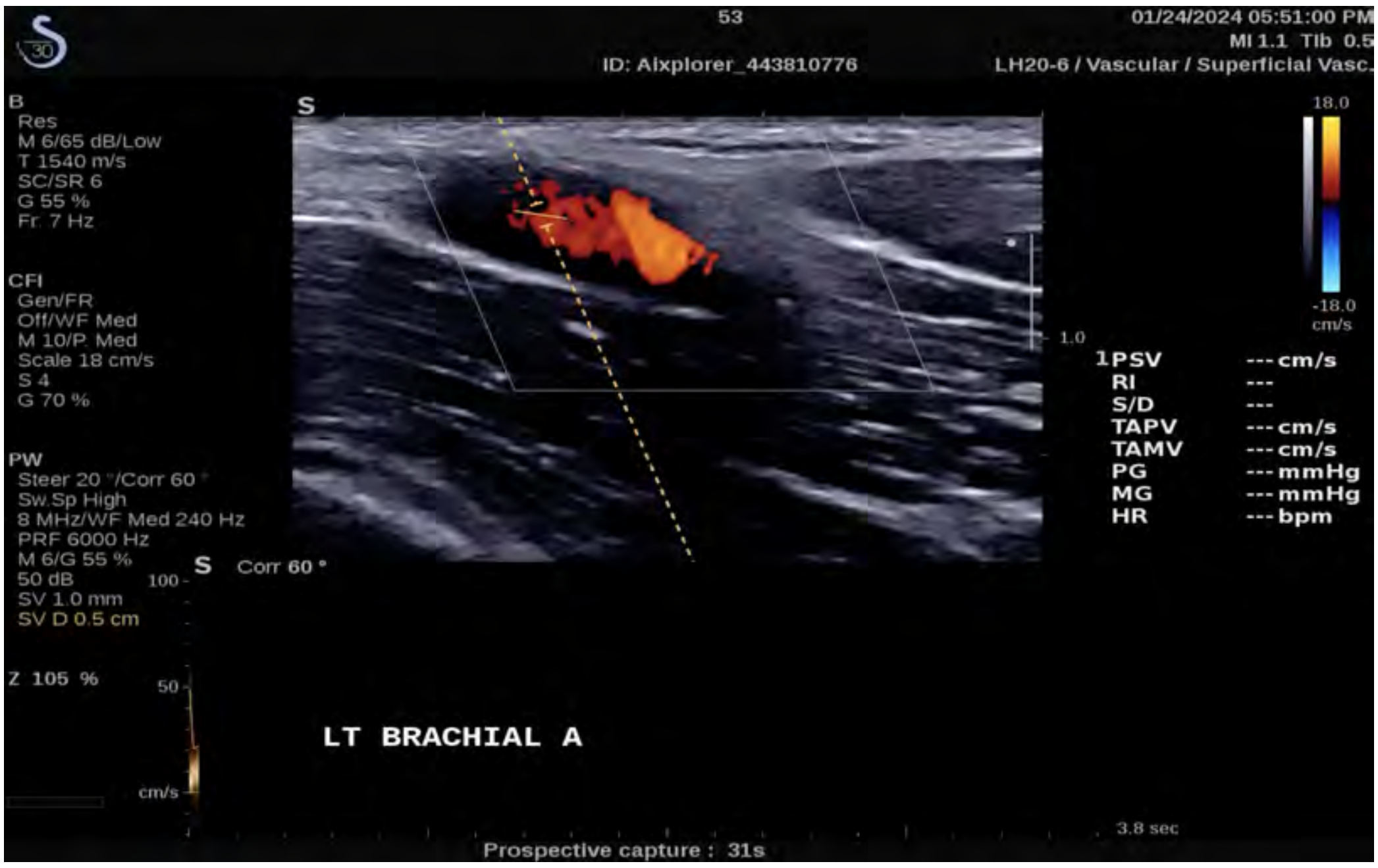

# 66 Supplementary Video 10. Doppler flowmetry of the radial artery

Doppler flowmetry of the left radial artery showing a peak systolic velocity of 57 cm/s.

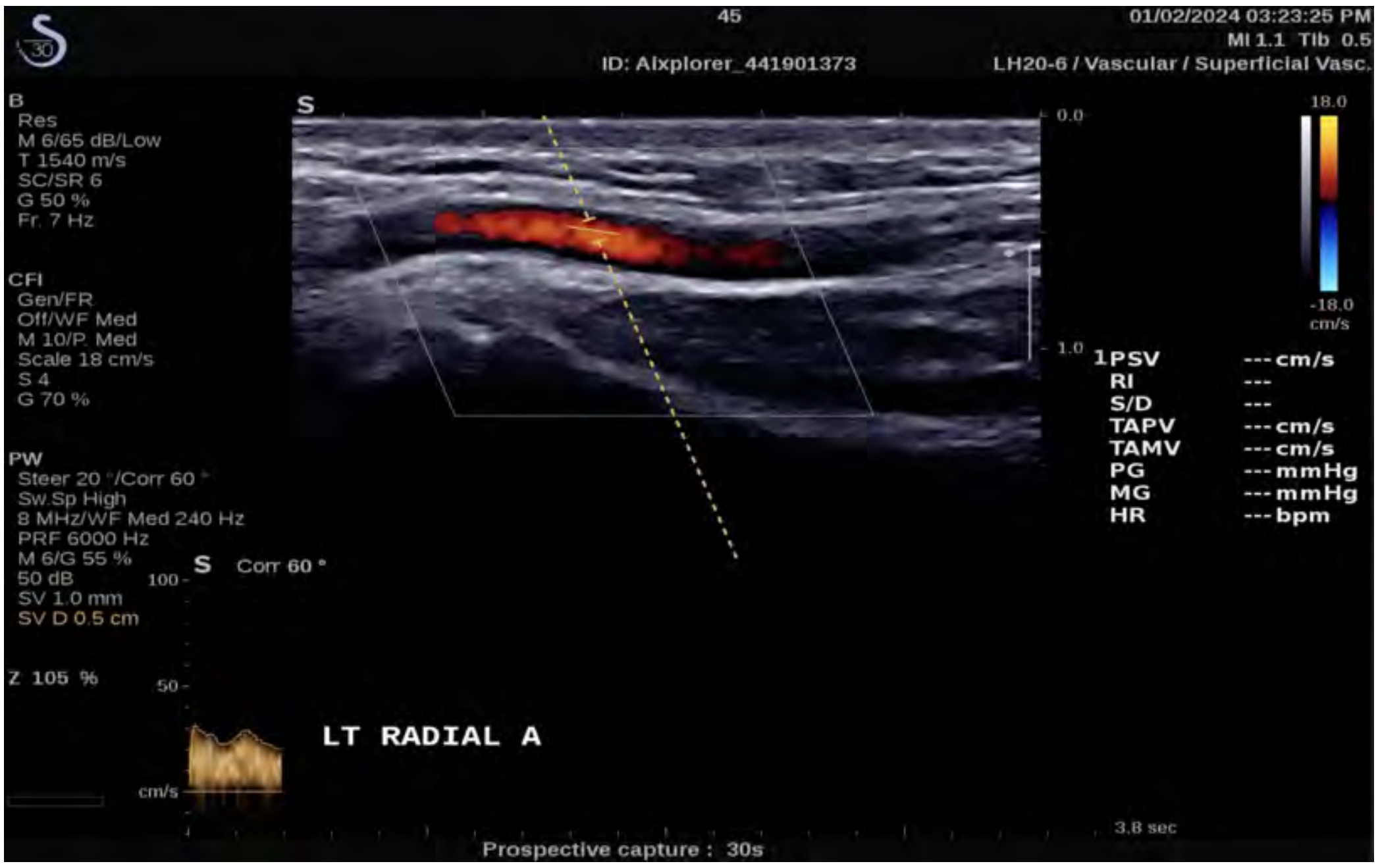

# 67 Supplementary Video 11. Doppler flowmetry of the ulnar artery

Doppler flowmetry of the left ulnar artery showing a peak systolic velocity of 59 cm/s.

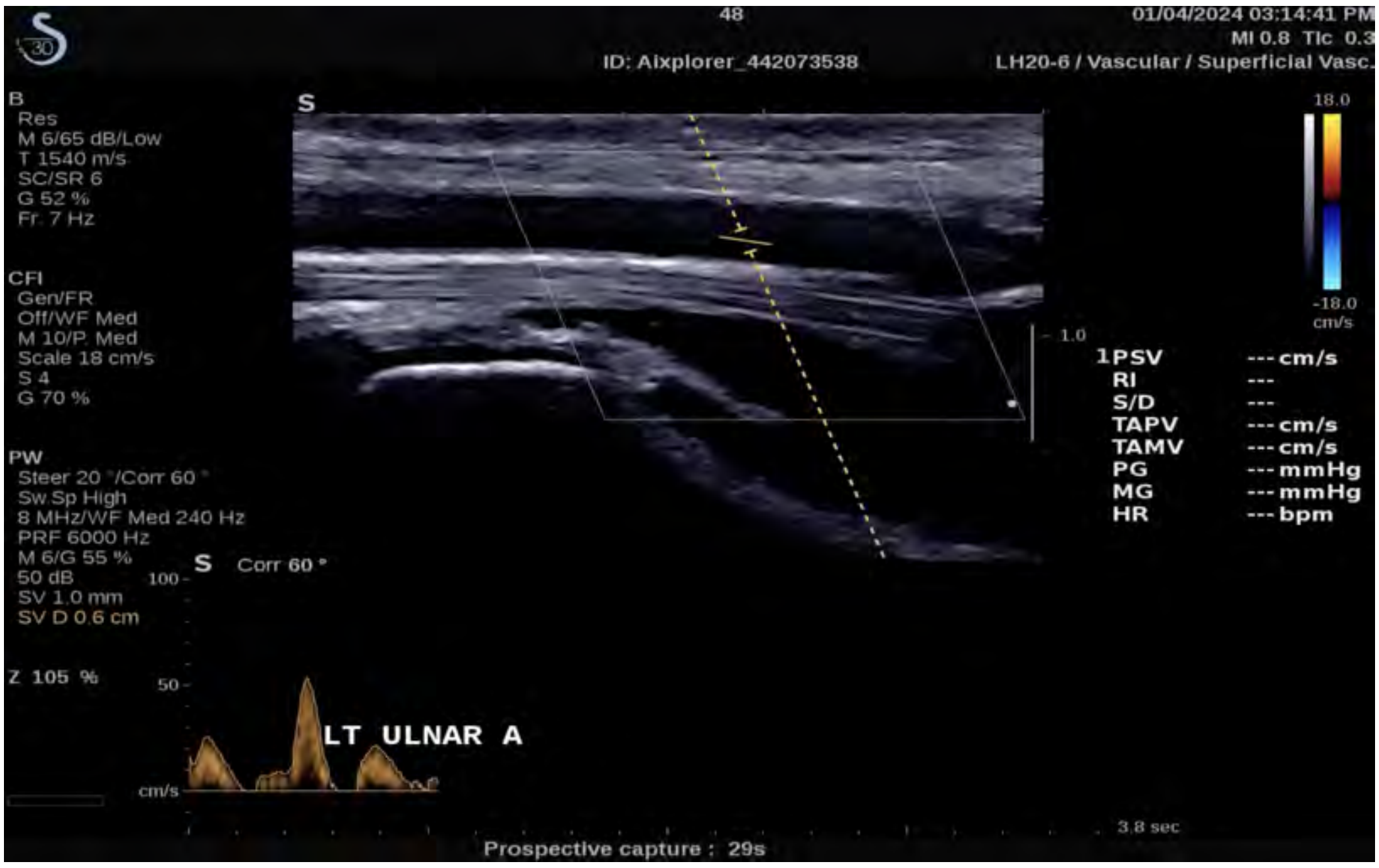

# Supplementary bibliography

| Figure (panel) | Comparison | N | Statistical Test | P value |
|---|---|---|---|---|
| **Figure 3** | | | | |
| 3 (c) | Radial artery radius | 100 | F(1, 98) = 949.6 | <0.0001 |
| | Radial blood conductivity | 100 | F(1, 98) = 1852 | <0.0001 |
| 3 (j) | nominal electrode position vs -10 mm | 4<br>4 | Dunn’s multiple comparisons test | 0.1829 |
| | nominal electrode position vs -7.5 mm | 4<br>4 | | 0.1177 |
| | nominal electrode position vs -5 mm | 4<br>4 | | 0.154 |
| | nominal electrode position vs -2.5 mm | 4<br>4 | | 0.2947 |
| | nominal electrode position vs 2.5 mm | 4<br>4 | | 0.9479 |
| | nominal electrode position vs 5 mm | 4<br>4 | | 0.9561 |
| | nominal electrode position vs 7.5 mm | 4<br>4 | | 0.7974 |
| | nominal electrode position vs 10 mm | 4<br>4 | | 0.9853 |
| 3 (l) | Radial artery radius | 17 | F(1, 15) = 1306 | <0.0001 |
| 3 (m) | Radial blood conductivity | 25 | F(1, 23) = 2610 | <0.0001 |

| Figure (panel) | Comparison | N | Statistical Test |
|---|---|---|---|
| **Extended data figure 1** | | | |
| 1 (c) | Radial artery radius | 17 | $F(1, 15) = 13.45$ |
| 1 (d) | Radial blood conductivity | 25 | $F(1, 23) = 617.5$ |

Considered the mean value of y-points

| P value |
| --- |
| |
| 0.0023 |
| <0.0001 |

| Figure (panel) | Comparison | N |
|---|---|---|
| **Extended data figure 2. Ultrasound results** | | |
| 3 (b i) Radial artery depth | Male | 39 |
| | Female | 36 |
| | Male and female | 75 |
| | Male vs<br>Female | 39<br>36 |
| 3 (b ii) Radial artery major diameter | Male | 39 |
| | Female | 36 |
| | Male and female | 75 |
| | Male vs<br>Female | 39<br>36 |
| 3 (b iii) Radial artery minor diameter | Male | 39 |
| | Female | 36 |
| | Male and female | 75 |
| | Male vs<br>Female | 39<br>36 |
| 3 (b iv) Radial artery peak systolic velocity | Male | 39 |
| | Female | 36 |
| | Male and female | 75 |
| | Male vs<br>Female | 39<br>36 |
| 3 (c i) Ulnar artery depth | Male | 39 |
| | Female | 36 |
| | Male and female | 75 |
| | Male vs<br>Female | 39<br>36 |
| 3 (c ii) Ulnar artery major diameter | Male | 39 |
| | Female | 36 |
| | Male and female | 75 |
| | Male vs<br>Female | 39<br>36 |
| 3 (c iii) Ulnar artery minor | Male | 39 |
| | Female | 36 |
| | Male and female | 75 |
| diameter | Male vs<br>Female | 39<br>36 |
| 3 (c iv) Ulnar artery peak systolic velocity | Male | 38 |
| | Female | 36 |
| | Male and female | 74 |
| | Male vs<br>Female | 38<br>36 |
| 3 (d i) Brachial artery depth | Male | 39 |
| | Female | 36 |
| | Male and female | 75 |
| | Male vs<br>Female | 39<br>36 |
| 3 (d ii) Brachial artery major diameter | Male | 39 |
| | Female | 36 |
| | Male and female | 75 |
| | Male vs<br>Female | 39<br>36 |
| 3 (d iii) Brachial artery minor diameter | Male | 39 |
| | Female | 36 |
| | Male and female | 75 |
| | Male vs<br>Female | 39<br>36 |
| 3 (d iv) Brachial artery peak systolic velocity | Male | 39 |
| | Female | 36 |
| | Male and female | 75 |
| | Male vs<br>Female | 39<br>36 |
| 3 (e i) Artery depth | Radial | 75 |
| | Ulnar | 75 |
| | Brachial | 75 |
| | Radial vs<br>Ulnar | 75<br>75 |
| | Radial vs<br>Brachial | 75<br>75 |
| | Ulnar vs<br>Brachial | 75<br>75 |
| | Radial | 75 |
| 3 (e i) Artery major diameter | Ulnar | 75 |
| | Brachial | 75 |
| | Radial vs<br>Ulnar | 75<br>75 |
| | Radial vs<br>Brachial | 75<br>75 |
| | Ulnar vs<br>Brachial | 75<br>75 |
| 3 (e i) Artery minor diameter | Radial | 75 |
| | Ulnar | 75 |
| | Brachial | 75 |
| | Radial vs<br>Ulnar | 75<br>75 |
| | Radial vs<br>Brachial | 75<br>75 |
| | Ulnar vs<br>Brachial | 75<br>75 |
| 3 (e i) Artery peak systolic velocity | Radial | 75 |
| | Ulnar | 74 |
| | Brachial | 75 |
| | Radial vs<br>Ulnar | 75<br>74 |
| | Radial vs<br>Brachial | 75<br>75 |
| | Ulnar vs<br>Brachial | 74<br>75 |

| Statistical Test | P value |
|---|---|
| | |
| Shapiro-Wilk normality test | 0.6573 |
| | 0.3342 |
| | 0.3191 |
| Unpaired T-test | 0.9554 |
| F(35, 38) = 1.334 | 0.3856 |
| Shapiro-Wilk normality test | 0.4858 |
| | 0.2249 |
| | 0.0713 |
| Welch's test | <0.0001 |
| F(35, 38) = 3.718 | 0.0002 |
| Shapiro-Wilk normality test | 0.629 |
| | 0.215 |
| | 0.16 |
| Unpaired T-test | <0.0001 |
| F(35, 38) = 3.718 | 0.0566 |
| Shapiro-Wilk normality test | 0.087 |
| | 0.0093 |
| | 0.0187 |
| Mann-Whitney test | 0.0101 |
| Shapiro-Wilk normality test | 0.4857 |
| | 0.3863 |
| | 0.1809 |
| Unpaired T-test | 0.0377 |
| F(35, 38) = 1.014 | 0.9632 |
| Shapiro-Wilk normality test | 0.961 |
| | 0.0157 |
| | 0.2862 |
| Mann-Whitney test | <0.0001 |
| Shapiro-Wilk normality test | 0.4413 |
| | 0.0114 |
| | 0.0976 |
| Mann-Whitney test | <0.0001 |
| Shapiro-Wilk normality test | 0.5429 |
| | 0.0466 |
| | 0.0397 |
| Mann-Whitney test | 0.0447 |
| Shapiro-Wilk normality test | 0.2221 |
| | 0.0002 |
| | 0.0004 |
| Mann-Whitney test | 0.4208 |
| Shapiro-Wilk normality test | 0.1936 |
| | 0.1151 |
| | 0.1559 |
| Unpaired T-test | <0.0001 |
| F(35, 38) = 1.849 | 0.0694 |
| Shapiro-Wilk normality test | 0.257 |
| | 0.6028 |
| | 0.2763 |
| Unpaired T-test | <0.0001 |
| F(35, 38) = 1.726 | 0.1063 |
| Shapiro-Wilk normality test | 0.4119 |
| | 0.7099 |
| | 0.6474 |
| Unpaired T-test | 0.0984 |
| F(38, 35) = 1.088 | 0.805 |
| Shapiro-Wilk normality test | 0.3191 |
| | 0.1809 |
| | 0.0004 |
| Dunn’s multiple comparisons test | 0.0059 |
| | <0.0001 |
| | <0.0001 |
| Shapiro-Wilk normality test | 0.0713 |
| | 0.2862 |
| | 0.1559 |
| Dunn's multiple comparisons | 0.00432 |
| | <0.0001 |
| | <0.0001 |
| Shapiro-Wilk normality test | 0.16 |
| | 0.0976 |
| | 0.2763 |
| Dunn’s multiple comparisons test | 0.001 |
| | <0.0001 |
| | <0.0001 |
| Shapiro-Wilk normality test | 0.0187 |
| | 0.0396 |
| | 0.6472 |
| Dunn’s multiple comparisons test | 0.5176 |
| | <0.0001 |
| | <0.0001 |

| Figure (panel) | Comparison | N |
|---|---|---|
| **Extended data figure 4. Experimental Group 1 resul** | | |
| 4 (a) Group 1 A Scaled heart rate | Bike recovery | 25 |
| | Treadmill recovery | 26 |
| | Bike recovery | 25 |
| | Treadmill recovery | 26 |
| 4 (a) Group 1 A Scaled systolic blood pressure | Bike recovery | 25 |
| | Treadmill recovery | 26 |
| | Bike recovery | 25 |
| | Treadmill recovery | 26 |
| 4 (a) Group 1 A Scaled diastolic blood pressure | Bike recovery | 25 |
| | Treadmill recovery | 26 |
| | Bike recovery | 25 |
| | Treadmill recovery | 26 |
| 4 (a) Group 1 A Scaled baseline resistance | Bike recovery | 25 |
| | Treadmill recovery | 26 |
| | Bike recovery | 25 |
| | Treadmill recovery | 26 |
| 4 (a) Group 1 A Scaled peak-to-peak resistance | Bike recovery | 25 |
| | Treadmill recovery | 26 |
| | Bike recovery | 25 |
| | Treadmill recovery | 26 |
| | | |
| 4 (a) Group 1 B Scaled heart rate | Cold pressor | 4 |
| | Valsalva | 4 |
| | Cold pressor | 4 |
| | Valsalva | 4 |
| 4 (a) Group 1 B Scaled systolic blood pressure | Cold pressor | 4 |
| | Valsalva | 4 |
| | Cold pressor | 4 |
| | Valsalva | 4 |
| 4 (a) Group 1 B Scaled diastolic blood pressure | Cold pressor | 4 |
| | Valsalva | 4 |
| | Cold pressor | 4 |
| | Valsalva | 4 |
| 4 (a) Group 1 B Scaled baseline resistance | Cold pressor | 4 |
| | Valsalva | 4 |
| | Cold pressor | 4 |
| resistance | Valsalva | 4 |
| 4 (a) Group 1 B Scaled peak-to-peak resistance | Cold pressor | 4 |
| | Valsalva | 4 |
| | Cold pressor | 4 |
| | Valsalva | 4 |
| | | |
| 4 (a) Group 1 C Scaled heart rate | Deep breathing | 15 |
| | Hand grip | 15 |
| | Valsalva | 16 |
| | Deep breathing | 15 |
| | Hand grip | 15 |
| | Valsalva | 16 |
| 4 (a) Group 1 C Scaled systolic blood pressure | Deep breathing | 15 |
| | Hand grip | 15 |
| | Valsalva | 16 |
| | Deep breathing | 15 |
| | Hand grip | 15 |
| | Valsalva | 16 |
| 4 (a) Group 1 Scaled diastolic blood pressure | Deep breathing | 15 |
| | Hand grip | 15 |
| | Valsalva | 16 |
| | Deep breathing | 15 |
| | Hand grip | 15 |
| | Valsalva | 16 |
| 4 (a) Group 1 C Scaled baseline resistance | Deep breathing | 15 |
| | Hand grip | 15 |
| | Valsalva | 16 |
| | Deep breathing | 15 |
| | Hand grip | 15 |
| | Valsalva | 16 |
| 4 (a) Group 1 C Scaled peak-to-peak resistance | Deep breathing | 15 |
| | Hand grip | 15 |
| | Valsalva | 16 |
| | Deep breathing | 15 |
| | Hand grip | 15 |
| | Valsalva | 16 |
| | | |
| 4 (a) Group 1 All Scaled heart rate | All | 65 |
| | Valsalva | 20 |
| | All | 65 |
| | Valsalva | 20 |
| 4 (a) Group 1 All | All | 65 |
| 4 (a) Group 1 All Scaled systolic blood pressure | Valsalva | 20 |
| | All | 65 |
| | Valsalva | 20 |
| 4 (a) Group 1 All Scaled diastolic blood pressure | All | 65 |
| | Valsalva | 20 |
| | All | 65 |
| | Valsalva | 20 |
| 4 (a) Group 1 All Scaled baseline resistance | All | 65 |
| | Valsalva | 20 |
| | All | 65 |
| | Valsalva | 20 |
| 4 (a) Group 1 All Scaled peak-to-peak resistance | All | 65 |
| | Valsalva | 20 |
| | All | 65 |
| | Valsalva | 20 |

| Statistical Test | P value |
|---|---|
| lts | |
| Shapiro-Wilk normality test | 0.4637 |
| | 0.781 |
| One sample t test; t(24)=6.597 | <0.0001 |
| One sample t test; t(25)=9.077 | <0.0001 |
| Shapiro-Wilk normality test | 0.1239 |
| | 0.9184 |
| One sample t test; t(24)=0.01430 | 0.1977 |
| One sample t test; t(25)=1.323 | 0.9887 |
| Shapiro-Wilk normality test | 0.2943 |
| | 0.0988 |
| One sample t test; t(24)=0.05564 | 0.9286 |
| One sample t test; t(25)=0.0905 | 0.9561 |
| Shapiro-Wilk normality test | 0.0016 |
| | <0.0001 |
| Wilcoxon test; W(25)=-49 | 0.5249 |
| Wilcoxon test; W(26)=55 | 0.4992 |
| Shapiro-Wilk normality test | 0.1889 |
| | 0.0006 |
| Wilcoxon test; W(25)=-107 | 0.1563 |
| Wilcoxon test; W(26)=-129 | 0.1048 |
| | |
| Shapiro-Wilk normality test | 0.5568 |
| | 0.6021 |
| One sample t test; t(3)=1.666 | 0.1942 |
| One sample t test; t(3)=0.8705 | 0.4481 |
| Shapiro-Wilk normality test | 0.4516 |
| | 0.1439 |
| One sample t test; t(3)=2.456 | 0.0912 |
| One sample t test; t(3)=0.3793 | 0.3793 |
| Shapiro-Wilk normality test | 0.4276 |
| | 0.3164 |
| One sample t test; t(3)=2.073 | 0.1299 |
| One sample t test; t(3)=0.5192 | 0.6395 |
| Shapiro-Wilk normality test | 0.1226 |
| | 0.0415 |
| Wilcoxon test; W(4)=10 | 0.125 |
| Wilcoxon test; W(4)=10 | 0.125 |
| Shapiro-Wilk normality test | 0.6252 |
| | 0.2497 |
| One sample t test; t(3)=2.335 | 0.1016 |
| One sample t test; t(3)=0.0006 | 0.9996 |
| | |
| Shapiro-Wilk normality test | 0.916 |
| | 0.098 |
| | 0.0328 |
| Wilcoxon test; W(15)=64 | 0.073 |
| Wilcoxon test; W(15)=120 | <0.0001 |
| Wilcoxon test; W(16)=136 | <0.0001 |
| Shapiro-Wilk normality test | 0.6357 |
| | 0.8083 |
| | 0.8465 |
| One sample t test; t(14)=3.280 | 0.0055 |
| One sample t test; t(14)=0.9709 | 0.3481 |
| One sample t test; t(15)=0.5451 | 0.5937 |
| Shapiro-Wilk normality test | 0.2915 |
| | 0.7082 |
| | 0.4195 |
| One sample t test; t(14)=3.631 | 0.0027 |
| One sample t test; t(14)=0.1027 | 0.9197 |
| One sample t test; t(15)=1.562 | 0.1392 |
| Shapiro-Wilk normality test | <0.0001 |
| | 0.0307 |
| | 0.0024 |
| Wilcoxon test; W(15)=-32 | 0.3894 |
| Wilcoxon test; W(15)=0 | >0.9999 |
| Wilcoxon test; W(16)=22 | 0.5966 |
| Shapiro-Wilk normality test | 0.107 |
| | 0.2913 |
| | 0.0726 |
| One sample t test; t(14)=1.190 | 0.2538 |
| One sample t test; t(14)=1.336 | 0.2029 |
| One sample t test; t(15)=0.1435 | 0.8878 |
| | |
| Shapiro-Wilk normality test | <0.0001 |
| | 0.0587 |
| Wilcoxon test; W(65)=1348 | <0.0001 |
| Wilcoxon test; W(20)=202 | <0.0001 |
| Shapiro-Wilk normality test | <0.0001 |
| | 0.5015 |
| Wilcoxon test; W(65)=-288 | 0.1929 |
| Wilcoxon test; W(20)=-4 | 0.9563 |
| Shapiro-Wilk normality test | 0.0006 |
| | 0.617 |
| Wilcoxon test; W(65)=-96 | 0.6679 |
| Wilcoxon test; W(20)=48 | 0.3884 |
| Shapiro-Wilk normality test | <0.0001 |
| | 0.0005 |
| Wilcoxon test; W(65)=-12 | 0.9604 |
| Wilcoxon test; W(20)=48 | 0.3884 |
| Shapiro-Wilk normality test | <0.0001 |
| | 0.0795 |
| Wilcoxon test; W(65)=-206 | 0.3536 |
| Wilcoxon test; W(20)=-24 | 0.6742 |

| Figure (panel) | Comparison | N |
|---|---|---|
| **Supplementary figure 17. Anthroprometric and demogra** | | |
| 21 (a) Age | Male | 39 |
| | Female | 36 |
| | Male and female | 75 |
| | Male vs<br>Female | 39<br>36 |
| 21 (b) Height | Male | 39 |
| | Female | 36 |
| | Male and female | 75 |
| | Male vs<br>Female | 39<br>36 |
| 21 (c) Weight | Male | 39 |
| | Female | 36 |
| | Male and female | 75 |
| | Male vs<br>Female | 39<br>36 |
| 21 (d) Body mass index | Male | 32 |
| | Female | 33 |
| | Male and female | 65 |
| | Male vs<br>Female | 32<br>33 |
| 21 (e) Wrist circumference | Male | 39 |
| | Female | 36 |
| | Male and female | 75 |
| | Male vs<br>Female | 39<br>36 |
| 21 (f) Forearm length | Male | 39 |
| | Female | 36 |
| | Male and female | 75 |
| | Male vs<br>Female | 39<br>36 |
| 21 (g) Upper arm length | Male | 39 |
| | Female | 36 |
| | Male and female | 75 |
| arm length | Male vs<br>Female | 39<br>36 |

| Statistical Test | P value |
|---|---|
| phic results | |
| Shapiro-Wilk normality test | 0.0007 |
| | <0.0001 |
| | <0.0001 |
| Mann-Whitney test | 0.9179 |
| Shapiro-Wilk normality test | 0.0888 |
| | 0.1082 |
| | 0.1157 |
| Unpaired T-test | <0.0001 |
| F(35, 38) = 1.049 | 0.8831 |
| Shapiro-Wilk normality test | 0.3267 |
| | 0.099 |
| | 0.3091 |
| Unpaired T-test | 0.0002 |
| F(35, 38) = 1.664 | 0.1264 |
| Shapiro-Wilk normality test | 0.5672 |
| | 0.0047 |
| | 0.0007 |
| Mann-Whitney test | 0.9811 |
| Shapiro-Wilk normality test | 0.3481 |
| | 0.8839 |
| | 0.4063 |
| Unpaired T-test | <0.0001 |
| F(35, 38) = 1.334 | 0.3856 |
| Shapiro-Wilk normality test | <0.0001 |
| | 0.0807 |
| | 0.0091 |
| Mann-Whitney test | <0.0001 |
| Shapiro-Wilk normality test | 0.2807 |
| | 0.1607 |
| | 0.4551 |
| Welch's T-test | 0.0027 |
| F(35, 38) = 2.049 | 0.0346 |